\documentclass[a4paper,fleqn,longmktitle]{cas-sc}

\usepackage[authoryear]{natbib}

\def\tsc#1{\csdef{#1}{\textsc{\lowercase{#1}}\xspace}}
\tsc{WGM}
\tsc{QE}

\def\spose#1{\hbox to 0pt{#1\hss}}
\def\deg{\ifmmode^\circ\else$\null^\circ$\fi}
\def\lta{\mathrel{\spose{\lower 3pt\hbox{$\mathchar "218$}}\raise 2.0pt\hbox{$\mathchar"13C$}}}
\def\gta{\mathrel{\spose{\lower 3pt\hbox{$\mathchar "218$}}\raise 2.0pt\hbox{$\mathchar"13E$}}}
\def\scinot#1.{\hbox{$\, \times \, 10^{#1}$}}
\def\CHfour{{\fam0 CH_4}}
\def\CHthree{{\fam0 CH_3}}

\def\COtwo{{\fam0 CO_2}}
\def\CO{{\fam0 CO}}
\def\CtwoHtwo{{\fam0 C_2H_2}}
\def\CtwoHthree{{\fam0 C_2H_3}}

\def\CtwoHfive{{\fam0 C_2H_5}}
\def\CtwoHsix{{\fam0 C_2H_6}}
\def\CthreeHfive{{\fam0 C_3H_5}}
\def\CthreeHsix{{\fam0 C_3H_6}}
\def\CthreeHseven{{\fam0 C_3H_7}}

\def\HCO{{\fam0 HCO}}
\def\HtwoCO{{\fam0 H_2CO}}

\def\HtwoO{{\fam0 H_2O}}
\def\Htwo{{\fam0 H_2}}
\def\Hatom{{\fam0 H}}

\def\Mthird{{\fam0 M}}

\def\Net{{\fam0 Net:}}
\def\OH{{\fam0 OH}}

\def\Oatom{{\fam0 O}}
\def\CHtwoNH{{\fam0 CH_2NH}}
\def\CHthreeNH{{\fam0 CH_3NH}}
\def\CHthreeNHtwo{{\fam0 CH_3NH_2}}
\def\NHtwo{{\fam0 NH_2}}
\def\NHthree{{\fam0 NH_3}}

\def\CN{{\fam0 CN}}
\def\HCN{{\fam0 HCN}}
\def\CthreeN{{\fam0 C_3N}}
\def\HCthreeN{{\fam0 HC_3N}}

\begin{document}
\let\WriteBookmarks\relax
\def\floatpagepagefraction{1}
\def\textpagefraction{.001}

\shorttitle{Saturn atmospheric response to the ring influx inferred from Cassini INMS}    

\shortauthors{J. I. Moses et al.}  

\title [mode = title]{Saturn's atmospheric response to the large influx of ring material inferred from Cassini INMS measurements}  



%

\author[1]{Julianne I. Moses}[orcid=0000-0002-8837-0035]

\cormark[1]


\ead{jmoses@spacescience.org}



\affiliation[1]{organization={Space Science Institute},
            city={Boulder},
            state={CO},
            country={USA}}

\author[2]{Zarah L. Brown}[orcid=0000-0002-4513-4876]





\affiliation[2]{organization={Lunar and Planetary Laboratory},
            addressline={University of Arizona}, 
            city={Tucson},
            state={AZ},
            country={USA}}

\author[2]{Tommi T. Koskinen}[orcid=0000-0003-3071-8358]


\author[3]{Leigh N. Fletcher}[orcid=0000-0001-5834-9588]


\affiliation[3]{organization={School of Physics and Astronomy},
            addressline={University of Leicester}, 
            city={Leicester},
            country={UK}}

\author[4]{Joseph Serigano}[orcid=0000-0003-3849-5064]


\affiliation[4]{organization={Department of Earth and Planetary Sciences},
            addressline={Johns Hopkins University}, 
            city={Baltimore},
            state={MD},
            country={USA}}

\author[5,12]{Sandrine Guerlet}[orcid=0000-0001-5019-899X]


\affiliation[5]{organization={Laboratoire de M{\'e}t{\'e}orologie Dynamique/Institut Pierre-Simon Laplace (LMD/IPSL)},
            addressline={Sorbonne Universit{\'e}, CNRS, {\'E}cole Polytechnique}, 
            addressline={Institut Polytechnique de Paris, {\'E}cole Normale Sup{\'e}rieure (ENS), PSL Research University}, 
            city={Paris},
            country={France}}

\author[6]{Luke Moore}[orcid=0000-0003-4481-9862]


\affiliation[6]{organization={Center for Space Physics},
            addressline={Boston University}, 
            city={Boston},
            state={MA},
            country={USA}}

\author[7]{J. Hunter Waite}[suffix=Jr,orcid=0000-0002-1978-1025]


\affiliation[7]{organization={Southwest Research Institute},
            city={San Antonio},
            state={TX},
            country={USA}}

\author[8]{Lotfi Ben-Jaffel}[orcid=0000-0003-4047-2793]


\affiliation[8]{organization={Institut d'Astrophysique de Paris},
            addressline={Sorbonne Universit{\'e}}, 
            addressline={UPMC \& CNRS}, 
            city={Paris},
            country={France}}

\author[9]{Marina Galand}[orcid=0000-0001-5797-914X]


\affiliation[9]{organization={Department of Physics},
            addressline={Imperial College London}, 
            city={London},
            country={UK}}

\author[9]{Joshua M. Chadney}[orcid=0000-0002-5174-2114]



\author[4]{Sarah M. H{\"o}rst}[orcid=0000-0003-4596-0702]


\author[10]{James A. Sinclair}[orcid=0000-0001-5374-4028]


\affiliation[10]{organization={Jet Propulsion Laboratory},
            addressline={California Institute of Technology}, 
            city={Pasadena},
            state={CA},
            country={USA}}

\author[11]{Veronique Vuitton}[orcid=0000-0001-7273-1898]


\affiliation[11]{organization={Univ. Grenoble Alpes},
            addressline={CNRS, CNES, IPAG}, 
            city={38000 Grenoble},
            country={France}}

\affiliation[12]{organization={LESIA, Observatoire de Paris, Université PSL, CNRS, Sorbonne Université, Université Paris Cit{\'e}},
            city={Meudon},
            country={France}}

\author[13]{Ingo M{\"u}ller-Wodarg}[orcid=0000-0001-6308-7826]


\affiliation[13]{organization={Blackett Laboratory},
            addressline={Imperial College London}, 
            city={London},
            country={UK}}

\cortext[1]{Corresponding author}



\begin{abstract}
During the Grand Finale stage of the \textit{Cassini} mission, organic-rich ring material was discovered to be flowing into Saturn's equatorial 
upper atmosphere at a surprisingly large rate.  Through a series of photochemical models, we have examined the consequences of this 
ring material on the chemistry of Saturn's neutral and ionized atmosphere.  We find that if a substantial fraction of this material 
enters the atmosphere as vapor or becomes vaporized as the solid ring particles ablate upon atmospheric entry, then the ring-derived vapor 
would strongly affect the composition of Saturn's ionosphere and neutral stratosphere.  Our surveys of \textit{Cassini} infrared and 
ultraviolet remote-sensing data from the final few years of the mission, however, reveal none of these predicted chemical consequences.  
We therefore conclude that either (1) the inferred ring influx represents an anomalous, transient situation that was triggered by some recent 
dynamical event in the ring system that occurred a few months to a few tens of years before the 2017 end of the \textit{Cassini} mission, or 
(2) a large fraction of the incoming material must have been entering the atmosphere as small dust particles less than $\sim$100 nm in radius, rather 
than as vapor or as large particles that are likely to ablate.  Future observations or upper limits for stratospheric neutral species such 
as HC$_3$N, HCN, and CO$_2$ at infrared wavelengths could shed light on the origin, timing, magnitude, and nature of a possible vapor-rich 
ring-inflow event.  
\end{abstract}




\maketitle


\section{Introduction}\label{sec:intro}

During the 2017 ``Grand Finale'' stage of the \textit{Cassini} mission, the spacecraft spent five months in a series 
of orbits with close-approach periapse distances between Saturn's rings and its upper atmosphere, culminating 
in the orbiter's planned final plunge into Saturn's atmosphere.  The spacecraft skimmed Saturn's equatorial
upper atmosphere at pressures less than $\sim$1 nbar in the last five of the Grand Finale orbits, as well as
during the ``Final Plunge'' trajectory itself.  The Ion Neutral Mass Spectrometer (INMS) instrument on the \textit{Cassini} orbiter 
directly sampled Saturn's atmospheric composition during these last orbital passes \citep{waite18}, in a manner similar to the earlier 
close flybys of Titan and the Enceladus plumes \citep[e.g.,][]{waite05,waite06,cravens06,cravens09enceladus,yelle06,vuitton06,cui09inms,magee09}.  
Grand Finale data from the Radio and Plasma Wave Science/Langmuir Probe (RPWS/LP) instrument, Magnetospheric Imaging Instrument (MIMI), 
and Cosmic Dust Analyzer (CDA) also provided complementary \textit{in situ} measurements of electron and ion densities, dust properties, and
other key pieces of information about the local atmosphere and its interaction with the rings and magnetosphere
\citep[e.g.,][]{mitchell18,wahlund18,hsu18sat,persoon19,hadid19,morooka19,johansson22}.  The \textit{in situ} results from these last Saturn 
orbital passes provided one of the most surprising and puzzling discoveries from the 13-year \textit{Cassini} mission.  

Saturn's uppermost neutral atmosphere was expected to be completely dominated by H$_2$, H, and He, because molecular diffusion confines 
heavier species that are intrinsic to the planet to the well-mixed homospheric region at deeper pressures ($P \gtrsim 0.1$ $\mu$bar).
However, along with the expected H$_2$ and He (H could not be measured by INMS at Saturn), the \textit{in situ} INMS measurements indicated the 
additional presence of significant quantities 
of heavier molecules, such as methane (CH$_4$), water (H$_2$O), ammonia (NH$_3$), carbon monoxide (CO), molecular nitrogen (N$_2$), and 
carbon dioxide (CO$_2$), along with more complex organic molecules, nanoparticles, and dust 
\citep{waite18,yelle18,perry18,mitchell18,miller20,serigano20,serigano22,koskinen22chap}.  The inferred vertical mixing-ratio profiles 
for these heavy molecular species were found to increase with altitude, indicating that the molecules are flowing in to the equatorial 
region from outside the planet and diffusing down through the atmosphere.  Some external material from the Enceladus plumes, interplanetary dust particles, cometary impacts, and 
Saturn's rings and satellites is expected (and observed) to pollute Saturn's upper atmosphere 
\citep[e.g.,][]{connerney84,feuchtgruber97,mosesbass00,moses00b,cavalie09,cavalie10,cavalie19,cassidy10,moore10,moore15,hartogh11,poppe12,odonoghue13,odonoghue19,ip16,poppe16,moses17poppe,hamil18}.  
However, the measured gas abundances and corresponding influx rates in the equatorial region of Saturn from the INMS data significantly exceed the 
inferred broader-scale influx of external material, as determined from global-average observations of stratospheric oxygen-bearing species and 
ionospheric electron densities on Saturn \citep[cf.][]{feuchtgruber97,moses00b,bergin00,moore10,cavalie10,fletcher12spire,abbas13,waite18,yelle18,miller20,serigano20}.  
The source of the heavy molecules for the INMS measurements appears to be atmospheric erosion of Saturn's rings --- particularly the inner 
D ring \citep{mitchell18,hadid19}, which can be resupplied with particles from the C ring \citep{waite18,yelle18,perry18,miller20}.  If the 
extremely large mass influx rate inferred from the INMS data \citep[$> 10^{4}$ kg s$^{-1}$ into a low-latitude band;][]{waite18} were maintained 
over time, the rings would be completely eroded within $\sim$50 Myr \citep[or $\sim$5000 years for the D ring alone;][]{koskinen22chap}.  

Several questions then arise.  Is the current mass influx rate inferred from the INMS \textit{in situ} measurements anomalously large and caused 
by some recent, transient disruptive event within the rings, or does it represent a more-or-less steady-state influx that has been operating in the 
equatorial region over larger time scales?  If a disruptive event is the culprit, how recently did it happen?  Conversely, if this 
ring-erosion process has been operating over long time scales, what are the consequences for Saturn's atmospheric chemistry, structure, 
and bulk elemental abundances, as well as for the evolution of the rings?  Can any other atmospheric observations be used to help evaluate the 
possible long- or short-term nature of this enormous ring influx rate seen by INMS?

Some of the consequences of this infalling ring-derived vapor and dust on Saturn's ionosphere have been discussed previously. Briefly, the 
inflow of neutral molecules from the rings causes heavy molecular ions with relatively short chemical lifetimes to supplant H$^+$ and H$_3$$\! ^+$ as 
the dominant ions near Saturn's main ionospheric peak at pressures $\sim$10$^{-6}$--10$^{-7}$ mbar 
\citep{moore18,waite18,cravens19ioncomp,dreyer21,chadney22,vigren22}, and incoming ring dust may influence ionospheric structure and charge balance 
\citep{mitchell18,hadid19,persoon19,morooka19,vigren22,johansson22}.  From model-data comparisons and
photochemical-equilibrium arguments, \citet{dreyer21} find that the dominant heavy ion at low latitudes near 1700 km altitude should 
have an effective electron-recombination rate coefficient of $\lesssim \, 3\scinot-7.$ at an electron temperature of 300 K, which makes HCO$^+$ a prime candidate 
for the dominant ion at the main peak \citep[see also][]{moore18}.  However, other heavy ions not considered in the \citet{moore18} model 
might also satisfy this constraint \citep{dreyer21}, and even H$_3$O$^+$ might satisfy existing constraints if the ion densities derived 
by \citet{morooka19} have been influenced by secondary electron emission currents \citep[see][]{johansson22}.

The consequences with respect to Saturn's deeper neutral atmosphere of this large influx of external material remain relatively unexplored, other 
than the general comment by \citet{waite18} that if this INMS-inferred influx rate were typical over long time scales extending into the past, 
then the inflowing ring material could have enhanced the atmospheric C/H ratio and other elemental abundances of Saturn's atmosphere over the 
lifetime of the rings.  

Here, we use a coupled ion-neutral photochemical model to explore the specific chemical consequences of the ring-derived vapor for 
both Saturn's ionosphere and neutral atmosphere.  We expand the number of ionic species considered by \citet{moore18}, and we track the chemical 
reactions and ultimate fate of the ring-derived molecules as they flow through the thermosphere and stratosphere, on their way down into the troposphere.  
We find that the large influx rate of ring vapor inferred from the \textit{Cassini} mission has significant consequences for the stratospheric 
composition --- consequences that are not currently observed by remote-sensing instruments from \textit{Cassini} or ground-based telescopes.  
Our results therefore have implications for the age of the perturbing event that is responsible for the apparent large influx of ring material 
during the \textit{Cassini} Grand Finale and/or for the possible dust-to-gas ratio of the incoming material.  Our results also have implications 
for the possibility of future observable changes to the composition of Saturn's stratosphere if the current large influx rate of material derives largely 
from vapor and were to continue at a similar rate into the next decade.

\section{Constraints on gas influx rates from Grand Finale measurements}\label{sec:obscon}

The trajectory of the \textit{Cassini} spacecraft in the last five complete Grand Finale orbits (orbit numbers 288--292) progressed from
north to south, with closest approach near $-5\deg$ planetocentric latitude at a thermospheric altitude of $\sim$1700 km above the 1-bar 
level \citep{waite18,yelle18}.  The Final Plunge trajectory (orbit number 293) never crossed the ring plane, and the last INMS data points were 
acquired near 9$\deg$ planetocentric latitude at an altitude of $\sim$1400 km \citep{waite18,yelle18}, well above the near-equatorial methane 
homopause altitude of $\sim$840--960 km \citep{vervack15,koskinen18,brown21}.  The \textit{in situ} measurements obtained at higher altitudes 
were generally also obtained at higher latitudes, such that disentangling effects caused by density variations with altitude vs.~latitude 
vs.~time can be complicated \citep{waite18}.  These complications are exacerbated by variations in thermospheric temperatures with latitude 
\citep{koskinen15sat,koskinen16,koskinen18,brown20,koskinen21empirical} and by variations in gravity with latitude caused by Saturn's rapid 
rotation and strong zonal winds that deform the planet's shape \citep{lindal85,anderson07,militzer19,strobel19}, such that the atmospheric 
density vs.~altitude (or vs.~radius) profile on Saturn is not uniform with latitude.  To compare density results acquired at different 
latitudes, investigators typically must convert local radii to a gravitational potential \citep[e.g.,][]{yelle18} or to an equivalent equatorial 
altitude \citep[e.g.,][]{koskinen22chap}.

With the Closed Source Neutral mode of the INMS instrument \citep[see][]{waite04} --- which is the mode used to determine the densities of 
non-reactive atmospheric gases such as H$_2$, He, CH$_4$, H$_2$O, NH$_3$, N$_2$, CO, CO$_2$, etc. --- the neutral atmospheric gas molecules and 
small dust particles encountered by the instrument during the spacecraft's $\sim$30 km s$^{-1}$ traverse through Saturn's upper atmosphere flow into a 
spherical antechamber, where they bounce off the walls, potentially fragment due to the high relative velocity, and become thermalized.  As described 
in \citet{waite04}, the inflowing gas ram pressure creates a density enhancement in the antechamber with this mode, increasing measurement precision 
and sensitivity.  Any dust particles impacting within the instrument at the high velocities experienced during the Saturn passes 
would most likely be vaporized and contribute to the ambient gas in the 
antechamber.  From there, the gas then flows to an ionization region, where a collimated electron beam ionizes it, and the resulting ions are 
deflected onto a quadrupole mass analyzer that filters the ions based on their mass-to-charge ($m/z$) ratio.  The instrument is sensitive to 
species with a mass range of 1--99 atomic mass units (amu), with a resolution of about 1 amu \citep{waite18}.  Derivations of the original 
neutral molecule densities require a simultaneous fit of their electron-impact fragmentation patterns in the mass spectra 
\citep[e.g.,][]{miller20,serigano20,serigano22}.  

During the Grand Finale passes through Saturn's atmosphere, the INMS detected not only the expected H$_2$ and He as major thermospheric 
constituents, but also significant quantities of heavier molecules.  The INMS mass spectra contained readily identifiable signatures of 
CH$_4$, H$_2$O, and NH$_3$, at inferred mixing ratios of 10$^{-5}$ to 10$^{-3}$ in the thermosphere, corresponding 
to maximum influx rates of a few $\times$ 10$^{9}$ molecules cm$^{-2}$ s$^{-1}$ at low latitudes
\citep{waite18,yelle18,miller20,serigano20,serigano22}.  The high abundances of these species, along with their nearly constant mixing ratios with 
altitude, indicated an external rather than internal source.  The presence of CO$_2$ was also inferred in the INMS spectra, along with a molecule 
with a mass peak at 28 amu --- corresponding to CO and/or N$_2$, with some contribution from C$_2$H$_4$ or impact fragments from heavier 
constituents \citep{waite18,moore18,miller20,serigano22,koskinen22chap}.  However, the latitude-altitude distributions of the heavier species were found to 
differ from that of CH$_4$ \citep{waite18,miller20}, indicating that the heavier species could have a potential contribution from impact 
vaporization and fragmentation of dust grains within the INMS antechamber.  A rich spectrum of other heavy organics was also apparent in the 
INMS spectra (again, with a possible contribution from dust impact vaporization and fragmentation), and \citet{waite18} 
estimated that heavy organics make up $\sim$37\% of the incoming material by mass.  All told, the INMS data suggest that the mass 
influx of external material to Saturn's equatorial atmosphere at the time of the Grand Finale ranged from 4800--74,000 kg s$^{-1}$ \citep{waite18,serigano22}.  
For numerous reasons discussed in \citet{waite18}, \citet{miller20}, and \citet{serigano22}, the rich spectrum of molecules identified by 
the Saturn Grand Finale measurements likely represents a real \textit{in situ} signal, rather than an instrument artifact, a signal from 
reactions within the chamber or chamber walls, or remnant adsorbed gases from the earlier Titan atmospheric flybys.  However, the degree to 
which high-velocity impact fragmentation could have affected the INMS mass spectra is still an open discussion \citep{miller20,serigano22}.

The fraction of the signal that is derived from impact vaporization and fragmentation of solid dust grains, rather than directly encountered as gas, 
is difficult to quantify accurately.  During the Grand Finale orbits, the CDA and MIMI instrument investigations \citep{hsu18sat,mitchell18} identified 
a prograde orbiting population of dust particles entering the atmosphere at the equator as the main dust component, presumably from atmospheric 
erosion of the D ring, as well as a population of charged grains being dynamically controlled by the magnetic field and entering the atmosphere 
over a wider spread of latitudes, consistent with the previously identified Saturn ``ring rain'' \citep[e.g.,][]{connerney84,odonoghue13}.  Depending 
on entry velocities, entry angles, and grain composition, these dust grains may or may not ablate to release vapor molecules within the atmosphere 
\citep[e.g.,][]{moses17poppe,hamil18}, but ablation at the higher altitudes relevant to the INMS measurements is relatively unlikely.  The fact 
that the INMS counts from the lightest $m/z$ 10--20 amu fragments greatly exceed (e.g., by three orders of magnitude) the counts from the heaviest 
$m/z$ 90--100 amu fragments suggests that smaller molecules, rather than larger molecules or grains, dominate the INMS input 
\citep[e.g.,][]{waite18,miller20}.  However, this observed variation with mass in the INMS measurements does not necessarily rule out the possibility 
that the smaller mass signals result from highly volatile components preferentially being fragmented or otherwise released from grains or 
macromolecules after impacts in the antechamber \citep[e.g.,][]{waite18,perry18,miller20}.  

\subsection{Constraints on dust versus gas fraction}\label{sec:dustvgas}

One of the strongest indicators of a direct gas influx component to the INMS mass spectra is the spatial distribution of 
the different $m/z$ signatures as a function of altitude, in comparison with the dust distribution inferred from the MIMI instrument.  At 
high altitudes, both the low-mass and high mass INMS signals are distributed such that the maximum abundance is centered at the equator, 
rather than at the closest-approach latitude where the atmospheric density is greatest \citep{waite18}.  This distribution is reminiscent of the 
narrow equatorial distribution of the nanograins that have been eroded from the D ring, as observed by MIMI \citep{mitchell18}, except that the 
low-mass INMS signal has a broader extent \citep{waite18}, as might be expected from diffusive spreading due to collisions of incoming 
volatile molecules with ambient atmospheric H$_2$, H, and He.  As the altitude decreases, the lower mass peaks (such as $m/z$ 15 amu, a 
dissociative ionization ionization fragment of $m/z$ 16 amu from methane, predominantly) 
spread more rapidly than the higher-mass peaks (e.g., $m/z$ 28 amu), while the nanograin dust component is still relatively narrowly 
confined to the equator \citep{waite18,mitchell18,perry18}.  Finally, at the lower altitudes probed during the spacecraft passage through the 
atmosphere, the INMS mass peaks are firmly centered at the closest-approach latitude, consistent with a population of gas-phase molecules that are 
diffusively coupled to the atmosphere. 

These altitude-dependent spatial distributions, along with a second piece of evidence, prompted \citet{waite18} and \citet{miller20} to suggest that 
only the most volatile species in the incoming D-ring material (such CH$_4$, CO, and N$_2$) enter the atmosphere in the gas phase or are sublimated 
at very high altitudes, whereas the more refractory species (e.g., H$_2$O, CO$_2$, NH$_3$, and heavier ``organics'') have contributions from impact 
vaporization of grains within the INMS antechamber.  The second piece of evidence supporting this hypothesis is that the mixing ratios CH$_4$ and 
the $m/z$ 28 species do not appear to change much with each of the Grand Finale orbital passes, whereas the mixing ratios of H$_2$O, NH$_3$, CO$_2$ and 
the more massive species are found to be depleted for the higher-latitude orbit 293 in comparison with the lower-latitude orbits 290-292.  If the 
more refractory species are derived from dust, then one might expect their distributions to be concentrated at the equator, while the gas molecules have 
longer residence times and can spread more readily in latitude.  
Based on the determination that the latitude distribution of the different mass signals is correlated with volatility, \citet{miller20} attempt to 
quantify the gas-to-dust ratio of the material that contributes to the INMS signals.  Assuming that CH$_4$ and the $m/z$ 28 species are derived 
entirely from gas, \citet{miller20} estimate an upper-limit gas-to-dust ratio 
for the ring material of $\sim$0.7--2, with the volatility of CO$_2$ representing the dividing line between species that are 
flowing in largely in the vapor phase vs.~within a dust component that vaporizes in the instrument.

\subsection{Constraints on gas influx rates}\label{sec:gasflux}

Because of uncertainties in the fraction of the contribution of the INMS signal caused by impact vaporization of dust grains that 
may not otherwise release vapor into Saturn's atmosphere, we examine separate possible cases for our atmospheric models.  First, we 
assume that all the molecules identified 
in the INMS mass-spectral analyses of \citet{serigano22} and/or \citet{miller20} derive from vapor species 
directly entering Saturn's atmosphere.  Second, we assume that only CH$_4$, CO, and N$_2$ --- as the most volatile species --- enter the 
atmosphere as vapor, whereas the other INMS mass signatures solely result from impact vaporization of dust within the instrument, and 
that these dust grains would not have otherwise ablated to contribute the more refractory gases to the atmosphere.  Based on the 
analysis of \citet{miller20}, these two cases could appropriately bracket the real situation.  

Because it is also unclear whether the vapor-phase equatorial ring material spreads efficiently with latitude before reaching the 
methane homopause, below which it can influence stratospheric photochemistry, we also examine separate cases that assume either no 
spreading beyond the latitude distribution already observed by INMS, or complete global spreading.  The latter case would decrease 
the ``effective'' vapor influx rate into the stratosphere by an order of magnitude, given that the INMS-inferred distribution is 
centered in a low-latitude band \citep{waite18,perry18} that is roughly 10\% of the surface area of the planet.  Based on the molecular 
diffusion coefficients of heavier species in an H$_2$-dominated atmosphere \citep[e.g.,][]{marrero72}, on inferred eddy diffusion 
coefficients for Saturn's upper atmosphere \citep[e.g.,][]{koskinen16}, and on the INMS-derived structure for the low-latitude 
thermosphere \citep[e.g.,][]{yelle18}, the ring-derived gas flowing in from the top of the atmosphere would take less than 10 hr to 
diffuse vertically down to the 1400 km altitude level that provided the last recorded INMS measurement during the Final Plunge (orbit 
293).  The observed nearly altitude-independent mixing ratios of CH$_4$ and other molecules identified by INMS through this altitude 
level, as well as the observed spatial distribution of these species, make it clear that the incoming ring material has not spread too 
widely up to that deepest altitude point --- i.e., global spreading above that point would have resulted in an apparent decrease in 
local equatorial mixing ratios with decreasing altitude, which is not observed in the INMS data \citep[e.g.,][their Fig.~7]{serigano22}.  
However, the gas would take roughly another month to diffuse vertically from that last recorded INMS point to the CH$_4$ homopause level, 
and horizontal spreading could have occurred in that intervening region.  

The possibility of 
effective meridional spreading between the INMS closest-approach altitudes and the CH$_4$ homopause is suggested by the fact that the 
thermospheric CH$_4$ mixing ratios inferred from the INMS data during the Grand Finale orbits \citep{yelle18,waite18,miller20,serigano20,serigano22} 
are typically larger than the CH$_4$ mixing ratios retrieved at the highest altitudes probed by the \textit{Cassini} Ultraviolet 
Imaging Spectrometer (UVIS) stellar and solar occultation observations \citep{koskinen15sat,koskinen16,koskinen18,brown21}.  Unless 
the thermospheric CH$_4$ spreads sufficiently in latitude before it reaches the homopause, or unless the CH$_4$ mixing ratio is reduced by chemical 
processes in the intervening region in the lower thermosphere/ionosphere, the UVIS and INMS CH$_4$ mixing ratios would appear to be mutually 
inconsistent.  The one-dimensional (1D) models we consider in this paper cannot accurately simulate horizontal spreading, so to account 
for this possibility, we consider additional cases that simply adopt an order-of-magnitude reduction in the inferred INMS influx rates 
to simulate global spreading of the ring-derived gases before they reach the stratosphere.  The results in the middle and upper 
thermosphere (where significant meridional spreading has not been observed) for these cases will be inaccurate, but if more widespread 
global redistribution is occurring deeper in the thermosphere, the stratospheric results for these cases may be more accurate than models 
that assume no spreading.  Note, however, that meridional wind speeds would need to be of order $\sim$30 m s$^{-1}$ in the lower thermosphere 
for widespread global transport of this vapor to occur before it reaches the homopause, and such large meridional wind speeds have not been 
predicted by thermospheric global circulation models (GCMs) at non-auroral latitudes \citep[e.g.,][]{mullerwodarg12,mullerwodarg19,smith12}. 
However, when wave drag is introduced to the Saturn Thermosphere Ionosphere Model (STIM) based on observed gravity-wave signatures in the 
UVIS-derived thermal structure, 
equatorward meridional winds approach $\sim$60 m s$^{-1}$ at mid-latitudes \citep{brown22}. These winds are convergent at the equator, and 
convergence would require a poleward return cell at some deeper pressure, potentially enabling the spreading of the ring material.  Horizontal 
diffusion could also be a factor.  Therefore, while we have no 
current evidence that suggests horizontal spreading away from the equator in the lower thermosphere would be occurring, this scenario is 
not out of the realm of possibility.

The influx rates inferred from the INMS Grand Finale data for the different model assumptions described above are shown in Table~\ref{tabinflux}.
The fluxes from \citet{serigano22} in Cases A and E relate to their top 10 most abundant incoming external species on average, plus C$_6$H$_6$.  
The values from \citet{miller20} pertain to the most abundant external species listed in their Table~2, with the corresponding fluxes determined 
from our photochemical models such that the mixing ratios from their Table~2 are reproduced.   

\begin{table}[!htb]
\caption{Model influx rates (molecules cm$^{-2}$ s$^{-1}$) based on different assumptions about the INMS measurements}\label{tabinflux}
\begin{tabular*}{\tblwidth}{@{}LCCCCCC@{}}
\toprule
Species & Case A$^a$ & Case B$^b$ & Case C$^c$ & Case D$^d$ & Case E$^e$ & Case F$^f$ \\ 
\midrule
CO         & 2.33$\scinot9.$ & 2.33$\scinot9.$ & 2.33$\scinot8.$ & 2.33$\scinot8.$ & 2.52$\scinot9.$ & 3.47$\scinot9.$ \\ 
H$_2$O     & 2.30$\scinot8.$ & --              & 2.30$\scinot7.$ & --              & 3.38$\scinot9.$ & 3.64$\scinot9.$ \\ 
N$_2$      & 2.14$\scinot9.$ & 2.14$\scinot9.$ & 2.14$\scinot8.$ & 2.14$\scinot8.$ & 2.21$\scinot9.$ & 3.45$\scinot9.$ \\ 
CH$_4$     & 1.46$\scinot9.$ & 1.46$\scinot9.$ & 1.46$\scinot8.$ & 1.46$\scinot8.$ & 1.95$\scinot9.$ & 1.74$\scinot9.$ \\ 
NH$_3$     & 8.24$\scinot7.$ & --              & 8.24$\scinot6.$ & --              & 1.29$\scinot9.$ & 2.00$\scinot9.$ \\ 
HCN        & 3.63$\scinot8.$ & --              & 3.63$\scinot7.$ & --              & 7.12$\scinot8.$ & 1.12$\scinot9.$ \\ 
CO$_2$     & 1.84$\scinot8.$ & --              & 1.84$\scinot7.$ & --              & 4.78$\scinot8.$ & --              \\ 
H$_2$CO    & 8.80$\scinot7.$ & --              & 8.80$\scinot6.$ & --              & 5.16$\scinot8.$ & 3.80$\scinot8.$ \\ 
C$_2$H$_2$ & 1.47$\scinot8.$ & --              & 1.47$\scinot7.$ & --              & 2.37$\scinot8.$ & --              \\ 
C$_2$H$_6$ & 1.14$\scinot8.$ & --              & 1.14$\scinot7.$ & --              & 2.33$\scinot8.$ & --              \\ 
C$_6$H$_6$ & 7.16$\scinot6.$ & --              & 7.16$\scinot5.$ & --              & 1.50$\scinot8.$ & --              \\ 
NO         & --              & --              & --              & --              & --              & 2.67$\scinot8.$ \\ 
\bottomrule
\end{tabular*}
\raggedright $^a$Final Plunge (Orbit 293) fluxes from the INMS analysis of \citet{serigano22}.  Case A is our standard model for 
9.2$\deg$ N latitude with ring-vapor influx. \\
$^b$Final Plunge (Orbit 293) fluxes from the INMS analysis of \citet{serigano22}, with the added assumption that only the most volatile 
molecules --- N$_2$, CO, CH$_4$ --- are entering the atmosphere as vapor, whereas the other INMS signals derive from impact vaporization of dust that never 
naturally ablates in the atmosphere. Case B is an alternative model for 9.2$\deg$ N latitude that assumes that much of the INMS signal is 
caused by vaporized dust rather than direct vapor.\\
$^c$Final Plunge (Orbit 293) INMS analysis of \citet{serigano22}, reduced by an order of magnitude.  This 9.2$\deg$ N latitude Case C is 
similar to Case A, except we assume global redistribution of the incoming vapor to simulate complete horizontal spreading of the ring vapor before it 
reaches the homopause.\\
$^d$Final Plunge (Orbit 293) INMS analysis of \citet{serigano22}, considering only the most volatile species (N$_2$, CO, CH$_4$), 
with fluxes reduced by an order of magnitude.  This 9.2$\deg$ N latitude Case D is similar to Case B, except we assume global redistribution of 
the incoming vapor to simulate complete horizontal spreading of the ring vapor before it reaches the homopause.\\
$^e$Orbit 290-292 average fluxes from the INMS analysis of \citet{serigano22}.  Case E is one of our standard models for $-5.5\deg$ N latitude, 
assuming no horizontal spreading and assuming INMS signals are caused predominantly by ring vapor.\\
$^f$Orbit 290-292 average, from the auto-fit mixing ratios derived from the INMS analysis of \citep[][see their Table 2]{miller20}. Case F is 
the second of our standard models for $-5.5\deg$ N latitude, assuming no horizontal spreading and assuming INMS signals are caused predominantly by ring vapor.\\
\end{table}

\section{Photochemical model}\label{sec:model}

We use the fully implicit finite-difference \texttt{KINETICS} photochemical model \citep{allen81,yung84} to solve the coupled set of continuity 
equations to  predict the 1D transport and chemical kinetics of Saturn's near-equatorial upper atmosphere.  Two different specific latitudes are 
examined: 9.2$\deg$ planetocentric latitude, relevant to the last recorded position of \textit{Cassini}\/'s Final Plunge trajectory for orbit 293, 
and $-5.5\deg$ planetocentric latitude, near closest approach during Grand Finale orbit 290 \citep{yelle18}.  The latter model also has 
relevance to orbits 288--292, given that the spacecraft trajectories were similar during these final full orbits. 
Both the 9.2$\deg$ and $-5.5\deg$ latitude 
models extend down to $\sim$5 bar at the bottom boundary.  The 9.2$\deg$ latitude model extends to 10$^{-11}$ mbar at the top boundary, with 260 vertical 
grid points equally spaced in log pressure.  The $-5.5\deg$ latitude model has a top boundary at a $\sim$10$^{-10}$ mbar, with 242 vertical grid 
points. This latter model has a warmer thermosphere, and the top boundary has been placed at a greater pressure to avoid extending into a region in 
which H would surpass H$_2$ as the dominant atmospheric constituent.  Note that H does not begin to compete with H$_2$ until extremely high altitudes, 
and neither of these models violates the atomic H upper limits derived from \textit{Cassini} UVIS solar occultations \citep{koskinen13sat}.

\subsection{Background atmosphere}\label{sec:backgroundatm}

As an input to \texttt{KINETICS}, we specify a background atmosphere in which the vertical variations in temperature, pressure, density, and altitude 
are predefined.  We adopt pressure-temperature profiles from \textit{Cassini} Composite Infrared Spectrometer (CIRS) limb retrievals for the stratosphere 
at pressures greater than $\sim$3$\scinot-3.$ mbar \citep[see][for a description of the data processing and retrieval 
procedures]{fouchet08,guerlet09,guerlet10,guerlet15,sylvestre15,koskinen18}, from CIRS nadir retrievals for tropospheric pressures less than 500 mbar 
\citep{fletcher19chap,fletcher21rev}, and Voyager radio occultations profiles at tropospheric pressures greater than 500 mbar \citep{lindal85,lindal92nep}.  
There were no \textit{Cassini} CIRS limb observations 
exactly within the Grand Finale closest-approach time frame, so we have adopted the pressure-temperature retrievals for the nearest latitude and time.  
For the 9.2$\deg$ N planetocentric latitude model, we use the CIRS limb observations from 10$\deg$ N planetographic (8.2$\deg$ planetocentric) latitude 
acquired in March 2017.  For the $-5.5\deg$ planetographic model, we use the CIRS limb observations from $-5\deg$ planetographic ($-4.1\deg$ 
planetocentric) latitude acquired in January 2017.  For the upper thermosphere, we use the INMS H$_2$ density measurements as a function of 
planetary radius (and latitude) from \citet{yelle18} to define the thermal structure in that region of the upper atmosphere.  In the intervening 
pressure region where there is a gap between the CIRS retrievals and the INMS data, we connect these regions with a Bates-type pressure-temperature
profile \citep[e.g.,][]{bates59,gladstone96}.  The full background atmospheric grid is then defined using a forward-modeling procedure that solves the 
hydrostatic equilibrium equation, assuming a planetary shape, rotation rate, and gravity parameters ($J_2$, $J_4$, $J_6$) described in 
\citet{anderson07}.  The forward-modeling procedure is iterated until the H$_2$ density at closest approach from the INMS measurements is 
reproduced to within some average specified percent (e.g., $<$ 5\%).

\begin{figure*}[!htb]
\vspace{-14pt}
\begin{tabular}{ll}
{\includegraphics[clip=t,scale=0.30]{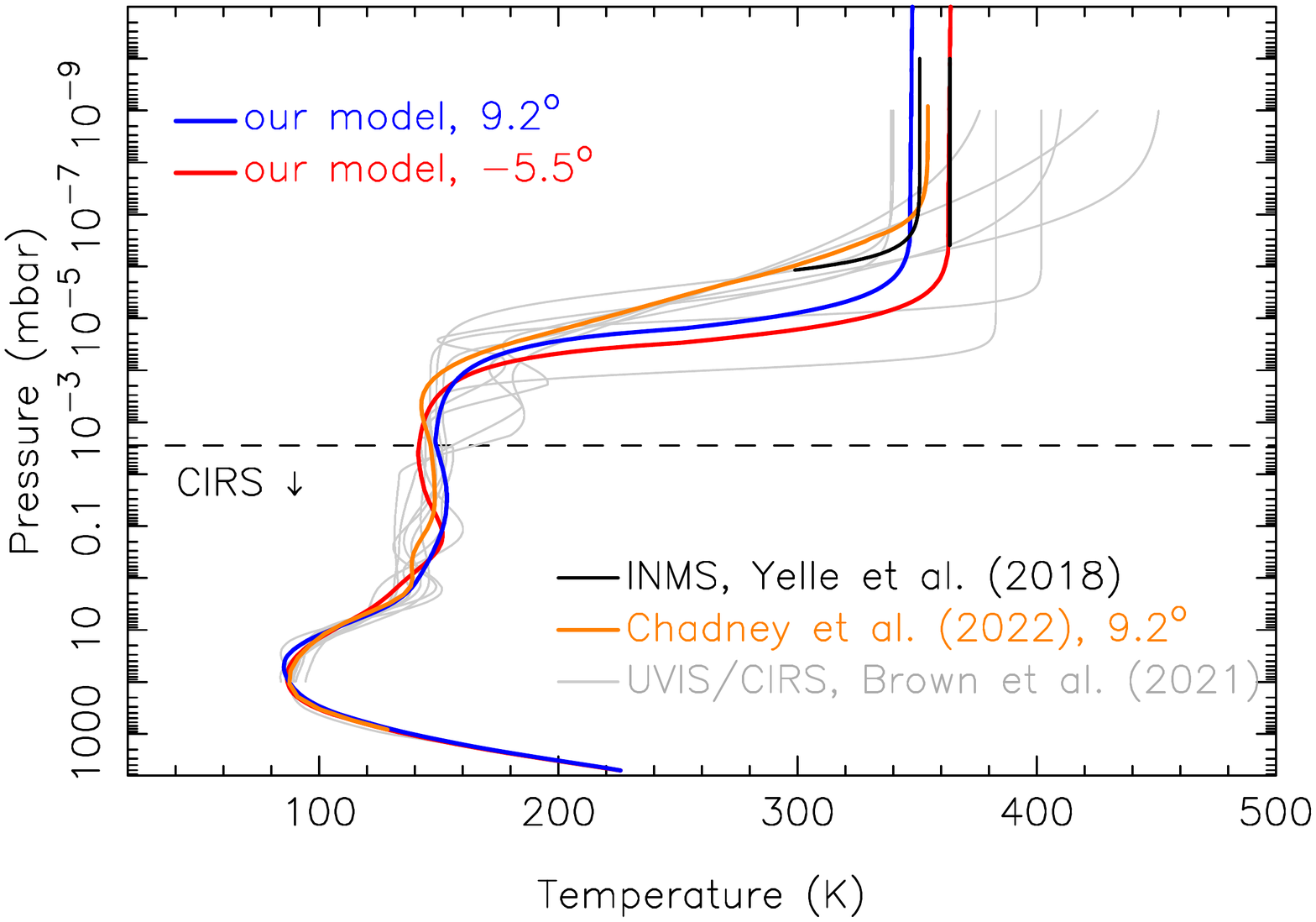}}
&
{\includegraphics[clip=t,scale=0.30]{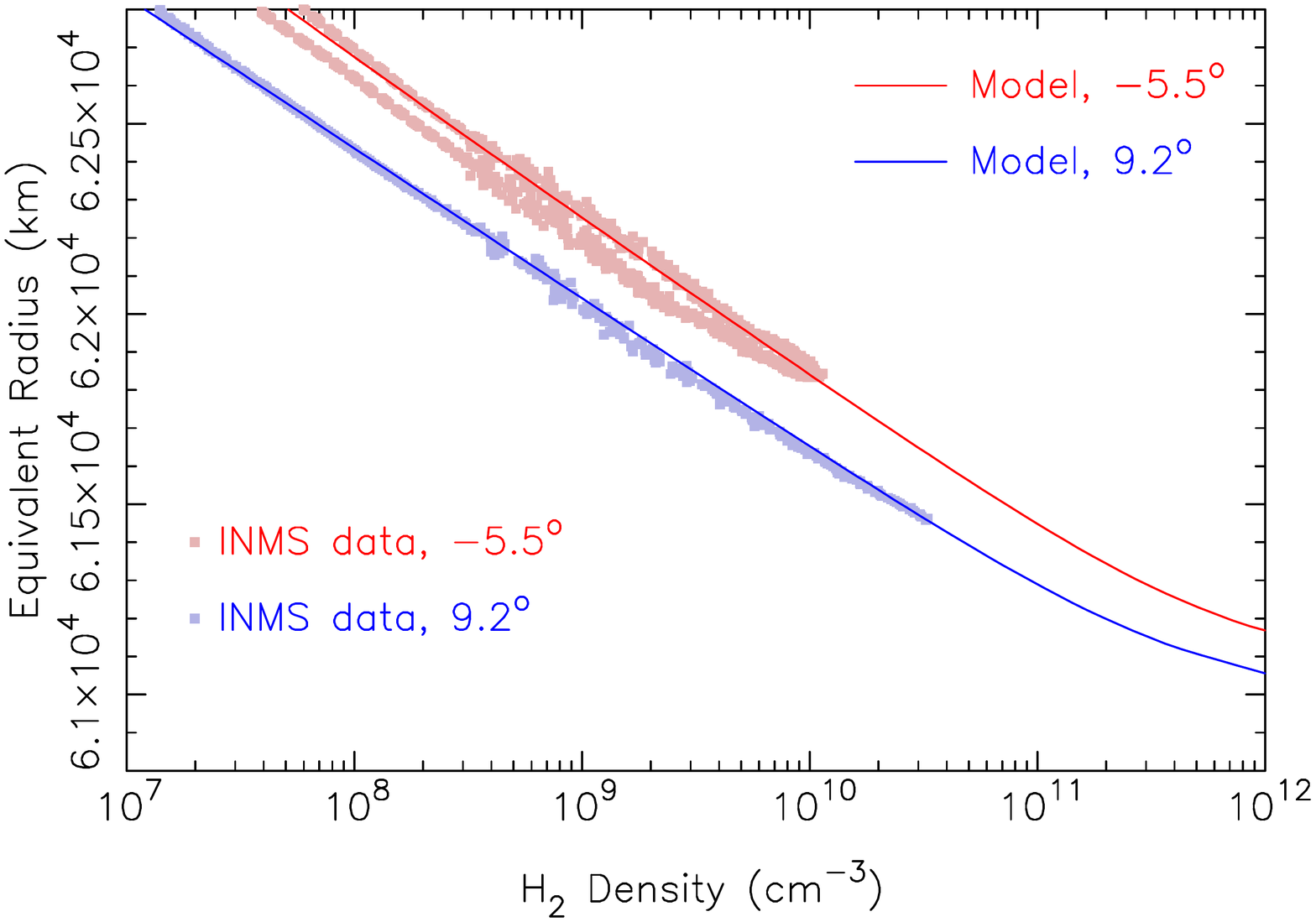}}
\end{tabular}
\caption{(Left) Temperature profiles adopted in our model for 9.2$\deg$ (blue) and $-5.5\deg$ (red) planetocentric latitude, in comparison 
with the INMS thermal structure (black) inferred for orbits 293 and 290 by \citet{yelle18}.  The orange curve is the \citet{chadney22} 9.2$\deg$ latitude model, 
and the gray curves are various low-latitude ($\pm \, $35$\deg$) temperature profiles derived from a combined analysis of 
UVIS occultations and CIRS limb retrievals from \citet{brown21}.  Temperatures for our red and blue models at pressures 
greater than the horizontal dashed line are derived from our combined \textit{Cassini} CIRS limb and nadir retrievals.  
(Right) Resulting model H$_2$ densities at 9.2$\deg$ (blue) and $-5.5\deg$ (red) planetocentric latitude, in comparison with INMS-derived 
H$_2$ densities \citep{yelle18} as a function of the equivalent radius at 9.2$\deg$ (gray-blue dots) and $-5.5\deg$ (gray-red dots) planetocentric 
latitude.  See text for an explanation of equivalent radii. The bifurcation in the measured densities at $-5.5\deg$ results from differences 
between the inbound and outbound measurements in orbit 290; the outbound leg has the greater H$_2$ densities.}
\label{figtempden}
\end{figure*}

Figure \ref{figtempden} illustrates the resulting pressure-temperature profiles and the H$_2$ density profiles as a function of ``equivalent 
radius'' in comparison with the INMS data for orbits 290 and 293.  The densities of H$_2$ and He from orbits 288--292 agree to 
within tens of percent \citep{yelle18}, and orbit 290 is simply used as a representative for the closest-approach $-5.5\deg$ latitude region here.  
\textit{Cassini} spacecraft trajectories were such that the measurement latitudes were constantly changing along with the spacecraft's radial position.  
As mentioned previously, because of Saturn's oblate 
shape and strong variation in gravity with latitude, the radial distance at which a specific H$_2$ density would be found at one latitude differs from that
at another latitude.  To account for this effect, we convert the actual radius to an equivalent radius at which this measured H$_2$ density would 
be located for the terminal latitude (for orbit 293) or for the closest-approach latitude (for orbit 290) via 
\begin{equation}
R_{\rm e} \ = \ R_{\rm 1bar,ref} \ + \ \left( \frac{g_{\rm i}}{g_{\rm ref}} \right) \left( R_{\rm i} - R_{\rm 1bar,i} \right) \quad ,
\end{equation}
where $R_{\rm e}$ is the equivalent radius at the reference latitude (in our case, at planetocentric latitude 9.2$\deg$ for the orbit 293 model 
or $-5.5\deg$ for the orbit 290 model), $R_{\rm 1bar,ref}$ is the 1-bar radius at the reference latitude, $g_{\rm i}$ is the gravitational 
acceleration at measurement latitude $i$ and measurement radius $R_{\rm i}$, $g_{\rm ref}$ is the gravitational acceleration at the reference latitude 
at a radius corresponding to that same measured H$_2$ density (derived from the hydrostatic-equilibrium model), and $R_{\rm 1bar,i}$ is the 
1-bar radius at measurement latitude $i$.  The 1-bar radii are derived from Eq.~(8a) of \citet{strobel19}.

The above expression assumes that the atmospheric temperature and mean molecular mass are constant with radius over the limited radial 
range that is being used to determine the exospheric temperature from the INMS data --- an assumption that is not strictly valid but 
that provides a reasonable structure model for our purposes.  An alternative method would be to convert the vertical coordinate to gravitational 
potential \citep{yelle18,chadney22,serigano22}. As can be seen, our background structure 
models have exospheric temperatures within a few degrees of previous analyses \citep{yelle18,waite18,chadney22} and consider a thermal structure in 
the lower-thermospheric ``gap'' region that connects the CIRS limb thermal structure appropriately to the H$_2$ densities measured by INMS.  
Equation (1) also assumes that thermospheric temperatures do not vary significantly with latitude.  The fact that the inbound and 
outbound H$_2$ density measurements for orbit 290 shown in light gray-red in Fig.~\ref{figtempden}b do not exactly align on the equivalent-radius grid 
for $-5.5\deg$ planetocentric latitude suggests that variations with latitude do exist, a conclusion that is also consistent with the UVIS 
occultation analyses of \citet{koskinen18} and \citet{brown20,brown21}.  Note that our model solutions for the temperature profile in the gap region are 
non-unique, and other selections of the $T$ profile within this gap region could produce equally good fits to the H$_2$ densities 
determined by INMS, but the assumptions here will have negligible effect on the resulting chemistry.  Note also that the H$_2$ densities reported 
by INMS are typically a factor of 
$\lesssim$2 larger than the H$_2$ densities derived from UVIS occultations at similar latitudes and slightly different times \citep{brown21}, as 
is discussed in \citet{koskinen22chap} and \citet{strobel22}.  This apparent INMS-UVIS inconsistency with respect to H$_2$ densities might have a slight 
influence on inferred thermospheric transport time scales and the radial structure determined from our model, but it would have little effect on the predicted 
constituent abundances as a function of pressure, particularly for the stratosphere, and thus would not affect our main conclusions for the 
neutral atmosphere. 

\subsection{Chemistry inputs}\label{sec:cheminputs}

Although our main focus in this paper is on the stratospheric neutral chemistry resulting from a large influx of ring material, we also 
examine how the ionospheric composition can change for different assumptions of the ring vapor molecular influx rates at the top of the 
atmosphere (see Table~\ref{tabinflux}).  
Our reaction list for the photochemical model includes 164 neutral and 118 ionized species containing H, He, C, N, and/or O atoms that interact 
with each other through 2531 reactions.  Neutral hydrocarbons with up to 18 carbon atoms are considered, including polycyclic aromatic hydrocarbons (PAHs), 
although the production and loss reactions for molecules with more than six carbon atoms are incomplete.  Neutral molecules with up to two 
nitrogen or oxygen atoms are also considered.  Our full reaction list is included in the Supplementary Material.  The neutral hydrocarbon and 
oxygen reactions and associated rate coefficients and photolysis cross sections derive largely from \citet{moses17poppe} and \citet{moses18}, with 
the neutral nitrogen reaction rate coefficients deriving from giant-planet models such as \citet{moses10,moses11} and Titan models such as 
\citet{loison15} and \citet{vuitton12}.  The ion reaction list also derives largely from Titan studies, particularly \citet{vuitton19}.  We 
neglect the coupled NH$_3$-PH$_3$ photochemistry that is expected to occur in Saturn's troposphere at pressures greater than a few hundred mbar 
\citep[e.g.,][]{kaye84}, so the results are most relevant to Saturn's stratosphere and thermosphere at pressures less than 100 mbar.

We include the reaction H$^+$ + H$_2 (v \ge 4)$ $\rightarrow$ H$_2$$\! ^+$ + H, which is exothermic for molecular hydrogen residing in the 4th or higher 
vibrational state \citep[e.g.,][]{mcelroy73,moore19chap}.  The rate coefficient for this reaction is assumed to be 1$\scinot-9.$ cm$^{-3}$ s$^{-1}$ 
\citep{huestis08hcoll}.  Based on \citet{majeed91fluor} and \citet{hallett05}, we approximate [H$_2 (v \ge 4)$ ], the density of H$_2$ in vibrational 
states $\ge$ 4, as [H$_2 (v \ge 4)$] = 10$^4$ cm$^{-3}$ for total H$_2$ densities [H$_2$] $>$ 10$^9$ cm$^{-3}$, and [H$_2 (v \ge 4)$] = 
10$^{-5} \cdot$[H$_2$] for [H$_2$] $\le$ 10$^9$ cm$^{-3}$.

Our model does not take grain charging into account, such that the electron density is simply assumed to be equal to the sum of the ion densities.  
Early analyses of the RPWS Langmuir Probe data seemed to indicate significant electron depletion due to grain charging \citep[e.g.][]{morooka19}. 
However, \citet{johansson22} recently demonstrated that secondary electron emission current can affect the Langmuir Probe bias voltage sweeps, and 
when such effects are considered, the data show no evidence for dust playing a significant role in charge balance in Saturn's ionosphere.  
\citet{johansson22} also suggest that electron temperatures may have been overestimated by earlier analyses \citep[e.g.,][]{morooka19}.  We assume 
the electron and neutral temperatures are equal in our models \citep[for a justification of this assumption, see][]{moore08}.  The influence of 
grain charging and greater electron temperatures is considered 
in the models of \citet{vigren22}; briefly, some electron depletion and grain charging is predicted if electron temperatures are as high as indicated 
by \citet{morooka19} but these effects become much less significant if electron and ion temperatures are assumed to be equal. 

To speed up the calculations, we consider the condensation of H$_2$O, CO$_2$, C$_4$H$_2$, C$_6$H$_6$, HCN, CH$_3$CN, and HC$_3$N only, although 
several additional species such as C$_4$H$_{10}$, PAHs, other refractory high-molecular-weight hydrocarbons and nitrogen species become abundant enough 
to condense, depending on ring-vapor influx assumptions.  
Our model will therefore over-estimate the vapor-phase mixing ratios of these species at pressures $P$ greater than a few mbar.  Condensation and evaporation 
are considered as separate reactions, in a manner described in \citet{moses00b}.  Aerosol scattering and absorption are not considered in the model, given 
that Saturn's equatorial stratospheric haze is optically thin at ultraviolet wavelengths \citep{pryor91,karkoschka05}, particularly at the higher altitudes 
($P$ $\lesssim$ 5 mbar) where the bulk of the interesting photochemistry is occurring in these models.

\subsection{Boundary conditions}\label{sec:bound}

At the model lower boundary, we assume fixed mixing ratios of He \citep[11\%, see][]{koskinen18} and CH$_4$ \citep[4.7$\scinot-3.$, see][]{fletcher09}, 
which have an internal source from the deep atmosphere.  An internal source of oxygen and nitrogen species is ignored, as the dominant O- and N-bearing 
constituents --- H$_2$O and NH$_3$ --- condense in the troposphere and thus do not affect the stratospheric/thermospheric composition.  One exception is 
CO, which is a disequilibrium quenched product from the deep atmosphere that is transported to the upper troposphere faster than it can be chemically destroyed 
\citep{prinn77,visscher05,visscher10co,wang16}, and for which we assume a fixed mixing ratio of 1$\scinot-9.$ at the lower boundary in our nominal model 
\citep{noll90,cavalie09}.  However, we omit this lower boundary condition for CO for the INMS-inspired high ring-influx models, as the influx source 
of CO then becomes more important than the interior source.  All other species in the model are assumed to have a zero concentration gradient at the 
lower boundary, such that they flow through the bottom of the model at a maximum possible velocity given by the diffusion coefficient divided by the 
atmospheric scale height.  At the top boundary in our nominal models --- that is, those models without the INMS-derived ring influx --- we include a 
smaller influx rate of H$_2$O, CO, and CO$_2$ from external sources such as Enceladus plumes or interplanetary dust particles \citep[for further details, 
see section~\ref{sec:resnom} and][]{moses17poppe}.  For the INMS-inspired ring-influx models, we consider vapor influx rates at the top boundary of the model as described 
in Table~\ref{tabinflux} (from \citealt{serigano22}, and \citealt{miller20}).

The steady-state solutions will always conserve mass, such that the flux of carbon, nitrogen, and oxygen enterring the atmosphere will balance the 
flux leaving the atmosphere.  In the nominal model, the hydrocarbons that are produced photochemically in the stratosphere diffuse down through 
the atmosphere and flow out the lower boundary; this loss of carbon is then balanced by an upward flux of methane from the interior to maintain the 
fixed (and observed) mixing ratio of CH$_4$ at the lower boundary.  In the ring-influx models, the loss of C, N, and O species through the lower 
boundary is balanced by both the influx of ring material from the top and the influx of methane from below.  The ring molecules take roughly a 
thousand years to flow vertically through the model, although achieving true steady state takes longer, as chemical time constants can be longer than 
the diffusion time scales.  The amount of carbon flowing in from the rings does not compete with the amount of carbon already in the atmosphere, so 
the ring influx does not significantly perturb the tropospheric or lower-stratospheric CH$_4$ profile, and the CH$_4$ flux at the lower boundary still 
ends up positive (upward) to counteract the loss of other carbon species through the lower boundary.

\begin{figure*}[!htb]
\vspace{-14pt}
\begin{tabular}{ll}
{\includegraphics[clip=t,scale=0.30]{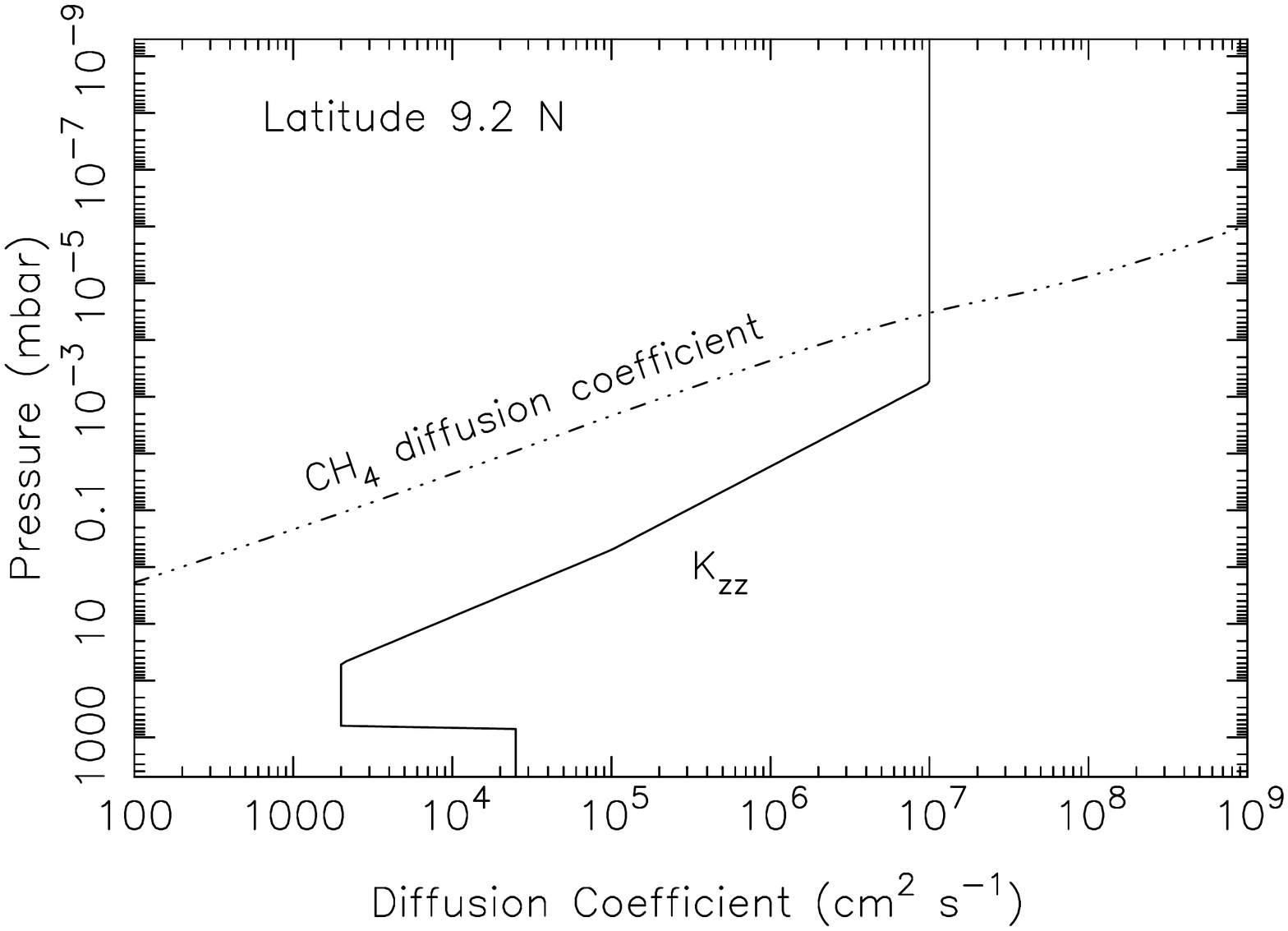}}
&
{\includegraphics[clip=t,scale=0.30]{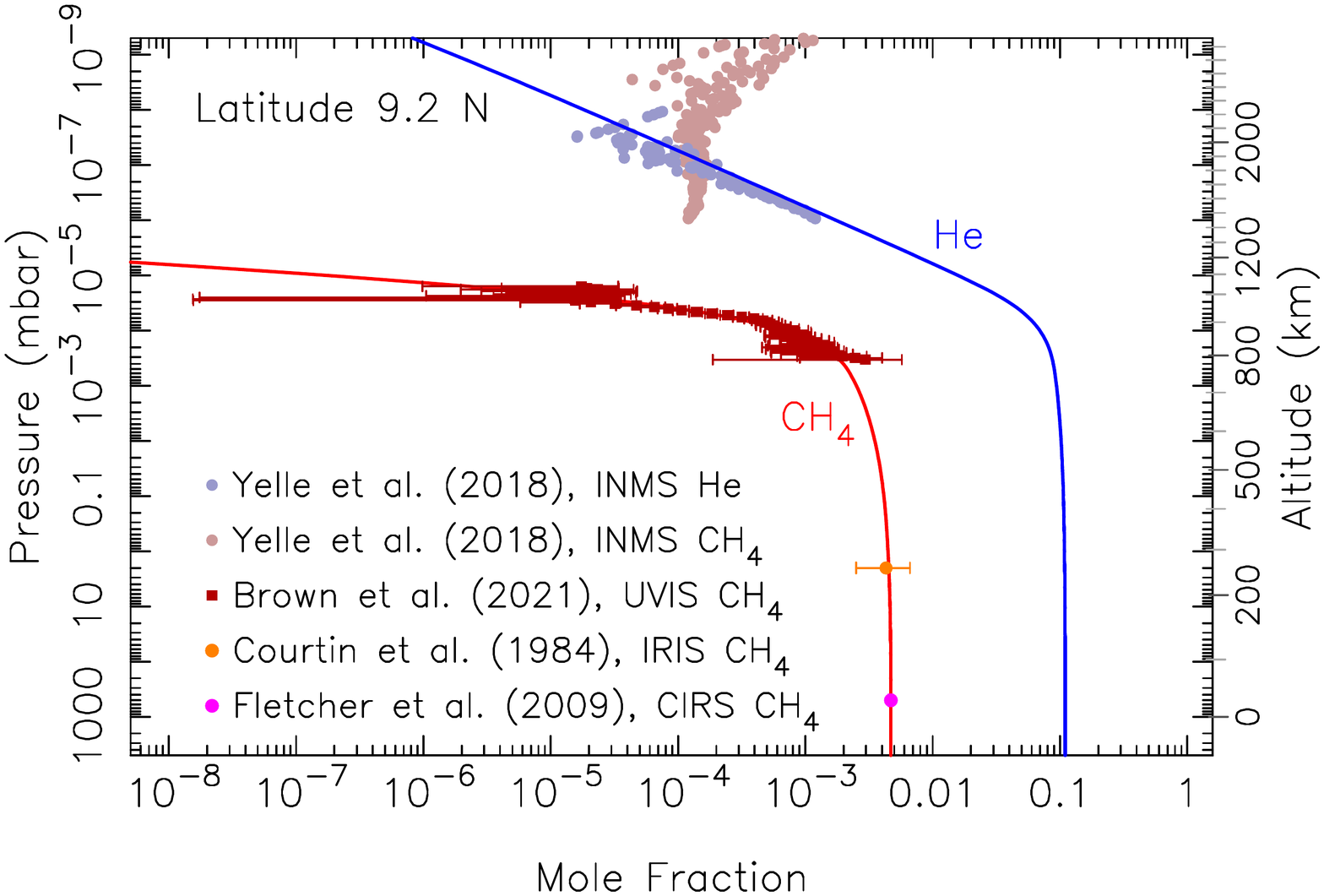}} \\
\vspace{-14pt}
{\includegraphics[clip=t,scale=0.30]{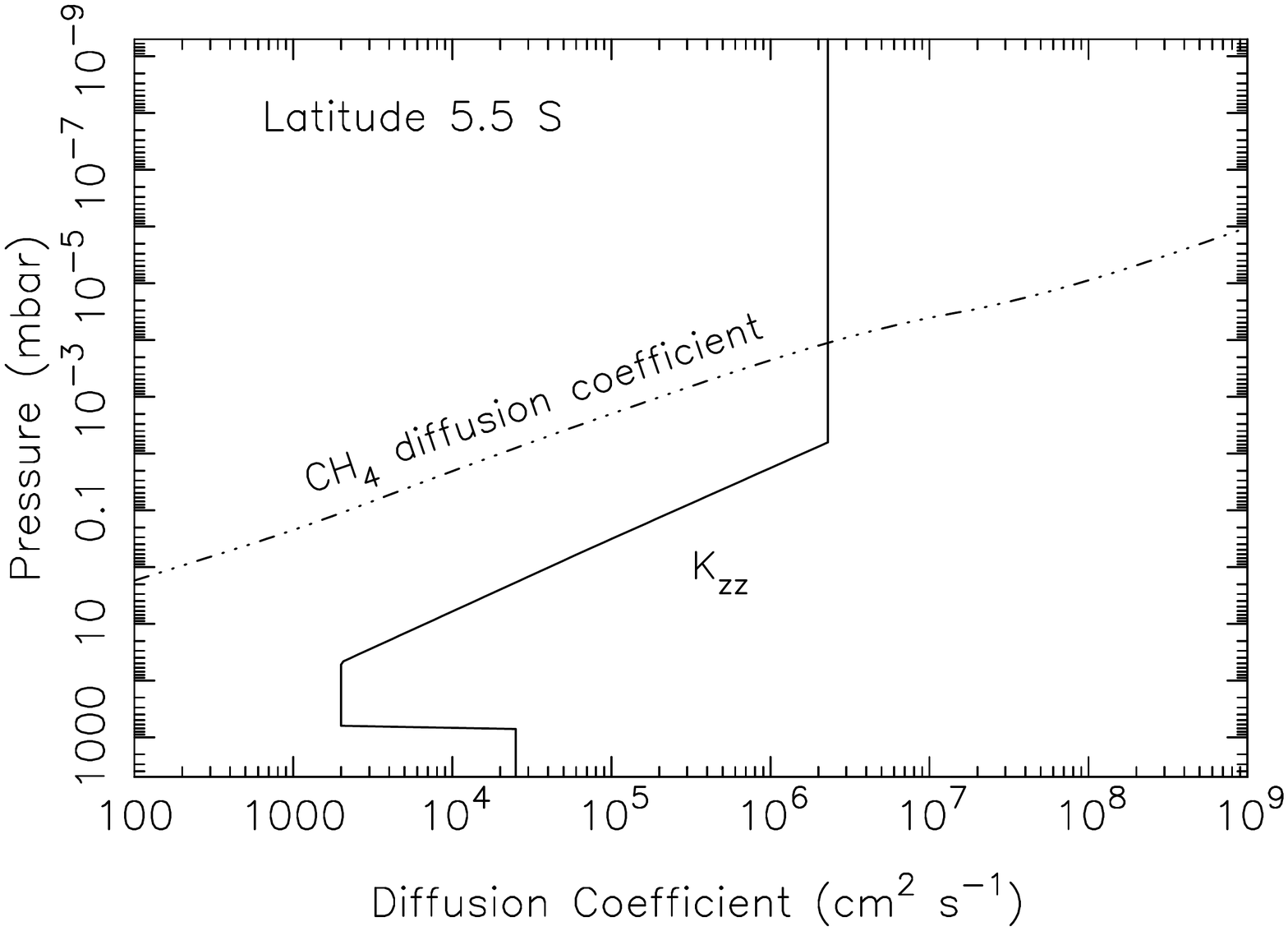}}
&
{\includegraphics[clip=t,scale=0.30]{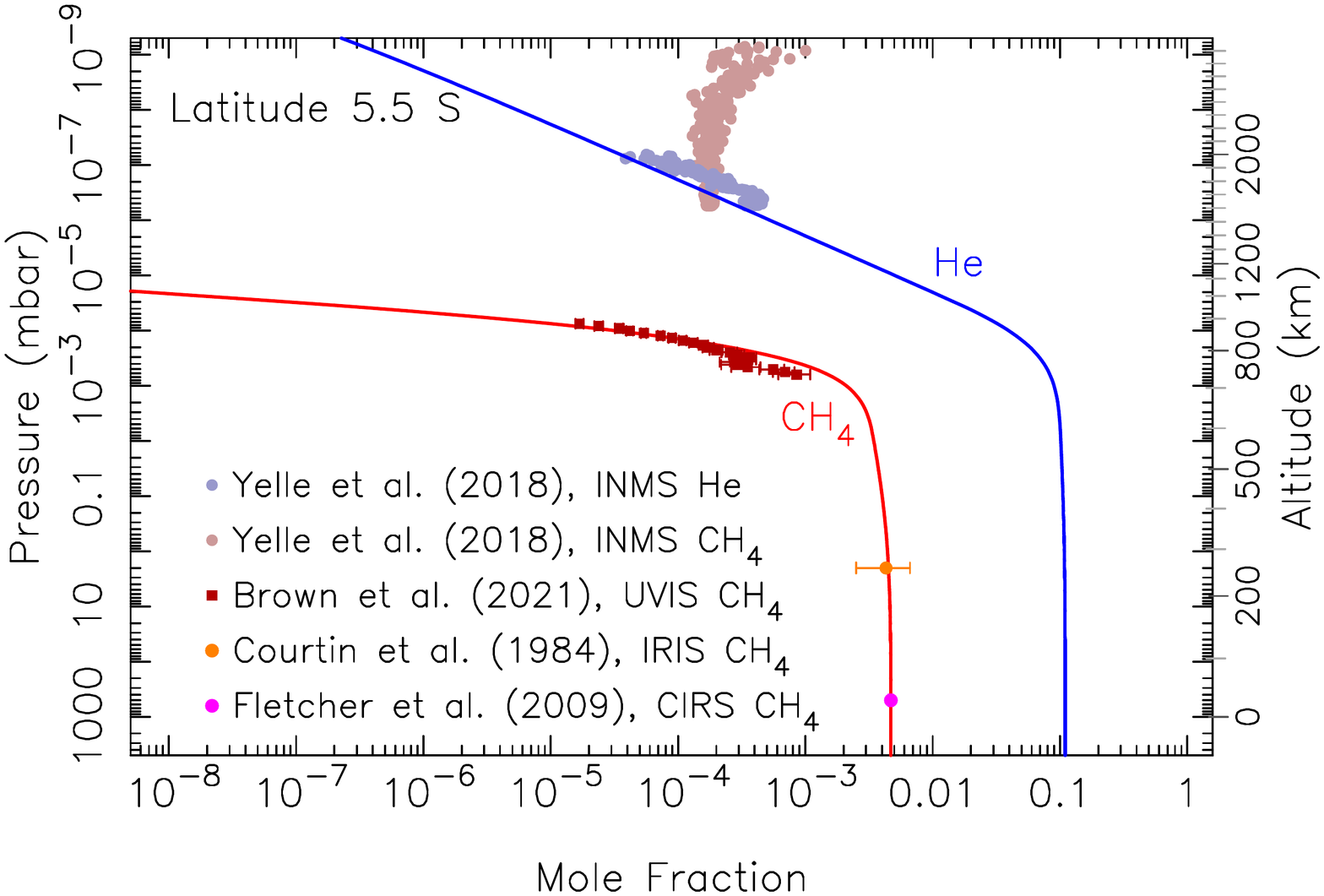}}
\end{tabular}
\caption{(Left panels) Eddy diffusion coefficient ($K_{zz}$) profile (solid black line) adopted in our model, in comparison 
with the CH$_4$ molecular diffusion coefficient (dot-dashed line) that corresponds to the adopted thermal structure shown in Fig.~\ref{figtempden}, 
for (Top Left) 9.2$\deg$ and (Bottom Left) $-5.5\deg$ planetocentric latitude.  
The methane homopause is located where $K_{zz}$ equals the CH$_4$ molecular diffusion coefficient, and the CH$_4$ mixing ratio drops off steeply 
above that pressure region. 
(Right panels) Resulting CH$_4$ (red) and He (blue) mixing ratios from our nominal model without the INMS-derived ring influx, assuming the $K_{zz}$ 
profile shown in the left panels for (Top Right) 9.2$\deg$ and (Bottom Right) $-5.5\deg$ planetocentric latitude.  The predicted model abundances are 
compared to various \textit{Voyager} Infrared Interferometer Spectrometer (IRIS) data and \textit{Cassini} CIRS, UVIS, and INMS data (for orbit 
293 in the top panel and orbit 290 in the bottom panel), as labeled in the legend
\citep{courtin84,fletcher09,yelle18,brown21}.}
\label{figeddy9n}
\end{figure*}

\subsection{Other model inputs and assumptions}\label{sec:modinputs}

Vertical transport in the 1D photochemical models occurs through molecular and eddy diffusion for the neutral species, with ambipolar diffusion 
included for the ions \citep[see][for a discussion of the latter]{mosesbass00}.  
At high altitudes, molecular diffusion dominates over eddy diffusion for the neutral constituents.  The mutual molecular diffusion 
coefficients of various species in a hydrogen atmosphere depend on 
atmospheric density, temperature, and properties of the species, and Appendix A describes our assumptions in detail.
The eddy diffusion coefficient ($K_{zz}$) profile is a free 
parameter in the models that is adjusted through forward modeling to fit various observational constraints.  In the upper-stratosphere, $K_{zz}$ is 
adjusted to fit the CH$_4$ density retrievals from the UVIS occultation observations of \citet{brown21} at the nearest latitude and time to the 
last Grande Finale orbits, which in the case of the 9.2$\deg$ planetocentric latitude model is the UVIS occultation of $\epsilon$ Orionis acquired 
at 12$\deg$ planetocentric latitude on June 25, 2017.  For the $-5.5\deg$ planetocentric latitude model, there were no UVIS occultations obtained near this 
latitude in 2017, so we use the CH$_4$ retrievals from the $6\deg$ planetocentric latitude occultation of $\zeta$ Puppis acquired October 3, 2014.  
In the lower stratosphere, $K_{zz}$ is adjusted to fit the C$_2$H$_6$ and C$_2$H$_2$ retrievals from \textit{Cassini} CIRS limb and nadir observations
\citep[e.g.,][]{guerlet09,guerlet10,guerlet15,koskinen18,fletcher10,fletcher15satpole,fletcher19chap,fletcher21rev,sinclair13} at latitudes and times closest to 
the Grand Finale orbits.  In the troposphere at pressures $\gtrsim$1 bar, $K_{zz}$ is assumed to increase again as a result of atmospheric 
convection, and we have chosen a tropospheric value of 2.5$\scinot4.$ cm$^{2}$ s$^{-1}$ to be consistent with the Jupiter study of 
\citet{edgington98}.  The adopted tropospheric $K_{zz}$ value has no effect on the results higher up in the stratosphere, but the minimum 
$K_{zz}$ value in the tropopause region can affect long-lived species such as C$_2$H$_6$ and CO \citep[e.g.,][]{bezard02}.  

Figure~\ref{figeddy9n} shows our resulting adopted $K_{zz}$ profiles for the 9.2$\deg$ and $-5.5\deg$ latitude models, 
along with the resulting nominal model (i.e., no INMS-inspired ring influx) profiles for CH$_4$ and He, which are strongly affected by $K_{zz}$.  
In Appendix B, we describe the sensitivity of the model results to changes in the adopted $K_{zz}$ profile.
Although the $K_{zz}$ profiles shown in Fig.~\ref{figeddy9n} were formulated based on stratospheric hydrocarbon remote-sensing observations 
alone, without the consideration of the INMS results, the modeled thermospheric He mixing-ratio profiles are reasonably consistent with the INMS He 
measurements of \citet{yelle18}, despite uncertainties in the deep helium abundance on Saturn \citep[e.g.,][]{koskinen18}, despite uncertainties in the 
thermal structure of Saturn in the upper-stratospheric ``gap'' region between the CIRS and UVIS temperature retrievals than can affect the He scale 
height, and despite the fact that the CH$_4$ UVIS occultation measurements were at a slightly different latitude and time than the INMS Grand Finale 
measurements.  The latter two effects, especially, are likely to contribute to the over-estimation of the He mixing ratio in the 9.2$\deg$ latitude 
model and the under-estimation of the He mixing ratio in the $-5.5\deg$ latitude model compared to INMS measurements, while the first two effects 
emphasize the degeneracies that complicate the determination of the deep He abundance using the INMS measurements alone or even in combination with 
the UVIS observations \citep[e.g.,][]{koskinen18}.  Note also that the UVIS CH$_4$ retrievals show clear evidence for molecular diffusion in the 
methane homopause region, with no evidence in these occultations (from October 2014 and June 2017) or any other UVIS occultations for the apparent 
large and nearly constant CH$_4$ mixing ratios extending into the thermosphere that are indicated by the INMS measurements. In fact, the CH$_4$ mixing 
ratios retrieved at the highest altitudes determined from the UVIS occultations (maroon squares in Fig.~\ref{figeddy9n}) are roughly an order of 
magnitude smaller than those determined by the INMS measurements at higher altitudes in the thermosphere, which illustrates why we 
consider some model cases that include efficient global spreading of the ring-derived vapor before the ring vapor reaches the CH$_4$ homopause,
as discussed earlier in section~\ref{sec:gasflux}.  

We adopt a solar-cycle average ultraviolet flux \citep[from][]{woods02} and fixed-season equinox geometry for the calculations to better predict the 
observable consequences of the inflowing ring material for Saturn's middle and lower stratosphere (section \ref{sec:obs}), given that time 
constants in this region of Saturn's atmosphere are longer than a Saturn year and are thus controlled by long-term average quantities \citep{mosesgr05}.  
The radiative-transfer calculations assume diurnally averaged transmission, except for a few specific cases where we consider diurnal variation 
and look at results near local noon to better compare the ionospheric results with previous models \citep[e.g.,][]{moore18,chadney22}.  

Our models do not consider the very high spectral resolution H$_2$ cross sections and solar spectrum that \citet{kim14} and \citet{chadney22} demonstrate 
can affect the structure of Saturn's ionosphere via the presence of ``micro-windows'' in the $\sim$80-110 nm wavelength region that allow some ionizing 
photons to penetrate down to the methane homopause region near $\sim$0.1 $\mu$bar.  Based on the discussion of \citet{kim14}, we increase the CH$_4$$\! ^+$ 
production rate in our model by a factor of 22 to partially account for our neglect of this extra CH$_4$ ionization source; however, \citet{chadney22} 
illustrate that this adjustment will not correctly capture the overall altitude dependence of the CH$_4$$\! ^+$ production profile, with lower-resolution 
models such as the ones presented here having a production peak at higher altitudes than the high-resolution models.  Our low-resolution approximation 
here will therefore affect the overall predicted vertical structure of the ionosphere and the ion densities within the secondary ionospheric peak, but 
it should not change our main conclusions with respect to the dominant ions throughout the ionosphere for the different assumed influx rates shown in 
Table~\ref{tabinflux}.  Similarly, we do not explicitly calculate the transport of photoelectrons and secondary ion production, but instead use the 
pressure-dependent parameterizations described in \citet{moore09} to approximate their effect.  Future models that aim to more realistically track the 
details of ionospheric chemistry and structure resulting from equatorial ring-vapor influx on Saturn should consider these processes without such 
approximations.  








\section{Results}\label{sec:results}

If we ignore the ring material flowing into the equatorial region, the chemistry of the stratosphere and thermosphere of Saturn is fairly straighforward.  
Methane is the only abundant non-H$_2$-and-He species intrinsic to the planet that is volatile enough to make it past the cold trap at the tropopause 
temperature minimum to be transported throughout the stratosphere.  Because CH$_4$ is heavier than the background hydrogen gas, there is a limit to 
how high up in altitude the CH$_4$ can be transported in the stratosphere before molecular diffusion starts to compete with atmospheric dynamical mixing, 
at which point the CH$_4$ mixing ratio drops off with a scale height much smaller than the background gas.  The region where molecular diffusion starts to 
dominate is called the methane homopause, and not coincidently, that region marks the base of the thermosphere, because CH$_4$ and its photochemical 
products efficiently radiate away the heat conducted down from higher thermospheric altitudes.  Solar ultraviolet photons with wavelengths 
less than $\sim$145 nm that penetrate to the methane homopause can dissociate CH$_4$ and trigger a cascade of photochemical reactions that produce 
more complex hydrocarbons \citep{strobel75,atreya86,yung99,moses00a,moses05}.  Hydrocarbons in this region also contribute ions to the lower ionosphere 
\citep{atreya75hc,mosesbass00,kim14}.  In the stratosphere below the homopause, methane photolysis and the resulting hydrocarbon neutral reactions dominate 
the chemistry.  In the thermosphere above the homopause, hydrogen reactions dominate the chemistry.  The main ionospheric peak is dominated by 
H$_3$$\! ^+$ and H$^+$, although incoming external water molecules may affect both the chemistry and structure of this main peak \citep{waite87,moore09}.
If we do not ignore the equatorial ring influx, the chemistry of the stratosphere and thermosphere/ionosphere becomes more complicated, as described in 
sections \ref{sec:casea}--\ref{sec:casef}.

\subsection{Nominal photochemical models without a large ring vapor influx rate}\label{sec:resnom}

\begin{figure*}[!htb]
\vspace{-14pt}
\begin{tabular}{ll}
{\includegraphics[clip=t,scale=0.30]{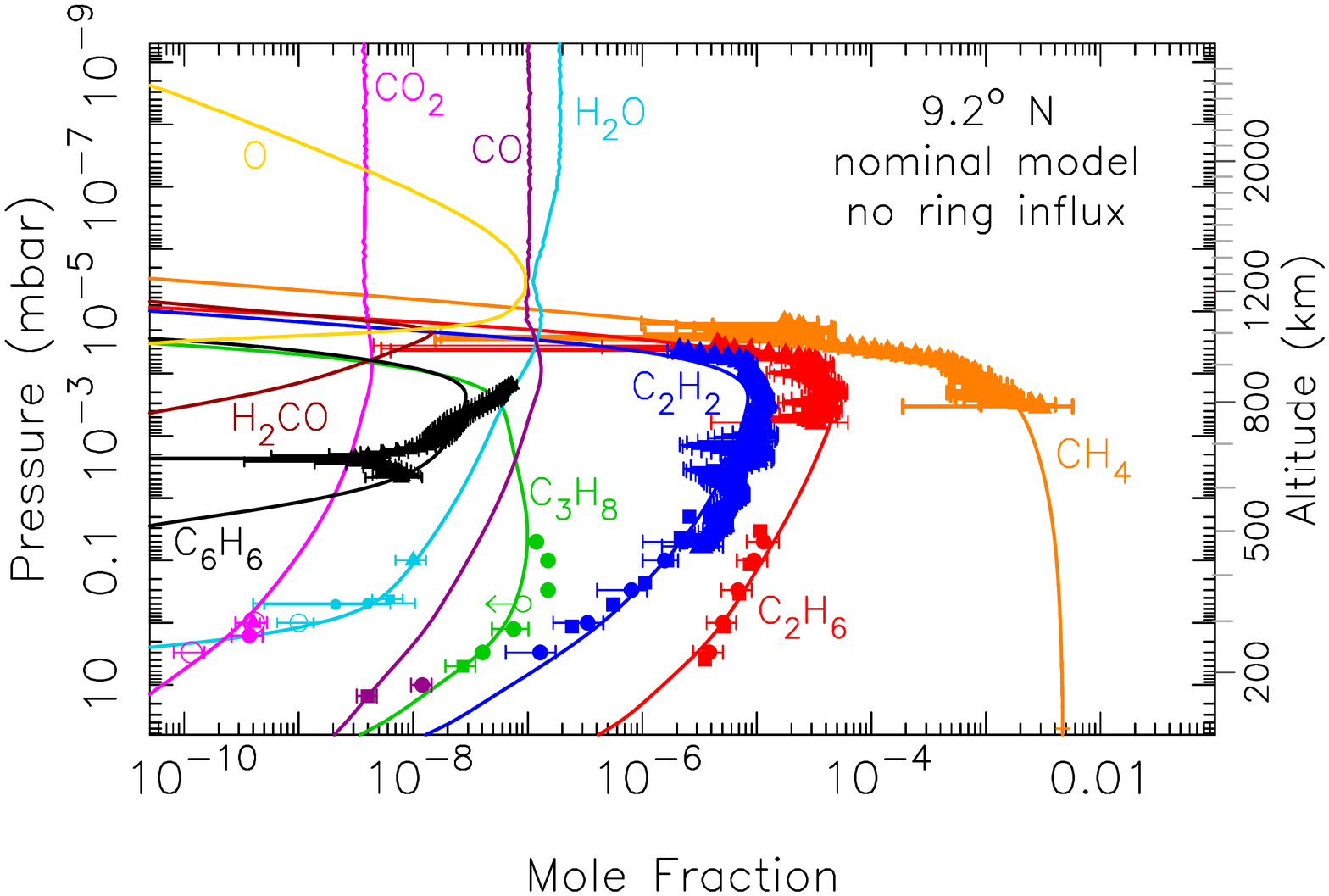}}
&
{\includegraphics[clip=t,scale=0.30]{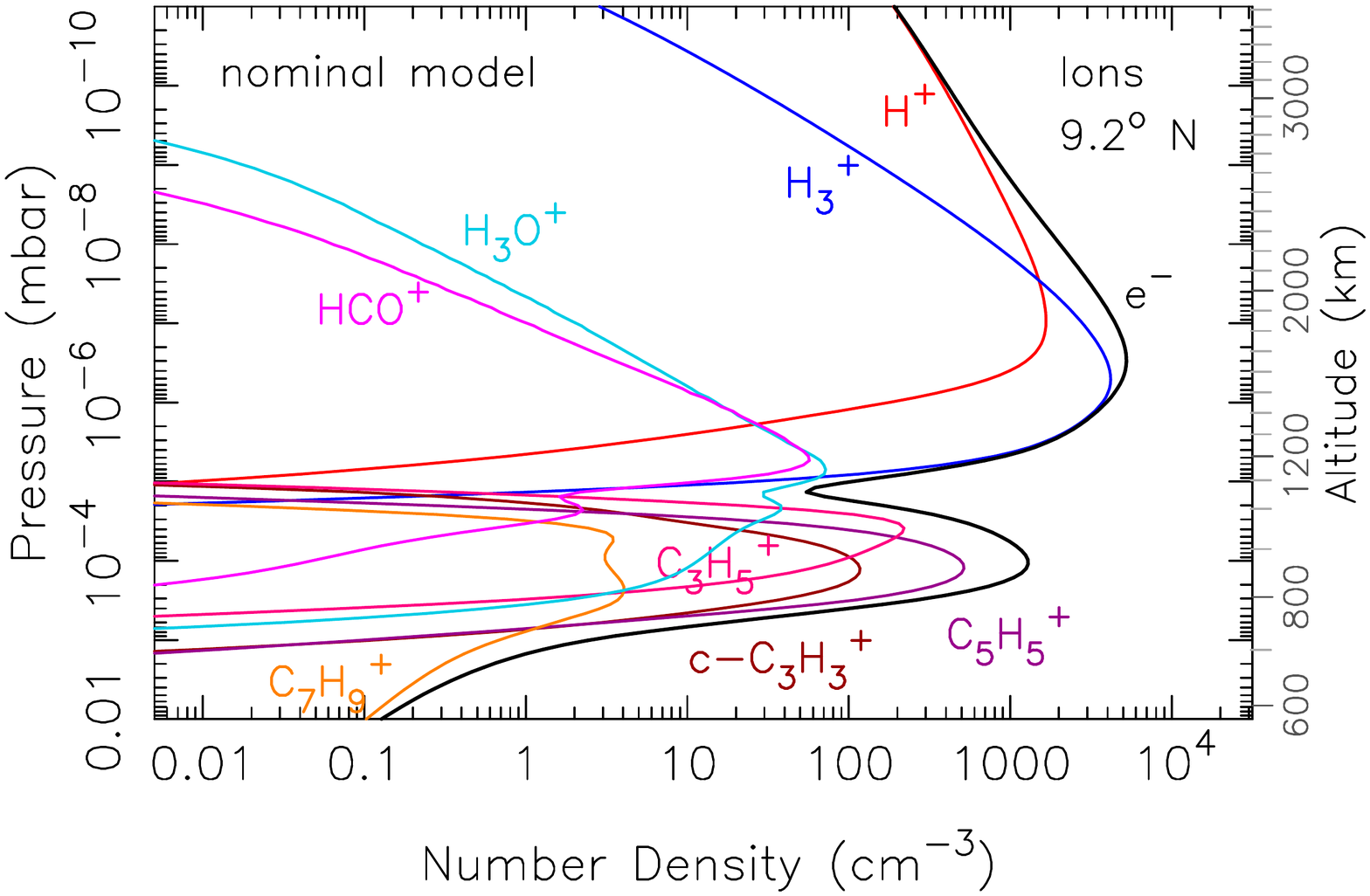}}
\end{tabular}
\caption{(Left) Mixing-ratio profiles for several hydrocarbon and oxygen species in our nominal model for 9.2$\deg$ planetocentric latitude (colored lines, 
as labeled), compared to hydrocarbon data from \textit{Cassini} CIRS and INMS instruments for the closest latitude and time (\citealt{brown21}, 
\citealt{fletcher21rev}, and this work), as well as to other broader-scale or global-average observations 
\citep[colored data points][]{fletcher09,fletcher12spire,guerlet10,moses00a,moses00b,bergin00,noll90,cavalie10,cavalie19,abbas13}.  
This nominal photochemical model does not consider the large influx of ring material derived from INMS Grand Finale measurements, 
but it does include a smaller external influx of H$_2$O, CO, and CO$_2$ to remain consistent with the observed global-average mixing ratios of oxygen 
species in Saturn's stratosphere \citep[see][]{moses17poppe}.
(Right) Densities for several ions in our nominal model (assuming diurnal-average transmission) for 9.2$\deg$ planetocentric latitude 
(colored lines, as labeled).  Note that H$_3$$\! ^+$ and H$^+$ dominate at the main peak and above, various hydrocarbon ions dominate the lower secondary 
peak, and H$_3$O$^+$ dominates in between the two peaks. Many more hydrocarbon ions contribute to the lower peak than are shown here.}
\label{fignominal9n}
\end{figure*}

The predicted abundance profiles for a few key neutral and ionized species from our nominal photochemical model for 9.2$\deg$ N planetocentric latitude 
are shown in Fig.~\ref{fignominal9n}.  This nominal model does not include the large influx of ring material derived from INMS Grand Finale measurements; 
instead, it serves as a baseline model for further comparisons into the chemical consequences of larger ring influx rates.  This nominal model does, however, 
include a smaller influx rate of three oxygen-bearing species at the upper boundary (e.g., from Enceladus) to justify the presence and global-average 
abundances of H$_2$O (assumed influx rate of 2$\scinot6.$ H$_2$O molecules cm$^{-2}$ s$^{-1}$), CO (assumed influx rate of 2$\scinot6.$ CO molecules cm$^{-2}$ 
s$^{-1}$), and CO$_2$ (assumed influx rate of 1.2$\scinot5.$ CO$_2$ molecules cm$^{-2}$ s$^{-1}$) that would otherwise not be present in such large 
quantities in Saturn's stratosphere because of condensation of intrinsic H$_2$O deeper in the troposphere.  
We introduce these external oxygen species through a flux boundary condition at the top of the model for simplicity, although depending on its source, 
the vertical source profile in the actual atmosphere may be more complicated \citep[e.g.,][]{moses00b,cavalie10,moses17poppe,hamil18}.
No nitrogen species have been identified in Saturn's stratosphere (e.g., intrinsic NH$_3$ condenses in the troposphere), so we have not included an 
influx of any nitrogen species in this nominal model.

Figure~\ref{fignominal9n} shows that CH$_4$ and its C$_2$H$_x$ photochemical products dominate the composition of Saturn's stratosphere, aside 
from the main background H$_2$ and He gas; however, even when the INMS-derived large influx rates are ignored, external oxygen species appear to be
abundant enough in a global-average sense that they can compete with C$_3$H$_x$ and heavier hydrocarbons to influence the chemistry and structure of 
the upper atmosphere.  Some coupling of oxygen and hydrocarbon photochemistry does occur, which can affect the abundance of unsaturated hydrocarbons
\citep[see][for further details of coupled oxygen-hydrocarbon photochemistry on Saturn]{moses00b,moses15,moses17poppe}.  Signatures of that 
coupled chemistry are readily apparent in Fig.~\ref{fignominal9n}, such as the decrease in H$_2$O mixing ratio with decreasing altitude in the thermosphere 
and the H$_2$O minimum near $\sim$3$\scinot-6.$ mbar --- resulting from H$_2$O photodissociation, ionization, and charge-exchange reactions with H$_3$$\! ^+$ 
and other ions ---  and the corresponding increase in species such as O, H$_3$O$^+$, and eventually CO and CO$_2$.  Water is recycled efficiently in this 
H$_2$-dominated atmosphere, but a small net loss can end up transferring some of the oxygen from H$_2$O into CO and CO$_2$ within the stratosphere.  The 
dominant pathway for that transfer within and just above the methane homopause is reaction of the H$_2$O molecules with H$_3$$\! ^+$ ions, forming H$_3$O$^+$ ions, 
which then recombine with an electron to form O + H$_2$ + H as a minor recombination channel \citep{mcewan07,neau00}.  The atomic O can react with CH$_3$ 
produced from methane photolysis near the homopause to form H$_2$CO \citep{atkinson92,preses00}, which then reacts with atomic H to form HCO and then CO 
\citep{baulch05}.  The CO can react with OH, which is a water photolysis product, to form CO$_2$ \citep{lissianski95,atkinson06}.  This net loss of H$_2$O 
leads to a slight increase in CO and CO$_2$ near $\sim$10$^{-4}$ mbar, which is visible in 
Fig.~\ref{fignominal9n}.  Both CO and CO$_2$ are relatively stable photochemically and diffuse down into the troposphere.  Water condenses in the lower 
stratosphere, contributing to haze layers in the region \citep{moses00b}.  Further details of neutral oxygen and hydrocarbon photochemistry on Saturn are 
discussed in \citet{moses00a,moses00b,moses05,moses15}.  

Figure~\ref{fignominal9n} also shows some key ionospheric species in the nominal model.  As has been discussed by \citet{majeed91jupsat}, \cite{mosesbass00}, 
\citet{moore04}, and \citet{kim14}, theory predicts that Saturn's ionospheric structure should be characterized by two peaks --- a main electron-density 
peak at higher altitudes that is dominated by H$^+$ and H$_3$$\! ^+$ ions, and a deeper, smaller, secondary peak that is dominated by hydrocarbon ions, with a 
possible contribution from metal ions from meteoric sources \citep[e.g.,][]{mosesbass00,kim01}.  A small-to-moderate influx of external oxygen species, such 
as is assumed in our nominal model, is expected to lead to a reduction in the overall electron density of the main peak, through reactions such as H$^+$ + H$_2$O 
$\rightarrow$ H$_2$O$^+$ + H, followed by H$_2$O$^+$ + H$_2$ $\rightarrow$ H$_3$O$^+$ + H, which partially converts the long-lived H$^+$ ions into molecular 
ions that recombine more rapidly with electrons \citep[e.g.,][]{connerney84}.  Spacecraft radio occultations of Saturn's ionosphere show that the electon-density 
structure is much more complicated than this simple theoretical picture, with an ionosphere that is strongly variable with latitude and time, perhaps due to 
complicated dynamical and electrodynamical effects, combined with a latitude-dependent influx of external species \citep[e.g.,][]{nagy06,moore10,moore15,barrow13}.  
However, as reviewed by \citet{moore19chap}, the standard theoretical picture with a moderate water influx concentrated at low latitudes is able to explain the 
gross ionospheric structure of Saturn, and our nominal model follows this standard view.  

The chemistry of Saturn's ionosphere has been reviewed recently by \citet{moore19chap}, where the reactions that dominate the main peak are 
described in detail.  One difference from previous water-only influx models is that we find in our nominal model that HCO$^+$ has a peak mixing ratio 
almost as great as that of H$_3$O$^+$, due to the assumed similar influx rates of external CO and H$_2$O, although the column abundance of H$_3$O$^+$ 
ends up becoming larger.  The HCO$^+$ ions are produced predominantly by H$_3$$\! ^+$ + CO $\rightarrow$ HCO$^+$ + H$_2$ \citep{mcewan07}, and lost through 
electron recombination to reform CO \citep{ganguli88,geppert04}.  The lower ionospheric peak in our nominal model forms near the methane homopause, 
where charge-exchange reactions with neutral 
hydrocarbons lead to a rich complexity of hydrocarbon ions that all contribute to the total ion density at these lower altitudes.  No single 
hydrocarbon ion dominates in this region in our model.  Except for nitrogen species, the chemistry in the lower ionosphere of Saturn in our nominal 
model has much in common with that of Titan, which is not surprising given the origin of the ion-chemistry reaction list; a full discussion of the
hydrocarbon ion chemistry on Titan can be found in \citet{vuitton19}.  One key difference between the ion chemistry on Saturn and Titan is the prevalence
of reactions of hydrocarbon ions with H$_2$ and and H on Saturn.  Further discussions of hydrocarbon ion chemistry on the giant planets can be found in 
\citet{kim94}, \citet{mosesbass00}, and \citet{kim14}.

\begin{figure*}[!htb]
\vspace{-14pt}
\begin{tabular}{ll}
{\includegraphics[clip=t,scale=0.30]{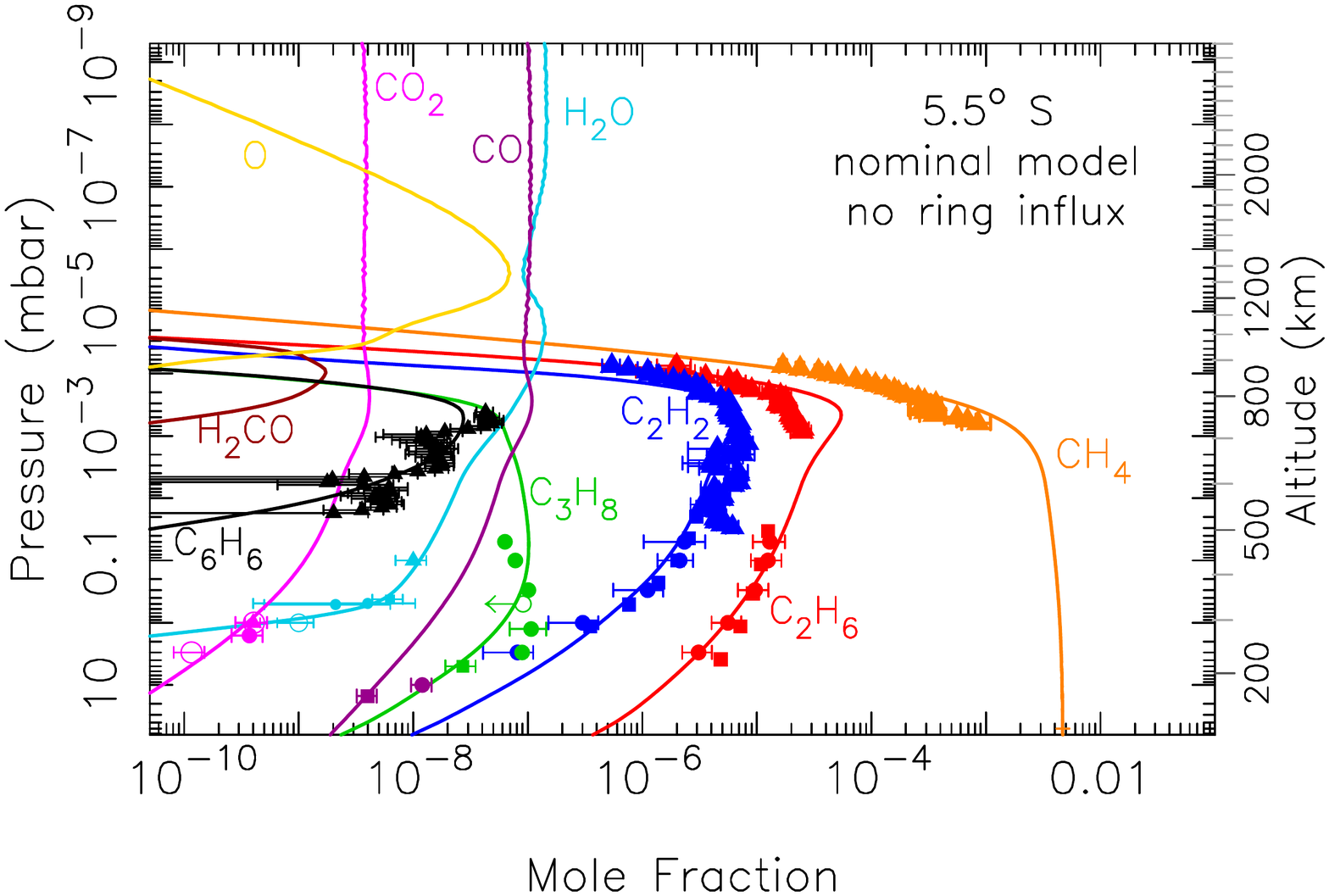}}
&
{\includegraphics[clip=t,scale=0.30]{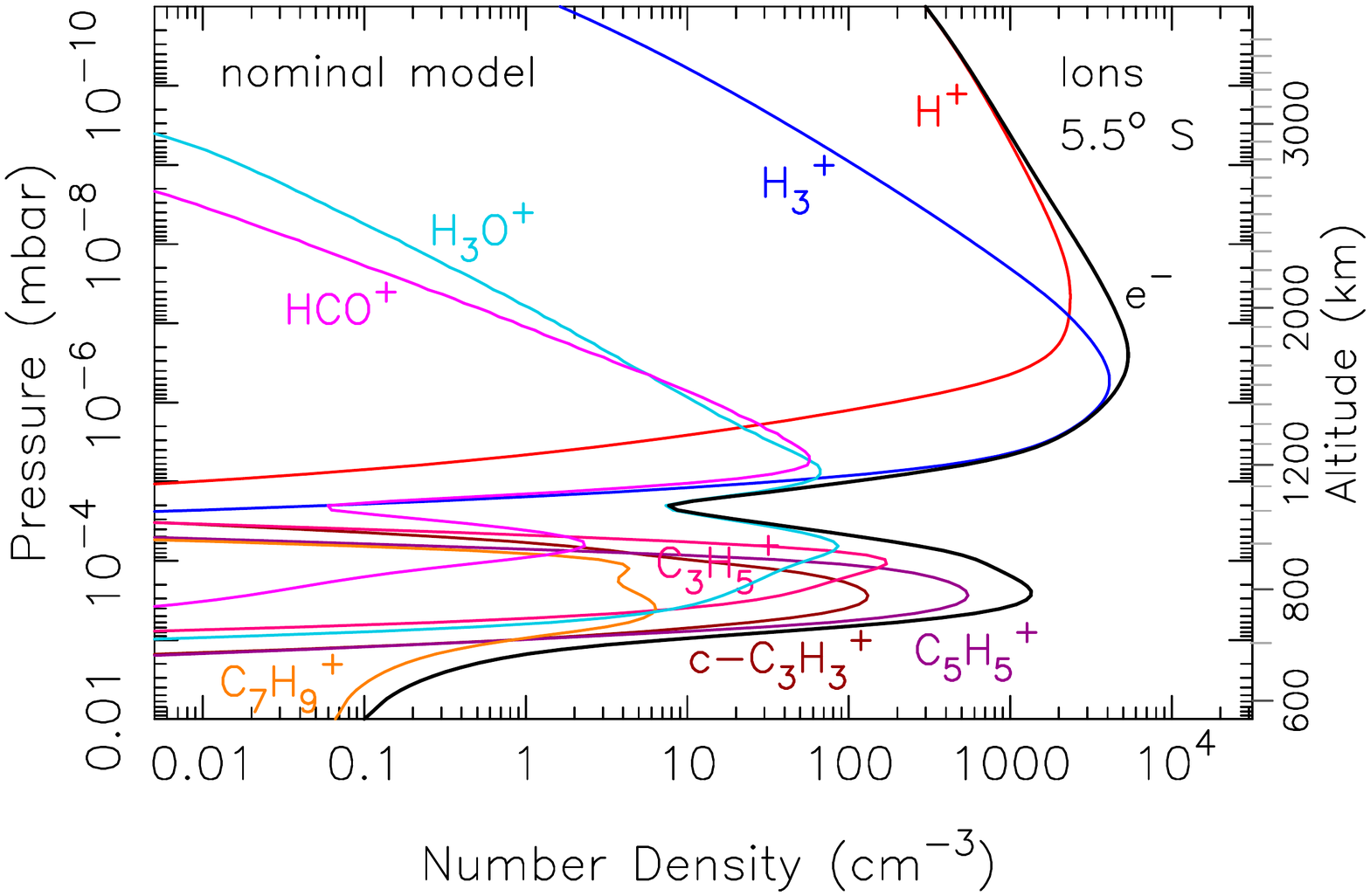}}
\end{tabular}
\caption{Same as Fig.~\ref{fignominal9n}, except for our nominal model at $-5.5\deg$ planetocentric latitude.}
\label{fignominal5s}
\end{figure*}

Corresponding results for the $-5.5\deg$ planetocentric latitude nominal model are shown in Fig.~\ref{fignominal5s}.  These results are very similar 
to the 9.2$\deg$ latitude case, with the main differences being caused by a slightly higher incident solar ultraviolet flux (recall that these models are 
calculated for northern vernal equinox conditions, so the $-5.5\deg$ case has a smaller effective zenith angle), a slightly different eddy diffusion coefficient 
profile and a correspondingly greater methane 
homopause pressure (see Fig.~\ref{figeddy9n}), and a slightly smaller assumed H$_2$O influx rate (1.5$\scinot6.$ 
molecules cm$^{-2}$ s$^{-1}$) that was adjusted to remain consistent with the global-average H$_2$O observations.  Aside from these minor differences, 
the same neutral and ion chemistry described above for 9.2$\deg$ N latitude still operates at $5.5\deg$ S latitude. 

As was suggested by order-of-magnitude estimates in \citet{koskinen16}, we find that ion chemistry provides a major 
source of benzene and PAHs in our nominal Saturn models, resulting in $\sim$4-6 times more upper-stratospheric C$_6$H$_6$ than the models that 
solely consider neutral chemistry.  However, this ion chemistry occurs predominantly at altitudes near the methane homopause.  At the higher 
thermospheric altitudes relevant to the INMS measurements, vertical diffusion is much faster than C$_6$H$_6$ chemical production, such that ion 
chemistry cannot account for the C$_6$H$_6$ signal detected by INMS at these altitudes.  The C$_6$H$_6$ detected in the thermosphere by \citet{serigano22} 
must either derive from direct influx of ring vapor or from impact vaporization and fragmentation of dust grains.

\subsection{Case A: Serigano orbit 293 influx, no horizontal spreading, all species gas phase}\label{sec:casea}

\begin{figure*}[!htb]
\vspace{-14pt}
\begin{tabular}{ll}
{\includegraphics[clip=t,scale=0.30]{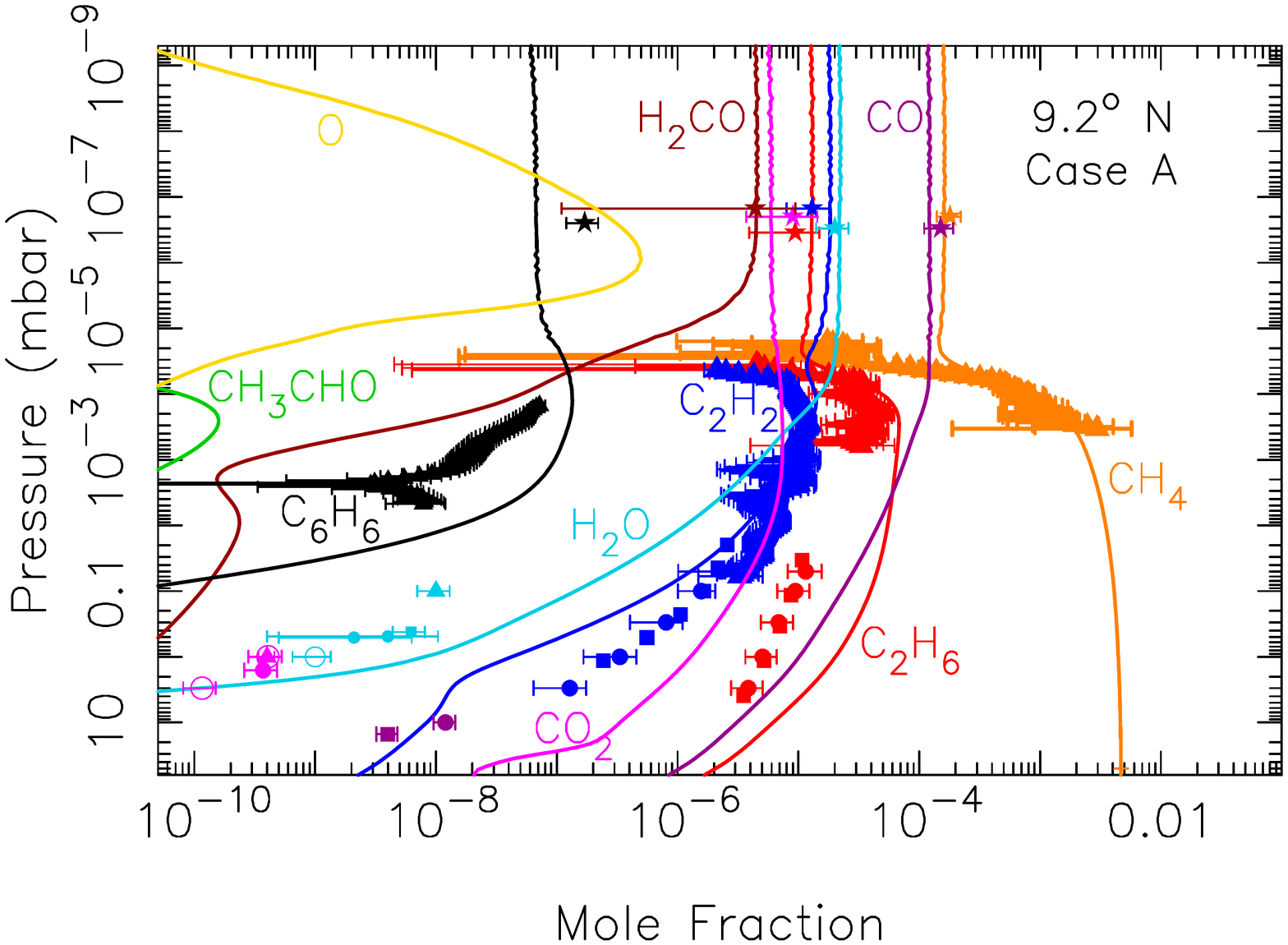}}
&
{\includegraphics[clip=t,scale=0.30]{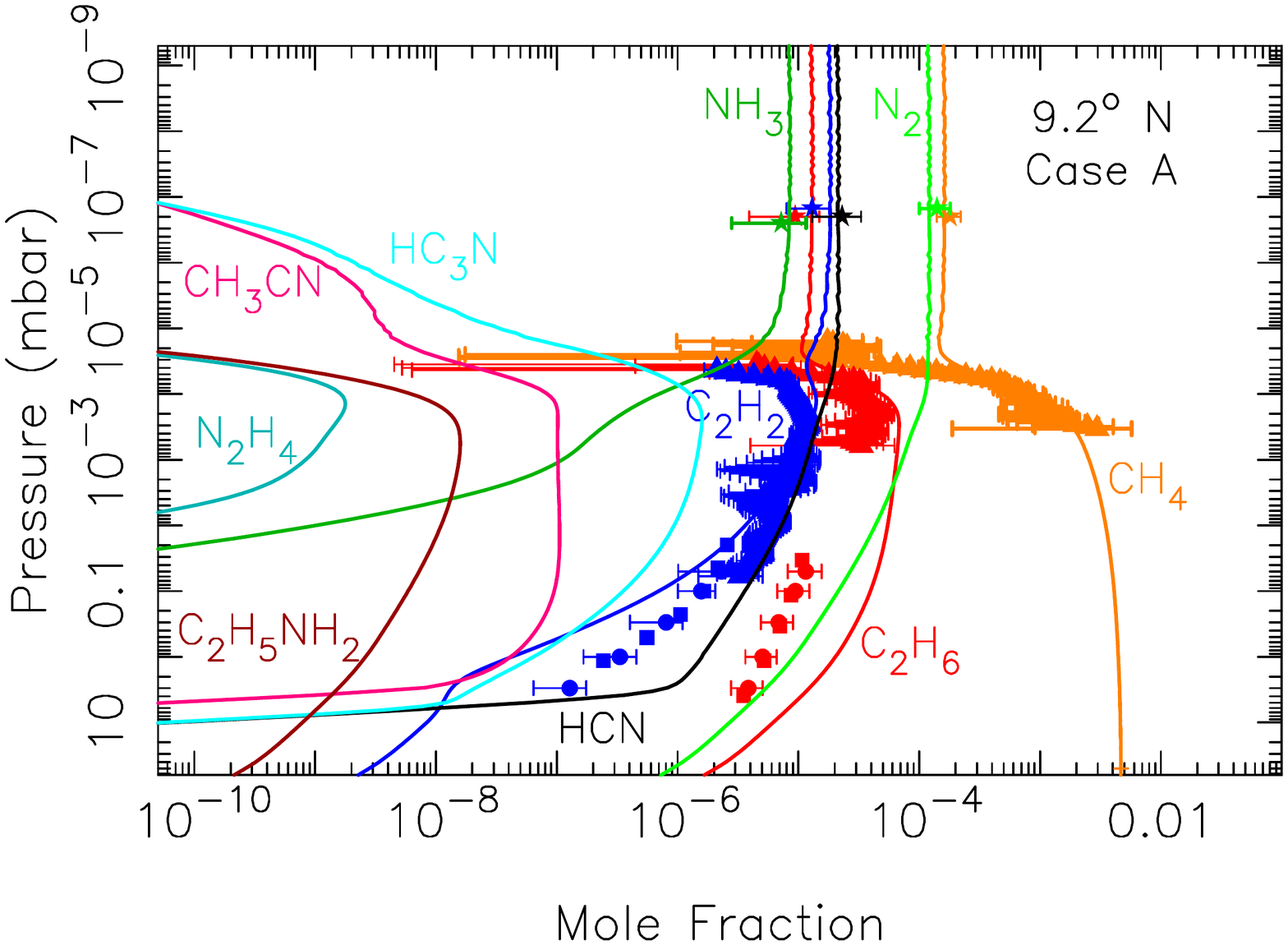}}
\end{tabular}
\caption{Mixing-ratio profiles for some neutral hydrocarbons, along with several key oxygen-bearing species (Left) and nitrogen-bearing species (Right) in 
our Case A model that adopts the INMS-derived, orbit 293, ring-vapor influx rates from \citet{serigano22} as upper boundary conditions.  Model results 
are indicated by solid colored lines, and are compared to various observational data points, including the INMS measurements (stars, from 
\citealt{serigano22}).  The incoming ring vapor notably changes the neutral composition of Saturn's stratosphere compared to the nominal 
model (cf. Fig.~\ref{fignominal9n}), through the addition of nitrogen species that would not otherwise be present, an increase in the 
abundance of several oxygen-bearing species, and modifications to the abundance of observable hydrocarbons, such as C$_2$H$_2$ and C$_2$H$_6$.}
\label{figcasea}
\end{figure*}

The results under the assumption of a strong influx of vapor from the rings, as indicated by the INMS Grand Finale data, are quite different 
from our nominal models, both in terms of the resulting stratospheric photochemistry/composition and the ionospheric chemistry.  The first model we 
explore is ``Case A'' from Table~\ref{tabinflux}, where the vapor influx rates at the upper boundary are taken from the \citet{serigano22} analysis of 
the Final Plunge orbit 293 at a terminal planetocentric latitude of 9.2$\deg$.  For this model, we include influx rates from the top 10 most abundant 
species from the \citet{serigano22} analysis, along with C$_6$H$_6$.  Results for key neutral stratospheric species are presented in Fig.~\ref{figcasea}, 
with key oxygen species shown on the left and key nitrogen species on the right.  The results for CH$_4$, C$_2$H$_2$, and C$_2$H$_6$ are shown in both 
figures to help provide a visual scale for comparison purposes.  Note that because we adopt a different background model atmosphere and different molecular 
diffusion coefficients for many species than the \citet{serigano22} model, our resulting thermospheric mixing ratios for the ring species end up 
differing from theirs.  Most of the thermospheric mixing ratios from our model fall within the quoted error bars from the \citet{serigano22} INMS analysis, 
but C$_6$H$_6$ is a notable exception.  The mismatch in C$_6$H$_6$ appears to be caused by a difference in our H$_2$--C$_6$H$_6$ mutual diffusion 
coefficients, for which we used a simple mass-scaling argument tied to the measured H$_2$--CH$_4$ diffusion coefficients \citep{gladstone96,moses00a} 
that may be a poor approximation for heavier species (see also Appendix A).

\subsubsection{Case A neutral oxygen species}\label{sec:caseaoxy}
Figure~\ref{figcasea} shows several interesting results.  First, the inferred large influx rate of CO, and to a lesser extent H$_2$O, CO$_2$, and 
H$_2$CO at the top of the atmosphere lead to significant increases in the column abundances of these and other oxygen species in Saturn's 
stratosphere.  With Case A, the stratospheric steady-state column abundances of H$_2$O, H$_2$CO, CO, and CO$_2$ increase by factors of 15, 120, 600, and 
$\sim$4000, respectively, in comparison with the nominal model, leading to significant over-predictions of the abundance of these species in comparison
to observations.  The resulting stratospheric column abundances of oxygen-bearing photochemical products 
(not all shown in the figure) are also increased accordingly.  For example, short-lived O and OH radicals are increased by a factor of 2.4 and 25, 
respectively.  Longer-lived methanol (CH$_3$OH), ketene (H$_2$CCO), and acetaldehyde (CH$_3$CHO) molecules are increased by factors of 60, 370, and 
$\sim$3000, respectively.  

Carbon monoxide (CO) survives relatively unscathed throughout the thermosphere and stratosphere, and CO is ultimately lost through diffusion down into 
the deeper troposphere.  Direct photolysis of CO is only important at very high altitudes, because the strong C$\equiv$O bond can only be broken by 
high-energy photons with wavelengths shorter than 1118 \AA, and so CO quickly becomes shielded by H$_2$, CH$_4$, and other atmospheric gases, including 
itself.  Although some CO is converted (by reaction with H$_3$$\! ^+$) to HCO$^+$ in the thermosphere/ionosphere \citep{mcewan07} or lost through three-body 
recombination with atomic H to form HCO in the stratosphere \citep{baulch05}, these species efficiently recycle back to CO through electron recombination 
or reaction with H, respectively \citep{ganguli88,geppert04,baulch05}.  In fact, when the CO production rate is integrated over the whole atmosphere 
above the tropopause, the net column production rate exceeds its loss rate in the Case A model, with the additional CO being produced from direct 
photolysis of H$_2$CO and CO$_2$ and from indirect pathways initiated by H$_2$O photolysis followed by reaction of O and OH with CH$_3$ and other 
hydrocarbons and nitrogen species \citep[e.g.,][]{atkinson92,atkinson06,preses00,nguyen06,senosiain06,jasper07}.  Carbon dioxide (CO$_2$) also has a net 
positive column production rate integrated through the stratosphere and thermosphere, with its production largely due to the reaction of CO with the OH that 
derives from water photolysis.  The main loss process for CO$_2$ is photolysis, which produces CO and can lead to CO$_2$ recycling.  

The high photochemical stability of CO and CO$_2$, combined with their large influx rates from the top of the atmosphere in our Case A model, 
leads to their high abundance in the lower stratosphere seen in Fig.~\ref{figcasea}.  In fact, the resulting CO column abundance above 200 mbar 
of 5.8$\scinot19.$ cm$^{-2}$ is more than two and a half orders of magnitude larger than the global-average abundance inferred from 
infrared \citep{noll90,moses00b} and (sub)millimeter observations \citep{cavalie10}.  The CO$_2$ column abundance above 10 mbar 
in the Case A model --- 2$\scinot18.$ cm$^{-2}$ --- is more than 3000 times the global-average value inferred from the Infrared Space 
Observatory observations of \citet{feuchtgruber97,feuchtgruber99} and \citet{moses00b} and more than 2000 times the equatorial values 
inferred from \textit{Cassini}/CIRS nadir observations \citep{abbas13}.  Immediately we can see one challenge to the idea that the ring vapor 
influx derived from \textit{Cassini}/INMS represents a steady, long-term inflow rate that remains concentrated at low latitudes --- the 
stratospheric abundances of CO and CO$_2$, at least, would then be much greater than is actually observed on Saturn.

The situation regarding the H$_2$O column abundance for Case A is less severe but still greatly over-predicted by the model.  The stratospheric 
column abundance of H$_2$O predicted by the Case A model is 2.7$\scinot16.$ cm$^{-2}$, which is $\sim$10--20 times larger than that inferred from 
global-average observations \citep{feuchtgruber99,moses00b,bergin00} and $\gtrsim$10 times greater than the equatorial column abundance inferred 
from the spatially resolved \textit{Herschel} observations of \citet{cavalie19}.  Water is readily photolyzed in Saturn's stratosphere, and although 
the OH can react back with H$_2$ or various hydrocarbons to reform the H$_2$O \citep{baulch05,jasper07,gierczak97}, our Case A model also contains 
some effective schemes to transfer some of the oxygen initially available in water over to CO and/or CO$_2$.  One dominant scheme is:
\begin{eqnarray}
\HtwoO \, + \, h\nu \, & \rightarrow & \, \OH \, + \, \Hatom \nonumber \\
\CHfour \, + \, h\nu \, & \rightarrow & \, \CHthree \, + \, \Hatom \nonumber \\
\OH \, + \, \CO \, & \rightarrow & \, \COtwo \, + \, \Hatom \nonumber \\
\COtwo \, + \, h\nu \, & \rightarrow & \, \CO \, + \, \Oatom \nonumber \\
\Oatom \, + \, \CHthree \, & \rightarrow & \, \HtwoCO \, + \, \Hatom \nonumber \\
\HtwoCO \, + \, h\nu \, & \rightarrow & \, \CO \, + \, \Htwo \nonumber \\
\noalign{\vglue -10pt}
\multispan3\hrulefill \nonumber \cr
\Net \ \ \HtwoO \, + \, \CHfour \, & \rightarrow & \, \CO \, + \, \Htwo \, + \, 4\,\Hatom , \\
\end{eqnarray}
\citep[see][]{lissianski95,atkinson92,preses00,atkinson06},  
where $h\nu$ represents an ultraviolet photon.
Note that CO$_2$ is an intermediary in this scheme, unlike a similar scheme in our nominal model that does not assume such large CO and CO$_2$ influx 
rates.  Carbon dioxide can help speed up the conversion from the O-H bond found in H$_2$O into the carbon-oxygen bond found in CO, because CO$_2$ 
photolysis releases an O atom, whereas H$_2$O photolysis more readily produces OH radicals.  Atomic O is more likely than OH radicals to react 
with hydrocarbons to form C-O bonded species under these conditions, whereas the OH more efficiently reacts with H$_2$ and hydrocarbons to 
simply recycle back to H$_2$O.  Water condenses in the lower stratosphere in our Case A model at pressures greater than $\sim$1 mbar, which 
represents a significant sink of the external oxygen.  

Although formaldehyde (H$_2$CO) flows into the atmosphere from the rings in the Case A model, it is relatively unstable once it reaches the 
stratosphere and so does not build up to observable abundances.  The column abundance of H$_2$CO above 10 mbar in our nominal model is just
$\sim$7$\scinot13.$ cm$^{-3}$.  Formaldehyde can be photolyzed by longer-wavelength UV radiation, and it is not shielded effectively by any 
more abundant molecules.  While some reactions such as $2\, \HCO \, \rightarrow \, \CO \, + \, \HtwoCO$ allow the H$_2$CO to be recycled \citep{baulch05}, 
there is a significant permanent conversion of the H$_2$CO into CO in our models (e.g., see the last reaction in scheme (2) above). 

Other stable oxygen-bearing molecules such as CH$_3$CHO are produced in non-trivial quantities in the stratosphere of our Case A model (see 
Fig.~\ref{figcasea}) --- predominantly through HCO + CH$_3$ + M $\rightarrow$ CH$_3$CHO + M \citep{callear90}, where M is any third species --- but the large 
abundance of CH$_3$CHO here relies on the very large column abundance of CO.  As discussed previously, the predicted CO column abundance from 
Case A is significantly higher than what is observed in Saturn's stratosphere, so the predicted CH$_3$CHO stratospheric column abundance from 
Case A is also likely to be an over-estimate.

\subsubsection{Case A neutral nitrogen species}\label{sec:caseanit}
Our Case A model predicts significant quantities of neutral nitrogen species in Saturn's stratosphere (see Fig.~\ref{figcasea}), which derive 
from the ring-vapor influx of molecular nitrogen (N$_2$), hydrogen cyanide (HCN), and ammonia (NH$_3$).  
Molecular nitrogen, like CO, possesses a strong triple bond and is very 
stable in the stratosphere, with photolysis occurring mostly in the thermosphere.  Neutral reactions that can destroy N$_2$ tend to have 
either very high energy barriers and so are ineffective at cool stratospheric temperatures or involve highly radical species (such as CH) that 
have short lifetimes and are only present in small quantities in the stratosphere.  Molecular nitrogen is therefore very stable in Saturn's stratosphere 
and will diffuse downward intact into the troposphere.  The only notable loss process for N$_2$, other than high-altitude photolysis, involves 
ion-neutral reactions in the thermosphere/ionosphere.  From a column-integrated standpoint in our Case A model, N$_2$ is lost mainly through 
the reaction of H$_3$$\! ^+$ + N$_2$ $\rightarrow$ N$_2$H$^+$ + H$_2$ \citep{mcewan07}.  However, most of the 
loss reactions for N$_2$H$^+$ end up just recycling the N$_2$, so the main ``permanent'' loss process for N$_2$ is photolysis to produce 
N atoms that can then react with hydrocarbons such as CH$_3$ to produce H$_2$CN \citep{marston89a,marston89b} and eventually HCN \citep[e.g.,][]{nesbitt90}.  
The atomic N can also react with NH$_2$ (from NH$_3$ photolysis) to produce N$_2$H \citep[based on][]{dransfeld87}, which eventually recycles the 
N$_2$ \citep{dean00}, or the N can react with OH to form minor amounts of NO \citep{smith94stewart}.   
Molecular nitrogen is a homonuclear molecule that is difficult to detect by remote-sensing methods, so although the 
Case A stratospheric column abundance of 3.3$\scinot19.$ N$_2$ molecules cm$^{-2}$ s$^{-1}$ is quite large, it is not surprising that 
N$_2$ has not been detected in Saturn's stratosphere to date.  

Hydrogen cyanide (HCN) is also quite stable under Saturn stratospheric conditions, and HCN even benefits from some net column 
production due to coupled hydrocarbon-nitrogen photochemistry.  Although photolysis of HCN continues well down into the stratosphere, the 
CN thus produced will very efficiently recycle the HCN via reactions with stable hydrocarbons and H$_2$ \citep[e.g.,][]{baulch94,sims93,yang93}.  
Some fraction of the HCN in our 
Case A model is lost through relatively inefficient reactions with hydrocarbon radicals, to eventually form heavier nitriles, or lost 
through more efficient charge-exchange reactions with various ions (e.g., H$_3$O$^+$, H$_3$$\! ^+$, C$_2$H$_5$$\! ^+$, CH$_5$$\! ^+$, HCO$^+$, and 
N$_2$H$^+$ being most important) to form HCNH$^+$.  The loss processes for the latter ion mostly lead to the recycling of HCN, although some 
some heavier $R\cdot$CNH$^+$ ions can form (with $R$ being a radical, such as CH$_3$, C$_2$H$_5$, or C$_2$H$_3$) to eventually produce 
heavier neutral nitrile species.  Some excess HCN production also occurs in the stratosphere through coupled nitrogen-hydrocarbon 
photochemistry, such as the following scheme \citep{kaye83ch3nh2,jodkowski95,dean00,moses10}:
\begin{eqnarray}
\NHthree \, + \, h\nu \, & \rightarrow & \, \NHtwo \, + \, \Hatom \nonumber \\
\CHfour \, + \, h\nu \, & \rightarrow & \, \CHthree \, + \, \Hatom \nonumber \\
\NHtwo \, + \, \CHthree \, + \, \Mthird & \rightarrow & \, \CHthreeNHtwo \, + \, \Mthird \nonumber \\
\CHthreeNHtwo \, + \, h\nu \, & \rightarrow & \, \CHthreeNH \, + \, \Hatom \nonumber \\
\CHthreeNH \, + \, \Hatom \, & \rightarrow & \, \CHtwoNH \, + \, \Htwo \nonumber \\
\CHtwoNH \, + \, h\nu \, & \rightarrow & \, \HCN \, + \, \Htwo \nonumber \\
\noalign{\vglue -10pt}
\multispan3\hrulefill \nonumber \cr
\Net \ \ \NHthree \, + \, \CHfour \, & \rightarrow & \, \HCN \, + \, 2\,\Htwo \, + \, 2\,\Hatom , \\
\end{eqnarray}
along with reactions such as N + CH$_3$ $\rightarrow$ H$_2$CN + H, followed by H$_2$CN + H $\rightarrow$ HCN + H$_2$, or direct production
from N + CH$_3$ \citep{marston89a,marston89b,nesbitt90}.  The Case A abundance of HCN is sufficient enough that it condenses in the lower stratosphere 
at pressures near $\sim$3 mbar, with a total stratospheric column abundance of 1.7$\scinot18.$ cm$^{-2}$.  That amount of HCN in Saturn's 
stratosphere should be readily observable in emission from infrared, millimeter, and sub-millimeter observations (see section~\ref{sec:obs}).  
Our predicted 1-mbar mixing 
ratio from Case A is $\sim$70 times greater than the HCN upper limit reported by \citet{fletcher12spire}, again casting doubt on the long-term nature 
of the ring-vapor influx rates currently inferred for Case A from INMS, and/or suggesting significant global spreading at altitudes below the INMS 
measurements, and/or suggesting a significant fraction of the INMS measurements are caused by dust particles impacting the instrument.

Ammonia is much more photochemically fragile than N$_2$ and HCN in Saturn's stratosphere. Photons with wavelengths less than $\sim$220 nm are responsible 
for dissociating NH$_3$ throughout Saturn's stratosphere.  The main photolysis product, NH$_2$, can react with CH$_3$ to form CH$_3$NH$_2$ 
and eventually HCN (see scheme (3) above), can react with H and H$_2$ to recycle the NH$_3$ \citep{pagsberg79,espinosa94,moses11}, 
and can react with N and NH to form N$_2$H$_x$ species that eventually produce N$_2$ \citep{klippenstein09,dean00,dransfeld87}.  
Some hydrazine (N$_2$H$_4$) can also form from NH$_2$ + NH$_2$ + M $\rightarrow$ N$_2$H$_4$ + M \citep{klippenstein09}, 
as it is predicted to do in the tropospheres of Jupiter and Saturn \citep[e.g.,][]{strobel73,atreya80}.  However, the N$_2$H$_4$ itself is 
unstable in Saturn's stratosphere, and its formation does not constitute a significant loss for NH$_3$.  Instead, NH$_3$ is lost through 
coupled NH$_3$-hydrocarbon photochemistry, with the N ending up predominantly in HCN, CH$_3$CN, C$_2$H$_5$NH$_2$, HC$_3$N, and other 
organo-nitrogen compounds, along with N$_2$.  The stratospheric column abundance of NH$_3$ in our Case A model is relatively small, at 
1.9$\scinot14.$ cm$^{-2}$, which would be difficult to detect in infrared and ultraviolet observations.

Acetonitrile (CH$_3$CN, also called cyanomethane) is a significant photochemical product in our Case A model whose production and loss schemes 
are somewhat speculative due to a lack of available rate coefficients for key reactions \citep[see also][]{loison15,vuitton19}.  Acetonitrile 
is produced in our model predominantly from CH$_2$CN + H + M $\rightarrow$ CH$_3$CN + M (rate coefficient estimated) and from CH$_3$CNH$^+$ + 
H$_2$O $\rightarrow$ H$_3$O$^+$ + CH$_3$CN (rate coefficient estimated).  The CH$_3$CNH$^+$ derives from HCNH$^+$ + CH$_3$ $\rightarrow$ 
CH$_3$CNH$^+$ + H \citep{vuitton19}.  The source of CH$_2$CN in the model is also speculative, involving triplet ground-state methylene 
($^3$CH$_2$) reactions with HCN or C$_2$H$_3$CN that likely have energy barriers of uncertain magnitude \citep[see also][]{hebrard09}.  The 
production reactions for CH$_3$CN involving coupled C$_2$H$_2$-NH$_3$ photochemistry that were speculated to be important on Jupiter by 
\citet{kaye83hcn}, \citet{ferris88}, \citet{keane96}, and \citet{moses10} are included and are of moderate importance but never dominate.  
The abundance of CH$_3$CN in the Case A model is sufficient enough that it condenses in the lower stratosphere in the 2--1000 mbar region.  
The column abundance of CH$_3$CN above 10 mbar in our Case A model is 3.2$\scinot16.$ cm$^{-2}$, which makes it another interesting 
nitrogen-bearing product that might be observable on Saturn \citep{lellouch85} if ring-vapor influx is prevalent; notably, the lower-stratospheric 
CH$_3$CN mixing ratio in Case A is predicted to be greater than that on Titan, where CH$_3$CN has been observed at millimeter wavelengths 
\citep[e.g.,][]{marten02}. 

The main (non-recycling) production mechanism for cyanoacetylene (HC$_3$N, also called propiolonitrile) is 
CN + C$_2$H$_2$ $\rightarrow$ HC$_3$N + H \citep{sims93}, where the CN derives from HCN photolysis, and both HCN and C$_2$H$_2$ are ring-inflow 
species.  The dominant scheme is the following:
\begin{eqnarray}
\HCN \, + \, h\nu \, & \rightarrow & \, \CN \, + \, \Hatom \nonumber \\
\CN \, + \, \CtwoHtwo \, & \rightarrow & \, \HCthreeN \, + \, \Hatom \nonumber \\
\noalign{\vglue -10pt}
\multispan3\hrulefill \nonumber \cr
\Net \ \ \HCN \, + \, \CtwoHtwo \, & \rightarrow & \, \HCthreeN \, + \, 2\,\Hatom , \\
\end{eqnarray}
The HC$_3$N mixing ratio peaks just below the methane homopause, and then decreases with decreasing 
altitude in the stratosphere due to photolysis and subsequent reactions that lead to HCN formation, despite a tendency for the main 
photolysis product C$_3$N to recycle the HC$_3$N through reaction with CH$_4$ (see Appendix B). 
Some HC$_3$N production and loss also occurs in the ionosphere, typically involving 
HC$_3$NH$^+$ ions \citep{vuitton19}.  Condensation of HC$_3$N occurs at pressures greater than a few mbar in the 
Case A model, and the total stratospheric column abundance of HC$_3$N is 8.6$\scinot16.$ cm$^{-2}$.  That amount of 
HC$_3$N in Saturn's stratosphere should be readily observable at mid-infrared wavelengths (see section~\ref{sec:obs}).

Ethylamine (C$_2$H$_5$NH$_2$) is produced in our model through addition reactions of NH$_2$ and C$_2$H$_5$.  Little is known about its photochemistry, and 
its production and loss reaction rates have been estimated based on simulations of complex laboratory experiments \citep[e.g.,][]{moses10}.  

\subsubsection{Case A hydrocarbons}\label{sec:caseahyd}
The large influx rate of ring vapor also affects the abundance of stratospheric hydrocarbons.  As can be seen from Fig.~\ref{figcasea}, the 
abundance of C$_2$H$_6$ significantly increases, while C$_2$H$_2$ significantly decreases, with Case A in comparison to the nominal model 
shown in Fig.~\ref{fignominal9n}.  The increase in C$_2$H$_6$ results from an increase in CH$_3$ from several sources, including photolysis 
of HC$_3$N, followed by its recycling and the net destruction of CH$_4$ through the following scheme \citep[see][]{vuitton12,vuitton19,fournier14}:
\begin{eqnarray}
2\, ( \HCthreeN \, + \, h\nu \, & \rightarrow & \, \CthreeN \, + \, \Hatom ) \nonumber \\
2\, ( \CthreeN \, + \, \CHfour \, & \rightarrow & \, \HCthreeN \, + \, \CHthree ) \nonumber \\
2\, \CHthree \, + \, \Mthird & \rightarrow & \, \CtwoHsix \, + \, \Mthird \nonumber \\
\noalign{\vglue -10pt}
\multispan3\hrulefill \nonumber \cr
\Net \ \ 2\, \CHfour \, & \rightarrow & \, \CtwoHsix \, + \, 2\, \Hatom . \\
\end{eqnarray}
Ethane is also better 
shielded from photolysis in the middle and lower stratosphere with the presence of the additional ring-influx species from the Case A model.  
The increase in atomic H in the Case A model, resulting from the above scheme and from HCN, NH$_3$, and H$_2$O photolysis, promotes the conversion of 
C$_2$H$_2$ into C$_2$H$_6$, through schemes such at the following:
\begin{eqnarray}
\CtwoHtwo \, + \, \Hatom \, + \, \Mthird \, & \rightarrow & \, \CtwoHthree \, + \, \Mthird \nonumber \\
\CHthree \, + \, \CtwoHthree \, & \rightarrow & \, \CthreeHfive \, + \, \Hatom \nonumber \\
\Hatom \, + \, \CthreeHfive \, + \, \Mthird & \rightarrow & \, \CthreeHsix \, + \, \Mthird \nonumber \\
\Hatom \, + \, \CthreeHsix \, + \, \Mthird \, & \rightarrow & \, \CthreeHseven \, + \, \Mthird \nonumber \\
\Hatom \, + \, \CthreeHseven \, & \rightarrow & \, \CtwoHfive \, + \, \CHthree \nonumber \\
\Hatom \, + \, \CtwoHfive \, & \rightarrow & \, 2\, \CHthree \nonumber \\
2\, \CHthree \, + \, \Mthird \, & \rightarrow & \, \CtwoHsix \, + \, \Mthird \nonumber \\
\noalign{\vglue -10pt}
\multispan3\hrulefill \nonumber \cr
\Net \ \ \CtwoHtwo \, + \, 4\, \Hatom \, & \rightarrow & \, \CtwoHsix  . \\
\end{eqnarray}
In fact, the higher stratospheric production rate of CH$_3$ and H in the Case A model leads to greater abundances of saturated hydrocarbons 
(i.e., alkanes such as C$_2$H$_6$, C$_3$H$_8$, and C$_4$H$_{10}$) at the expense of unsaturated hydrocarbons (such as C$_2$H$_2$, 
C$_2$H$_4$, CH$_3$C$_2$H, and C$_4$H$_2$), so the resulting saturated-to-unsaturated hydrocarbon ratios are much greater in the 
Case A model than the nominal no-ring-influx model.  The column abundance of C$_2$H$_6$ above 10 mbar in the Case A model 
is a factor of $\gtrsim$4 greater than that in the nominal model, and the C$_2$H$_6$ abundance is no longer consistent with the stratospheric 
retrievals from CIRS limb observations, nadir observations, and UVIS occultations (see the red circles, squares, and triangles in comparison to 
red curve in Fig.~\ref{figcasea}).  By the same token, the Case A model predicts a C$_2$H$_2$ column abundance above 10 mbar that is a factor 
of $\lesssim$5 smaller than the predictions of the nominal model without the ring-vapor influx.  Again, the large ring influx from Case A causes a 
significantly degraded fit to the lower-stratospheric C$_2$H$_2$ observations from \textit{Cassini} CIRS and UVIS.  

\begin{figure*}[!htb]
\vspace{-14pt}
\begin{center}
\includegraphics[clip=t,scale=0.4]{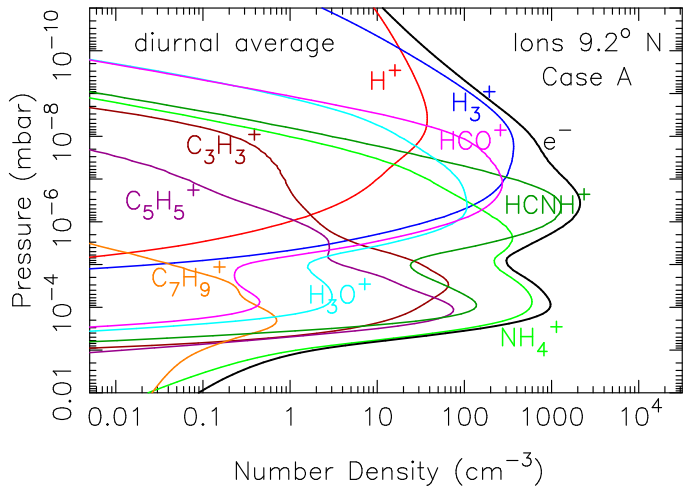}
\end{center}
\caption{Ion and electron number densities for our Case A model, assuming diurnally averaged transmission 
at 9.2$\deg$ planetocentric latitude.  Note the prevalence of nitrogen-bearing ions, in contrast to the nominal model (cf. Fig.~\ref{fignominal9n}).  
The incoming ring vapor significantly changes the ion composition of Saturn's stratosphere compared to the nominal model.} 
\label{figioncasea}
\end{figure*}

\subsubsection{Case A ion chemistry}\label{sec:caseaion}

The ion densities in our Case A model are shown in Fig.~\ref{figioncasea} for our standard assumption of diurnally averaged transmission; results from 
the diurnally variable models at times near local noon are presented later in section~\ref{sec:iono}.  The large influx of ring vapor --- and the 
presence of nitrogen-bearing species in particular --- 
significantly affects the predicted ionospheric composition compared to our nominal model without any ring influx (cf.~Figs.~\ref{fignominal9n} \& 
\ref{figioncasea}).  Instead of a main ionospheric peak dominated by H$_3$$\! ^+$ and H$^+$, the dominant ion at the main peak in the Case A model is 
HCNH$^+$.  Instead of a deeper, secondary peak dominated by hydrocarbon ions, the secondary peak in the Case A model is dominated by NH$_4$$\! ^+$ and 
HCNH$^+$.  As is discussed by \citet{vuitton06} for the case of Titan, protons ``flow'' through various molecular species via proton-transfer reactions 
and terminate at molecules with the greatest proton affinity.  The prevalence of these nitrogen-bearing ions in our Case A model is therefore 
simply the result of the presence of key nitrogen parent species such as HCN and NH$_3$ entering the atmosphere from the rings, combined with 
the high proton affinities of these species.  Other nitrogen ions not shown here, such as CH$_3$CNH$^+$ and HC$_3$NH$^+$, are also non-trivial 
components of the Case A ionosphere.

At the main peak, the HCNH$^+$ ions form through proton-transfer reactions from abundant local ions, such as H$_3$$\! ^+$, HCO$^+$, 
CH$_5$$\! ^+$, H$_3$O$^+$, and N$_2$H$^+$, to HCN \citep{mcewan07,vuitton19}, and are lost through electron recombination (to form HCN + H; see \citealt{semaniak01}) 
and through proton transfer to NH$_3$ (which has a higher proton affinity than HCN) to form HCN + NH$_4$$\! ^+$ \citep{vuitton19}.  At the deeper secondary peak, 
proton transfers from H$_3$O$^+$, C$_2$H$_5$$\! ^+$, CH$_5$$\! ^+$, HCO$^+$, and C$_2$H$_7$$\! ^+$ to HCN dominate production, and loss is primarily due to 
HCNH$^+$ + NH$_3$ $\rightarrow$ NH$_4$$\! ^+$ + HCN, with contributions from HCNH$^+$ + CH$_3$ $\rightarrow$ CH$_3$CNH$^+$ and HCNH$^+$ + C$_6$H$_6$ 
$\rightarrow$ C$_6$H$_7$$\! ^+$ + HCN \citep{vuitton19}.

The NH$_4$$\! ^+$ ions are formed via HCNH$^+$ + NH$_3$ $\rightarrow$ NH$_4$$\! ^+$ + HCN \citep{vuitton19}, with contributions from proton-transfer reactions of 
H$_3$$\! ^+$, HCO$^+$, H$_3$O$^+$, CH$_5$$\! ^+$, and C$_2$H$_3$$\! ^+$ to NH$_3$ to form H$_2$, CO, H$_2$O, CH$_4$, and C$_2$H$_2$, respectively, along with 
the NH$_4$$\! ^+$ \citep{mcewan07,vuitton19}.  The dominant loss process for NH$_4$$\! ^+$ is electron recombination \citep{alge83,ojekull04}.

Note that the electron density (i.e., assumed to be equal to the total ion density) at the main peak in our Case A model shown in Fig.~\ref{figioncasea} 
is smaller than that from our nominal model shown in Fig.~\ref{fignominal9n}, and the gap between the main and secondary peaks is less pronounced.  
Ion-neutral reactions between the neutral 
ring vapor species and H$^+$ and H$_3$$\! ^+$ help reduce the main peak magnitude by converting some portion of these ions to heavier molecular ions 
with larger electron recombination rate coefficients and thus shorter lifetimes. Similarly, the presence of large amounts of ring vapor species 
above the methane homopause in the Case A model provides a source of ions below the main peak and above the secondary peak that is not present 
to such an extent in the nominal model.

We can compare our Case A ionospheric model results with some other INMS-based models in the literature.  \citet{moore18} include INMS-inspired 
constant thermospheric mixing ratios of CH$_4$, CO$_2$, H$_2$O, NH$_3$, and a mass 28 species in their ion chemistry model, and they examine how the 
mass 28 constituent affects the ionospheric composition based on whether that species is N$_2$, CO, or C$_2$H$_4$.  In all three of their different 
mass-28 models, H$^+$ dominates the topside 
ionosphere, H$_3$$\! ^+$ dominates and forms the main peak at some point below H$^+$, and then various heavier ions take over below the H$_3$$\! ^+$, followed by 
a deeper region that is always dominated by H$_3$O$^+$ despite H$_2$O being the least abundant of the heavier INMS-included neutrals.  In the 
$\sim$1700--1800 km region in their $-6\deg$ latitude model (which would correspond to $\sim$2--3$\scinot-7.$ mbar in our $-5.5\deg$ latitude model and 
$\sim$1--2$\scinot-7.$ mbar in our 9.2$\deg$ latitude model), ions heavier than H$_3$$\! ^+$ contribute significantly to the main peak (including 
H$_3$O$^+$), but the ions aside from H$_3$O$^+$ that populate the region depend on the mass 28 neutral constituent.  If the mass 28 species is N$_2$, then 
N$_2$H$^+$ is important in that region of their model but never dominant; if the mass 28 species is C$_2$H$_4$, then C$_2$H$_3$$\! ^+$ is important but 
never dominant; if the mass 28 species is CO, then HCO$^+$ becomes dominant for a range of altitudes in between the H$_3$$\! ^+$ and H$_3$O$^+$ 
regions.  Their model stops at 1500 km altitude (which would correspond to $\sim$1$\scinot-6.$ mbar in our $-5.5\deg$ model and $\sim$7$\scinot-7.$ 
mbar in our 9.2$\deg$ model), and so never reaches the deeper secondary-peak region that we predict in our model.  

Our Case A main-peak location and peak electron density are similar to that of \citet{moore18}.  
Our H$_3$$\! ^+$ profiles are quite similar, too, and our Case A HCO$^+$ peak has a similar shape to theirs for the model in which 
CO is the mass 28 species, although our peak HCO$^+$ concentration is slightly smaller than theirs.  The details of the hydrocarbon ions are 
clearly different between our models. However, the most important difference between our models involves our Case A predicted dominance 
of HCNH$^+$ at the main peak and our Case A dominance of NH$_4$$\! ^+$ below the main peak, whereas \citet{moore18} expect H$_3$O$^+$ 
to dominate below the main peak.  \citet{moore18} do not include HCN as a ring-influx 
species, so their lack of HCNH$^+$ at the main peak is not a surprise. Similarly, although 
ammonia is included in the \citet{moore18} model, NH$_4$$\! ^+$ ions are not.  
Given the greater proton affinity of NH$_3$ in comparison to H$_2$O, the H$_3$O$^+$ 
ions in our Case A model drop off in favor of NH$_4$$\! ^+$ ions below the main peak, despite the greater influx rate of H$_2$O over NH$_3$ 
in Case A.  If NH$_3$ vapor is flowing in from the rings, then NH$_4$$\! ^+$ will likely be an important component of the lower ionosphere 
on Saturn, simply due to the high proton affinity of NH$_3$, and we would not then expect the H$_3$O$^+$ dominance predicted by \citet{moore18}.
However, if NH$_3$ and HCN are not flowing in as vapor, then our ion-chemistry results would end up similar to those of \citet{moore18}.

\citet{chadney22} also include an INMS-inspired influx of CH$_4$ in their ion-production model, and although they do not include other 
possible ring vapor molecules, they do show that the presence of inflowing CH$_4$ from the rings substantially affects the production 
rates of CH$_4$$\! ^+$, CH$_3$$\! ^+$, CH$_2$$\! ^+$, and CH$^+$ from methane photoionization and dissociative photoionization at altitudes above 
the methane CH$_4$ homopause.  \citet{chadney22} do not present modeled ion densities, so we cannot directly compare our chemistry and 
structure predictions with theirs.  However, Chadney et al.~do show the influence of high-resolution H$_2$ cross sections on the ion 
production rates.  As already mentioned in section~\ref{sec:modinputs}, our procedure that uses low-resolution cross sections and related 
approximations (including boosting CH$_4$$\! ^+$ production by a factor of 22 to partially atone for the low-resolution approximation) 
leads to a similar integrated CH$_4$$\! ^+$ production rate as Chadney et al., but the vertical distribution in our models is 
different from the high-resolution predictions \citep[e.g.,][]{kim14,chadney22}. Our peak CH$_4$$\! ^+$ production rate is slightly smaller 
and overall vertical distribution broader such that 
our models predict a greater CH$_4$$\! ^+$ production rate at higher altitudes than would be indicated with more realistic high-resolution 
H$_2$ cross sections.  Our use of low-resolution cross sections may affect the predicted structure of the ionosphere but is unlikely to 
affect the main conclusions with respect to the dominant ions, given the prevalence of ion-neutral reactions.  That is, regardless of 
the species that are originally ionized, ion-neutral reactions will act to quickly settle on a dominant ionized constituent.

The dominance of HCNH$^+$ at the main peak in our Case A model is consistent with the combined INMS and RPWS data analysis that points to a heavy 
ion being the dominant species at the main peak \citep{cravens19ioncomp}; it also satisfies the combined RPWS/LP and INMS analysis that 
suggests that the main-peak ion should have a recombination rate coefficient $\lesssim$3$\scinot-7.$ cm$^3$ s$^{-1}$ if the 
ion densities inferred by \citet{morooka19} are correct \citep{dreyer21} (but see \citealt{johansson22} for an alternate interpretation).  However, 
the H$^+$/H$_3$$\! ^+$ ratio predicted by the model at 3$\scinot-7.$ mbar, the deepest point in the trajectory of orbit 292, is much smaller than 
is observed \citep{waite18,moore18}, and the observed ion and electron densities at this pressure level are inconsistent with the INMS-inferred 
abundance of neutral species that are able to react with H$^+$ and maintain photochemical equilibrium \citep{vigren22}.  These issues are not 
limited to Case A and are discussed more completely in section~\ref{sec:iono}.

\subsubsection{Observational implications of Case A influx}\label{sec:caseaimp}
The fact that the Case A model greatly over-predicts the stratospheric column abundance of CO$_2$, CO, H$_2$O, C$_2$H$_6$, HCN, HC$_3$N (and probably 
CH$_3$CN), and under-predicts the C$_2$H$_2$ column abundance in comparison with previous assessments of mixing-ratio profiles that fit remote-sensing 
observations, suggests (1) that the large Case A ring-vapor influx rates cannot have been in operation over long time scales, (2) that horizontal 
spreading at altitudes between the INMS measurements 
and the methane homopause could be diluting the ring vapor species, and/or (3) that much of the material detected by INMS derives from solid 
particles impacting the instrument that would not normally ablate in Saturn's atmosphere.  Because Case A clearly does not represent a 
reasonable long-term average for the equatorial influx, we examine some of these other scenarios below.

We should also note that the column abundance for all the ions in the Case A model, as well as for the other cases discussed below, is relatively 
small --- for example, at best $\sim$1$\scinot11.$ cm$^{-2}$ for HCNH$^+$ in Case A at noon local time --- which could pose a challenge for the 
idea of using remote-sensing observations to distinguish between different ring-vapor inflow scenarios.  Stratospheric neutral species will 
likely provide the best clues to the long-term average ring influx rates.   

\subsection{Case B: Serigano orbit 293 influx, no horizontal spreading, only CH$_4$, CO, N$_2$ entering atmosphere in gas phase}\label{sec:caseb}

Given that the abundances of several neutral stratospheric species on Saturn are over-estimated with the Case A model, we next examine a situation in 
which only the most volatile species enter the atmosphere as vapor \citep[e.g.,][]{miller20}, hypothesizing that the other mass signatures 
from the INMS measurements result from impact vaporization of small dust particles that would not otherwise ablate in Saturn's atmosphere 
\citep[e.g., because the particles are too small or come in too slowly;][]{hamil18}.  This ``Case B'' scenario has the advantage that it reduces 
the variety of highly photochemically reactive species 
coming in from the rings that can affect neutral stratospheric chemistry; however, the volatile species that do come in --- CH$_4$, N$_2$, 
and CO --- still have extremely large influx rates, and the overall mass influx to the atmosphere from vapor plus dust \citep{waite18,perry18,serigano22} 
is still large from a ring-stability standpoint \citep[e.g.,][]{koskinen22chap}.  The Case B model flux boundary conditions for CH$_4$, N$_2$, 
and CO are kept the same as they were for Case A (see Table~\ref{tabinflux}), and all the other species are assumed to have a zero-flux upper boundary 
condition.  Although Enceladus plumes could conceivably also be contributing an additional source of water or other external material to Saturn's upper
atmosphere, as we assumed in the nominal models, we do not include this extra external source in the Case B model, so that we can better investigate 
the chemistry differences that occur when H$_2$O vapor is not explicitly flowing into the atmosphere.
As with Case A, our Case B is designed to represent the situation at 9.2$\deg$ planetocentric latitude, relevant to the \textit{Cassini} Final 
Plunge orbit 293.  Note that based on the \citet{serigano22} analysis, CO, N$_2$, and CH$_4$ together account for $\sim$65\% of the 
mass of the incoming ring material for orbit 293, and so this Case B scenario is still consistent with the \citet{miller20} upper limit for the 
gas-to-dust ratio of the incoming ring debris of $\sim$0.7--2.  

\begin{figure*}[!htb]
\vspace{-14pt}
\begin{tabular}{ll}
{\includegraphics[clip=t,scale=0.30]{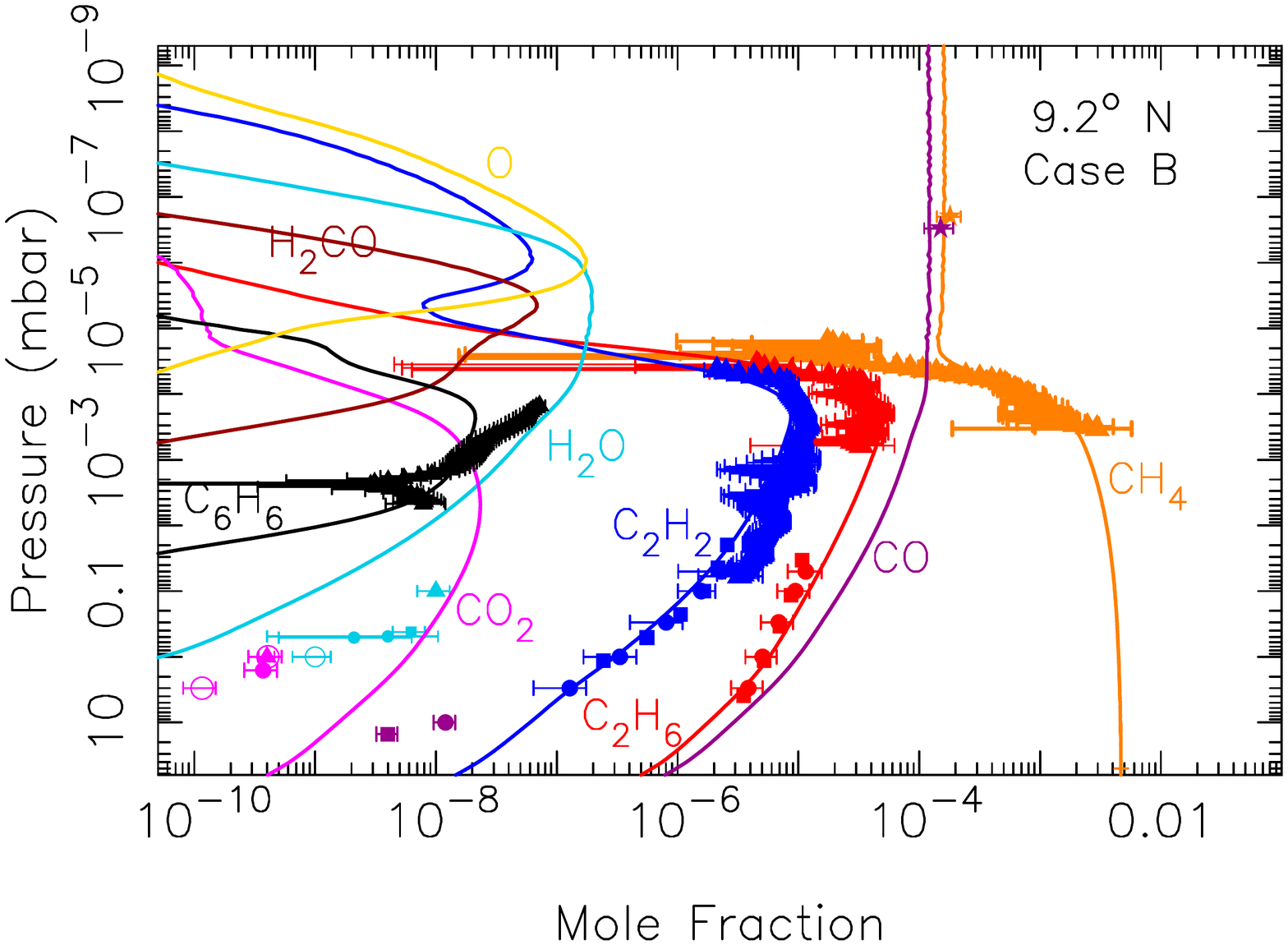}}
&
{\includegraphics[clip=t,scale=0.30]{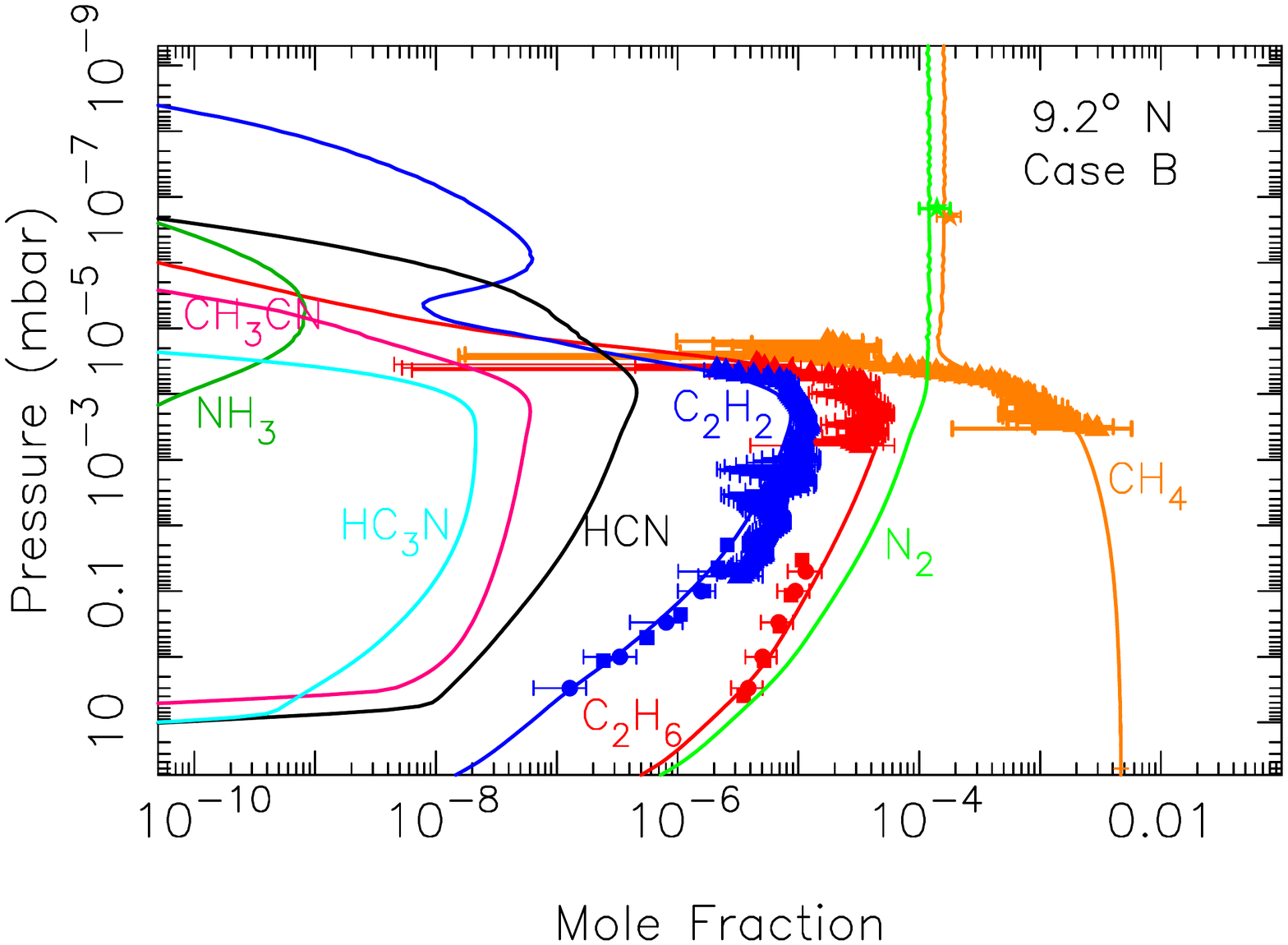}}
\end{tabular}
\caption{Mixing-ratio profiles for some neutral hydrocarbons, along with several key oxygen-bearing species (Left) and nitrogen-bearing species (Right) in 
our Case B model that adopts the INMS-derived, orbit 293, ring-vapor influx rates from \citet{serigano22} as upper boundary conditions, but assumes that 
only the most volatile species --- CH$_4$, CO, and N$_2$ --- are coming in as vapor.  All the other molecular species identified by INMS are presumed to 
derive from vaporization of dust grains that impact within the instrument.}
\label{figcaseb}
\end{figure*}

Results for key neutral stratospheric species in Case B are presented in Fig.~\ref{figcaseb}, which can be directly compared with the results 
from Case A in Fig.~\ref{figcasea} and the nominal no-ring-inflow model in Fig.~\ref{fignominal9n}.  When only CH$_4$, CO, and N$_2$ are 
flowing in through the top of the atmosphere, the stratospheric photochemistry results are different from Case A.  All three of these molecules 
are relatively stable photochemically in Saturn's stratosphere, so that both oxygen and nitrogen photochemical reactions are suppressed in this 
Case B.  

While CO can be photolyzed at high altitudes in the thermosphere, much of the interesting oxygen chemistry is actually initiated by 
ionospheric reactions.  For example, the incoming CO can accept a proton from CH$_5$$\! ^+$, H$_3$$\! ^+$, C$_2$H$_3$$\! ^+$, N$_2$H$^+$, and other 
ions to produce HCO$^+$ \citep{mcewan07,vuitton19}.  Electron recombination of HCO$^+$ predominantly recycles the CO, but some OH can form 
through the HCO$^+$ + electron $\rightarrow$ OH + C pathway \citep{geppert04}.  The OH then reacts with H$_2$ to form H$_2$O \citep{baulch05} 
and with CO to form CO$_2$ \citep{lissianski95,atkinson06,moses11}.  In that manner, H$_2$O and CO$_2$ are both photochemically produced 
and are predicted to be present in the stratosphere despite the absence of ring-vapor inflow of either species, albeit at reduced abundances 
compared to Case A.  The atomic O from CO photolysis and from another HCO$^+$ 
electron-recombination pathway that produces O + CH can react with CH$_3$ to produce H$_2$CO and eventually recycle the CO 
\citep[see][]{geppert04,atkinson92,preses00}.

Most of the oxygen species are greatly reduced in Case B as compared with Case A.  The stratospheric column abundances of CO and CO$_2$ 
in Case B, however, are still much greater than remote-sensing observations indicate (see Fig.~\ref{figcaseb}).  Case B predicts 
a column above 200 mbar of 5.5$\scinot19.$ CO molecules cm$^{-2}$, which is 400-800 times the CO column determined by \citet{moses00a} 
from the infrared observations of \citet{noll90}; the predicted CO mixing ratio at 15 mbar in Case B is 90-600 times that determined 
from (sub)millimeter observations \citep{cavalie09,cavalie10}.  Case B predicts a CO$_2$ column abundance above 10 mbar of 
9.5$\scinot15.$ cm$^{-2}$, which is 15 times the column abundance derived from global-average infrared observations \citep{moses00b} 
and 11 times the maximum equatorial value determined from \textit{Cassini}/CIRS observations \citep{abbas13}.  In contrast, the Case B 
H$_2$O stratospheric column abundance of 1.9$\scinot14.$ cm$^{-2}$ is a factor of 7 times smaller than is observed from globally averaged 
infrared observations \citep{moses00b} and a factor of $\sim$10 smaller than the equatorial column observed from \textit{Herschel} 
\citep{cavalie19}.  This under-estimate could be eliminated by including a source of H$_2$O from Enceladus plumes \citep{cavalie19} 

The only volatile nitrogen species flowing into the atmosphere in Case B is N$_2$, with all other nitrogen presumably tied up in more 
refractory grains.  The photochemical stability of N$_2$ reduces the importance of neutral nitrogen chemistry in the 
Case B model stratosphere in comparison to Case A.  The dominant column-integrated loss process for N$_2$ is the ion-neutral reaction H$_3$$\! ^+$ + N$_2$ 
$\rightarrow$ N$_2$H$^+$ + H$_2$ \citep{mcewan07}, but the loss processes for N$_2$H$^+$ mostly recycle the N$_2$.  One exception is the very minor 
recombination pathway N$_2$H$^+$ + electron $\rightarrow$ NH + N \citep{molek07,vigren12}.  Direct photolysis of N$_2$ is a more important source 
of atomic N, and the N so 
produced can react with CH$_3$ to form H$_2$CN and (directly, or eventually) HCN \citep{marston89a,marston89b,nesbitt90}.  Some CH$_3$CN is produced 
from CH$_2$CN via reactions of CH$_2$ with HCN and C$_2$H$_3$CN, as discussed for Case A in section~\ref{sec:caseanit}, and HC$_3$N is produced through 
CN + C$_2$H$_2$ $\rightarrow$ HC$_3$N + H, where the CN derives from HCN photolysis, as well as produced through recycling reactions involving C$_3$N.  
Condensation of HCN, CH$_3$CN, and HC$_3$N occurs under Case B conditions, limiting the stratospheric column abundances to 
3.2$\scinot16.$ HCN molecules cm$^{-2}$, 1.1$\scinot16.$ CH$_3$CN molecules cm$^{-2}$, and 3.2$\scinot15.$ HC$_3$N molecules cm$^{-2}$, 
which should be observable (see section~\ref{sec:obs}).  

The over-prediction of C$_2$H$_6$ and under-prediction of C$_2$H$_2$ that was seen with the Case A model has disappeared with the 
Case B model, and the stratospheric hydrocarbon abundances are more in line with what has been observed by CIRS 
\citep[e.g.,][]{fletcher19chap,fletcher21rev}.  

\begin{figure*}[!htb]
\vspace{-14pt}
\begin{center}
\includegraphics[clip=t,scale=0.40]{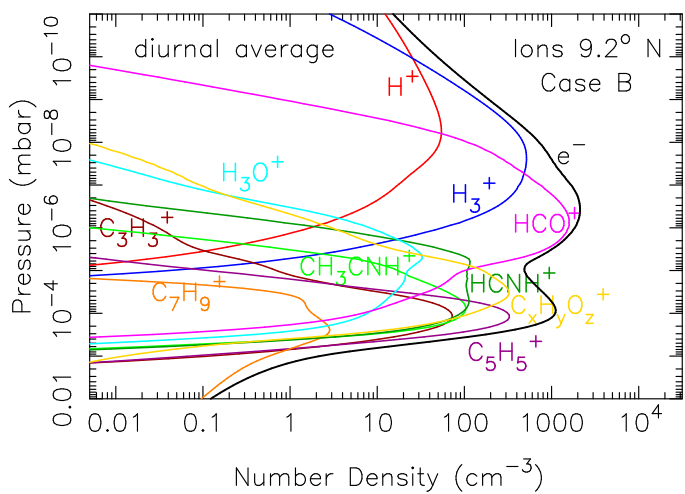}
\end{center}
\caption{Ion and electron number densities for our Case B model, assuming diurnally averaged transmission 
at 9.2$\deg$ planetocentric latitude.  Note that HCO$^+$ dominates at the main peak.}
\label{figioncaseb}
\end{figure*}

In terms of the ion composition with the Case B model, the nitrogen species no longer dominate the ionospheric chemistry now that 
HCN and NH$_3$ do not flow in from the top of the atmosphere (see Fig.~\ref{figioncaseb}).  The main ionospheric peak in Case B is dominated 
by HCO$^+$, which results from CO having the largest proton affinity in comparison to the other ring-inflow species (CH$_4$ and N$_2$) 
included in this model. As with Case A, complex hydrocarbon ions and HCNH$^+$ are still important in the secondary-peak region in the Case B 
model, but NH$_4$$\! ^+$ is no longer a major player in this region because of the low abundance of photochemically produced NH$_3$.  One 
additional major ion in the secondary peak region of the Case B model is ``C$_x$H$_y$O$_z$$\! ^+$,''  which is a catchall species we have introduced 
for heavy ions that contain C, H, and O but for which we no longer have the lab data to track their specific chemistry.  By far the largest production 
pathway for C$_x$H$_y$O$_z$$\! ^+$  in the Case B model is reaction of C$_2$H$_3$$\! ^+$ with CO to form a C$_2$H$_3$CO$^+$ adduct 
\citep{mcewan07}, where the rate coefficient for this reaction is taken from \citet{vuitton19}.  Unlike with the \citet{moore18} model, 
H$_3$O$^+$ ions are not dominant at any point below the main peak because photochemically produced H$_2$O has an abundance smaller than 
other key neutral constituents.  The dominance of HCO$^+$ at the main peak is consistent with joint analyses of INMS and RPWS/LP 
\citep{cravens19ioncomp,dreyer21}.

In all, the Case B model does a much better job of predicting neutral stratospheric 
abundances that stay within the bounds of what has been observed.  The main problems remain HCN, HC$_3$N, and the oxygen species, with a significant 
Case B over-prediction of CO, CO$_2$, HCN, and HC$_3$N, and an under-prediction of H$_2$O.  Adding a source of water from the Enceladus plumes 
\citep[e.g.,][]{cavalie19} to fix the water problem would just make the CO and CO$_2$ problem worse.  This CO and CO$_2$ over-prediction
could potentially be resolved by assuming that the mass 28 species seen by INMS is predominantly N$_2$, rather than CO, but then nitrogen 
species would be an even bigger component of Saturn's neutral stratosphere.  With the current Case B scenario, HCN and HC$_3$N would have been
readily detectable from \textit{Cassini} CIRS (see section~\ref{sec:obs}).

\subsection{Case C: Serigano orbit 293, complete horizontal spreading, all species gas phase}\label{sec:casec}

Because Case B does not completely resolve the remaining problems with the over-prediction of neutral stratospheric molecular abundances 
caused by the enormous influx of ring material detected by the INMS, we now examine a ``Case C'' in which the incoming ring vapor experiences 
efficient horizontal spreading across the whole globe at some altitude region between the lowest recorded measurements of INMS and the methane 
homopause near 10$^{-4}$ mbar.  This postulated global spreading would not only help reduce the amount of ring vapor flowing into the 
stratosphere to help limit stratospheric neutral abundances, but it could potentially also help explain the apparent roughly order-of-magnitude 
discrepancy between the large CH$_4$ mixing ratio determined by INMS in the high-altitude thermosphere, compared to smaller CH$_4$ mixing ratios 
determined at deeper levels just above the methane homopause from analyses of the UVIS occultation observations \citep[][see also 
Fig.~\ref{figcaseb}]{koskinen16,koskinen18,brown21}.  Our models do not predict efficient chemical loss of CH$_4$ in this intervening region, 
so horizontal spreading remains one possible explanation.

Our 1D photochemical model cannot handle horizontal spreading, so we instead simply reduce the INMS-derived influx rates at the top of the 
model to approximate the flux into the stratosphere that would result from spreading.  This model then does not accurately predict the 
ionospheric chemistry in the main-peak region, which is above the putative spreading region.  We therefore ignore the ionospheric results here 
and only report the neutral stratospheric results for Case C in this section.  
The INMS measurements from the Grand Finale orbits demonstrate that the low-latitude region over which the 
incoming ring material is concentrated encompasses at least $\pm$6$\deg$ in latitude, so allowing that material to spread over the whole surface 
area of the globe can at best reduce the flux into the stratosphere by a factor of $\sim$10.  Our Case C model therefore simply adopts 
the \citet{serigano22} orbit 293 fluxes that have been reduced by a factor of 10 (see Table~\ref{tabinflux}).  

\begin{figure*}[!htb]
\vspace{-14pt}
\begin{tabular}{ll}
{\includegraphics[clip=t,scale=0.30]{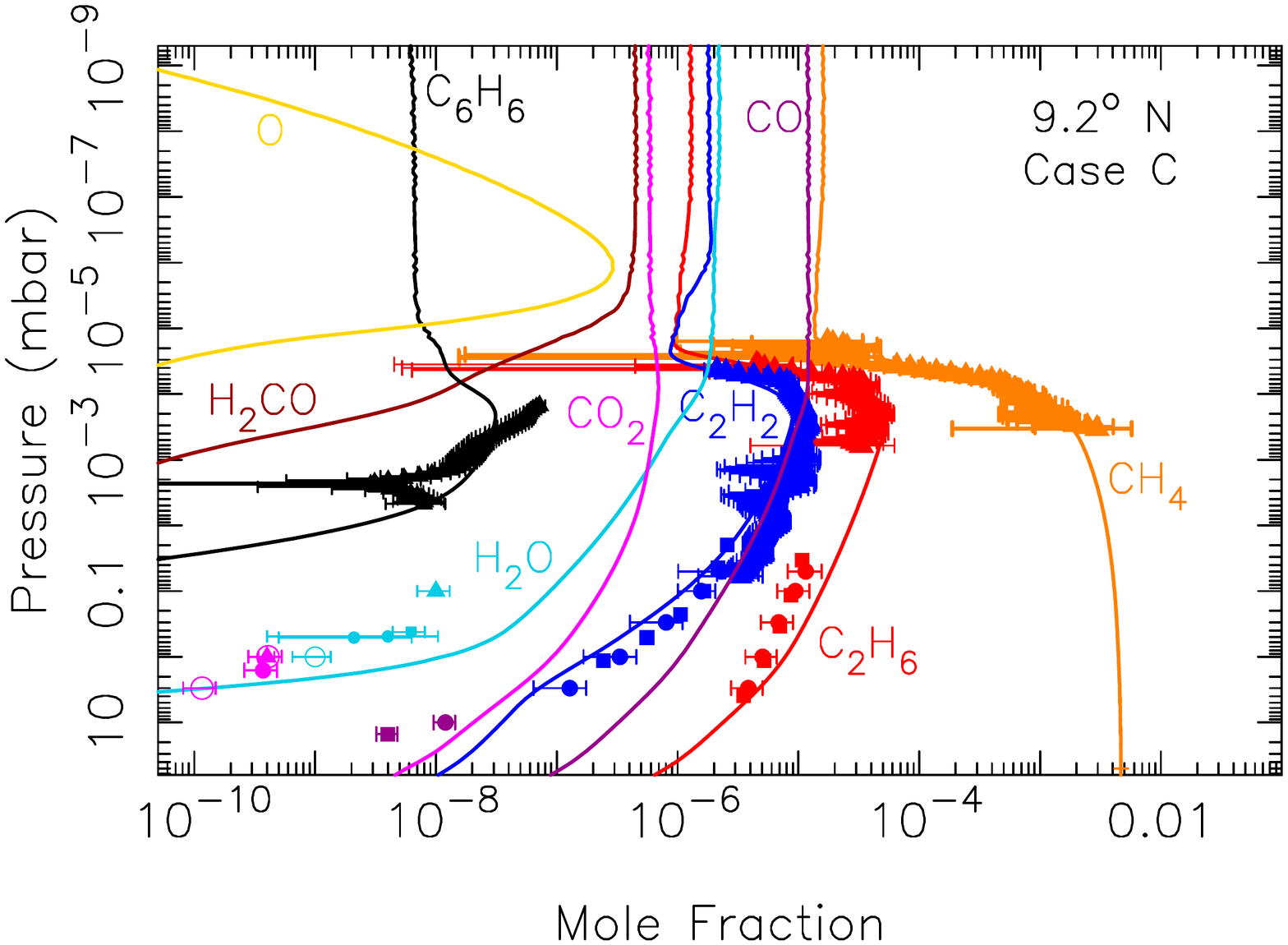}}
&
{\includegraphics[clip=t,scale=0.30]{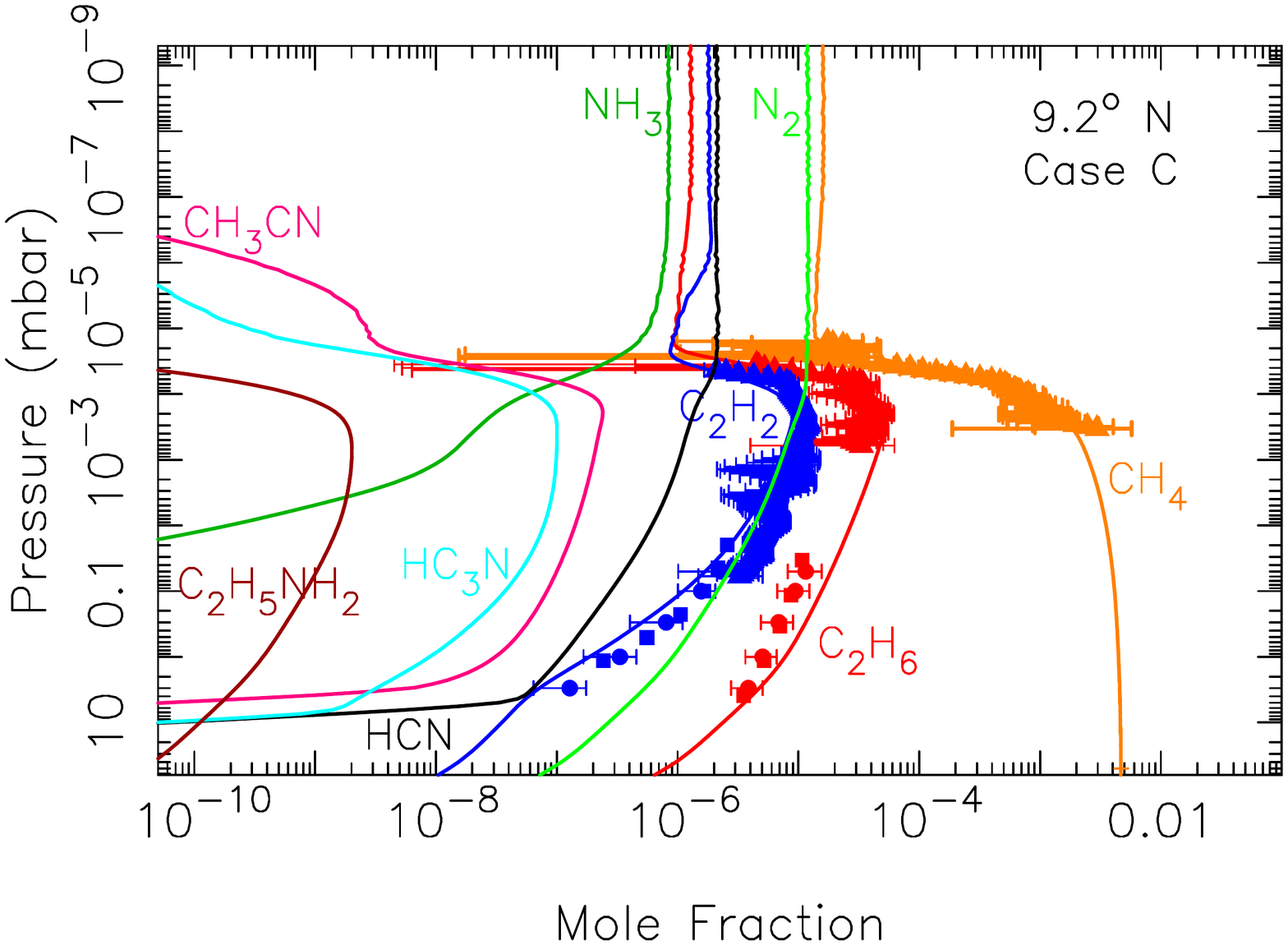}}
\end{tabular}
\caption{Mixing-ratio profiles for some neutral hydrocarbons, along with several key oxygen-bearing species (Left) and nitrogen-bearing species (Right) in 
our Case C model that adopts the INMS-derived, orbit 293, ring-vapor influx rates from \citet{serigano22} as upper boundary conditions, reduced by 
a factor of 10 to simulate complete global spreading before that vapor reaches the stratosphere.}
\label{figcasec}
\end{figure*}

The Case C results for neutral stratospheric species (see Fig.~\ref{figcasec}) are similar overall to Case A, except for a reduction in the abundances 
of many key species. The stable incoming N$_2$ and CO are still very abundant in the stratosphere, exceeding the column abundance of all the 
hydrocarbon photochemical products except C$_2$H$_6$.  The photochemistry of H$_2$O, CO$_2$, and other oxygen species is similar to Case A, as is 
the photochemistry of HCN, CH$_3$CN, HC$_3$N, but the oxygen and nitrogen photochemical products have reduced abundances.  Because less 
photochemically produced CH$_3$ and H are available, the Case C model over-prediction of C$_2$H$_6$ and under-prediction of C$_2$H$_2$ is not as 
severe as it was in Case A.

The stratospheric column abundances of CO$_2$, CO, and H$_2$O are still too high in comparison with observations.  With Case C, the CO$_2$ column 
abundance above 10 mbar is 1.4$\scinot17.$ cm$^{-2}$, which is $\sim$200 times greater than observations \citep{moses00a,abbas13}; the CO column 
abundance above 200 mbar is 6.3$\scinot18.$ cm$^{-2}$ and above 10 mbar is 1.8$\scinot18.$ cm$^{-2}$, which are $\sim$20--100 times greater than 
observations \citep{moses00b,cavalie09,cavalie10}; the stratospheric H$_2$O column abundance is 1.5$\scinot16.$ cm$^{-2}$, which is about an 
order of magnitude greater than observations \citep{moses00b,bergin00,fletcher12spire,cavalie19}.  The abundance of potentially observable 
neutral nitrogen species in the stratosphere is also still large.  With Case C, the stratospheric column abundance of HCN is 1.4$\scinot17.$ cm$^{-2}$, 
CH$_3$CN is 4.9$\scinot16.$ cm$^{-2}$, and HC$_3$N is 1.1$\scinot16.$ cm$^{-2}$.  These column abundances exceed those of minor hydrocarbon 
photochemical products that have already been observed on Saturn \citep[e.g.,][]{fouchet09,guerlet10}, and the Case C amount of HCN and HC$_3$N should be 
observable at infrared wavelengths (see section~\ref{sec:obs}).

\subsection{Case D: Serigano orbit 293, complete horizontal spreading, only CH$_4$, CO, N$_2$ gas phase}\label{sec:cased}

The stratospheric column abundances of CO, CO$_2$, and H$_2$O are still too high in Case C to be consistent with infrared and (sub)millimeter 
observations, so we examine a ``Case D'' where we not only assume complete global spreading of the ring material at some altitude above the methane 
homopause, but we also assume that only CH$_4$, CO, and N$_2$ are flowing in as vapor from the rings, whereas the other INMS mass signatures from 
less volatile species are caused by vaporization of solid ring particles as they impact within the instrument.  Case D is therefore similar to 
Case B, except that the influx rates of CH$_4$, CO, and N$_2$ have been reduced by an order of magnitude.

\begin{figure*}[!htb]
\vspace{-14pt}
\begin{tabular}{ll}
{\includegraphics[clip=t,scale=0.30]{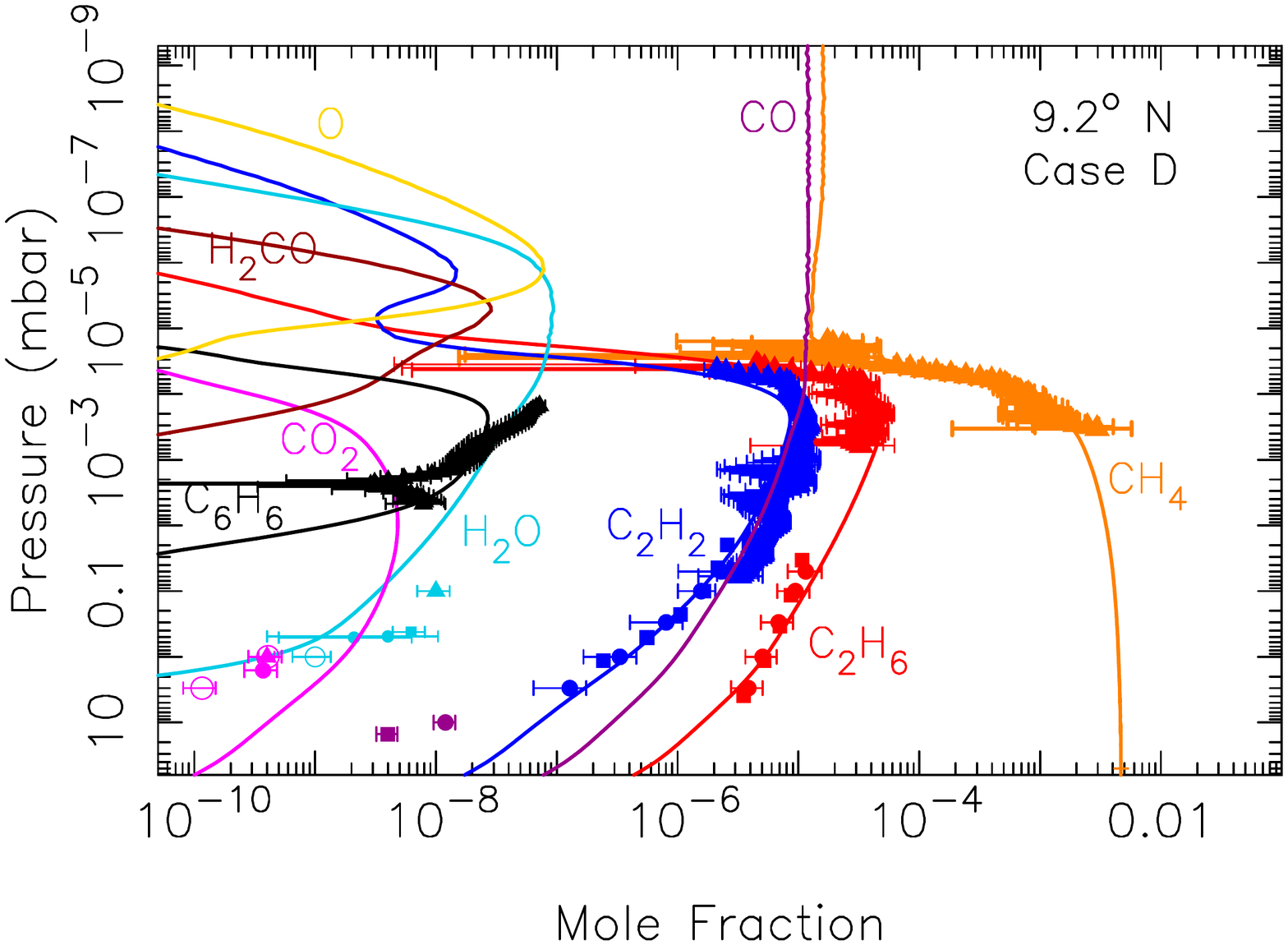}}
&
{\includegraphics[clip=t,scale=0.30]{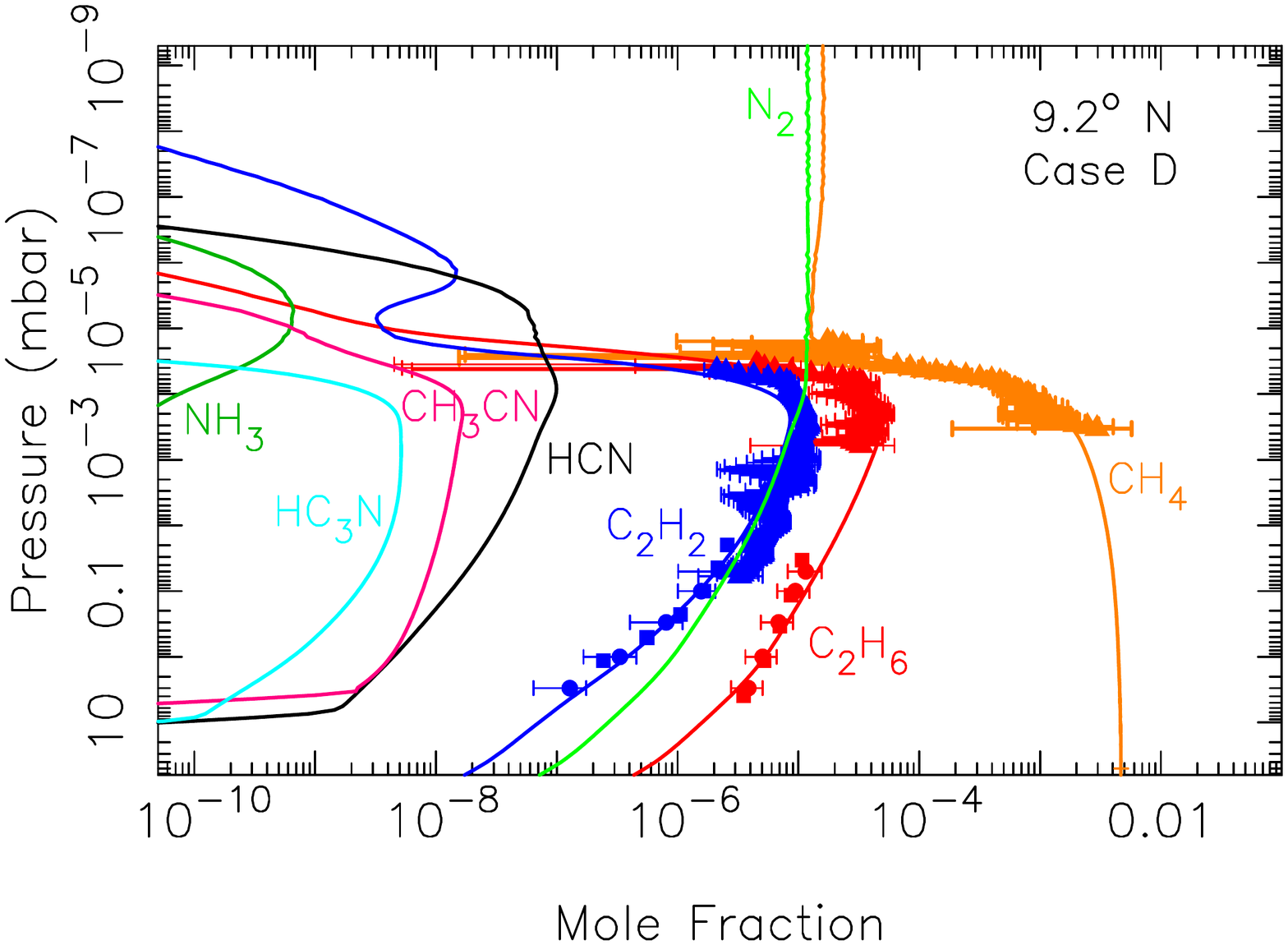}}
\end{tabular}
\caption{Mixing-ratio profiles for some neutral hydrocarbons, along with several key oxygen-bearing species (Left) and nitrogen-bearing species (Right) in 
our Case D model that assumes that only CH$_4$, CO, and N$_2$ flow in as vapor from the rings and that adopts the INMS-derived, orbit 293, ring-vapor 
influx rates from \citet{serigano22} for these species, but reduced by a factor of 10 to simulate complete global spreading before the ring vapor 
reaches the stratosphere.}
\label{figcased}
\end{figure*}

The Case D results for neutral stratospheric species (see Fig.~\ref{figcased}) are similar overall to Case B, except for a reduction in the abundances 
of many key species.  N$_2$ and CO are still photochemically stable species that end up very abundant in the stratosphere, exceeding the column abundance 
of all the hydrocarbon photochemical products except C$_2$H$_6$.  The photochemistry of oxygen and nitrogen species in Case D is qualitatively similar 
to Case B, except for the fact that the O- and N-bearing photochemical products have reduced abundances.  The inflowing species now have less effect on 
the hydrocarbon photochemical product abundances, which now remain consistent with the CIRS lower-stratospheric retrievals.  

The stratospheric column abundances of CO$_2$ and CO in Case D, however, are still greater than is observed on Saturn.  For example, the CO column abundance 
in the Case D model is 5.6$\scinot18.$ cm$^{-2}$ above 200 mbar and 1.6$\scinot18.$ cm$^{-2}$ above 10 mbar, which is $\sim$10--60 
times greater than observations \citep{moses00b,cavalie09,cavalie10}.  The Case D CO$_2$ column abundance of 2.6$\scinot15.$ cm$^{-2}$ above 10 mbar 
is $\sim$3--4 times the observed abundance \citep{moses00b,abbas13}.  On the other hand, the Case D stratospheric H$_2$O column abundance of 
5.6$\scinot14.$ cm$^{-2}$ is 2--4 times smaller than indicated by observations \citep{moses00b,cavalie19}.  With Case D, nitrogen species are notably
less abundant --- stratospheric column abundances are 6.7$\scinot15.$ HCN molecules cm$^{-2}$, 4.0$\scinot15.$ CH$_3$CN molecules
cm$^{-2}$, and 9.1$\scinot14.$ HC$_3$CN molecules cm$^{-2}$.  The observability of the nitrogen species under these conditions is discussed 
in section~\ref{sec:obs} --- both HCN and HC$_3$N have infrared signatures that should be detectable in CIRS spectra at Case D abundances.  
Therefore, even Case D with its lower influx rates of ring species has problems explaining the absence of observations of certain N- and O-bearing 
neutral species in Saturn's stratosphere.

\subsection{Case E: Serigano orbit 290--292, no horizontal spreading, all species gas phase}\label{sec:casee}

\begin{figure*}[!htb]
\vspace{-14pt}
\begin{tabular}{ll}
{\includegraphics[clip=t,scale=0.30]{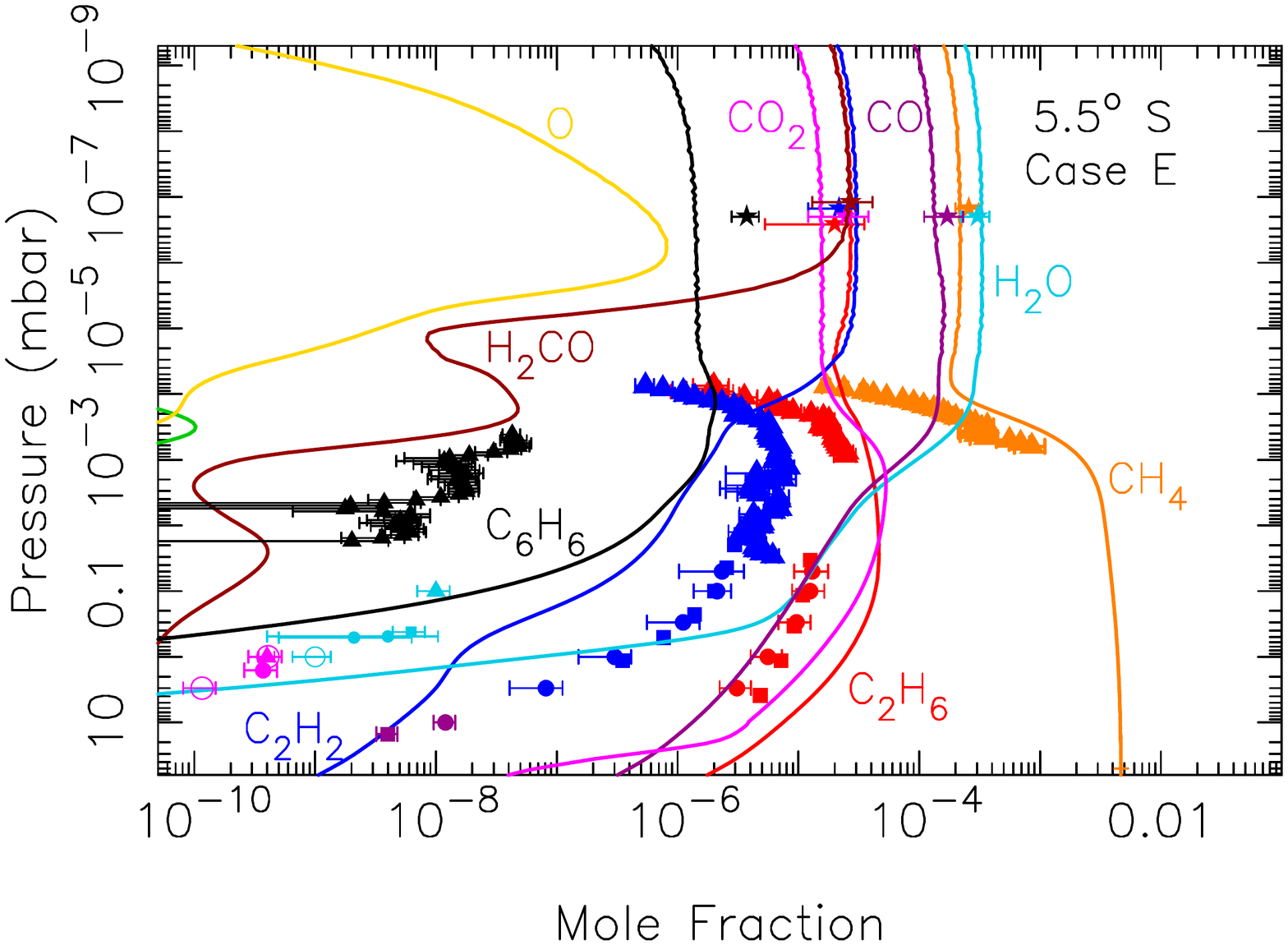}}
&
{\includegraphics[clip=t,scale=0.30]{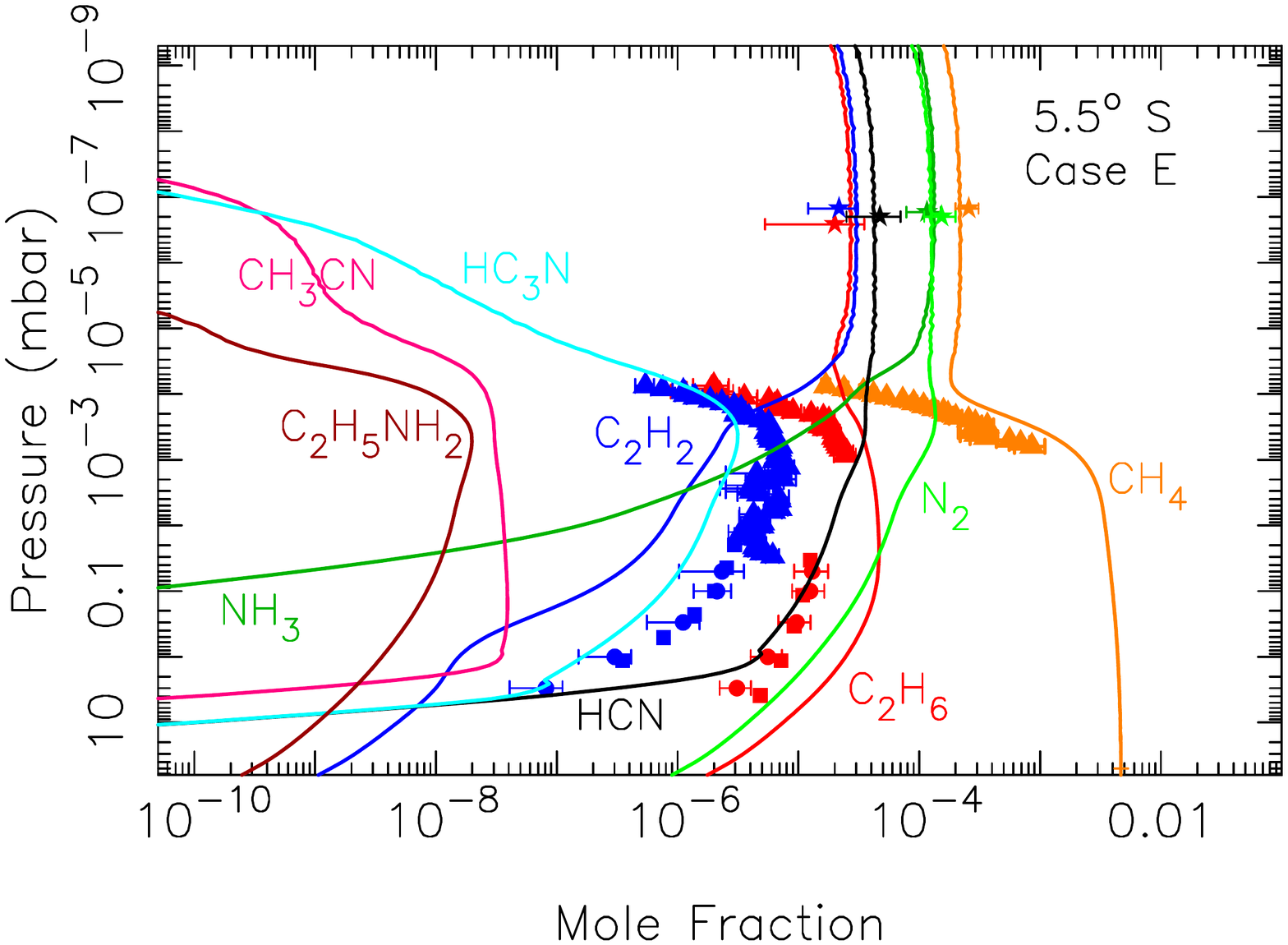}}
\end{tabular}
\caption{Mixing-ratio profiles for some neutral hydrocarbons, along with several key oxygen-bearing species (Left) and nitrogen-bearing species (Right) in 
our Case E model that considers the average orbit 290--292 abundances of incoming species from \citet[][]{serigano22}.  
The increase in ring-vapor abundances with decreasing altitude at the top of the model is simply a numerical boundary effect caused by rapid diffusion and 
the higher-pressure top boundary in this $-5.5\deg$ latitude model.}  
\label{figcasee}
\end{figure*}

All four previous Case A--D models were relevant to 9.2$\deg$ N planetocentric latitude, which was the terminal latitude of the INMS measurements 
from the Final Plunge orbit 293.  For completeness, we also examine models at 5.5$\deg$ S planetocentric latitude, near the closest approach latitude from 
orbits 288 and 290--292 (orbit 289 was not optimized for INMS and its analysis has been omitted).  The mixing ratios of the ring-derived molecules from 
orbits 290--292 were found to 
be higher, on average, than orbit 293, either because of the lower periapse latitude for these orbits or because of time variability in the 
source region.  Here we present the results from the higher-flux ``Case E'' model (Fig.~\ref{figcasee}), for which we adopt upper boundary conditions 
that represent an average of the derived fluxes from orbits 290--292 for the 10 most abundant incoming vapor species plus C$_6$H$_6$ in the INMS analysis 
of \citet{serigano22}, as listed in Table~\ref{tabinflux}.

The results for the oxygen species have qualitative similarities to those from Case A shown in Fig.~\ref{figcasea}, but the much higher influx rate of 
H$_2$O in Case E leads to a larger overall water abundance, as well as a much larger CO$_2$ abundance due to OH radicals from H$_2$O photolysis 
reacting with CO to form CO$_2$ + H.  Note that both the CO and H$_2$O mixing ratios are reduced in the middle and lower stratosphere by this process, with CO$_2$ 
eventually becoming the dominant oxygen-bearing constituent at pressures greater than a few microbars.  The column density of CO$_2$ above 10 mbar
is 2.4$\scinot19.$ cm$^{-2}$, which is a whopping 30,000--40,000 times greater than observations \citep{moses00b,abbas13}.  Water condensation, which is 
included in the model, helps ultimately remove H$_2$O in the lower stratosphere.  In fact, some H$_2$O condensation also occurs in the upper stratosphere 
in our Case E model; however, condensation is inefficient at these altitudes because the low atmospheric densities and relatively low encounter rates 
of the water molecules with existing condensation nuclei make condensation less effective in comparison with vertical transport, such that H$_2$O 
supersaturations can build up in the upper stratosphere.  The H$_2$O stratospheric column abundance in Case E is 1.6$\scinot18.$ cm$^{-2}$, which is 
three orders of magnitude greater than is observed \citep{moses00b,cavalie19}.  The CO column abundance in the Case E model is 2.4$\scinot19.$ cm$^{-2}$ 
above 200 mbar and 7$\scinot18.$ cm$^{-2}$ above 10 mbar, which is $\sim$100--300 times what is observed \citep{moses00b,cavalie09,cavalie10}.

The results for the nitrogen species are also qualitatively similar to Case A, except the large influx rates lead to greater abundances of nitrogen-bearing 
constituents in the stratosphere.  Although NH$_3$ is lost to some degree near the homopause, with a corresponding clear increase in N$_2$ and organo-nitrogen 
species in Fig.~\ref{figcasee} (see the chemistry discussion in section~\ref{sec:caseanit}), the NH$_3$ survives deeper into the stratosphere in Case E 
thanks in part to more efficient 
self shielding, but more importantly, to the enhanced H production due to the greater incoming abundance of H$_2$O and NH$_3$, which allows the ammonia to be 
more efficiently recycled once it is photolyzed.  Several nitriles are produced through HCN photolysis or through coupled nitrogen-hydrocarbon photochemistry, 
as discussed in section~\ref{sec:caseanit}.  The stratospheric column abundances of HCN, CH$_3$CN, and HC$_3$N in the Case E model are 3.6$\scinot18.$ 
HCN molecules cm$^{-2}$, 2.2$\scinot16.$ CH$_3$CN molecules cm$^{-2}$, and 1.8$\scinot17.$ HC$_3$N molecules cm$^{-2}$.  These species would be readily 
observed under Case E conditions (see section~\ref{sec:obs}).

\begin{figure*}[!htb]
\vspace{-14pt}
\begin{center}
\includegraphics[clip=t,scale=0.40]{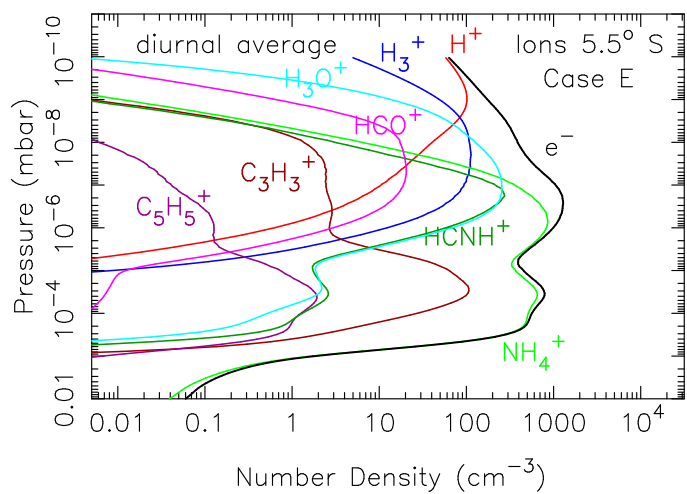}
\end{center}
\caption{Ion and electron number densities for our Case E model, assuming diurnally averaged transmission 
at $-5.5\deg$ planetocentric latitude.  Note that NH$_4$$\! ^+$ dominates at both the main peak and secondary peak.}
\label{figioncasee}
\end{figure*}

The stratospheric neutral hydrocarbons are strongly affected by the large influx rate of ring species in Case E.  The greater H$_2$O abundance in the thermosphere 
in Case E allows the atmosphere to become optically thick to Lyman alpha photons at relatively high altitudes near the methane homopause, which helps 
shield CH$_4$ from photolysis and causes the CH$_4$ mixing ratio near the homopause to be elevated over what it was in the nominal $-5.5\deg$ latitude model 
shown in Fig.~\ref{fignominal5s}.  The incoming hydrocarbons and HCN trigger a significant increase in CH$_3$ production, which augments the C$_2$H$_6$ 
abundance (see schemes (4) \& (5) above); the increase in atomic H from the photolysis of many of the ring-derived species helps convert unsaturated 
hydrocarbons into saturated ones, thus depleting the abundance of C$_2$H$_2$, C$_3$H$_4$, and C$_4$H$_2$ below what has been observed at these latitudes 
\citep[e.g.,][]{guerlet09,guerlet10,sinclair13,fletcher10,fletcher19chap,fletcher21rev}.  The column abundance of C$_6$H$_6$ is also greatly enhanced as a 
result of a larger derived C$_6$H$_6$ influx rate for orbits 290--292, of efficient ionospheric and upper-stratospheric recycling, and of enhanced stratospheric 
production resulting from the enhanced abundances of radicals such as C and CH$_3$.  The abundance of C$_6$H$_6$ in Case E greatly exceeds the observed 
upper-stratospheric abundances from the UVIS occultation observations \citep{brown21}.  PAHs (not shown in Fig.~\ref{figcasee}) are also generally 
enhanced with Case E.  In all, the predicted abundances of stratospheric neutral species in the Case E model are significantly in conflict with 
abundances retrieved from CIRS spectra.

The ionospheric results for Case E are shown in Fig.~\ref{figioncasee}.  Because of the large influx rate of NH$_3$ in the Case E model, and 
the large proton affinity of NH$_3$, the dominant ion becomes NH$_4$$\! ^+$ at both the main ionospheric peak and deeper secondary peak.  The 
higher influx rate of neutral ring vapor in Case E leads to a smaller main-peak electron density than in Cases A or B, resulting in a slightly 
worse fit to the majority of the electron-density measurements (see section \ref{sec:iono}).

\subsection{Case F: Miller orbit 290-292, no horizontal spreading}\label{sec:casef}

Different groups adopt different assumptions and analysis procedures when fitting the INMS mass spectra \citep[cf.][]{miller20,serigano22}, 
resulting in different conclusions with regard to the incoming ring vapor composition and influx rates.  These differences may affect the 
subsequent stratospheric chemistry and composition.  We therefore examine an additional ``Case F'' model that adopts upper boundary fluxes (see 
Table~\ref{tabinflux}) that allow our $-5.5\deg$ latitude model to reproduce the orbit 290--292 average mixing ratios for the dominant ring species 
determined from the auto-fit procedures of (\citealt{miller20}; see their Table~2).  Several of the same incoming ring molecules are included in 
this model as compared to the previous \citet{serigano22}-based Case A, C, \& E models (i.e., CO, H$_2$O, N$_2$, CH$_4$, NH$_3$, HCN, and H$_2$CO); 
however, \citet{miller20} do not include CO$_2$, C$_2$H$_6$, C$_2$H$_2$, or C$_6$H$_6$ in the list of dominant species in their Table 2, but they do
include NO, which was not in the \citet{serigano22} list of top species.  The inferred influx rates for the molecules the two groups have in 
common also differ for orbits 290--292 (cf.~Case E \& F in Table~\ref{tabinflux}), albeit not by large factors.  Comparisons between Case E and F 
predictions may therefore provide a useful illustration of ``systematic'' uncertainties related to different analysis assumptions and procedures.  

\begin{figure*}[!htb]
\vspace{-14pt}
\begin{tabular}{ll}
{\includegraphics[clip=t,scale=0.30]{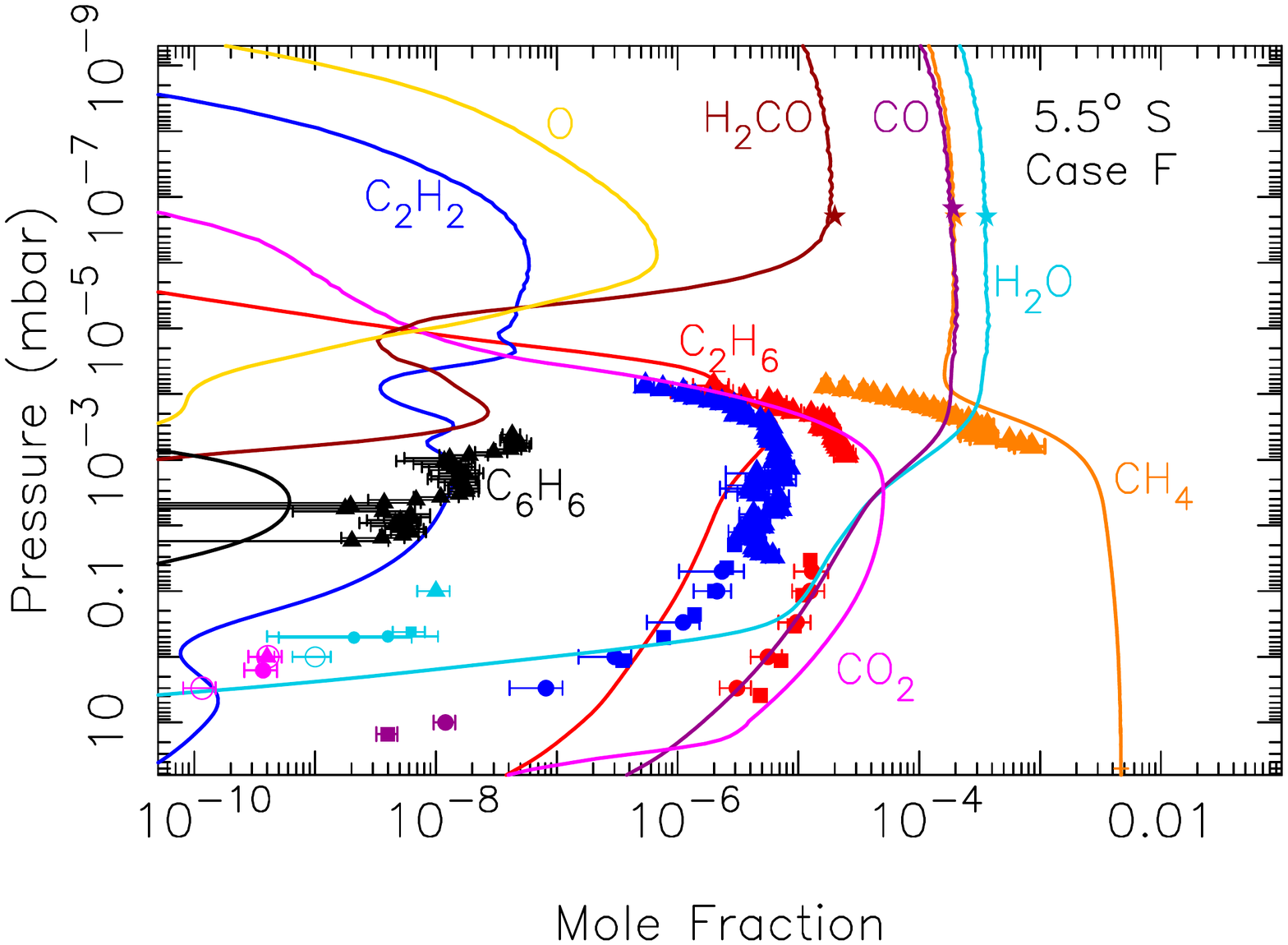}}
&
{\includegraphics[clip=t,scale=0.30]{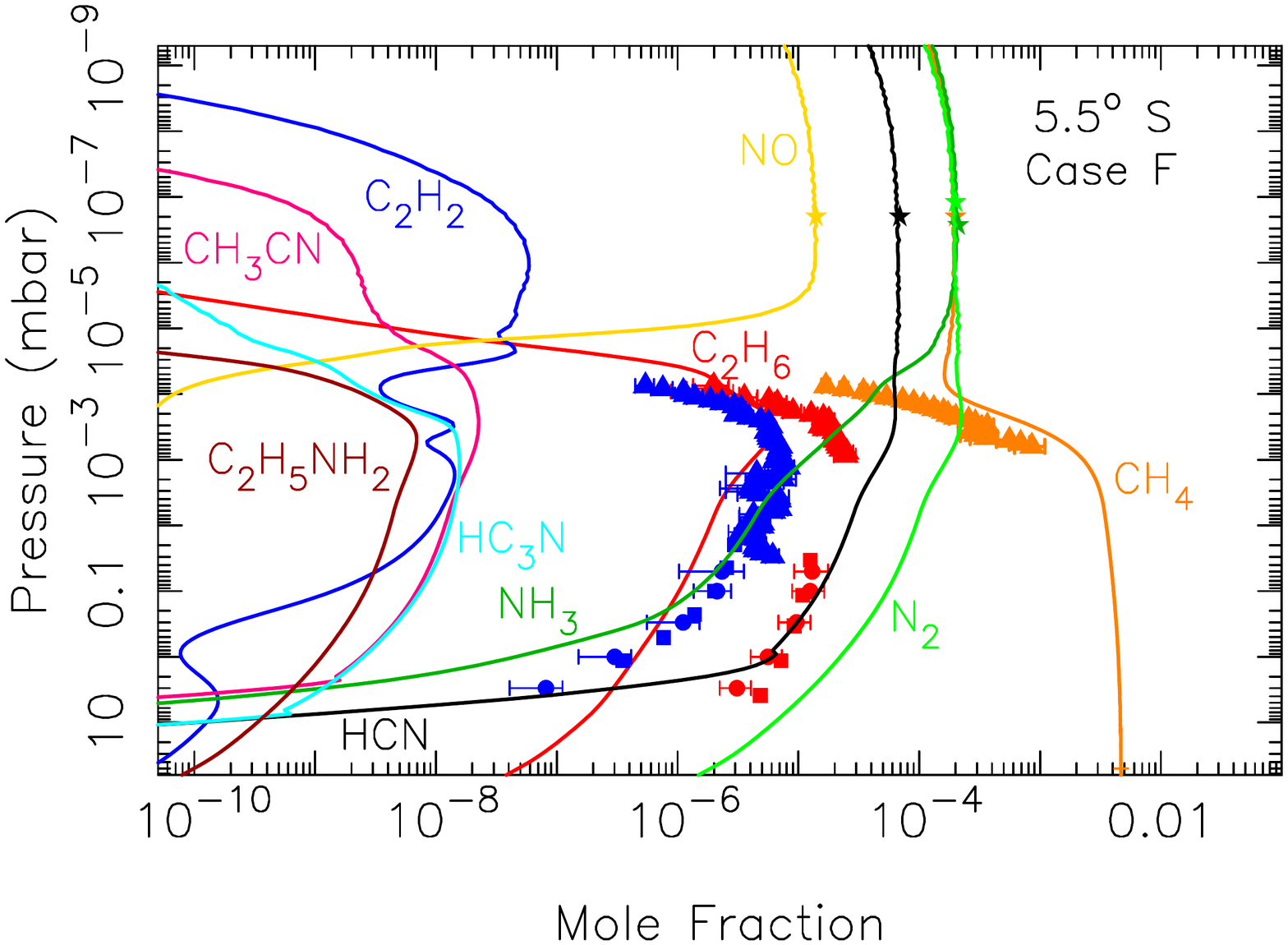}}
\end{tabular}
\caption{Mixing-ratio profiles for some neutral hydrocarbons, along with several key oxygen-bearing species (Left) and nitrogen-bearing species (Right) in 
our Case F model that considers the average orbit 290--292 abundances of incoming species from the auto-fit procedures of \citet[][;their Table 2]{miller20}.}  
\label{figcasef}
\end{figure*}

Some neutral stratospheric results for Case F are shown in Fig.~\ref{figcasef}.  Despite seemingly minor differences in the influx rates of the 
incoming ring species in common between Cases E \& F, the overall results are surprisingly different in terms of the stratospheric neutral 
composition.  First, we point out what the two models have in common.  The vertical profiles of H$_2$O, CH$_4$, CO, N$_2$, and HCN are similar in 
Cases E \& F, with differences being clearly related to differences in the assumed incoming ring-vapor flux.  The profile of CO$_2$ below the methane 
homopause is also very similar in the two models, despite no CO$_2$ coming in from the rings in Case F: the coupled H$_2$O and CO photochemistry in the 
stratosphere efficiently accounts for the CO$_2$ production in Case F.  

The Case F results for hydrocarbons, however, are significantly different from those for Case E.  The large incoming flux and thermospheric column 
abundance of H$_2$O help shield upper-stratospheric CH$_4$ from photolysis by Lyman photons in both models.  The main difference in the hydrocarbon 
results is due to the fact that heavier hydrocarbons (especially C$_2$H$_2$) are included as ring-vapor species in the \citet{serigano22} model 
(Case E) but not in the \citet{miller20} model (Case F).  The lack of these hydrocarbon in Case F throttles efficient production of CH$_3$ from 
schemes (4) and (5), removing an efficient C$_2$H$_6$ production pathway that is found in Case E.  Moreover, with atomic H being so abundant from 
H$_2$O and NH$_3$ photolysis, the few CH$_3$ and other hydrocarbon radicals that are produced in Case F are more likely to recombine with H to recycle 
methane than to produce C$_2$H$_6$ and other hydrocarbons.  Without these stratospheric hydrocarbon photochemical products in Case F, the coupled 
NH$_3$-hydrocarbon photochemistry is weaker, and fewer CH$_3$CN, HC$_3$N, and other nitriles and organo-nitrogen species are produced.  The 
NH$_3$ in Case F therefore survives to deeper altitudes than in Case E.  

\begin{figure*}[!htb]
\vspace{-14pt}
\begin{center}
\includegraphics[clip=t,scale=0.40]{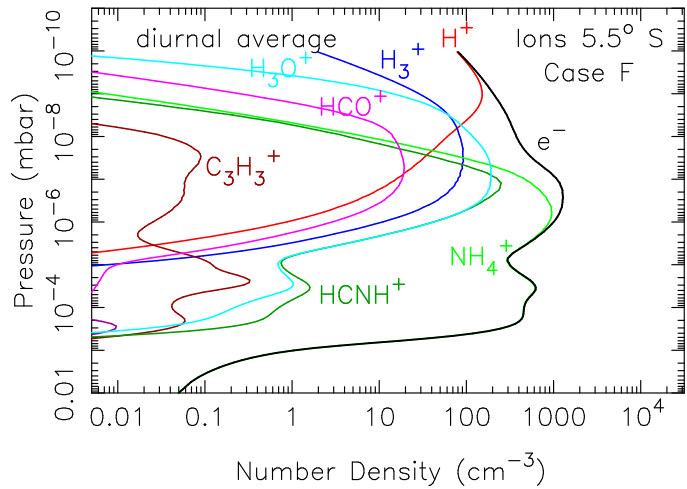}
\end{center}
\caption{Ion and electron number densities for our Case F model, assuming diurnally averaged transmission 
at $-5.5\deg$ planetocentric latitude.  Note that NH$_4$$\! ^+$ dominates at both the main peak and secondary peak, and hydrocarbon 
ions are not prevalent.}
\label{figioncasef}
\end{figure*}

The H$_2$O stratospheric column abundance in Case F is 1.9$\scinot18.$ cm$^{-2}$, which as with Case E, is three orders of magnitude greater than is 
observed \citep{moses00b,cavalie19}.  The CO column abundance in Case F is 2.8$\scinot19.$ cm$^{-2}$ above 200 mbar and 9$\scinot18.$ cm$^{-2}$ 
above 10 mbar, which again is $\sim$100--400 times what is observed \citep{moses00b,cavalie09,cavalie10}.  The column density of CO$_2$ above 10 mbar
is 2.4$\scinot19.$ cm$^{-2}$, which again is 30,000--40,000 times greater than observations \citep{moses00b,abbas13}.  
For the nitrogen species, the stratospheric column abundances of HCN, CH$_3$CN, and HC$_3$N 
in Case F are 4.7$\scinot18.$ HCN molecules cm$^{-2}$, 3.5$\scinot15.$ CH$_3$CN molecules cm$^{-2}$, and 3.9$\scinot15.$ HC$_3$N molecules cm$^{-2}$.  
This much HCN and HC$_3$N, at least, should be observable (see section~\ref{sec:obs}).  Unlike the other models, the stratospheric NH$_3$ abundance 
in Case F builds up to an observable column abundance of 1.8$\scinot17.$ cm$^{-2}$, such that emission line cores should be observed in the infrared.  
The photochemically produced hydrocarbons in Case F, on the other hand, are significantly depleted compared to the other models, and in fact are very 
inconsistent with observations from \textit{Cassini} CIRS.

The ionospheric results for Case F are shown in Fig.~\ref{figioncasef}.  Because of the lack of incoming hydrocarbons in the ring vapor, as well 
as the reduced hydrocarbon photochemical production below the methane homopause, Case F contains fewer hydrocarbon ions, and the 
ionospheric chemistry is less complex than in the other models.  The large influx rate and large proton affinity of NH$_3$ again lead to
NH$_4$$\! ^+$ becoming the dominant ion at both the main ionospheric peak and the deeper secondary peak.   

\subsection{Time-variable model}\label{sec:timevar}

The differences in the predicted stratospheric composition from Cases A--F are large and demonstrate that the assumptions about 
the incoming ring vapor can significantly affect the results.  However, one conclusion common to all cases is a very significant model 
over-prediction of the stratospheric abundance of the oxygen-bearing constituents CO and CO$_2$ and the nitrogen-bearing species HCN and 
HC$_3$N in comparison to observations.  Many of our 
ring-influx models also predict too much H$_2$O compared to observations, and the ring species often wreak havoc with the 
hydrocarbon abundances in ways that are not observed during the \textit{Cassini} era or at any other time.  This 
model-data mismatch suggests that the ring vapor could not have not penetrated very far into the stratosphere by the end of 
the \textit{Cassini} mission, or CIRS would have seen evidence for the oxygen-bearing species and HCN and HC$_3$N, if not additional nitrogen species 
and hydrocarbon perturbations.  On the other hand, 
the INMS measurements produce clear signals of heavy molecules in Saturn's thermosphere that do not appear to be an instrumental artifact 
\citep{waite18,miller20,serigano22}, and the INMS-inferred ring vapor influx rates are consistent to first order with ion and electron 
densities inferred for the ionosphere \citep[e.g., see the models above and section~\ref{sec:iono}, as well 
as][]{moore18,cravens19ioncomp,persoon19,hadid19,morooka19,dreyer21,vigren22}.  
One possibility that could potentially reconcile the stratospheric results with the thermospheric results is if the inferred ring vapor 
during the Grand Finale were the result of a recent disruptive event in the rings that generated a recent, transient inflow of dust and 
vapor.  How ``recent'' does this event have to be to prevent observable consequences to the stratosphere by the end of the \textit{Cassini} mission 
in 2017?

To answer this question, we have constructed a time-variable model that starts with the nominal 9.2$\deg$ latitude model 
results (i.e., no ring influx) and then introduces a constant flux of ring vapor from Case A (see Table~1) at the top of the model.  Rather than 
simply evolving the model to a long-term steady state, we track the intervening variation in atmospheric composition with time as the ring 
vapor diffuses down through the atmosphere, chemically reacting with the background atmosphere and other ring species as it flows
downward.  Vertical diffusion time scales are short in the thermosphere but increase with increasing 
atmospheric density.  Model time steps are correspondingly variable, being a small fraction of a second initially but grow with 
increasing run time as the ring vapor diffuses through the atmosphere; the choice of time-step size does not significantly affect the results. 
The model again is one dimensional and does not consider possible horizontal spreading of the incoming ring 
material.  Figure~\ref{figtimevara} shows the results from this time-variable model at specific times after the ring influx 
is triggered.

Figure~\ref{figtimevara} shows that it takes less than a day for the ring vapor species to diffuse to $\sim$1 nbar, where the vapor species are 
observed by INMS.  At this point, the CH$_4$ mixing ratio near the homopause is fully consistent with the UVIS occultation retrievals of \citet{brown21}, 
and the ring vapor is confined to the thermosphere.  By the 4-month mark, the ring vapor has made it past the CH$_4$ homopause level and into the upper 
stratosphere.  As we will show in section~\ref{sec:obs}, evidence for some neutral ring species should already appear in mid-infrared spectra of low 
latitudes after 4 months.  After 3 years, the 0.1--0.01 mbar region has been affected by the ring vapor, and observational consequences are readily 
apparent.  By 10 years, much of the vapor has penetrated past millibar pressure levels, with very strong emission signatures in infrared spectra.
The fact that infrared signatures of HCN, HC$_3$N, and excess CO$_2$ are not seen in CIRS spectra from 2017 severely limits the age of the ring-perturbing 
event, as we discuss more fully in section~\ref{sec:event}.

\begin{figure*}[!htb]
\vspace{-14pt}
\begin{center}
\includegraphics[clip=t,scale=0.60]{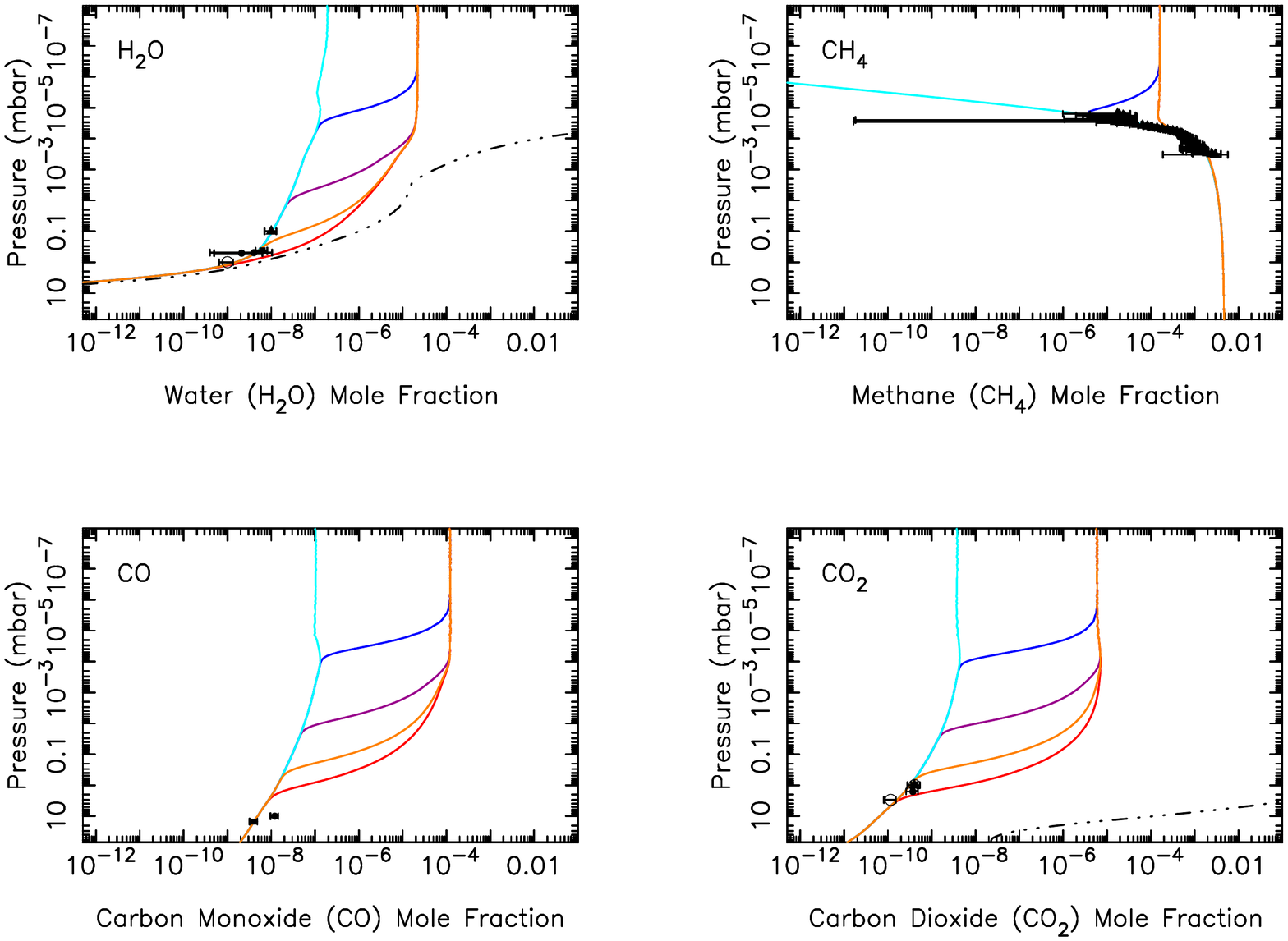}
\end{center}
\noindent{\textit{(continued).}}
\end{figure*}

\clearpage

\begin{figure*}[!htb]
\vspace{-14pt}
\begin{center}
\includegraphics[clip=t,scale=0.60]{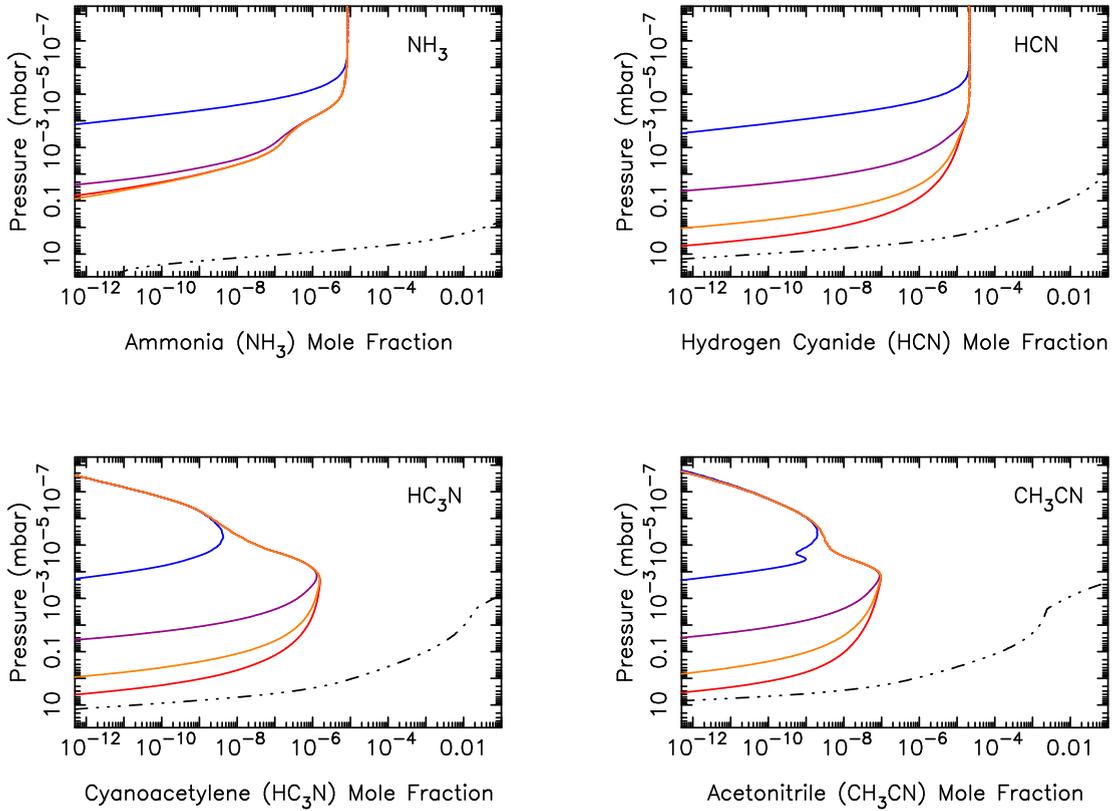}
\end{center}
\caption{The mixing-ratio profiles of several species (as labeled) in a time-variable model that starts with the nominal no-ring-influx 9.2$\deg$ N 
model results as an initial condition but then adopts a steady Case A ring-vapor flux upper boundary condition starting at time zero.  The colored 
curves show the initial state of the atmosphere (light blue/cyan), and the results after 1 day (darker blue), 4 months (purple), 3 years (orange) 
and 10 years (red).  The ring vapor species flow quickly through the thermosphere and upper stratosphere but more slowly deeper in the stratosphere. 
The dash-triple-dot lines show the saturation vapor pressure curves for the molecules in question.}
\label{figtimevara}
\end{figure*}

We also ran a time-variable model starting with the steady-state Case A solution but then terminating the 
ring-vapor inflow at the upper boundary (but keeping a smaller background influx of H$_2$O, CO, and CO$_2$ from interplanetary dust and Enceladus, 
as in the nominal model with no ring influx).  We determined that excess nitrogen species, CO, and CO$_2$ still 
remained in the stratosphere in observable quantities after more than a thousand years, indicating that it takes $\gtrsim$4000 years for the stratosphere to 
fully ``reset'' after a significant ring-influx event ceases to actively feed vapor to Saturn.  If the large ring-inflow rates were instead short-lived 
and not in steady state to begin with, then the time it would take to remove the evidence for this inflow would be correspondingly reduced and would 
depend on the total mass of material delivered to Saturn. 

\section{Discussion}\label{sec:discussion}

Comparisons of our model results with observations allow us to identify certain consequences of the ring-atmosphere interactions on Saturn that 
have implications for the planet's current ionospheric chemistry and structure, neutral stratospheric composition, potential temporal 
variability, timing of any ring-influx event, and the exact nature of the ring interaction.  We now discuss these implications.

\subsection{Implications for the ionosphere}\label{sec:iono}

\begin{figure*}[!htb]
\vspace{-14pt}
\begin{tabular}{ll}
{\includegraphics[clip=t,scale=0.30]{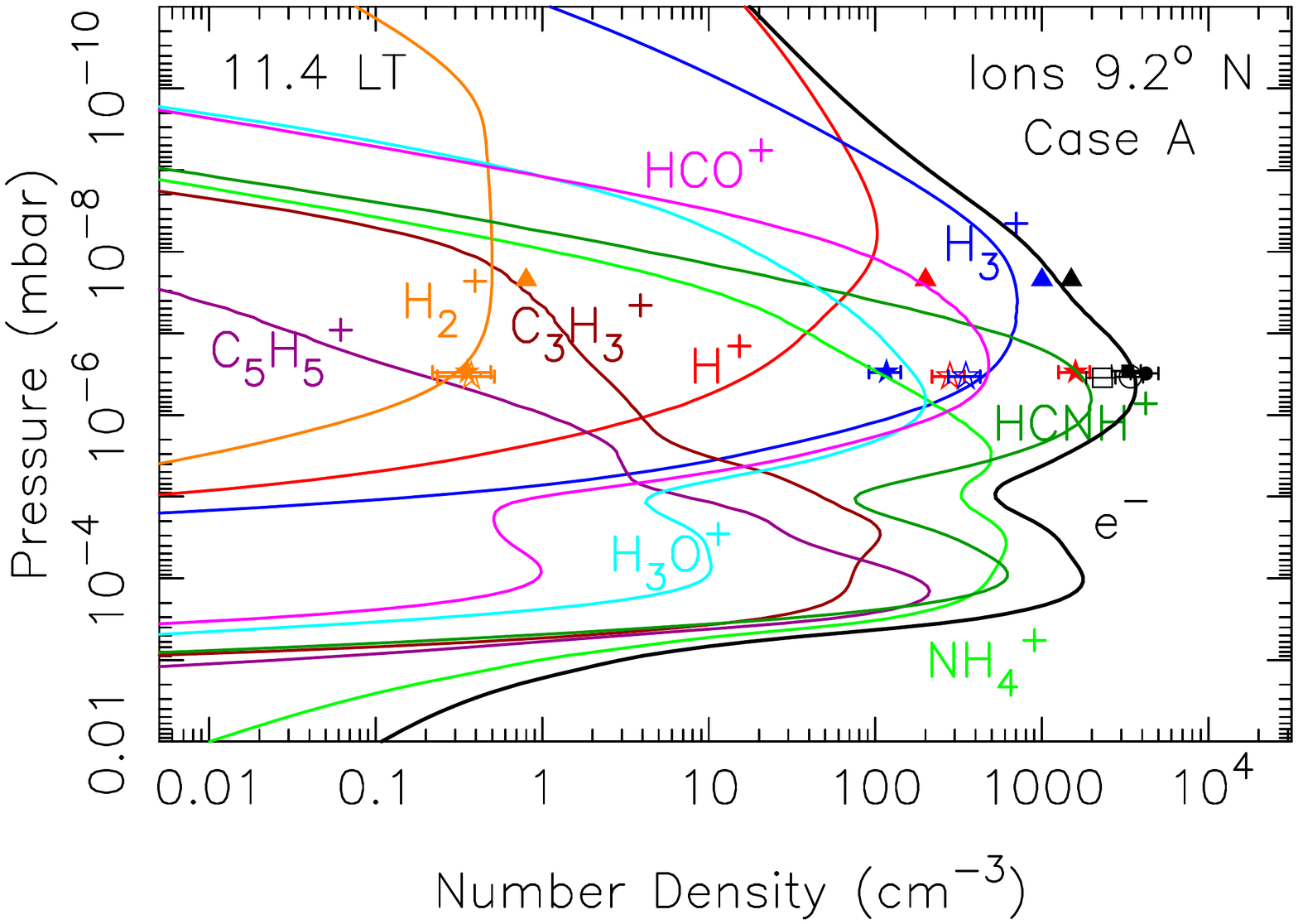}}
&
{\includegraphics[clip=t,scale=0.30]{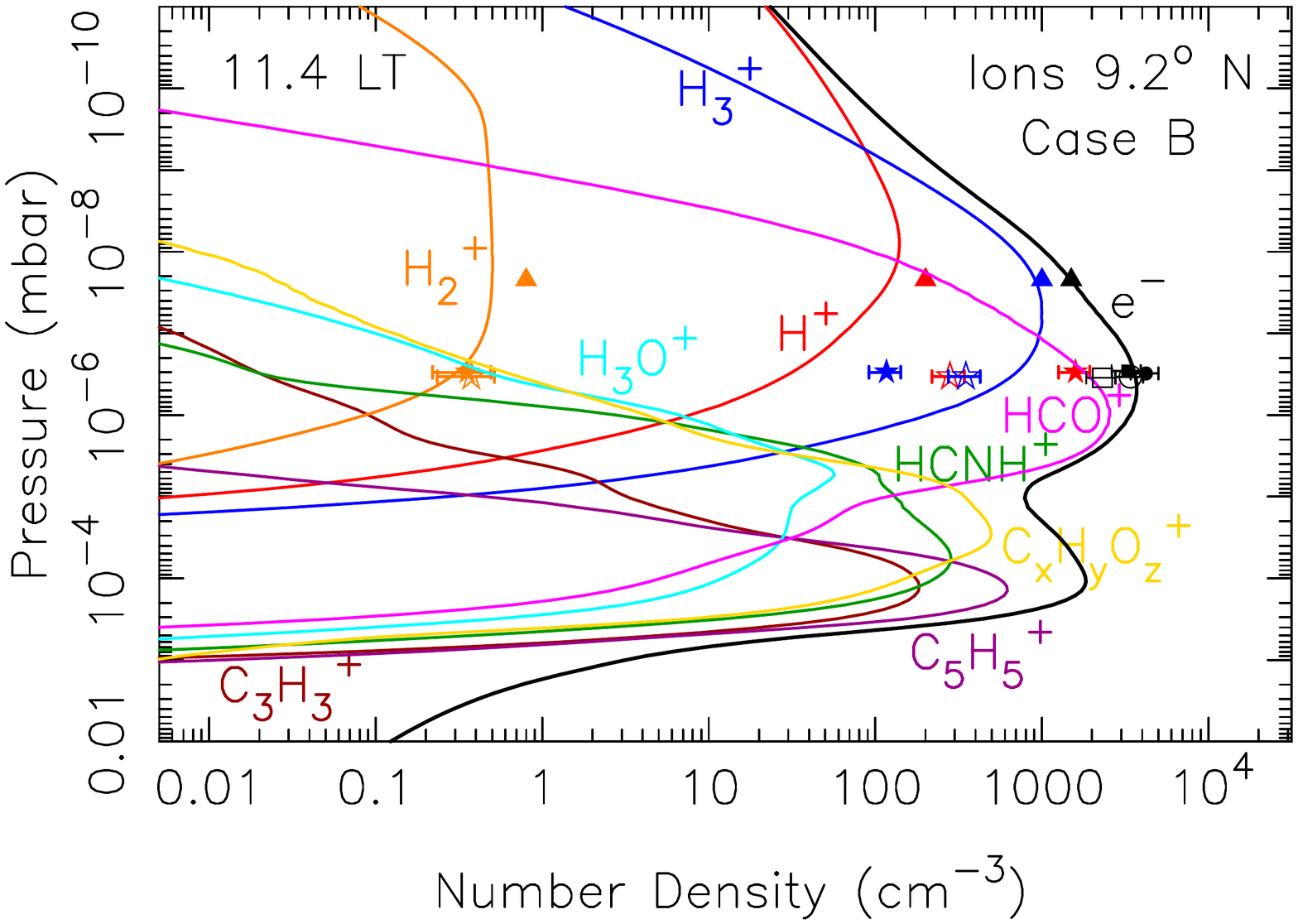}} \\
\vspace{-14pt}
{\includegraphics[clip=t,scale=0.30]{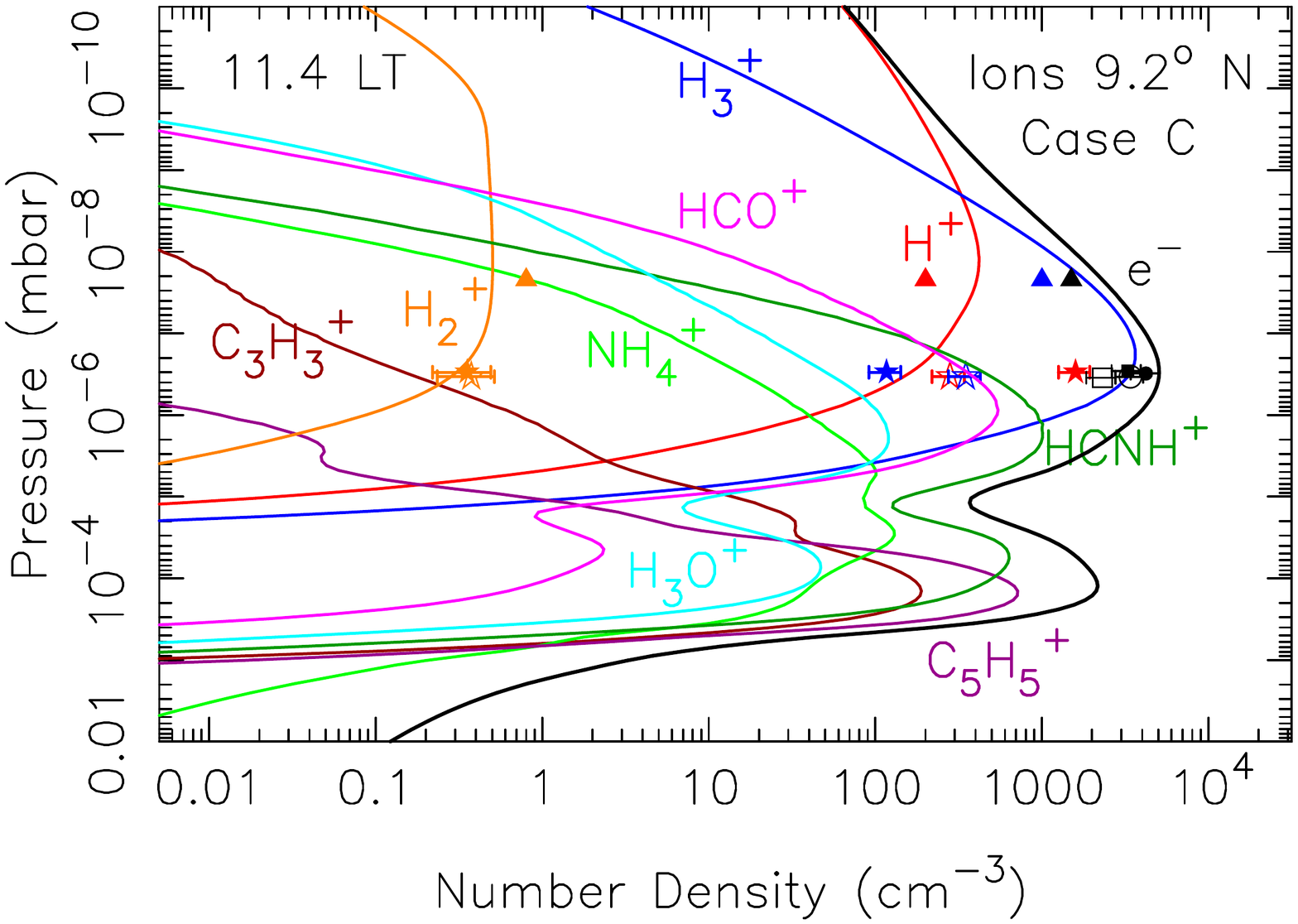}}
&
{\includegraphics[clip=t,scale=0.30]{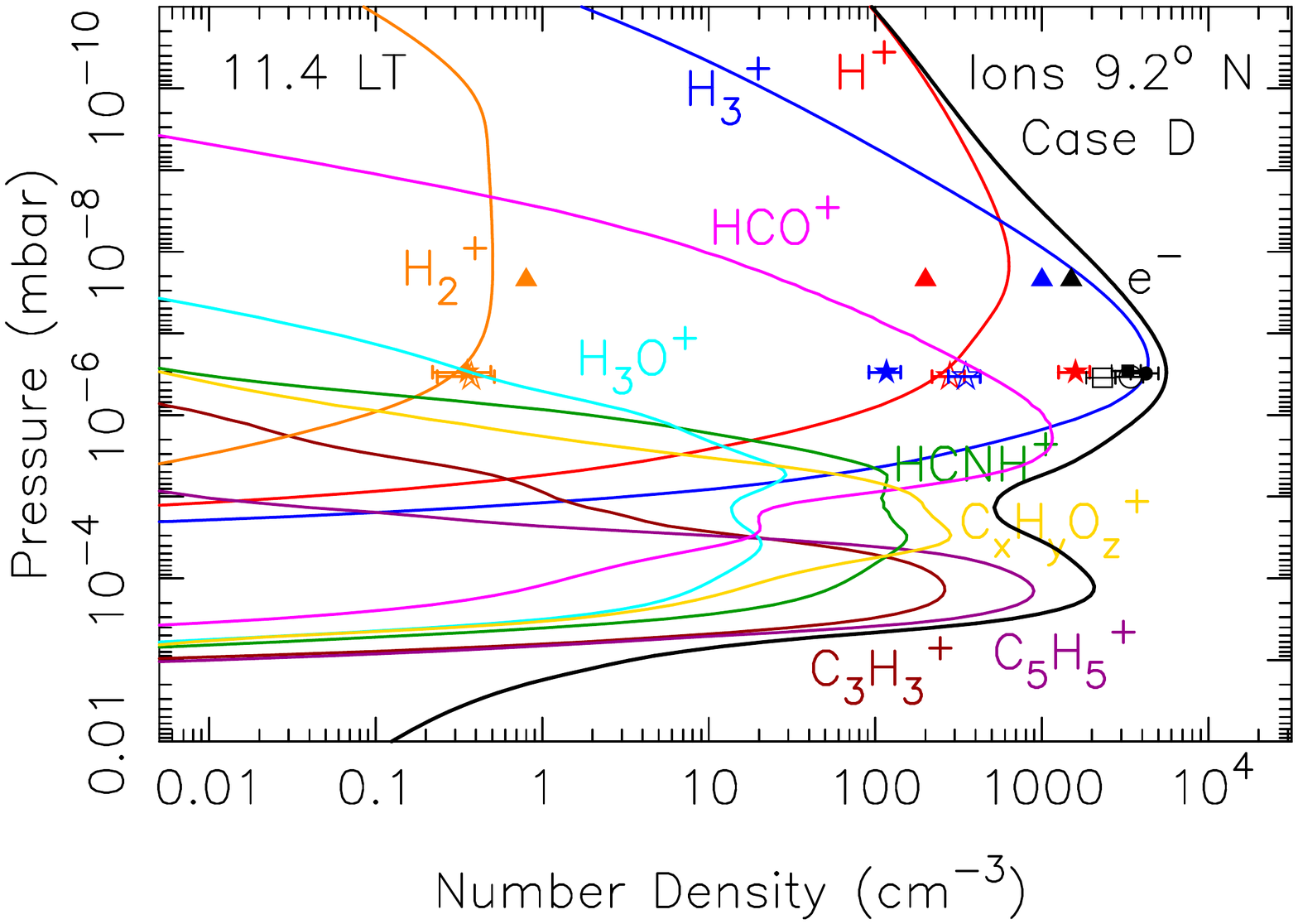}} 
\end{tabular}
\caption{Concentrations of key ions in our 9.2$\deg$ planetocentric latitude model for ring-influx assumptions from Case A (Top Left), 
Case B (Top Right), Case C (Bottom Left), and Case D (Bottom Right) from our diurnally variable model at a local time of 11.4 (just before 
noon), relevant to the \textit{Cassini} Grand Finale orbits.  Also shown are INMS ion-density measurements from orbit 292 (solid symbols) 
and orbit 288 (open symbols) at two specific locations along \textit{Cassini}'s orbital path; the points for the upper location (triangles) 
are from the ``rounded'' values given in \citet{vigren22} that were acquired near 2--4$\deg$ latitude, whereas the deeper points with error 
bars are taken from \citet{waite18} and \citet{moore18} at the closest-approach altitude near $-5\deg$ latitude.  Data points are plotted at 
the pressures corresponding to the measured H$_2$ density at those locations \citep[also from][]{waite18,moore18}.  The colors correspond to 
H$_2$$\! ^+$ (orange), H$_3$$\! ^+$ (blue), H$^+$ (red), with various other modeled ion profiles in other colors, as labeled.  The number density of 
free electrons from the models (black curves) and from RPWS measurements \citep[black points; ][]{morooka19,persoon19} derived from sweep data 
(squares) and upper hybrid resonance emissions (circles) are also presented.}   
\label{figion9ncases}
\end{figure*}

\begin{figure*}[!htb]
\vspace{-14pt}
\begin{tabular}{ll}
{\includegraphics[clip=t,scale=0.30]{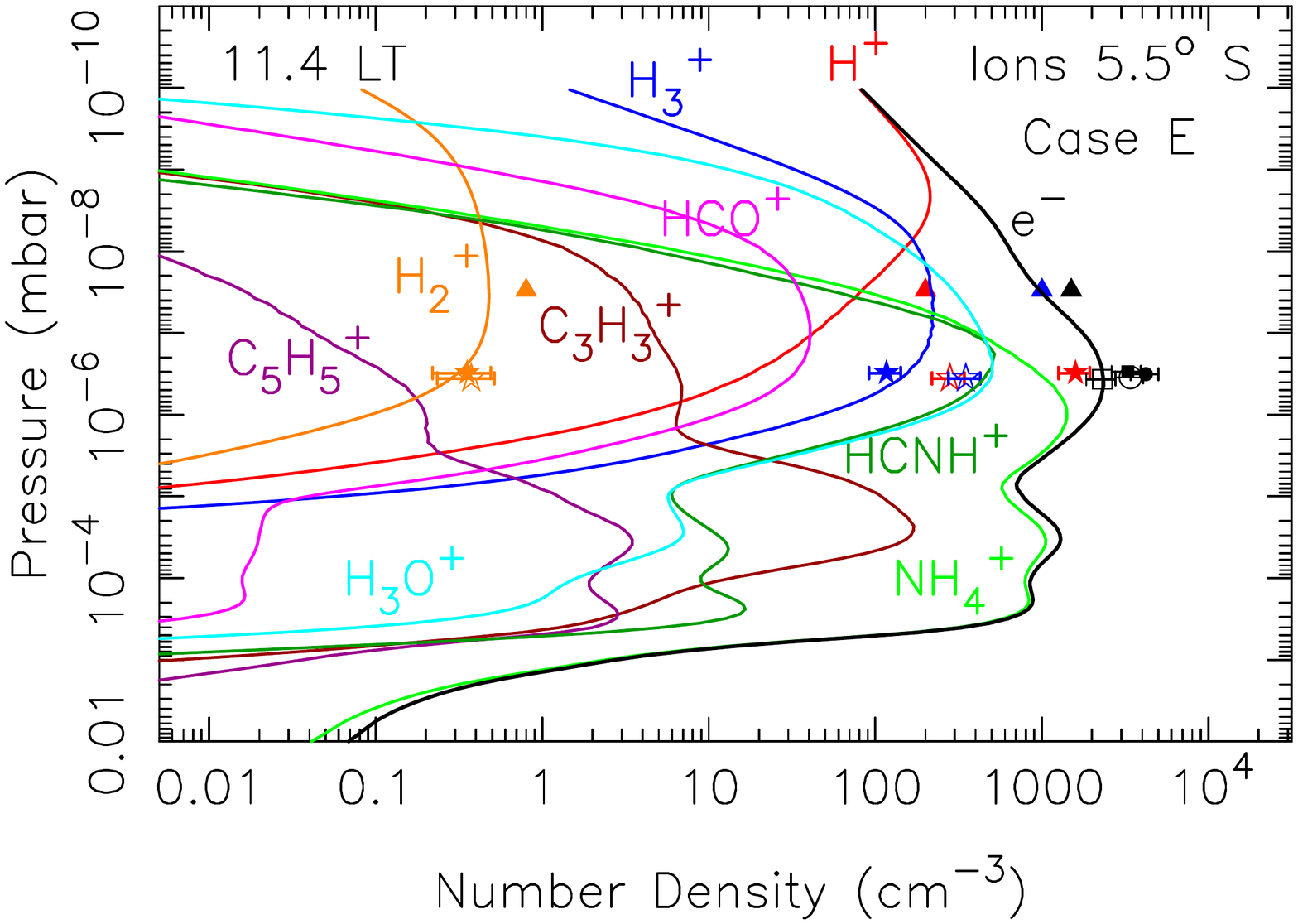}}
&
{\includegraphics[clip=t,scale=0.30]{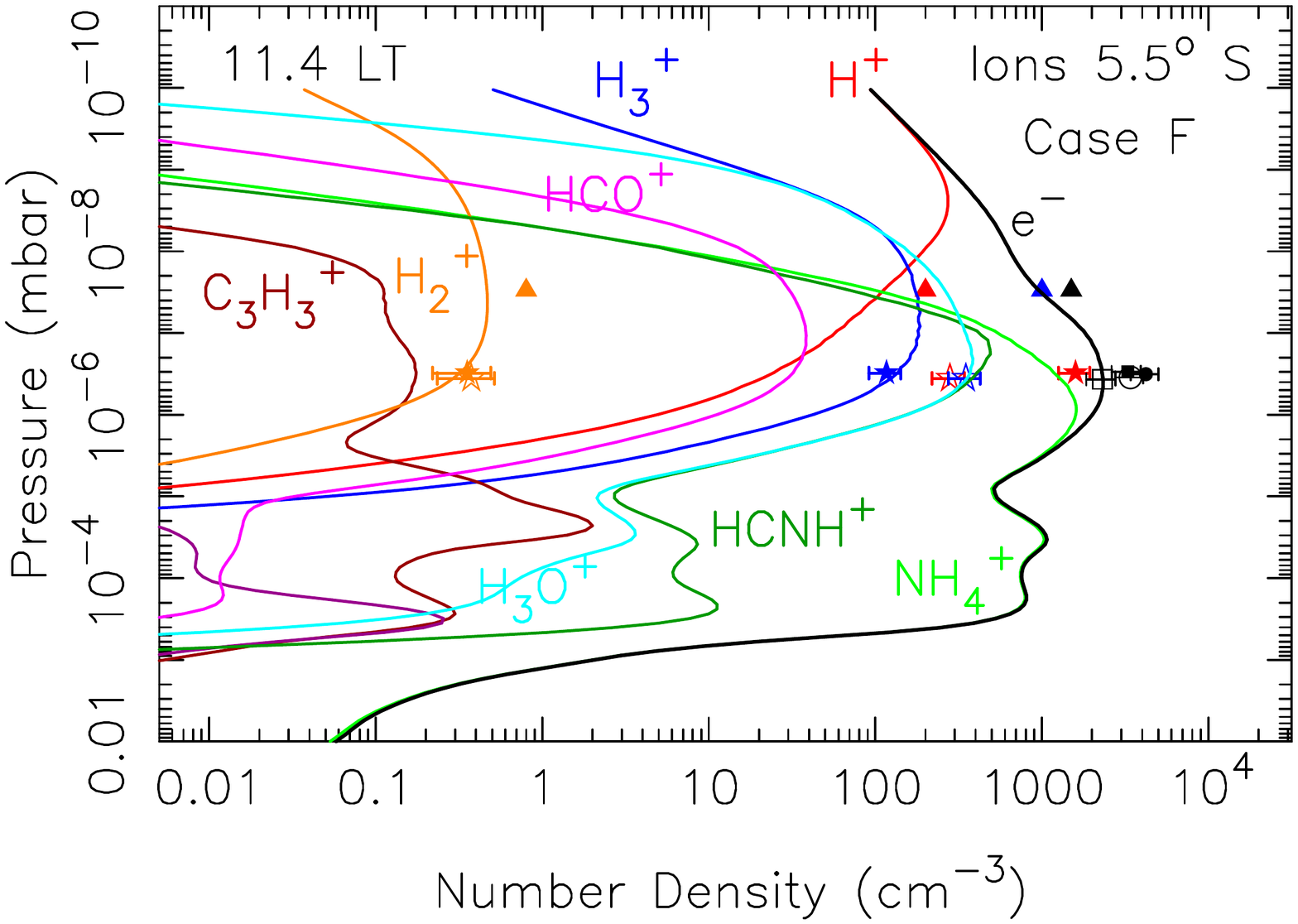}} 
\end{tabular}
\caption{Same as Fig.~\ref{figion9ncases}, except for the $-5.5\deg$ planetocentric latitude model for ring-influx assumptions from Case E (Left) 
and Case F (Right).}
\label{figion5scases}
\end{figure*}

Our diurnally varying model results for the ionosphere at a local time near noon are presented in Figs.~\ref{figion9ncases} 
and \ref{figion5scases} and compared with number densities of various ions measured by INMS \citep{waite18,moore18,vigren22} and number densities 
of free electrons determined from RPWS/LP observations \citep{wahlund18,morooka19,persoon19,hadid19} at two different atmospheric pressures.  
Because of the high encounter velocity of the spacecraft with respect to Saturn's atmosphere, only the densities of the lightest ions 
H$^+$, H$_2$$\! ^+$, H$_3$$\! ^+$, and He$^+$ were recorded by INMS \citep{waite18}.  

Although photoionization of H$_2$ to produce H$_2$$\! ^+$ dominates the initial ion production rate from a column-integrated standpoint 
in the models \citep[e.g., see also][]{kim14,chadney22}, the H$_2$$\! ^+$ then reacts efficiently with H$_2$ to form H$_3$$\! ^+$ + H, as well as 
reacts with atomic H to form H$^+$ + H$_2$ \citep{mcewan07}; thus, H$_2$$\! ^+$ is expected to become a relatively minor species in 
Saturn's ionosphere.  The INMS measurements confirm this expectation \citep[see][and Figs.~\ref{figion9ncases} and \ref{figion5scases}]{waite18,moore18}.  
The H$_3$$\! ^+$ ions formed in this manner become a major component of the ionosphere, with electron recombination and proton-transfer reactions with 
incoming neutral ring molecules (e.g., H$_2$O, CO, CH$_4$) and their photochemical products (e.g., O, OH) controlling H$_3$$\! ^+$ loss \citep{mcewan07,milligan00}.  
Direct photoionization of H, photoionization of H$_2$ to form H$^+$ + H + $e^-$, and reaction of H$_2$$\! ^+$ with H are responsible for producing 
H$^+$ in the models.  The H$^+$ reacts efficiently with incoming CH$_4$, HCN, H$_2$O, NH$_3$, CO$_2$, and (when available) C$_2$H$_6$ and 
C$_2$H$_2$ in our models \citep{mcewan07,smith92}, such that H$^+$ only dominates the total ion density at high altitudes but not at the main peak.  Instead, 
H$_3$$\! ^+$ dominates at the main peak if neutral ring-vapor influx rates are in the $\sim$10$^6$--10$^9$ molecules cm$^{-2}$ s$^{-1}$ range (such as 
with Cases C, D, and our nominal model) or H$_3$$\! ^+$ is supplanted by heavier ions at the main peak if ring-molecule influx rates are $\gtrsim$10$^9$ molecules 
cm$^{-2}$ s$^{-1}$ (such as with Cases A, B, E, F).  The relative abundances of H$^+$ and H$_3$$\! ^+$ in Cases A \& B are qualitatively consistent with 
the INMS ion measurements at altitudes above the main peak, but these models predict a much stronger drop off in H$^+$ density and a weaker drop off 
in H$_3$$\! ^+$ density with decreasing altitude toward the main peak than is borne out by the INMS measurements.  In fact, the INMS data suggest that 
H$_3$$\! ^+$ becomes notably depleted at the lowest altitudes and latitudes probed during orbits 288 and 292, whereas the number density of H$^+$ either 
increases (orbit 288) or remains roughly constant (orbit 292) in this closest-approach region of the orbits.  This surprising observed behavior is 
not borne out by any models that consider INMS-derived influx rates of neutral species (this work and \citealt{moore18}) and has interesting 
implications that are discussed by \citet{cravens19ioncomp} and \citet{vigren22} and examined further here.

The models shown in Figs.~\ref{figion9ncases} \& \ref{figion5scases} demonstrate that the composition of Saturn's ionosphere will be strongly 
sensitive to assumptions about the influx rates and composition of the incoming ring vapor.  The nitrogen-bearing ions NH$_4$$\! ^+$ and HCNH$^+$ 
can dominate the total ion density at and below the main ionospheric peak if NH$_3$ and HCN vapor molecules flowing in from the rings are 
abundant enough (e.g., Cases A, C, E, \& F).  Both HCN and NH$_3$ have large proton affinities and can outcompete other species in 
proton-transfer reactions even if the influx rates of NH$_3$ and HCN are smaller than that of other species such as CH$_4$, H$_2$O, 
CO, and N$_2$.  In cases in which only the most volatile molecules CO, N$_2$, and CH$_4$ are flowing in from the rings, HCO$^+$ can dominate the 
total ion density at the main peak when the CO flux is high enough (Case B), or at least become important below the main peak if the CO influx rate 
is too small to replace H$_3$$\! ^+$ at the main peak (e.g., Case D).  The number density of He$^+$ (not shown in the figures) is also found to be 
very sensitive to the assumed influx rate of neutral ring species.  The He$^+$ density at closest approach is over-estimated with Cases C \& D and 
under-estimated with cases A, B, E, and F, with Case B providing the closest match to observations.

Similarly, the models predict that the electron density, which is assumed to be equal to the sum of the ion densities in our models, has a peak abundance 
that is sensitive to the ring-vapor influx rates.  Cases E and F have the highest vapor influx rates and the lowest predicted peak electron 
densities.  Cases A and B, which are relevant to the Final Plunge terminal latitude at the end of the mission, have slightly reduced influx 
rates, either because of the higher terminal latitude (i.e., the ring material is concentrated at the equator) or because of time variability 
in the ring-influx source; the interaction of the ions with 
a smaller abundance of heavy neutrals in these cases results in a reduction in ion loss, such that the peak electron densities in Case A \& B 
end up being greater than in Cases E \& F.  The Case C and D models, which have fluxes that have been reduced from Cases A \& B by an order 
of magnitude to simulate global horizontal spreading, predict the greatest peak electron densities.  These latter models were not designed 
to truly simulate the current state of the ionosphere in the Grand Finale stage of the \textit{Cassini} mission because the INMS data exhibit 
no reduction in heavy neutral species' mixing ratios with decreasing altitude at low latitudes, as might be expected if horizontal spreading 
were effective.  The results, however, are still useful for gauging the sensitivity of the electron density to the influx rates, so we 
include them here.  The various 
techniques for deriving the electron density from the RPWS/LP experiment give slightly different results 
\citep{wahlund18,morooka19,persoon19,hadid19}, and the peak densities in Cases A, B, E, F are all quantitatively consistent with one or 
more of these results from the different techniques (see Figs.~\ref{figion9ncases} \& \ref{figion5scases}), whereas the predicted electron 
densities for Cases C \& D are too large.  The INMS-derived neutral influx rates for Cases A, B, E, F as derived from \citet{serigano22} and 
\citet{miller20} therefore appear consistent to first order with the total electron densities inferred from the RWPS/LP experiment.

However, there are two potential complications with this interpretation.  First, some analyses of the Langmuir Probe data from the RPWS instrument 
indicate that the total ion density at closest approach typically exceeds the electron density --- sometimes by as much as an order of magnitude --- 
suggesting that negatively charged grains are present \citep{morooka19}, in which case the electron densities should not be used as a proxy for 
the total ion density in comparisons of the models with the data.  However, this complication is negated by a more recent analysis that includes 
the influence of secondary-electron emission from impacts of gas molecules on the Langmuir Probe \citep{johansson22} and finds no convincing 
evidence for positive ion densities that exceed electron densities.  Second, although our models do a reasonable job of predicting the relative 
abundance of H$^+$ and H$_3$$\! ^+$ at higher altitudes, especially for Case B that assumes only the most volatile ring-derived species are 
flowing in as vapor (Fig.~\ref{figion9ncases}), the models significantly under-predict the H$^+$/H$_3$$\! ^+$ ratio at low altitudes/latitudes 
along the trajectory (see the deepest data points in Figs.~\ref{figion9ncases} \& \ref{figion5scases}).  
This second complication is a significant problem that is discussed further in \citet{vigren22} and below.  

To reproduce the high H$^+$ and low H$_3$$\! ^+$ abundances inferred for the closest-approach region, 
\citet{cravens19ioncomp} and \citet{vigren22} use photochemical equilibrium arguments to demonstrate that the data in this region 
imply that the abundance of neutral molecules ``$\mathcal{M}$'' that can react with both H$^+$ and H$_3$$\! ^+$ (e.g., CH$_4$, H$_2$O, 
NH$_3$, HCN, C$_2$H$_x$, CO$_2$) must be much smaller than the abundance of neutral molecules ``$\mathcal{R}$'' that react only 
with H$_3$$\! ^+$ (e.g., CO and N$_2$).  \citet{vigren22}, who have updated the general modeling technique of \citet{cravens19ioncomp} to 
include dust charging, demonstrate that the neutral abundances required to explain the closest-approach ion-density measurements are 
then inconsistent with the INMS measurements of neutral species, creating what they emphasize is a major conundrum.  Our modeling 
introduces some production and loss processes not considered in the \citet{vigren22} analysis, especially involving the abundant 
species atomic H, but our overall conclusions are similar --- the same conundrum still stands.
Not only are the neutral mixing ratios of H$_2$O, CH$_4$, HCN, CO$_2$, and other $\mathcal{M}$ species (and their photochemical products 
O and OH) too great to explain the large H$^+$ density at low altitudes, but the mixing ratios of CO and N$_2$ as the key 
$\mathcal{R}$ species are too low to account for the needed reduction in H$_3$$\! ^+$ at deep altitudes.  Moreover, the inferred 
mixing ratios of neutral $\mathcal{M}$ and $\mathcal{R}$ species needed to explain the low-altitude, low-latitude ion 
density data are inconsistent with the results inferred at higher altitudes/latitudes.

The H$^+$ density is the source of the largest mismatch with data.  At the altitudes probed during the closest approach in orbits 288 and 292, 
the dominant production and loss processes for H$^+$ are 
\begin{eqnarray}
\Hatom \, + \, h\nu \, & \rightarrow & \, \Hatom ^+ \, + \, e^{-} \\ 
\Htwo \, + \, h\nu \, & \rightarrow & \, \Hatom ^+ \, + \, \Hatom \, + \, e^{-}  \\
\Htwo ^+ \, + \, \Hatom \, & \rightarrow & \, \Hatom ^+ \, + \, \Htwo  \\
\Hatom ^+ \, + \, \mathcal{M} & \, \rightarrow & \, \mathcal{M} ^+ \, + \, \Hatom  , 
\end{eqnarray}
with $e^{-}$ being an electron and $\mathcal{M}$ being any of the ring-derived neutral species CH$_4$, HCN, H$_2$O, C$_2$H$_x$, NH$_3$, etc.~that react with 
H$^+$.  The abundance of atomic H derived from our models is greater than that 
assumed in \citet{chadney22}, and reaction (7) dominates slightly over reaction (8) in producing H$^+$ at the closest-approach altitude, with 
reaction (9) contributing $\sim$10\% of the total production.  The production rate of H$^+$ equals $J_{7} [\Hatom ] + J_{8} [ \Htwo ] + k_{9}
[\Htwo ^+ ] [\Hatom ]$, with $J_{7}$ and $J_{8}$ being the photoionization rate (s$^{-1}$) of reactions (7) and (8), respectively, $k_{9}$ being the 
rate coefficient for reaction (9), and the square bracket terminology indicating number density of that species (cm$^{-3}$).  The loss rate of H$^+$ is 
$ k_{10} [ \Hatom ^+ ] [ \mathcal{M} ]$, with $ [ \mathcal{M} ]$ being the sum of the densities of $\mathcal{M}$ = CH$_4$, HCN, H$_2$O, C$_2$H$_x$, NH$_3$, etc., 
and $k_{10}$ being the weighted average of the rate coefficients of the reaction of H$^+$ with 
$\mathcal{M}$.  If we assume photochemical equilibrium, such that the production and loss rates are 
equal, then one can solve for the H$^+$ density:
\begin{equation}
[ \Hatom ^+ ] \, = \, \frac{ [ \Hatom ] ( J_{7} + k_{9} [ \Htwo ^+ ] ) + J_{8} [ \Htwo ]}{ k_{10} [ \mathcal{M} ] } .
\end{equation}
To increase the H$^+$ number density to better match the INMS measurements at the lower altitudes and latitudes near closest approach, we would 
need to substantially decrease $ [ \mathcal{M} ]$ or substantially increase $J_7$ and/or $[ \Hatom ]$ at lower altitudes/latitudes, given that 
$J_{8}$, $k_9$, $[ \Htwo ]$, and $[ \Htwo ^+ ]$ seem more firmly established by theory, data, and/or the model-data comparisons.  Although 
$ [ \mathcal{M} ]$ may also seem firmly established by the INMS neutral-density measurements at the same location, there is always the 
possibility that some or much of the INMS signal is caused by impact vaporization of solid dust particles, such that the inferred mixing 
ratios of $\mathcal{M}$ species in the thermosphere have been over-estimated by INMS analyses.  

Is there a way to resolve this apparent discrepancy in the H$^+$ abundance in our models at low altitudes/latitudes by increasing the H$^+$ 
production or decreasing its loss?  First, we note that our adoption of low-resolution H$_2$ cross sections and solar-flux values causes us to 
under-estimate $J_7$ at low thermospheric altitudes \citep{chadney22}, and switching to high resolution would help increase H$^+$ production.  
However, the models of \citet{chadney22} predict that the largest increase in $J_7$ from the adoption of high wavelength resolution occurs 
near the methane homopause, whereas at the altitudes (and H$_2$ densities) relevant to the closest approach distance in orbits 288--292, the increase 
in $J_7$ is only a factor of $\sim$2.  The adoption of high-resolution cross sections and solar flux values is therefore unlikely to resolve the more 
than order-of-magnitude under-prediction in the H$^+$ density at closest approach that plagues our models in Cases A, B, E, and F (i.e., in the models 
that do not include horizontal spreading).  We also have adopted solar-cycle average ultraviolet solar flux in these models, which was chosen because 
of our great interest in the stratospheric chemistry, with its longer chemical and diffusion time constants.  We could gain another factor of a few 
if we were to adopt solar-cycle maximum conditions.  A greater extreme ultraviolet flux would also increase H$_2$$\! ^+$ abundances to better match 
observations.  To increase atomic H at the closest-approach altitude, we could adopt a higher eddy diffusion coefficient 
within and above the methane homopause region where the atomic H production is a maximum, such that more H is transported into the ionospheric 
main-peak region.  However, our current $K_{zz}$ profile is constrained by \textit{Cassini} UVIS occultation observations of CH$_4$ and other 
hydrocarbons at low latitudes \citep{koskinen16,koskinen18,brown21}, and increasing $K_{zz}$ in the upper stratosphere and thermosphere would 
lead to a higher-altitude methane homopause than is indicated by the UVIS observations.  The fact that the UVIS occultations were not simultaneous with 
the INMS measurements gives us some leeway to potentially adjust $K_{zz}$, but the required increase would be significantly different from what has been 
inferred at low latitudes at other times, and a physical explanation for this sudden change would need to be sought.  Moreover, $K_{zz}$ cannot be 
so high that the atomic H upper limits from \citet{koskinen13sat} are violated (i.e., the H mixing ratio must remain less than $\sim$11\% at [H$_2$] = 
8.9$\scinot7.$ cm$^{-3}$, which is located at $\sim$4.4$\scinot-9.$ mbar in our 9.2$\deg$ latitude model).

\begin{figure*}[!htb]
\vspace{-14pt}
\begin{center}
\includegraphics[clip=t,scale=0.40]{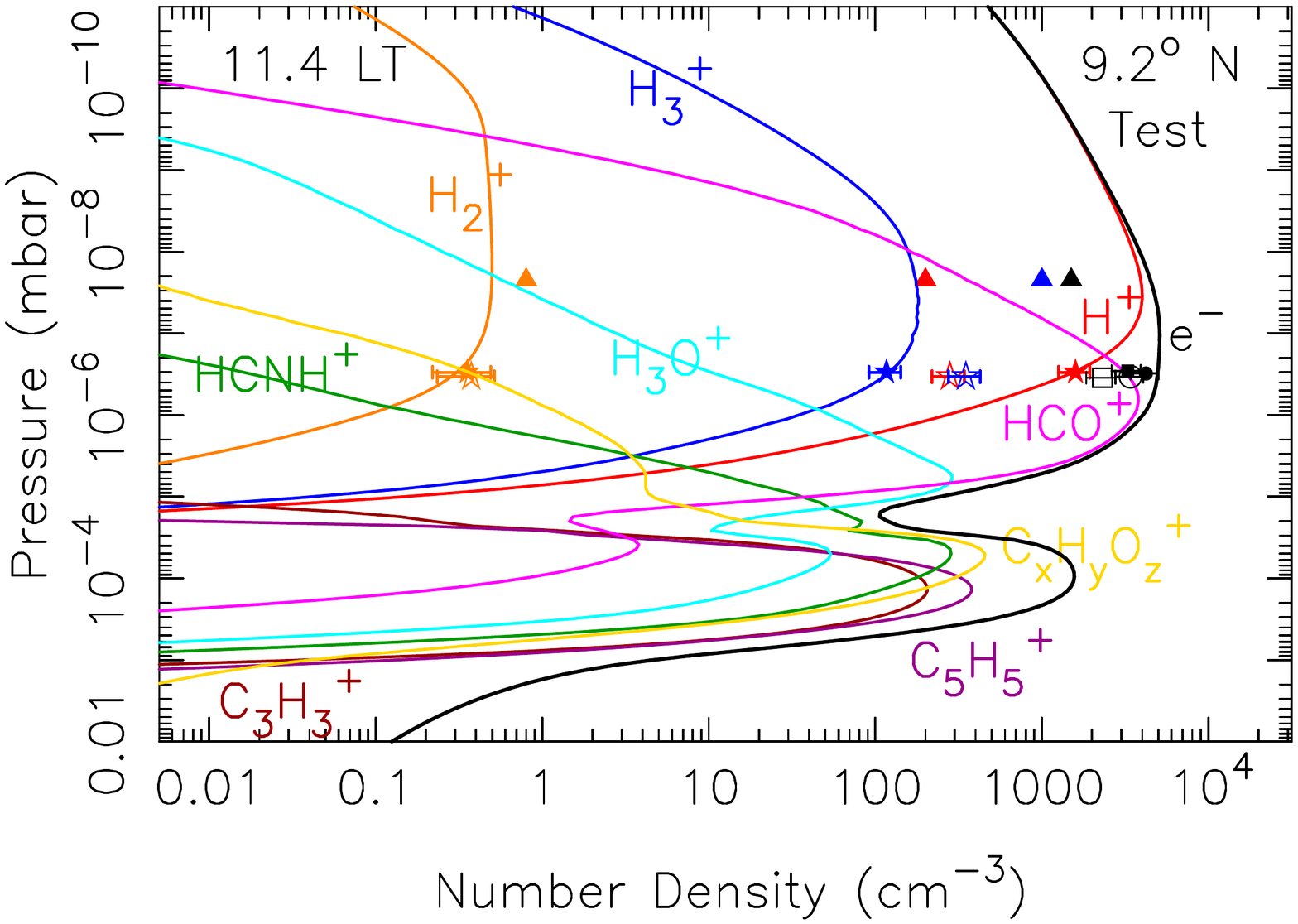}
\end{center}
\caption{Same as Fig.~\ref{figion9ncases}, except for a 9.2 N planetocentric latitude model in which the influx rate of neutral species 
at the top of the atmosphere is assumed to be 5$\scinot10.$ CO molecules cm$^{-2}$ s$^{-1}$, 2.14$\scinot9.$ N$_2$ molecules cm$^{-2}$ s$^{-1}$, 
3$\scinot7.$ CH$_4$ molecules cm$^{-2}$ s$^{-1}$, and 2$\scinot6.$ H$_2$O molecules cm$^{-2}$ s$^{-1}$.  The relatively low influx rates of 
CH$_4$ and H$_2$O help maintain the large observed H$^+$ density at low altitudes, whereas the high influx rates of CO and N$_2$ help reduce the H$_3$$\! ^+$ 
density at low altitudes, providing a good fit to the low-altitude, low-latitude ion densities determined by INMS at closest approach for 
orbit 292 (solid stars).  However, such a model fails to reproduce the mixing ratios of neutral species determined by the INMS 
measurements in the closest-approach region \citep{miller20,serigano22} and fails to reproduce the ion densities determined by INMS at 
higher altitudes and latitudes \citep{waite18}.} 
\label{figionfitdeep}
\end{figure*}

Decreasing the mixing ratios of the $\mathcal{M}$ species available to react with H$^+$ would seem to be the most straightforward way to increase H$^+$ 
to explain the closest-approach INMS ion measurements.  Figure \ref{figionfitdeep} shows a solution for 9.2 N planetocentric latitude 
similar to Case B but for which we have reduced the inflow of $\mathcal{M}$ species, such that the incoming CH$_4$ flux is 3$\scinot7.$ molecules 
cm$^{-2}$ s$^{-1}$ (in comparison to the 1.46$\scinot9.$ molecules cm$^{-2}$ s$^{-1}$ inferred for CH$_4$ for orbit 293 from \citealt{serigano22}), 
and we include a small background Enceladus H$_2$O influx of 2$\scinot6.$ molecules cm$^{-2}$ s$^{-1}$.  To help reduce the H$^+$ 
loss in the main-peak region in this model, we have also omitted the reaction of H$^+$ with vibrationally excited H$_2$ \citep[see][]{moore19chap} 
and limited the influence of the reaction H$^+$ + H$_2$ + M $\rightarrow$ H$_3$$\! ^+$ + M by capping the high-pressure limiting rate coefficient to 
be equal to the rate coefficient for radiative association of 1.3$\scinot-16.$ cm$^{3}$ s$^{-1}$ \citep{gerlich92,mcewan07}, although these 
reactions have less influence on H$^+$ loss than the reaction of H$^+$ with CH$_4$ and H$_2$O \citep{mcewan07}.  For the model presented in 
Fig.~\ref{figionfitdeep}, we 
found that we also needed to increase the influx of the $\mathcal{R}$ species CO and/or N$_2$ to fit the low H$_3$$\! ^+$ densities inferred by INMS 
at closest approach for orbit 292, as has previously been discussed by \citet{cravens19ioncomp} and \citet{vigren22}.  We have kept the influx rate for N$_2$ 
the same as in Cases A \& B (2.14$\scinot9.$ N$_2$ molecules cm$^{-2}$ s$^{-1}$) in this new model, while increasing the CO influx rate from the
2.33$\scinot9.$ molecules cm$^{-2}$ s$^{-1}$ determined by \citet{serigano22} to 4$\scinot10.$ molecules cm$^{-2}$ s$^{-1}$ --- other 
combinations of similar increases of N$_2$ and/or CO would have had a similar effect.  

This model with increased influx of $\mathcal{R}$ species and decreased influx of $\mathcal{M}$ species better reproduces the INMS ion measurements 
\citep{waite18,moore18} and the
RPWS/LP total ion estimates \citep{morooka19} at closest approach for orbit 292 (see Fig.~\ref{figionfitdeep}), but the resulting mixing ratio of CO is 
then significantly higher (and the mixing ratio of CH$_4$ and other $\mathcal{M}$ species significantly lower) than the mixing ratios derived 
from the simultaneous INMS closest-approach neutral measurements \citep{miller20,serigano22}, creating the major ``conundrum'' emphasized by 
\citet{vigren22}. We have no solution to this conundrum with our current chemistry inputs.  Larger icy particles coming in from the rings could begin to
ablate in this region of the atmosphere \citep{moses17poppe,hamil18}, potentially introducing H, CO, N$_2$ from thermochemistry within the meteor 
trails, but again, the INMS neutral measurements would have seen evidence for the excess CO and/or N$_2$.  Metals and SiO might also have been 
introduced to equatorial latitudes within this altitude region from the ablation of ring particles, thereby triggering metal-ion chemistry that 
that could be affecting the ionospheric structure and chemistry \citep[see][]{lyons95,mosesbass00,kim01}. Again, however, the INMS data show 
no evidence for silicon-bearing or metal ions \citep{serigano22}.   Dust particles themselves may be affecting the ion chemistry and structure at equatorial 
latitudes, but it is difficult to imagine how reactions with dust would preferentially decrease H$_3$$\! ^+$ and increase H$^+$, as is required 
by the ion measurements at low altitudes/latitudes.  \citet{vigren22} suggest that collision-induced dissociation of H$_3$$\! ^+$ 
with H$_2$ to form H$^+$ + 2$\,$H$_2$ could be one way to produce H$^+$ at the expense of H$_3$$\! ^+$, but they emphasize that this reaction is 
endoergic by $\sim$4.3 eV and point out other reasons for its infeasibility in resolving the H$^+$/H$_3$$\! ^+$ ratio dilemma in Saturn's ionosphere.

Although we currently offer no viable solution to this apparent neutral-vs-ion INMS inconsistency at equatorial low altitudes during closest 
approach, we should also point out the considerable variability in H$^+$ and H$_3$$\! ^+$ density measurements between orbits 288 and 292 
\citep[see][]{waite18,moore18}, as seen in Fig.~\ref{figionfitdeep}, and the significant variability in the electron-density profiles derived 
for Saturn from the \textit{Cassini} radio-occultation experiment \citep[e.g.,][]{nagy06}.  This variability suggests that chemistry is not 
alone in controlling processes in Saturn's ionosphere and that other dynamical or electrodynamical processes contribute to the 
observed ionospheric behavior \citep[e.g.,][]{mosesbass00,nagy06,matcheva12,barrow13}.  Moreover, there is some reason to be cautious about 
the inferred abundances of the lightest species species detected by the INMS.  Quadrupole mass spectrometers are hampered by a problem 
known as the ``zero-blast effect'' that makes it difficult to accurately measure the abundance of light species (1--4 amu) \citep{evans19,strobel22},  
although the INMS uses precision grinding of hyperbolic quadrupole rods to help mitigate this effect \citep{waite04}.  Scattering in high-dynamic-pressure 
situations, such as occurs at closest approach during the high-velocity atmospheric passes in the proximal orbits, can exacerbate the effect 
\citep[e.g.,][]{magee09}.  Best efforts were made to correct for these instrumental effects in the calibration and analysis of the INMS data,
but the results for the light species may retain higher inherent uncertainty \citep[see also][]{strobel22}.

\subsection{Implications for thermospheric neutral chemistry}\label{sec:other}

Our models indicate that the heavy molecules detected by INMS during the Grand Finale orbits are unlikely to have been formed from 
photochemical processes within Saturn's upper thermosphere.  In other words, the inflow of CH$_4$ or other ``parent'' molecules from the rings would not 
have triggered rapid formation of more complex organic gas-phase photochemical products that then are detected by INMS.  The vertical diffusion time scale of 
molecules through the upper thermosphere is much faster than local chemical production and loss rates, even when considering relatively rapid ion chemistry.  
Similarly, chemical loss of the inflowing vapor between the INMS measurement altitudes and the top of the stratosphere is relatively minor for most 
species, so photochemistry does not remove the incoming vapor before it reaches the stratosphere.

\subsection{Observational implications for the neutral stratosphere and for the timing of the ring-influx event}\label{sec:obs}

The equatorial inflow of ring vapor discovered by INMS not only affects the ionosphere but also introduces new molecules to Saturn's stratosphere 
that will alter the neutral chemistry and composition.  The differences between our Case A--F models illustrate that the ultimate fate of the 
ring-derived species depends not only on the total influx rate but also on the molecular makeup of the incoming vapor.  The new neutral molecules 
introduced to the stratosphere, either directly from the ring influx or indirectly from subsequent stratospheric photochemistry, have abundances 
that rival those of the hydrocarbon photochemical products that have already been observed on Saturn \citep[see the review of][]{fouchet09}.  We 
therefore must consider whether these ring-derived neutral species are currently observed and/or should be observable in Saturn's stratosphere.

\subsubsection{Comparisons to infrared observations}\label{sec:irobs}

\begin{figure*}[!htb]
\vspace{-14pt}
\begin{tabular}{ll}
{\includegraphics[clip=t,scale=0.30]{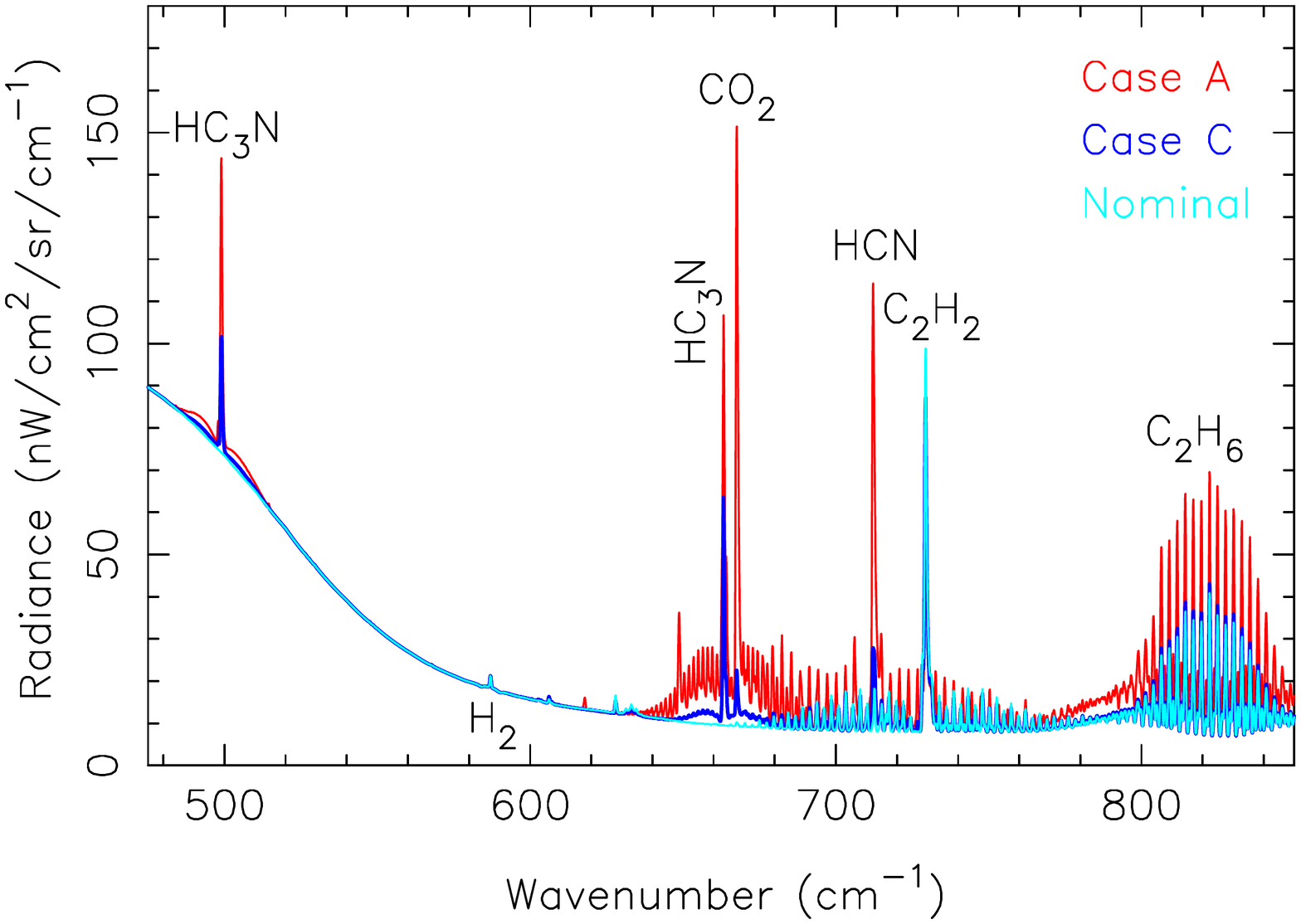}}
&
{\includegraphics[clip=t,scale=0.30]{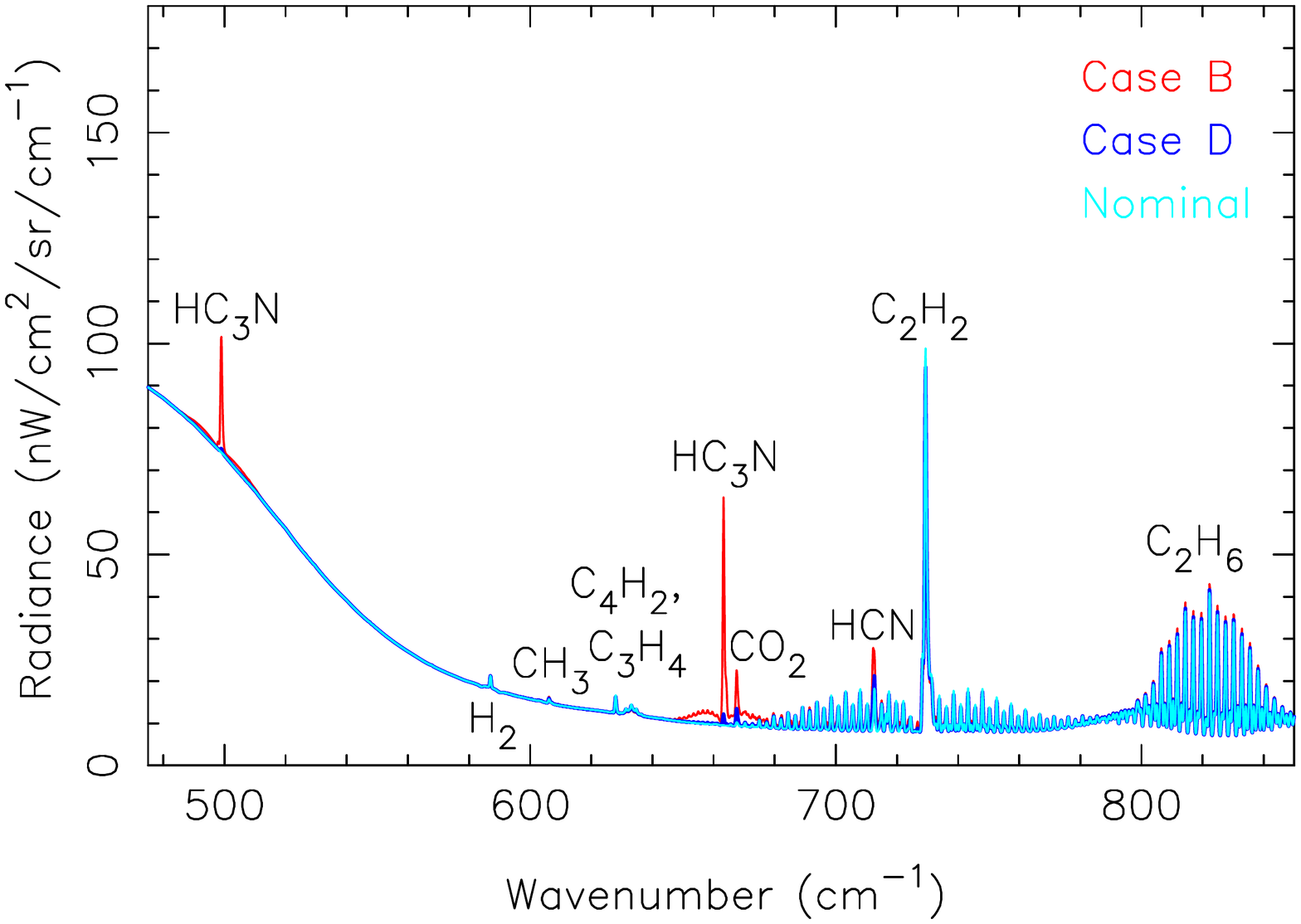}} \\
\vspace{-14pt}
{\includegraphics[clip=t,scale=0.30]{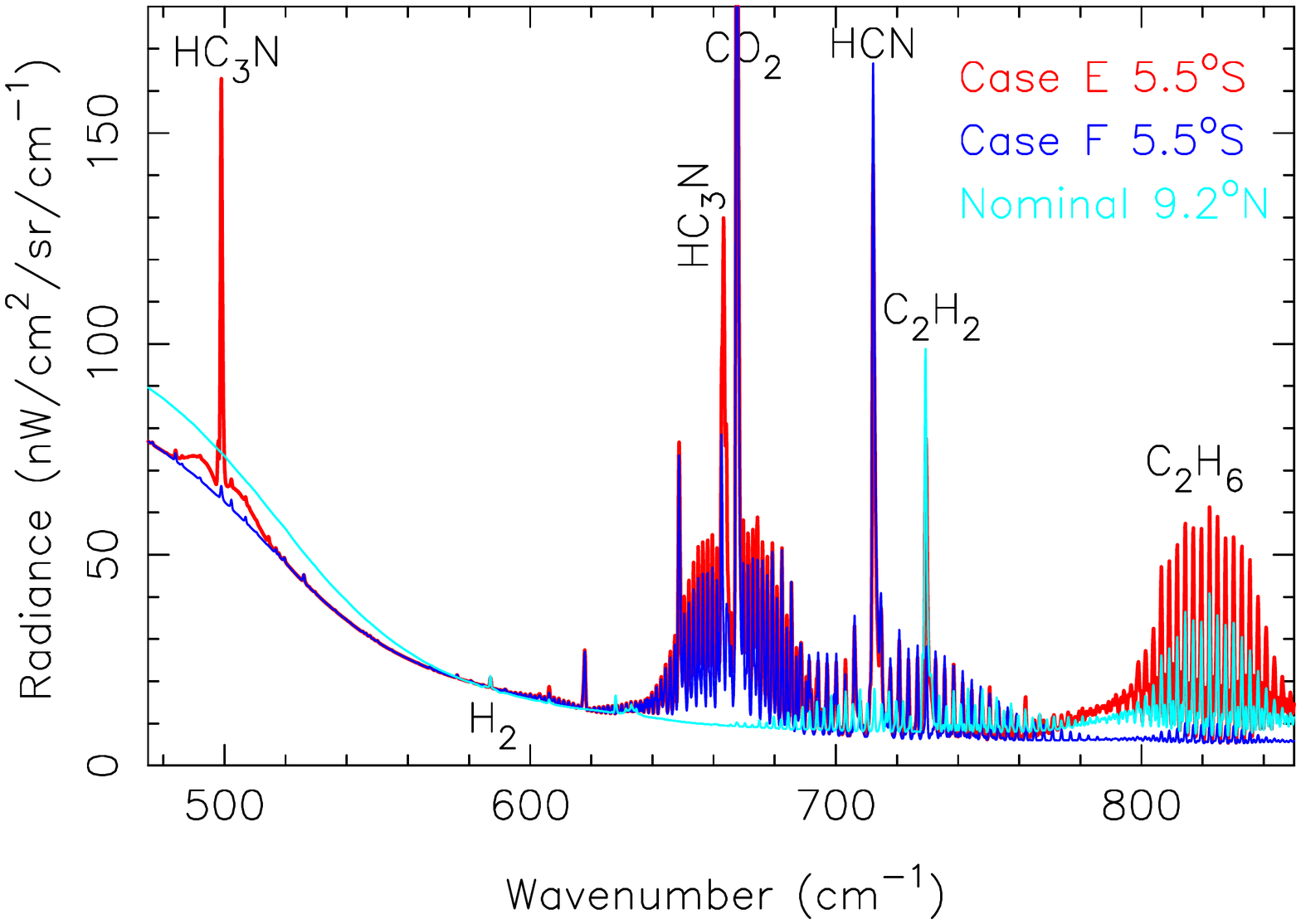}}
&
{\includegraphics[clip=t,scale=0.30]{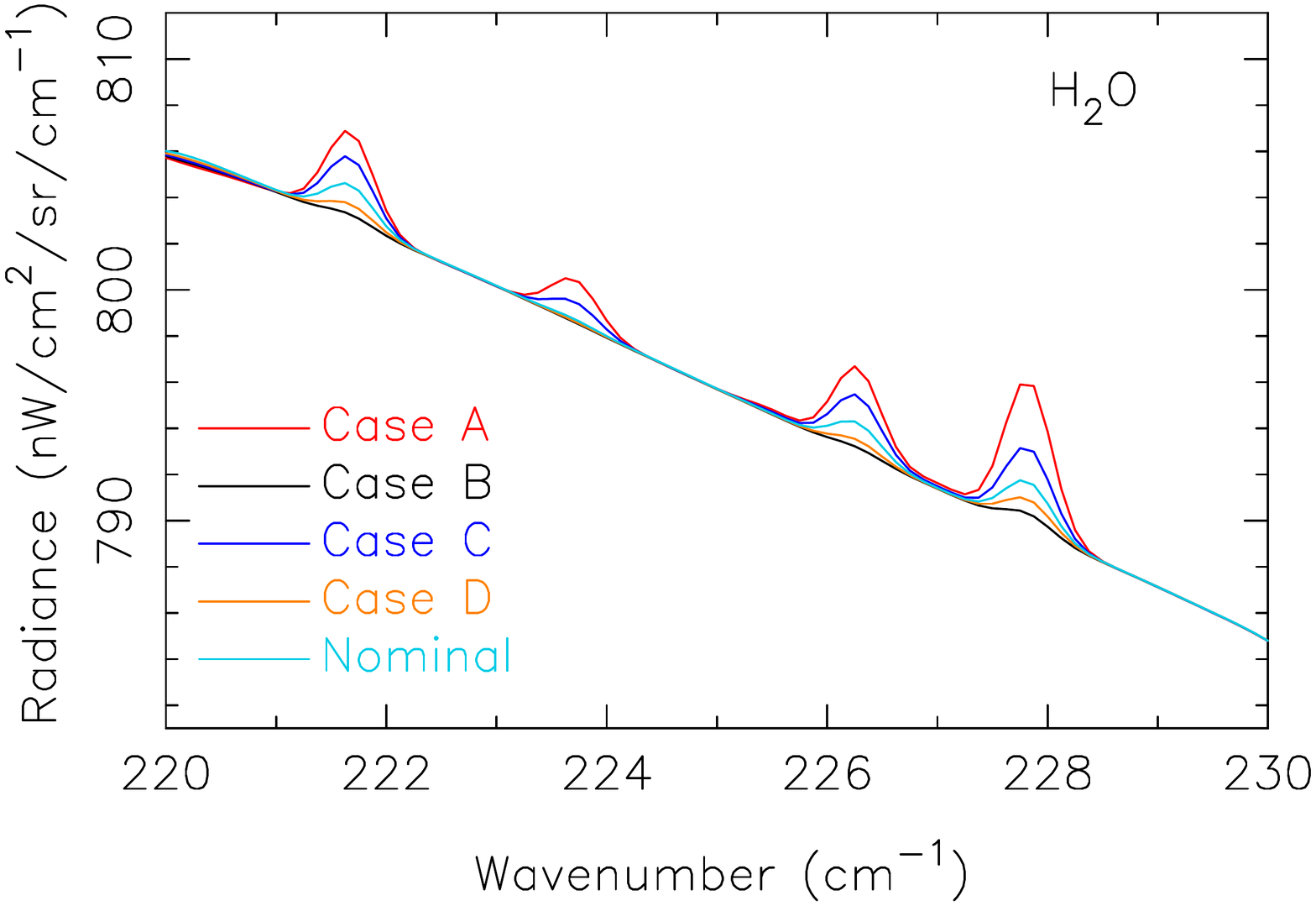}}
\end{tabular}
\caption{Synthetic mid-infrared emission spectra of Saturn at 0.5-cm$^{-1}$ resolution --- relevant to the highest-resolution mode of \textit{Cassini} CIRS --- 
from ring-vapor influx model Cases A \& C (Top Left), Cases B \& D (Top Right), Cases E \& F (Bottom Left), compared to the nominal 9.2$\deg$ N 
planetocentric latitude with no ring influx.  Note the prominent spectral features resulting from CO$_2$, HCN, and HC$_3$N in this spectral 
region.  Even if these molecules are not directly flowing in from the rings, they are produced photochemically within the stratosphere and can 
affect infrared observations.  Also presented (Bottom Right) are synthetic spectra for Cases A--D at 9.2$\deg$ N for a narrow far-infrared spectral 
region that contains water rotational lines.}
\label{figspecmidir}
\end{figure*}

\begin{figure*}[!htb]
\begin{tabular}{ll}
{\includegraphics[clip=t,scale=0.165]{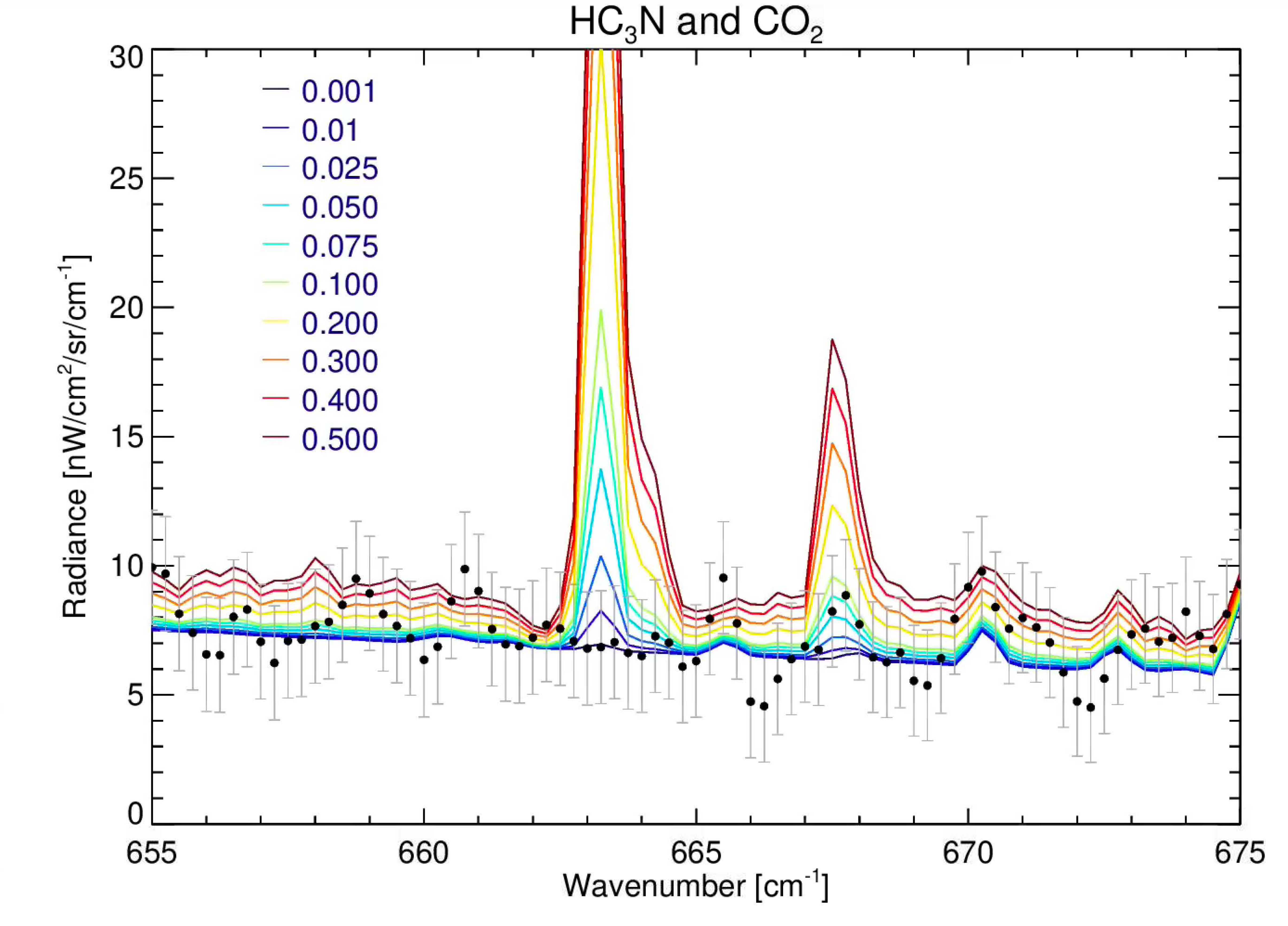}}
&
{\includegraphics[clip=t,scale=0.165]{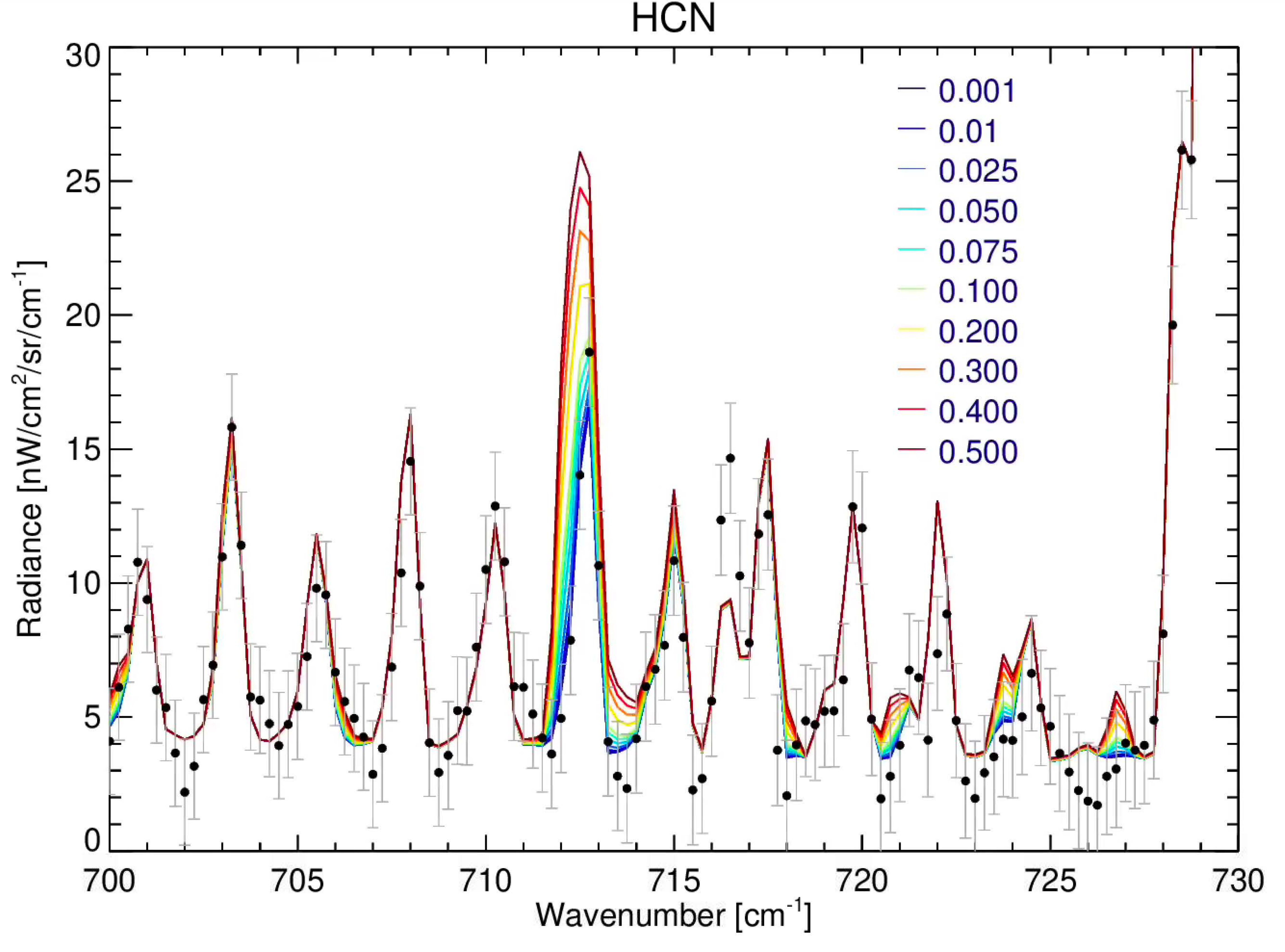}} 
\end{tabular}
\caption{\textit{Cassini}/CIRS nadir spectra (black data points with error bars) from April 28, 2017 at 3$\deg$ N planetocentric latitude, along with our
best-fit temperature and hydrocarbon retrieval, with the addition of HC$_3$N and CO$_2$ (Left) and HCN (Right) with Case B model abundances scaled by factors
of 0.001--0.5 (colored lines, as labeled).}
\label{figfletchcirs}
\end{figure*}

To address the question of observability, we have created synthetic nadir spectra from the photochemical model results using the \texttt{NEMESIS} 
radiative-transfer software package \citep{irwin08} in a line-by-line forward-modeling mode.  Figure~\ref{figspecmidir} shows the predicted mid-infrared 
emission spectra for the Case A--F models in nadir-looking geometry, along with the 
9.2$\deg$ N nominal model with no ring influx, at the 0.5-cm$^{-1}$ wavenumber resolution relevant to the highest-resolution mode of the \textit{Cassini} 
CIRS instrument.  If the ring-vapor influx rates determined by the INMS analyses of \citet{miller20} and \citet{serigano22} have been in operation 
over long time scales, then Fig.~\ref{figspecmidir} demonstrates that several strong bands of CO$_2$, HCN, and HC$_3$N would be prominently visible
in Saturn spectra in the mid-infrared region, at least at the low latitudes where the inflow is concentrated (see Fig.~\ref{figspecmidir}).  Other than 
the CO$_2$ feature near 667--668 cm$^{-1}$, these prominent bands have never been detected on Saturn (e.g., see Fig.~\ref{figfletchcirs}), and the observed 
intensity of the CO$_2$ feature is much smaller than is predicted by the INMS ring-inflow models.  

To directly compare our model predictions with low-latitude \textit{Cassini}/CIRS observations of Saturn near the Grand Finale stage of the mission, 
we have analyzed 0.5 cm$^{-1}$ resolution CIRS nadir emission at 600--860 cm$^{-1}$ (Focal Plane 3) and 1230-1375 cm$^{-1}$ (Focal Plane 4) acquired on 
April 28, 2017 from a low-latitude region on Saturn centered near 3$\deg$ N planetographic latitude, for which the CIRS footprint covered 0--7$\deg$ N 
planetographic latitude. Spectra from all longitudes were averaged together.  For this part of the analysis, we inverted the CIRS spectra using the 
optical-estimation approach of \texttt{NEMESIS} and using the correlated-k approximation to increase the speed of the radiative-transfer calculations,  
as is described in \citet{fletcher07,fletcher10,fletcher11beacon,fletcher15satpole}.  
Our Case B model temperature-pressure profile and hydrocarbon abundances (e.g., CH$_3$, CH$_4$, C$_2$H$_2$, C$_2$H$_4$, C$_2$H$_6$, C$_3$H$_4$, C$_4$H$_2$) were 
used as priors for the retrieval; the CH$_4$ profile was held fixed, whereas temperatures were allowed to vary freely in the retrieval, 
and the vertical profiles of the hydrocarbons were scaled by a constant factor until the synthetic spectra provided a statistical best fit to 
the observed spectra.  Resulting scale factors for the hydrocarbons were generally small (0.75--1.32), implying that the Case B hydrocarbon profiles 
provide a reasonable estimate of those needed to reproduce the April 2017 spectra.  Moreover, the retrieved temperatures remained within a 
few K of our 9.2 N latitude model profile.  However, we found that the CIRS spectra show no evidence for HCN or HC$_3$N emission features 
(Fig.~\ref{figfletchcirs}), and the Case B profiles of HCN, HC$_3$N, and CO$_2$ needed to be scaled down by orders of magnitude to remain consistent with the 
CIRS nadir spectra.  A simple visual comparison of Fig.~\ref{figfletchcirs} suggests that HC$_3$N mixing ratios in Saturn's low-latitude stratosphere are 
limited to $\lesssim$0.025 times that predicted from Case B, while HCN and CO$_2$ mixing ratios are limited to $\lesssim$0.1 times those from Case B.  Even for
Case D that assumes only incoming CO, N$_2$, and CH$_4$ and that the ring vapor spreads globally before reaching the stratosphere (not shown in 
Fig.~\ref{figfletchcirs}), the emission from HC$_3$N should have been apparent at intensities larger than is consistent with CIRS observations.  This 
HC$_3$N band, which is well separated from several other emission features and therefore clearly detectable above the surrounding hydrogen-helium continuum,
provides a particularly stringent test of the vapor influx rates.  The lack of observed HC$_3$N emission here suggests that the gas-to-dust 
ratio of the incoming ring material could be considerably smaller than the upper limit posed by \citet{miller20}; alternatively, the inflow must be recent 
and not represent steady state.

Saturn spectra of the H$_2$O rotational lines at far-infrared wavelengths may also be useful for constraining ring-inflow scenarios (Fig.~\ref{figspecmidir}), 
but firm conclusions are hampered by the 
fact that analyses of the CIRS far-infrared water emission lines have never been published in the refereed literature. However, clear detections of 
far-IR lines of H$_2$O have been reported in preliminary CIRS analyses \citep[e.g.,][]{hesman15,bjoraker19}, suggesting that our model predictions 
can ultimately be compared with the CIRS data.  Based on global-average observations of H$_2$O lines detected by the \textit{Infrared Space Observatory} 
\citep{feuchtgruber97,moses00b} and the \textit{Submillimeter Wave Astronomy Satellite} \citep{bergin00}, and based on spatially resolved H$_2$O lines 
detected by \textit{Herschel} \citep{cavalie19}, the H$_2$O line-to-continuum ratios predicted by Cases A, C, E, \& F would be too large in comparison 
with observations, whereas those for Cases B \& D that have no H$_2$O flowing in from the top of the atmosphere (and just H$_2$O produced photochemically 
from CO) would be too small.  

Although CH$_3$CN is abundant enough to potentially have spectral signatures in our Cases A--F models, the mid-infrared bands at 900--930 cm$^{-1}$ 
are weak and are likely to be obscured by other atmospheric emission or absorption bands (e.g., C$_2$H$_4$, NH$_3$, PH$_3$).  The best chance for detection 
may be at millimeter and sub-millimeter wavelengths \citep[e.g.,][]{lellouch85,marten02}.   

\begin{figure*}[!htb]
\vspace{-14pt}
\begin{tabular}{ll}
{\includegraphics[clip=t,scale=0.30]{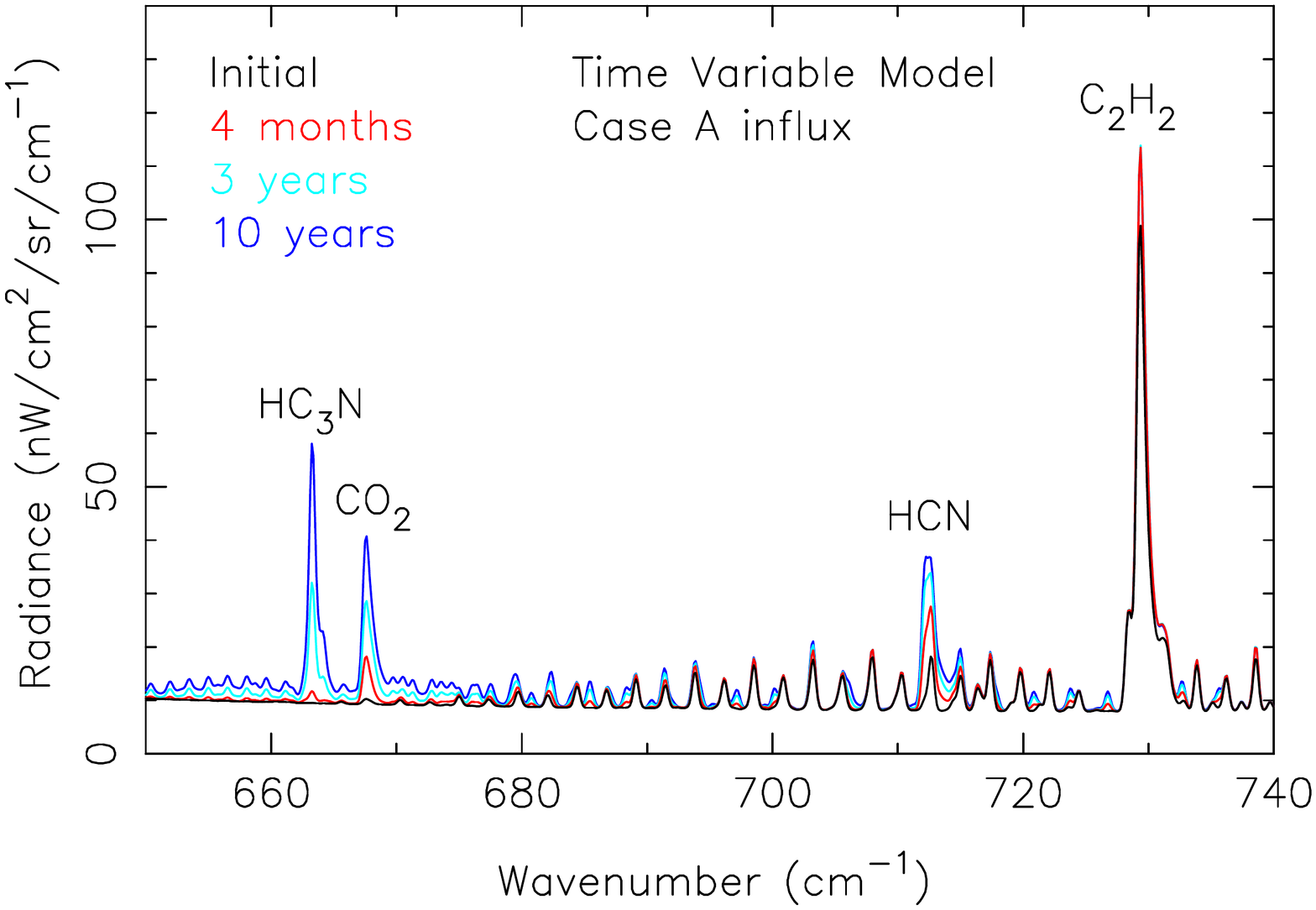}}
&
{\includegraphics[clip=t,scale=0.30]{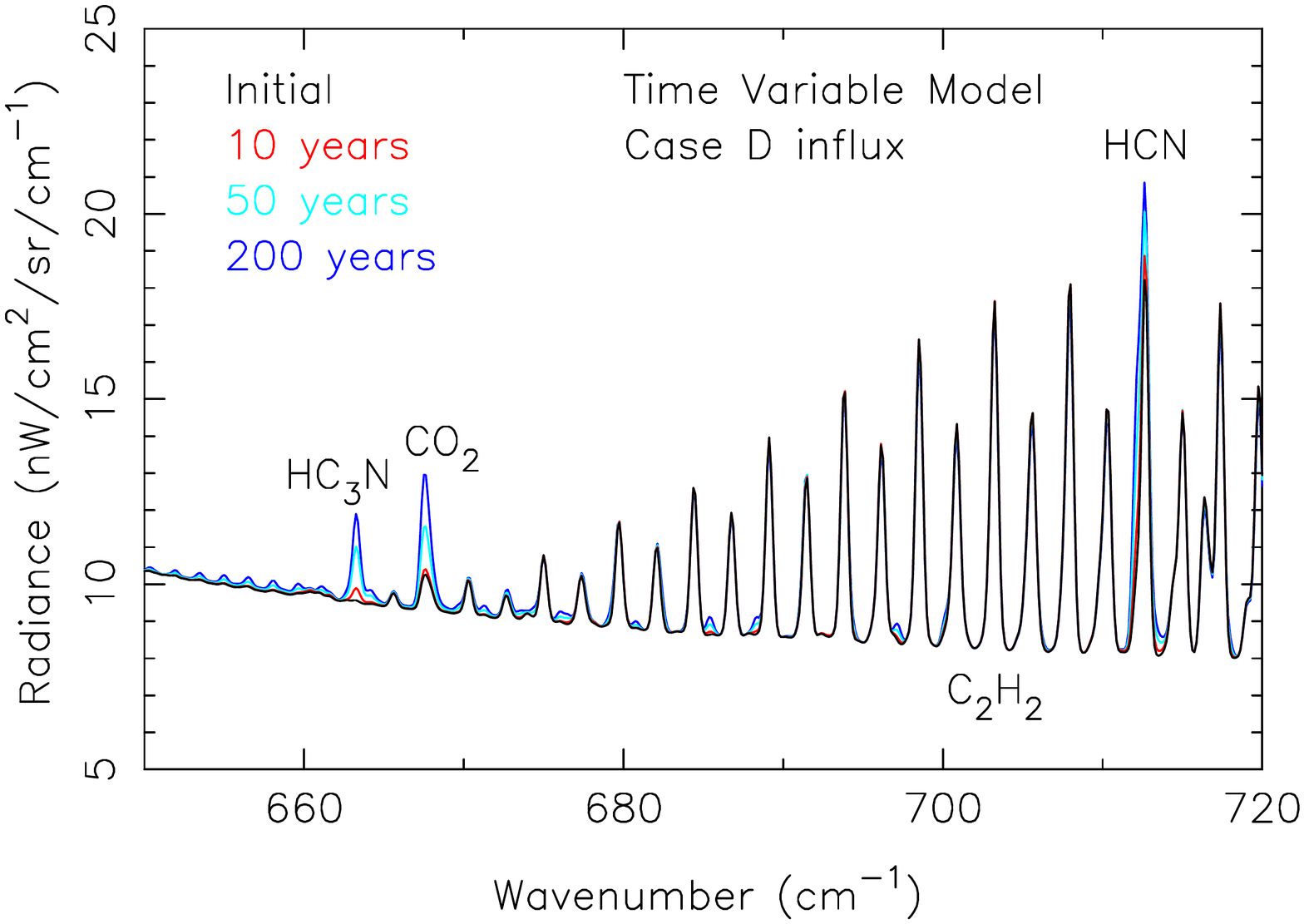}} 
\end{tabular}
\caption{Synthetic spectra from our time-variable models that start with the fully converged 9.2$\deg$ N nominal model with no ring influx and then 
march forward in time assuming a downward ring-vapor flux relevant to Case A (Left) and Case D (Right) --- note the different axis scales for the 
two panels.  
Results are shown for the initial nominal state (black) and various time periods after the onset of the ring-vapor influx (colored lines, as labeled).  
The HC$_3$N, CO$_2$, and HCN spectra features become more obvious with time, with consequences becoming obvious in as little as 4 months with Case A 
assumptions and in $<$50 years with Case D assumptions.}
\label{figspectime}
\end{figure*}

The above analysis assumes that the ring-vapor influx has been occurring for long time scales.  
If the large influx rates inferred from the INMS measurements derive from a transient ring-inflow event, our time-variable model described in 
section~\ref{sec:timevar} can give us some insight into how long it would take for ring-derived species to be observable in the stratosphere.  
The ring vapor that flows in from the top of the atmosphere takes a couple weeks to reach the methane homopause, and then about 4 months to 
reach microbar pressures.  At that point, ring vapor species such as HCN, CO$_2$, and HC$_3$N would be observable at mid-infrared wavelengths 
if the neutral vapor influx rates remain at constant Case-A levels for the entire 4-month period (Fig.~\ref{figspectime}).  It takes 500-1000 yrs 
for the ring vapor to flow down to the troposphere.  The Case E \& F 
influx results are similar to those of Case A, at least in terms of HCN and CO$_2$ --- spectral consequences should be apparent within a couple months.  
However, if only the most volatile species CO, N$_2$, and CH$_4$ are flowing in and those ring-derived molecules spread globally before reaching 
the homopause (e.g., Case D influx assumptions), then the spectral signatures of HCN, CO$_2$, and HC$_3$N could take tens of years to become apparent 
(Fig.~\ref{figspectime}).  The emission bands of these molecules become more and more prominent with time in the years following the onset of 
ring-vapor influx, yet such enhanced emission bands are not seen in CIRS spectra.  This interesting finding severely limits the age of any 
ring-perturbing event, suggesting that it must be contemporaneous with the $\sim$5-month Grand Finale stage of the mission if the INMS measurements 
truly represent a vapor inflow from the rings and global horizontal spreading is not diluting the ring species, or must have begun within the 
last few decades if efficient global spreading occurs and impact vaporization and fragmentation of dust contributes significantly to the INMS signals. 

\begin{figure*}[!htb]
\begin{tabular}{ll}
{\includegraphics[clip=t,scale=0.38]{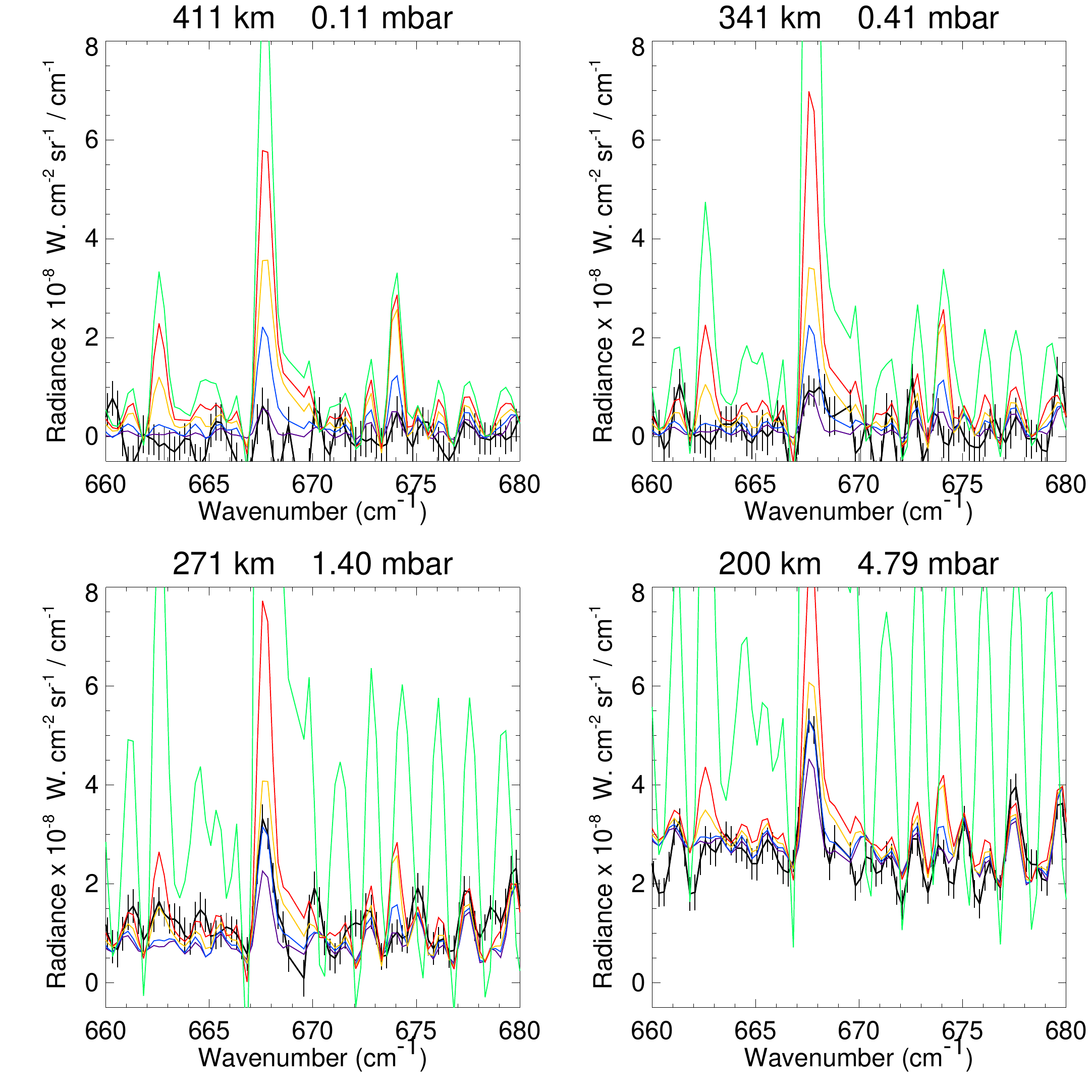}}
&
{\includegraphics[clip=t,scale=0.38]{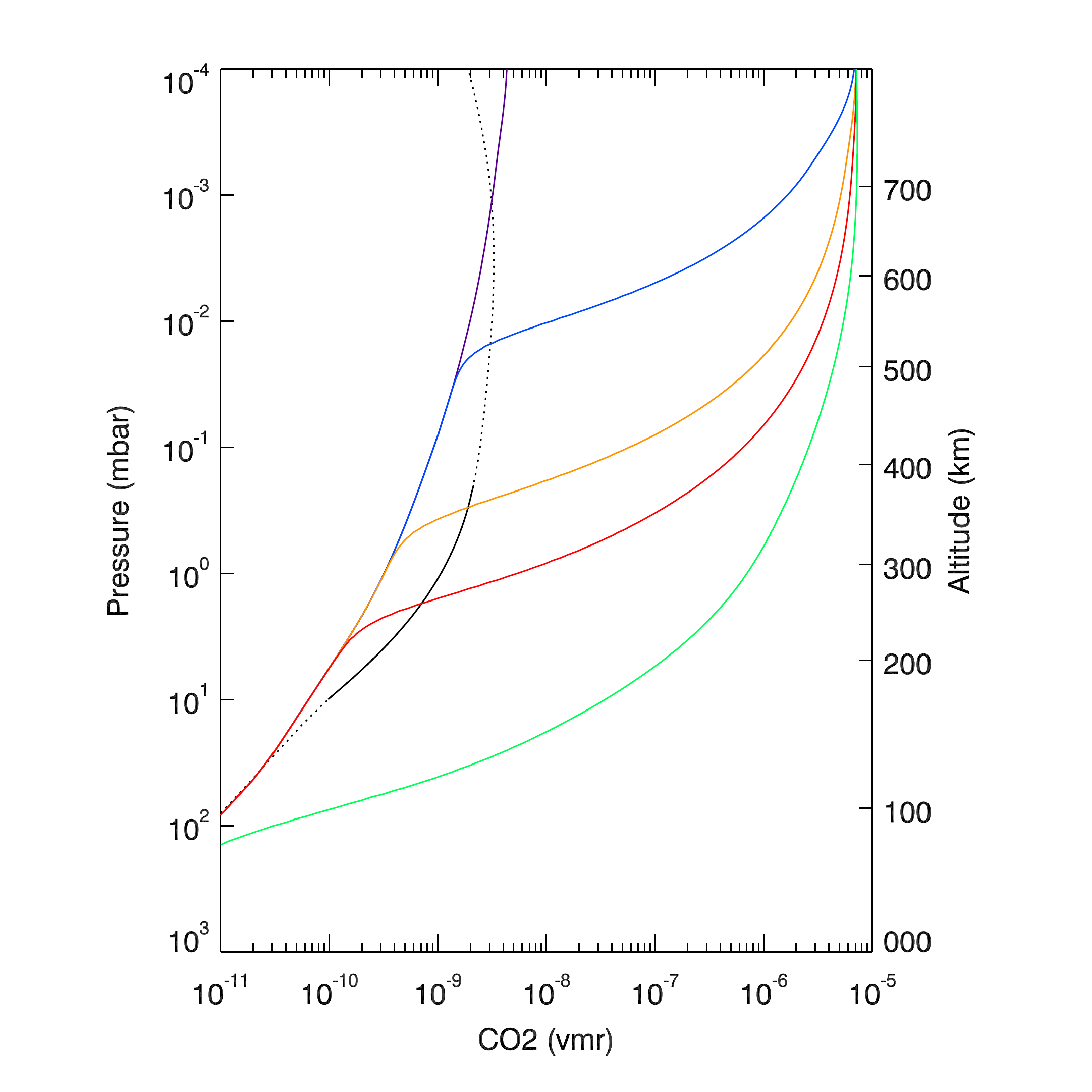}} 
\end{tabular}
\caption{(Left) \textit{Cassini} CIRS limb spectra acquired for the equatorial region in January 2008 (black) at 4 different tangent altitudes, showing 
the detection of CO$_2$, in comparison with synthetic spectra from the nominal 9.2$\deg$ latitude model (purple), and our time-variable model results at
4 months (blue), 3 years (orange), 10 years (red), and 100 years (green) after initiating ring-vapor influx at Case-A values.  (Right) The CO$_2$ 
mixing ratio profile retrieved from the same 2008 CIRS limb dataset (black solid line), compared to our nominal and time variable models, using the same 
color scheme as in the left panel.  Note that the nominal model under-estimates the CO$_2$ abundance in the $\sim$0.1--10 mbar equatorial region and that 
the time-variable model over-estimates the CO$_2$ emission at pressures $\lesssim$0.41 mbar in as little as 4 months of continuous inflow at the 
Case-A influx rates.}
\label{figsandrine2008}
\end{figure*}

\begin{figure*}[!htb]
\begin{center}
\includegraphics[clip=t,scale=0.50]{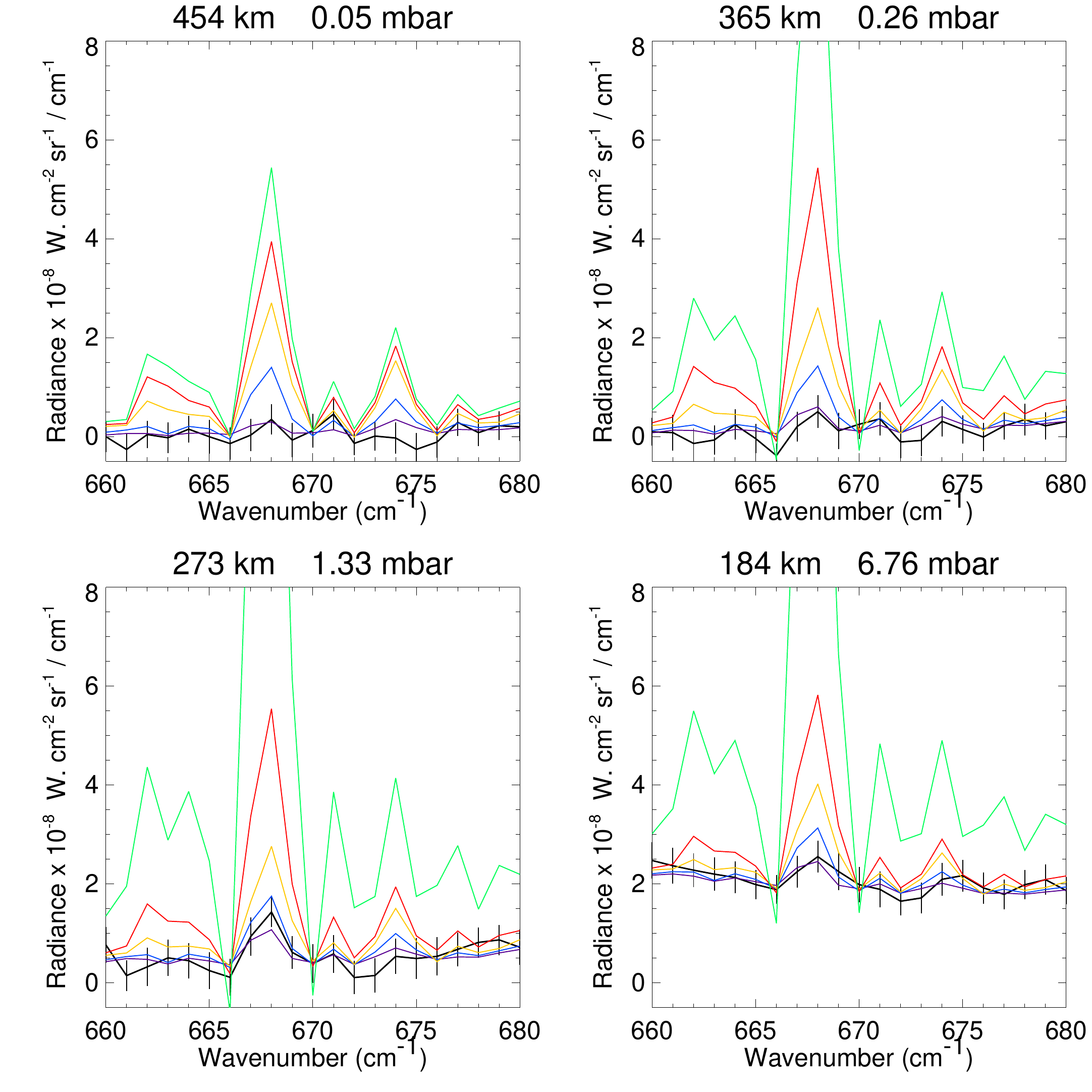}
\end{center}
\caption{\textit{Cassini} CIRS limb spectra showing the detection of CO$_2$ bands at $P$ $\gtrsim$ 1 mbar from 5$\deg$ N planetographic latitude in August 2017 (black) at 
4 different tangent altitudes, as labeled at the top of each image, compared with synthetic spectra from the nominal 9.2$\deg$ latitude model (purple), 
and our time-variable model 4 months (blue), 3 years (orange), 10 years (red), and 100 years (green) after initiating ring-vapor influx at Case A values.}
\label{figsandrine2017a}
\end{figure*}

\begin{figure*}[!htb]
\begin{center}
\includegraphics[clip=t,scale=0.50]{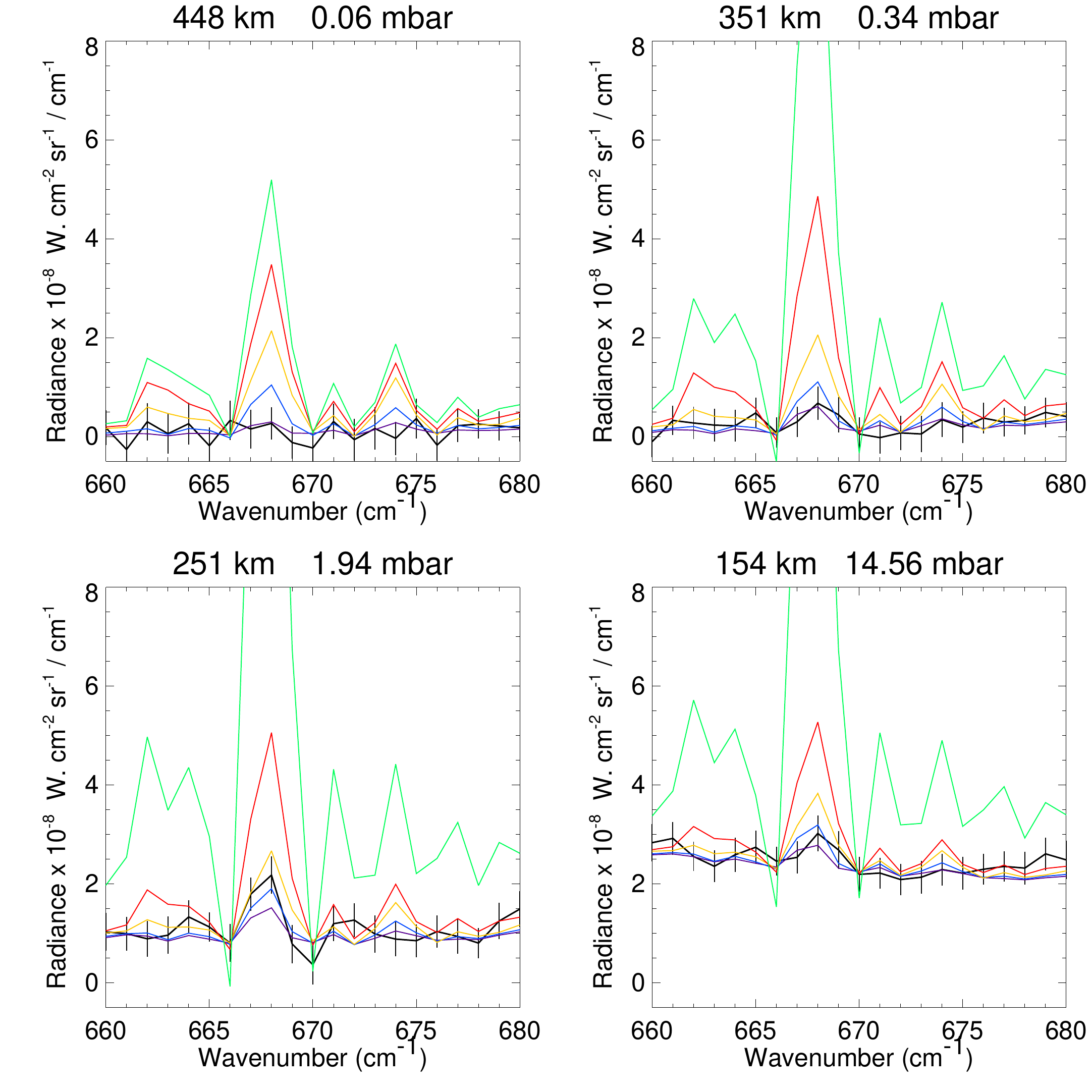}
\end{center}
\caption{Same as Fig.~\ref{figsandrine2017a}, except for \textit{Cassini} CIRS limb spectra from 10$\deg$ S planetographic latitude in August 2017.}
\label{figsandrine2017b}
\end{figure*}

Because CO$_2$ emission at 667-668 cm$^{-1}$ has already been detected in CIRS spectra \citep{abbas13}, that feature might provide particularly
good current constraints on the influx rates of oxygen-bearing ring vapor, especially for observations acquired at limb geometry that probe the 
highest stratospheric altitudes.  We therefore take a more in-depth look at CO$_2$ emission from the CIRS limb observations 
\citep[e.g.,][]{fouchet08,guerlet09,guerlet10,guerlet15}.  Most of the CIRS observations 
in limb geometry were acquired at the lowest 15 cm$^{-1}$ spectral resolution, which makes detecting narrow spectral bands such as the 667 
cm$^{-1}$ CO$_2$ feature more difficult.  A few higher-resolution CIRS limb datasets do exist, however, including the 1 cm$^{-1}$-resolution limb 
observations of Saturn's equatorial region from January, 2008, that has been described more fully in \citet{guerlet15}, and the 3 cm$^{-1}$-resolution 
limb observations of 5$\deg$ N and 10$\deg$ S planetographic latitude from August, 2017, that we examine further here.  The 2017 data are of lesser 
quality (lower signal-to-noise ratio) than the 2008 data because of a shorter acquisition time and a larger field of view, the latter of which yields 
a coarser vertical resolution.  Nevertheless, CO$_2$ is detected in all three of these 2008 and 2017 CIRS limb data sets (Figs.~\ref{figsandrine2008}, 
\ref{figsandrine2017a}, \& \ref{figsandrine2017b}).  

Figure~\ref{figsandrine2008} shows the 2008 equatorial CIRS limb observations in the relevant CO$_2$ spectral region, along with synthetic spectra 
from the Case A time-variable model at various points in time after initiation of the ring inflow.  Carbon dioxide is clearly detected in these data 
at abundances that are actually greater at $\sim$0.1-10 mbar than the initial nominal-model predictions with no ring inflow (see the right panel of 
Fig.~\ref{figsandrine2008}).  However, at higher tangent altitudes ($P$ $\lesssim$ 0.4 mbar), the CIRS data already become incompatible with the Case A 
inflow predictions 4 months after the ring inflow was initiated (blue curves in Fig.~\ref{figsandrine2008}).  The same conclusion can be made for 
the 2017 observations that are shown in Fig.~\ref{figsandrine2017a} and \ref{figsandrine2017b}.  
The CIRS limb observations therefore provide additional important constraints on the magnitude and/or timing of the influx of ring-derived oxygen-bearing 
species.  If the INMS-derived influx rates of CO (which is the main progenitor of CO$_2$ in the stratosphere, see sections \ref{sec:casea}--\ref{sec:casef} 
above) from \citet{serigano22} and \citet{miller20} are taken at face value, then a continuous CO influx at the inferred rate could not have been in 
operation at such a high magnitude for more than a few months, without there being evidence for larger CO$_2$ mixing ratios in the CIRS limb data.  
Based on the retrieved versus predicted column abundances of CO$_2$, observational consequences in the CIRS limb data would become apparent 
$\sim$50 years after any ring event in which the Case D influx scenario is initiated, such that only the most volatile species enter the atmosphere 
as vapor and complete global spreading occurs before the CO reaches the homopause.  These Case A and Case D scenarios effectively bracket the age 
of any ring-derived vapor inflow event.

\begin{figure*}[!htb]
\begin{center}
\includegraphics[clip=t,scale=0.50]{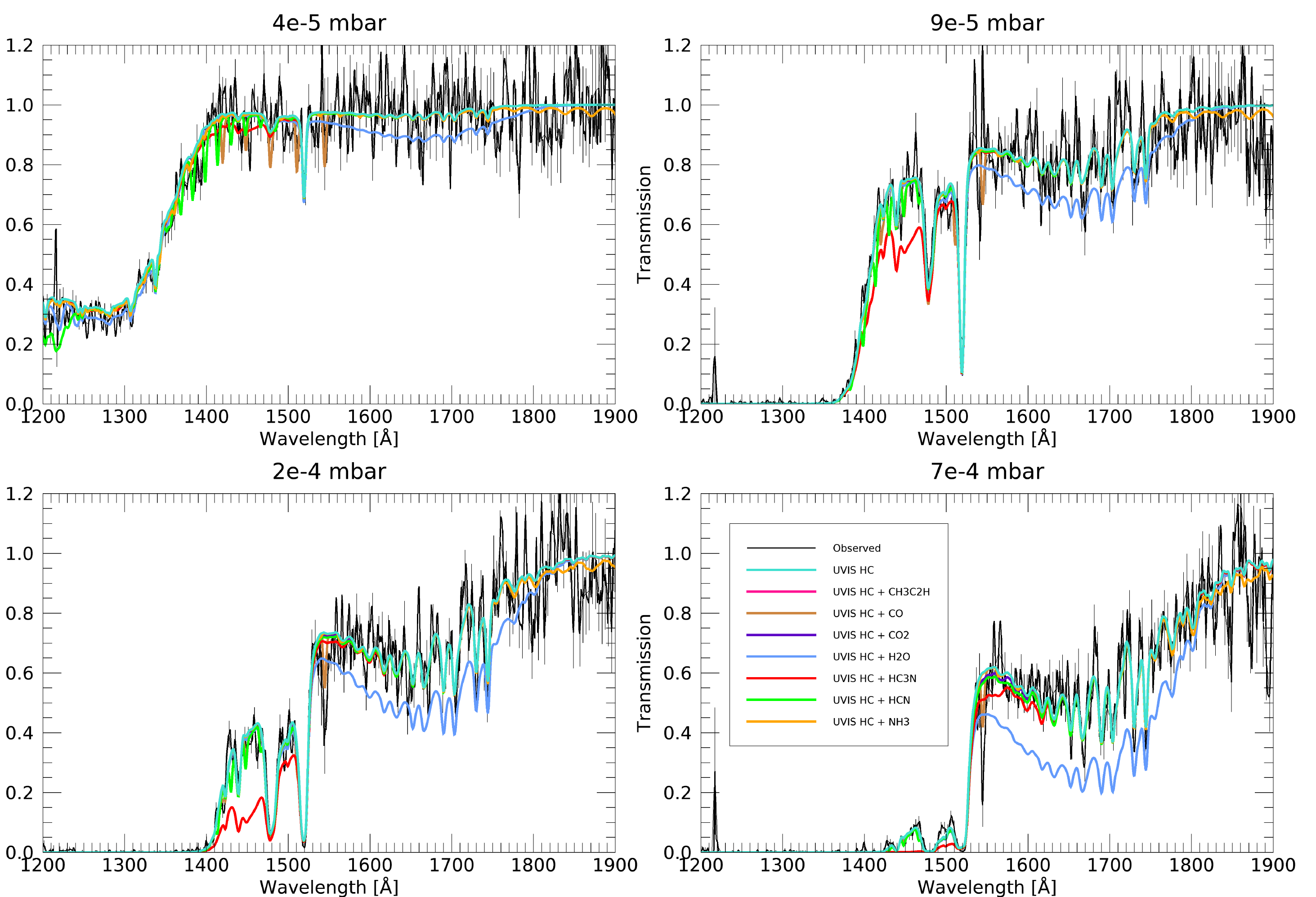}
\end{center}
\caption{Line-of-sight transmission of the stellar spectrum through Saturn's atmosphere from the \textit{Cassini} UVIS $\epsilon$ Ori stellar occultation 
performed on June 25, 2017 at 12$\deg$ planetocentric latitude (black curves) compared to the synthetic transmission predicted assuming the retrieved 
atmospheric vertical profiles of CH$_4$, C$_2$H$_2$, C$_2$H$_4$, and C$_2$H$_6$, along with the retrieved C$_6$H$_6$ profile at tangent 
pressures $P$ $>$ 1.38$\scinot-4.$ mbar --- where collectively these hydrocarbons are called ``HC'' in the legend --- along with the addition of various 
ring-derived vapor species from our Case A model (colored lines, as labeled), for tangent radii relevant to $P$ = 4$\scinot-5.$ 
mbar  (Top Left), 9$\scinot-5.$ mbar (Top Right), 2$\scinot-4.$ mbar (Bottom Left), and 7$\scinot-4.$ mbar (Bottom Right).  
Excess absorption from H$_2$O (blue curves), HC$_3$N (red curves), and to a lesser extent HCN (green curves) 
becomes apparent at $P$ $\gtrsim$ 9$\scinot-5.$ mbar, such that the predicted Case A abundances of these species are inconsistent 
with the UVIS occultation observations.}
\label{figuvis_casea}
\end{figure*}

\subsubsection{Comparisons to ultraviolet observations}\label{sec:uvobs}

Some of the inflowing ring vapor species are also abundant enough and have strong-enough ultraviolet absorption bands that they could be detected from 
ultraviolet solar and stellar occultations at Saturn.  We have therefore analyzed the \textit{Cassini}/UVIS occultation data from the $\epsilon$ Ori 
stellar occultation that was acquired on June 25, 2017 at 12$\deg$ N planetocentric latitude.  First, 
the background atmospheric structure and vertical profiles of CH$_4$, C$_2$H$_2$, C$_2$H$_4$, and C$_2$H$_6$ were derived from 
the UVIS occultation light curves through a retrieval process \citep[for details on the procedure, see][]{koskinen15sat,koskinen16,koskinen18,brown20,brown21}.  
For tangent altitudes less than 820 km above the 1-bar level (i.e., for $P$ $<$ 1.38$\scinot-4.$ mbar), vertical profiles for C$_6$H$_6$ were also 
retrieved, as the occultations show evidence for its presence at deeper stratospheric pressures.  Then, the Case A model results for the CH$_3$C$_2$H, CO, 
CO$_2$, H$_2$O, HC$_3$N, HCN, and NH$_3$ densities were interpolated in log pressure 
onto an altitude grid relevant to the 12$\deg$ N latitude occultation.  The species' column densities along the UVIS occultation line of sight were then 
calculated using the discretized method outlined in \citet{quemerais06} and multiplied by the species' ultraviolet cross sections to determine line-of-sight 
optical depth as a function of wavelength.  After summing the optical depths 
over different combinations of species, the transmission of the starlight through Saturn's atmosphere was calculated and convolved with the UVIS point-spread 
function and binned to the UVIS wavelength spacing.  These synthetic transmission spectra as a function of tangent pressure in the atmosphere are
compared to the observed transmission in Fig.~\ref{figuvis_casea}.  All model curves in this figure contain identical retrieved profiles for CH$_4$, 
C$_2$H$_2$, C$_2$H$_4$, and C$_2$H$_6$ (with C$_6$H$_6$ also included for the two deeper pressures); then, each of the colored curves contains a single 
additional species at its Case A vertical profile. 

Figure \ref{figuvis_casea} demonstrates that the steady-state Case A abundances of H$_2$O and HC$_3$N are inconsistent with the UVIS observations from this 
low-latitude occultation, with H$_2$O providing too much absorption in the $\sim$1520-1800 \AA\ region (see blue curves compared to observations) and 
HC$_3$N providing too much absorption in the $\sim$1400-1580 \AA\ region (see red curves compared to observations).  The predicted HCN abundance from Case A 
also leads to some excess absorption compared to observations (see green curves), but other species including CH$_3$C$_2$H, CO, CO$_2$, and NH$_3$ do not 
significantly affect the predicted transmission (see other colored curves that frequently plot on top of each other).  

\begin{figure*}[!htb]
\begin{center}
\includegraphics[clip=t,scale=0.50]{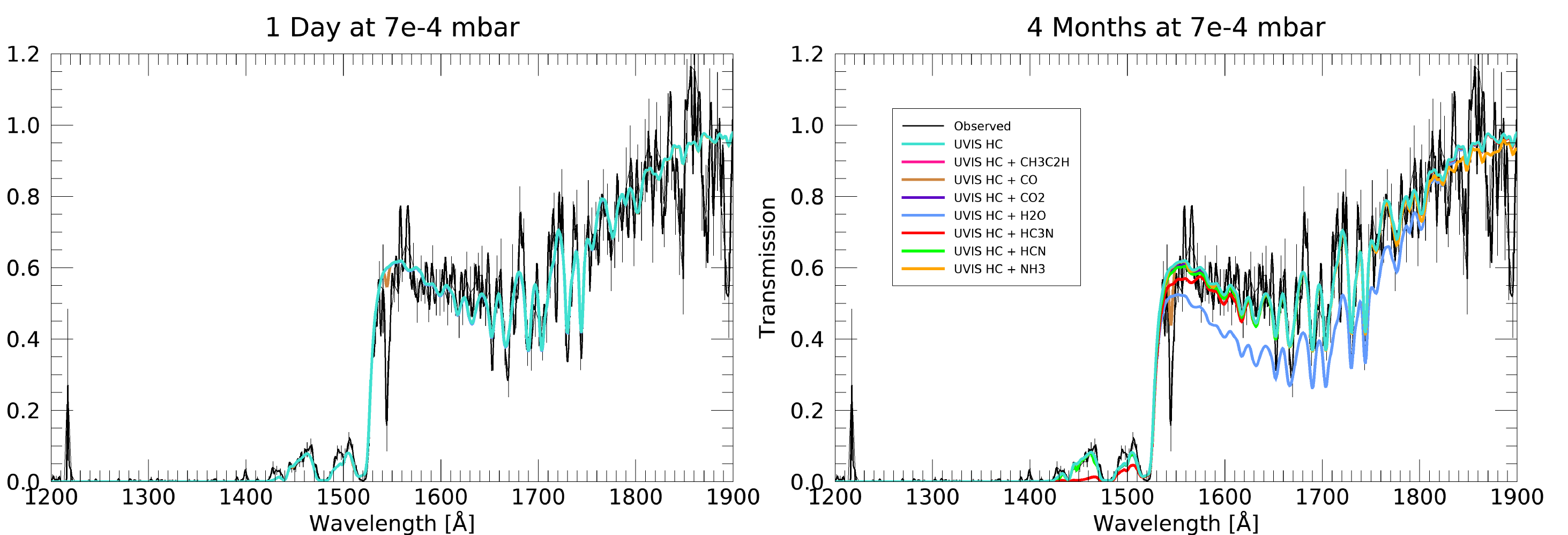}
\end{center}
\caption{Same as Fig.~\ref{figuvis_casea}, except the observations (black) are compared to the time-variable model simulations for the Case A influx 
after 1 day (Left) and 4 months (Right) at a tangent radius corresponding to a pressure of 7$\scinot-4.$ mbar.  Excessive absorption from H$_2$O, 
HC$_3$N, and C$_6$H$_6$ would appear in as little as a 4 months of continuous ring-vapor influx at Case A vapor influx rates.}
\label{figuvis_timevar}
\end{figure*}

In Fig.~\ref{figuvis_timevar}, we compare the UVIS occultation observations with simulations of our time-variable model that starts with no ring-vapor 
influx and then marches forward in time with constant Case A influx rates.  Here we can see that the predicted H$_2$O and HC$_3$N absorption notably 
exceeds that observed at a tangent pressure of 7$\scinot-4.$ mbar within 4 months of the onset of the ring-vapor inflow.  The same would be true of our other high 
H$_2$O- and ring-vapor inflow Cases E \& F.  However, we find that models that only consider the most volatile species N$_2$, CO, and CH$_4$ (i.e., Cases 
B \& D) remain consistent with the UVIS occultation data even 1000 years after the onset of the continuous inflow (not shown in Fig.~\ref{figuvis_timevar}.  
Under the Case B \& D inflow scenarios, less stratospheric H$_2$O and HC$_3$N are photochemically produced, and the line-of-sight column abundances do not 
grow large enough to affect the atmospheric transmission in the ultraviolet.

\begin{figure*}[!htb]
\begin{center}
\includegraphics[clip=t,scale=0.75]{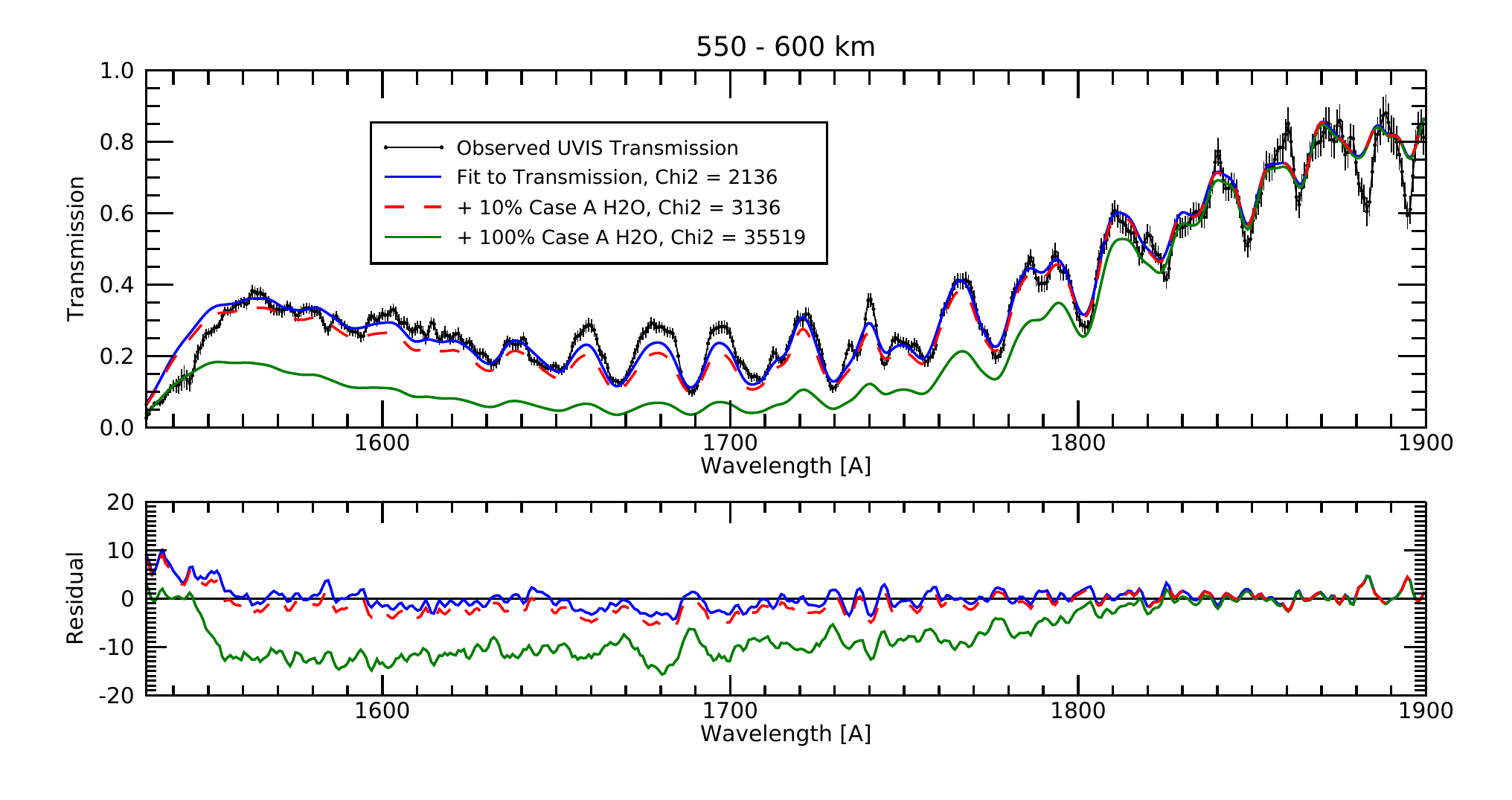} \\
\includegraphics[clip=t,scale=0.75]{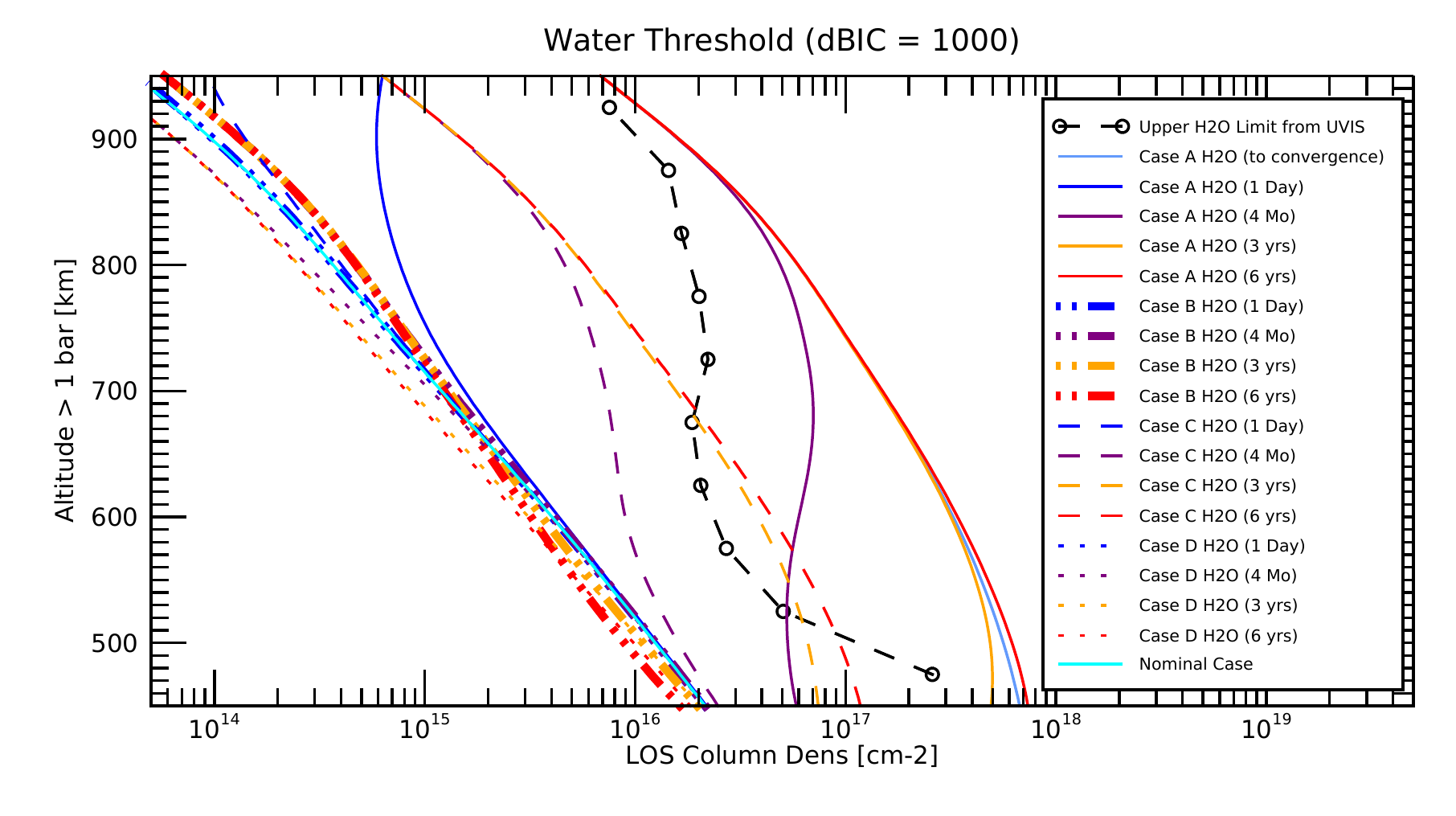}
\end{center}
\caption{(Top) UVIS transmission spectrum from the $\epsilon$ Ori occultation at 550--600 km altitude above the 1-bar level (0.01-0.005 mbar; black data points) 
in comparison to a model that includes the retrieved vertical profiles of CH$_4$, C$_2$H$_2$, C$_2$H$_4$, C$_2$H$_6$, and C$_6$H$_6$ (blue curve), as well as 
to the same model with the addition of our Case A vertical profile of H$_2$O (green curve), and to a model in which the H$_2$O profile is set to 10\% of the Case A 
abundance profile (red dashed).  (Middle) Residuals (in percent) for the three models shown in the top panel. (Bottom) The maximum allowable water abundance 
(open circles), in terms of line-of-sight column density, as a function of altitude above the 1-bar level, as determined from a dBIC value of 1000 (see text), 
compared to the water profiles from various of our models, as labeled.}
\label{figdbic}
\end{figure*}

The UVIS occultation observations allow constraints to be placed on the inflow of H$_2$O to Saturn's equatorial region.  To quantify these constraints, we 
use the Bayesian Information Criterion (BIC) to determine the amount of water that can be added before the fit to the data becomes statistically 
worse than our best-fit retrieval of the hydrocarbons CH$_4$, C$_2$H$_2$, C$_2$H$_4$, C$_2$H$_6$, and C$_6$H$_6$ from the UVIS occultations.  
Starting with the Case A model, the column abundance of H$_2$O is adjusted up or down until the change in BIC (delta BIC, or dBIC) becomes equal to 
1000, which we take to denote decisive evidence against the profile with water added \citep[e.g., adapted from][]{kass95}.  To improve the signal-to-noise 
ratio and to better define the 
threshhold H$_2$O value, we first binnned the UVIS transmission in 50-km altitude intervals.  Figure~\ref{figdbic} shows the results of this procedure.  The 
nominal model and Cases B and D all fit within the dBIC = 1000 threshold at 550-600 km altitude (0.01--0.005 mbar pressure), as
do the Cases A and C at early times after onset of the ring influx (e.g., 1 day and 4 months after 
ring-vapor influx was initiated).  At the 3-year point or later after the onset of ring-vapor influx, the H$_2$O profiles from influx rates relevant to 
Cases A and C can be ruled out from the UVIS occultation data.  The steady-state equatorial influx rate of water must be less than 2.3$\scinot7.$ molecules 
cm$^{-2}$ s$^{-1}$ (i.e., Case C) to remain consistent with the UVIS occultations, corresponding to a mass influx of 44 kg s$^{-1}$ if we 
assume the water influx is confined to $\pm$8$\deg$ latitude (i.e., 0.4--10\% of the water mass influx rate determined from INMS for the different Grand 
Finale orbits \citep{waite18,serigano22}). However, a higher transient H$_2$O 
influx rate can be accommodated if the inflow were triggered within a few months of the \textit{Cassini} Grand Finale orbits.

\subsubsection{Summary of observational implications}\label{sec:sumobs}

Based on the above comparisons with \textit{Cassini} infrared observations, we conclude that if vapor makes up a substantial fraction of 
the inflowing ring material, then the surprisingly large influx rates inferred by the INMS must represent a recent, transient event that was 
triggered a few months to a few tens of years before the 2017 INMS measurements, or there would be spectral consequences in the infrared that are 
not seen by \textit{Cassini}.  A similar conclusion, at least in a qualitative sense, was reached by \citet{cavalie19} when 
comparing the INMS-derived H$_2$O influx rates to the significantly smaller influx rates inferred from stratospheric H$_2$O observations on 
Saturn from the \textit{Herschel Space Telescope} acquired in 2010--2011.  
If horizontal spreading 
of the vapor away from low latitudes in the lower thermosphere is relatively ineffective, then 
the INMS-inferred ring inflow may have been triggered during the Grand Finale stage of the mission itself, which lasted 5 months.  Under that scenario, 
a 2017 ring-influx event, followed by an assumed continuous total influx rate of $\sim$5000--70,000 kg s$^{-1}$ \citep[as inferred from the INMS 
measurements,][]{waite18,perry18,serigano22} over the intervening half-decade since the end of the \textit{Cassini} mission, would lead to an equatorial 
stratosphere that should now show obvious signs of this inflow at mid-infrared wavelengths where HC$_3$N, CO$_2$, and HCN have emission bands.
On the other hand, if meridional spreading in the lower thermosphere is effective and if only the most 
volatile species detected by INMS are derived from incoming ring vapor, with the other mass signals deriving from impact vaporization and fragmentation 
of dust, then the inflow 
event could have been triggered at any time within the past $\sim$50 years, and infrared observational consequences might not be imminent, depending on when the 
actual event began.  Searches for spectral signatures of HCN, HC$_3$N, CH$_3$CN, and CO$_2$ and their variability over time with the \textit{James Webb 
Space Telescope} (JWST) or ground-based infrared and sub-millimeter facilities could be valuable in further constraining the nature, timing, and 
magnitude of this apparent transient ring-inflow event.

Alternatively, if the inflow of ring material is steady rather than deriving from some anomalous recent event, then the INMS measurements must be caused 
predominantly by impact vaporization and fragmentation of dust particles within the instrument during \textit{Cassini's} atmospheric passage, and those particles 
must not otherwise ablate in Saturn's atmosphere.  The CIRS observations place strong limits on the steady-state influx of external 
oxygen- and nitrogen-bearing species to Saturn, and even our Case D scenario would produce observable stratospheric consequences in the infrared. 
The HC$_3$N column-abundance results and model-data comparisons, in particular, limit the N$_2$ influx to at least a factor of $\sim$4--11 smaller 
than indicated by Case D.  Although uncertainties in the photochemistry of HC$_3$N limit firmer quantitative conclusions with respect to the N$_2$ 
influx (see Appendix B), spreading alone does not appear sufficient to dilute the nitrogen vapor enough to explain the lack of 
stratospheric HC$_3$N on Saturn.  The bulk of the nitrogen must therefore either be tied up in dust grains, or the inflow must be recent.

The \textit{Cassini} UVIS occultations provide additional constraints on the amount of H$_2$O in Saturn's equatorial stratosphere.  The resulting 
constraints are in line with those determined from comparisons of models with the \textit{Cassini} CIRS observations.  Continuous Case A, C, E, and F water 
influx rates are determined to be inconsistent with UVIS observations, such that the equatorial H$_2$O influx must be below 2.3$\scinot7.$ cm$^{-2}$ 
s$^{-1}$ (i.e., $\sim$44 kg s$^{-1}$ in a $\pm$8$\deg$ latitude band) from a steady-state standpoint.  However, if the incoming 
equatorial water spreads globally before reaching the stratosphere or if the 
inflow is triggered by an event that occurred within a few months of the Grand Finale \textit{in situ} measurements, then the larger equatorial H$_2$O influx 
rates that were inferred from INMS can be accommodated while still remaining consistent with the UVIS observations.  This observational 
upper limit of 2.3$\scinot7.$ H$_2$O molecules cm$^{-2}$ s$^{-1}$ flowing into Saturn in steady state is fully consistent with the previously identified 
more modest ``ring rain'' source \citep[e.g.,][]{connerney84,odonoghue13,odonoghue19} and previous inferrences of the global-average and low-latitude 
H$_2$O influx rates on Saturn \citep[e.g.,][]{moses00b,hartogh11,moore15,moses17poppe,cavalie19}, which are all inferred to be well below this upper limit.  

\subsection{Additional considerations with respect to a specific ring-inflow event}\label{sec:event}

As just discussed in section \ref{sec:obs}, the fact that none of these predicted spectral consequences of the ring-vapor inflow have been reported 
from \textit{Cassini} remote-sensing instruments to date suggests that the inflow might have been caused by a perturbation of the rings that occurred
relatively recently, within the past 4 months to 50 years.  Based on an analysis of visible-wavelength \textit{Cassini} Imaging Science Subsystem 
(ISS) D-ring observations, \citet{hedman19} report the 
detection of a series bright clumps in the narrow ``D68'' ringlet in Saturn's inner D ring that formed sometime in 2014 or early 2015, and whose morphology 
and brightness are observed to vary with time in the last 2-3 years of the mission.  \citet{hedman19} suggests that these clumps consist of fine-grained 
material generated by collisions of few-meter-sized or larger objects within the D68 ringlet, and that some event (most likely a dynamical event 
rather than an impact from an external body) occurred within the 2014--2015 time frame to generate this new collisional debris.  Small dust grains 
and vapor molecules released during the collisional event could make their way into Saturn's upper atmosphere, potentially providing a connection to the 
surprisingly large influx of ring material observed by the \textit{in situ} instruments during the Grand Finale.  However, unless the liberated 
vapor from this collisional event took a couple years to reach Saturn, then this 2014--2015 D68 ringlet perturbation is unlikely to have triggered 
the observed strong vapor inflow, or the stratospheric consequences shown in section~\ref{sec:obs} would have been already obvious to CIRS in 2017.  That 
statement assumes that the INMS measurements indeed came from ring-derived vapor present in the atmosphere and not from dust that was vaporized 
upon impact with the spacecraft or instrument.  If only the most volatile material is entering the atmosphere as vapor and if that vapor spreads 
globally before reaching the stratosphere (i.e., as in our Case D model), then this event in the D68 ringlet could indeed be responsible for the 
inflowing material seen by INMS.

However, if direct vapor were predominantly responsible for the INMS measurements, and if that vapor did not spread globally before reaching the stratosphere, 
then the influx event must have been triggered contemporaneously with 5-month Grand Finale stage of the mission itself --- that is, at some point between 
\textit{Cassini}'s first pass through the gap between the atmosphere and innermost rings in late April, 2017, and the first deep dive into 
Saturn's atmosphere for orbit 288 in mid-August 2017.  This apparent timing begs the question of whether \textit{Cassini} itself could have 
caused the inflow.  
While an intriguing possibility, this suggestion seems unlikely.  The spacecraft mass is too low to have generated gravitational perturbations or 
density waves within the rings.  The dust density and ring thickness 
within this gap region are small \citep{hsu18sat,ye18} and the spacecraft cross section is small (2--3$\scinot5.$ cm$^{2}$), so vaporization of dust 
particles in the ring plane during the earlier Grand Finale orbits falls far short of quantitatively accounting for the necessary influx of vapor.  
Moreover, impacts of the orbiter with larger particles that could have generated larger amounts of vapor would have been catastrophic to the spacecraft.  
A cascade of collisions similar to a space-debris event within the D ring could possibly generate the necessary amount of dust and/or vapor, but again, it is 
difficult to imagine how the \textit{Cassini} spacecraft could have interacted with ring dust in such a way as to trigger such an event.  So, we are left 
with either a coincidence in the timing of the vapor inflow to within the Grand-Finale orbital time period, or with the conclusion that the ring inflow 
identified by INMS consisted predominantly of dust, with little in the way of vapor.

\subsection{Other implications with respect to the nature of the incoming ring material}\label{sec:other}

The possibility that the incoming ring material consists largely of dust, rather than vapor, that then vaporizes when impacting within the INMS 
instrument can potentially reconcile the large abundance of heavy neutral molecules inferred by INMS with the lack of evidence of ring-derived vapor 
in the stratosphere, but only if that incoming dust never ablates in the atmosphere.  If a substantial fraction of the incoming ring dust ablates, 
then oxygen- and nitrogen-bearing vapor would be released into the atmosphere anyway, reintroducing the inconsistency between INMS and CIRS/UVIS
data.  Interplanetary dust grains would enter Saturn's atmosphere fast enough to fully ablate \citep{moses17poppe}. However, dust grains from the 
ring system would be coming in more slowly and would likely be smaller if their source is atmospheric erosion (\citealt{mitchell18,perry18}; see 
also \citealt{liu14,ip16}), which makes the incoming ring grains more likely to survive without ablating. \citet{hamil18} 
demonstrate that ring particles with radii less than a few 10's of nm entering the atmosphere at velocities less than 25 km s$^{-1}$, assuming an 
incidence angle with respect to the horizontal of 20$\deg$, would not experience much in the way of ablation.  Nanograins coming in to Saturn's upper 
atmosphere from the rings were detected and characterized by MIMI \citep{mitchell18} and CDA \citep{hsu18sat}, but the inferred influx rate of such 
grains was only 5 kg s$^{-1}$ \citep{mitchell18}, in contrast to the $\gtrsim$10,000 kg s$^{-1}$ of ring material detected by INMS 
\citep{waite18,perry18,serigano22}.  The differences in fluxes here led \citet{perry18} to suggest that much of the mass flux is entering the 
atmosphere at masses below the MIMI lower limit of $\sim$8000-10,000 amu, although we also suggest that larger grains above the MIMI upper limit of 
$\sim$40,000 amu \citep{mitchell18} and on up to a maximum of $\sim$3$\scinot8.$ amu (i.e., a radius of 50 nm, assuming spherical grains of 
density 1 g cm$^{-3}$) could be contributing to the mass inflow without adding significant amounts of ablated vapor to the atmosphere that cannot 
be accounted for by CIRS and UVIS.

Small dust grains would not directly affect neutral stratospheric chemistry in the major ways described in our photochemical models. However, the particles 
could indirectly influence the chemistry through facilitating aerosol nucleation, providing surfaces upon which hetereogeneous chemical reactions 
could occur, or enhancing absorption of solar radiation to potentially enhance local atmospheric heating.  Interactions of the ring particles with 
ionospheric electrons and ions could also potentially affect ionospheric structure and charge balance.  These possibilities are worth exploring in future 
work.

\section{Conclusions}\label{sec:conclusions}

During the 2017 Grand Finale stage of the \textit{Cassini} mission, ring debris was discovered to be flowing into Saturn's equatorial atmosphere
at a surprisingly large rate of $\gtrsim$10,000 kg s$^{-1}$ \citep{waite18,perry18,serigano22}.  We have developed a coupled ion-neutral photochemical model 
of Saturn's thermosphere, ionosphere, and stratosphere to investigate the extent to which this influx of ring material could affect atmospheric 
chemistry.  We find that if a substantial fraction of the ring debris identified by the \textit{Cassini} INMS instrument were to enter the 
atmosphere as vapor or become vaporized within the atmosphere --- an assumption that is consistent with current analyses of the INMS data 
\citep{waite18,miller20,serigano22} --- then our models predict that this ring-derived vapor would strongly affect both the ionospheric composition 
and neutral stratospheric composition of Saturn.  The fact that the predicted stratospheric consequences of the ring influx have not been observed 
by any remote-sensing instruments onboard \textit{Cassini} to date suggests (1) that the inferred influx of ring vapor represents an anomalous, very 
recent, transient condition, and/or (2) that much of the material detected by INMS results from solid particles impacting within the instrument and that 
these particles would not otherwise have ablated and released vapor into Saturn's atmosphere.  Global horizontal spreading of the ring vapor in the 
lower thermosphere could potentially reduce the effective influx of ring vapor into the stratosphere by an order of magnitude if meridional winds or 
diffusion are strong enough, but horizontal spreading alone cannot ``hide'' the ring vapor from stratospheric observations without either scenario 
(1) or (2) above being true.  Our modeling can therefore help place limits on the origin and timing of the apparent ring-inflow event and/or the 
nature and properties of the incoming ring material (i.e., solid grains vs.~vapor, particle-size constraints, vapor-composition constraints).

Based on our theoretical calculations and spectral predictions, including model-data comparisons with new retrievals from \textit{Cassini} 
CIRS and UVIS observations, we make the following conclusions with respect to the consequences of the \textit{in situ} Grand Finale results for 
Saturn's atmospheric composition:

\begin{itemize}

\item Photochemistry within Saturn's upper thermosphere cannot explain the origin of the complex molecules identified in the INMS mass spectra, 
as was originally hypothesized by \citet{serigano20,serigano22}.  Vertical diffusion time scales of vapor molecules in the upper thermosphere are much 
shorter than chemical production and loss time scales of the major ``parent'' molecules (e.g., CH$_4$, H$_2$O, NH$_3$, CO, N$_2$) that are responsible 
for generating other photochemical products in the thermosphere.  The parent molecules flow through the upper thermosphere relatively unscathed, and our models do 
not predict sufficient production of heavy organics needed to explain the higher-mass spectra ($m/z$ $\gtrsim$24 amu) recorded by INMS.  Therefore, the 
heavy species identified in the INMS data either must have already been flowing into the atmosphere (e.g., from vapor produced from collisions 
within the D-ring region) or must have formed from impact vaporization and fragmentation of dust particles within the INMS instrument during the 
high-speed spacecraft passage through the atmosphere.  

\item Photochemistry within Saturn's lower thermosphere and ionosphere cannot explain the apparent INMS-UVIS inconsistency in CH$_4$ mixing 
ratios between the thermosphere and upper stratosphere \citep[cf.][]{koskinen16,koskinen18,brown21,waite18,yelle18,miller20,serigano22}.  Although some 
photochemical loss of the aforementioned parent species does occur in the lower thermosphere/ionosphere at altitudes below the deepest INMS 
measurements, our models do not predict sufficient destruction of methane above its homopause region to account for the roughly order-of-magnitude 
decrease in the mixing ratio of CH$_4$ from the middle thermosphere sampled by the INMS measurements to the upper stratosphere probed by the 
UVIS occultations.  Efficient global horizontal spreading of the incoming CH$_4$ in the lower thermosphere could potentially resolve this 
INMS-UVIS discrepancy, or again, the discrepancy, could be resolved if the CH$_4$ identified in the INMS mass spectra were derived from 
impact vaporization and fragmentation of dust within the instrument.

\item We confirm the previous predictions of \citet{koskinen16} in that ion chemistry near the methane homopause on Saturn strongly increases 
the production rate of benzene and PAHs, a finding that is also consistent with models of Titan \citep[e.g.,][]{vuitton19} and Neptune \citep{dobrijevic20}.  
In our Saturn models without ring-vapor inflow, the coupled chemistry of ions and neutral species increases the equatorial upper-stratospheric benzene 
abundance by a factor of 4--6 times over models that consider neutral chemistry alone.

\item The inflow of ring vapor strongly affects Saturn's ionospheric chemistry and structure.  Our models predict that H$^+$, H$_3$$\! ^+$, and electron 
densities are notably reduced due to the ring-vapor inflow, and ions such as HCNH$^+$ and NH$_4$$\! ^+$ become dominant constituents within and below the 
main ionospheric peak if HCN and NH$_3$ are present in the incoming vapor.  Both HCN and NH$_3$ have high proton affinities and can outcompete other 
species (e.g., H$_2$O, CO, N$_2$, CO$_2$) in proton-transfer reactions even if the local abundance of HCN and NH$_3$ is smaller.  The predicted composition 
of the ionosphere is very sensitive to both the total influx rate of neutral vapor and to the assumed molecular speciation of that incoming vapor.  If 
HCN and NH$_3$ are not flowing in from the rings, ions such as HCO$^+$ can become correspondingly more important or dominant.  Note that HCNH$^+$, 
NH$_4$$\! ^+$, and HCO$^+$ all satisfy the requirement discussed by \citet{dreyer21} that the dominant ion at at equatorial altitudes below 2000 km 
should have an electron recombination rate coefficient $\lesssim$3$\scinot-7.$ cm$^{-3}$ s$^{-1}$ to explain the combined RPWS/LP and 
INMS data, if the inferred ion densities of \citet{morooka19} are correct.

\item Our INMS-based ring-inflow models do a reasonable job of reproducing the densities of H$^+$, H$_2$$\! ^+$, H$_3$$\! ^+$, and electrons above the main 
ionospheric peak \citep[see][]{waite18,moore18,wahlund18,morooka19,persoon19,hadid19}, but they are inconsistent with the H$^+$ and H$_3$$\! ^+$ 
densities and their relative abundances at the lower altitudes and latitudes relevant to the spacecraft closest approach.  As is discussed in 
\citet{cravens19ioncomp} and \citet{vigren22}, the inferred H$^+$ and H$_3$$\! ^+$ densities in this close-approach region require a very large concentration 
of a neutral species such as CO and/or N$_2$ that can react with H$_3$$\! ^+$ but not H$^+$ and a much smaller concentration of other neutral species such as 
CH$_4$, H$_2$O, NH$_3$, or HCN that react with both H$^+$ and H$_3$$\! ^+$.  These constraints are inconsistent with INMS measurements of neutral species 
at closest approach, and our models do not resolve this discrepancy.

\item If the INMS measurements of neutral molecules truly result from atmospheric vapor and not from impact vaporization and fragmentation of dust, then our 
models suggest that the inferred large inflow rates cannot represent a long-term average flux into Saturn but must reflect a recent, anomalously 
large, transient ring-inflow event, given that infrared spectral signatures of HCN, HC$_3$N, and excess CO$_2$ in Saturn's 
stratosphere become obvious within a couple months of the inflow onset.  If the incoming ring material contains a large vapor fraction that remains 
confined to low latitudes, then the transient event must have been contemporaneous with the Grand-Finale orbits in the last few months of the mission, 
or spectral consequences would have been seen in CIRS and UVIS observations from 2017.  Alternatively, assuming that only the most volatile molecules 
identified in the INMS $m/z$ $>$10 amu 
mass spectra (e.g., CO, N$_2$, CH$_4$) are entering the atmosphere as vapor and are then spreading globally before reaching the stratosphere (e.g., 
our model Case D), then the transient ring-inflow event could have occurred as long ago as 50 yrs before the 2017 end of the \textit{Cassini} mission, 
but no longer than $\sim$50 years or even Case D would have resulted in spectral consequences.  
This latter scenario is consistent with the suggestion \citep{waite18,perry18,mitchell18} that some dynamical event in ringlet D68 was triggered in 
2014-2015 time frame \citep{hedman19}, resulting in collisional debris that then entered Saturn's upper atmosphere and was detected during the Grand 
Finale \textit{in situ} orbital measurements.  

\item If, instead, the INMS measurements of neutral molecules result predominantly from impact vaporization and fragmentation of dust within the 
instrument, then that dust must be smaller than $\sim$50 nm if the grains enter the atmosphere at speeds of $\sim$25 km s$^{-1}$ (or smaller than 
$\sim$100 nm if the grains enter the atmosphere at speeds of 15 km s$^{-1}$), or the dust grains would completely ablate and release their vapor anyway 
\citep{hamil18}, becoming inconsistent with the lack of evidence of ring-derived vapor in the stratosphere.  
MIMI data acquired during the Grand Finale orbits provide evidence for nanograins of $\sim$8000--40,000 amu in Saturn's equatorial atmosphere 
that are consistent with such a population of small dust grains \citep{mitchell18}; however, the inferred influx rate of the grains detected 
by MIMI is only 5 kg s$^{-1}$ \citep{mitchell18}, falling far short of the $\gtrsim$10,000 kg s$^{-1}$ influx rate of material inferred from INMS 
\citep{waite18,perry18}.  The missing grain population must then have masses that reside outside the range detected by MIMI \citep[see][]{perry18}.

\item \textit{Cassini} UVIS stellar occultation observations near the equatorial region of Saturn in 2017 do not reveal absorption from H$_2$O in 
the upper stratosphere, limiting the average influx of water vapor to $\lesssim$2$\scinot7.$ molecules cm$^{-2}$ s$^{-1}$ at low latitudes during the 
Grand Finale stage of the mission.  This finding is in conflict with the 1--2 order of magnitude greater H$_2$O influx rates inferred from the INMS 
\textit{in situ} measurements.  The absence of large amounts of stratospheric water in Saturn's stratosphere again suggests that the H$_2$O signals 
from the INMS measurements derive predominantly from small dust particles that impact, vaporize, and fragment within the instrument and/or that 
the observed INMS inflow resulted from a transient event that occurred contemporaneously within the 5-month Grand Finale orbital time frame.  Global 
spreading of the incoming water vapor cannot resolve the model-data mismatch.

\item Infrared observations from JWST or ground-based infrared, millimeter, or sub-millimeter observations could be used to constrain long-term average 
ring-vapor influx rates on Saturn or to potentially monitor transient, short-term ring-influx events, given that stratospheric consequences appear 
on time scales of months to years.  Key stratospheric species that could be used for this purpose include CO$_2$, H$_2$O, CO, HCN, HC$_3$N, and 
potentially CH$_3$CN.  Determining the latitude variation of these potential ring-derived species would also help address the potential horizontal 
spreading away from the equatorial region.  Because the inflow of ring vapor also affects the stratospheric mixing ratios of photochemically produced 
C$_2$H$_2$, C$_2$H$_6$, and other hydrocarbons in ways that are not currently observed, mid-infrared observations of hydrocarbons might also be useful 
for monitoring short-term inflow events on $\lesssim$50-year time scales.  However, in practice, it might be difficult to separate any ring-inflow contribution from 
any stratospheric circulation patterns that are advecting the hydrocarbons.  Although we find that the ionospheric composition is very sensitive to changes 
in the incoming ring influx rates and exact molecular makeup, key ions such as HCNH$^+$, NH$_4$$\! ^+$, and HCO$^+$ are predicted to have relatively 
small vertical column densities that could be difficult to observe.  Neutral stratospheric species should therefore provide the best clues to 
long-term average ring-vapor influx rates or short-term, transient inflow events.  

\end{itemize}

Our current modeling cannot distinguish between the allowed possibilities of the INMS neutral measurements being caused (1) by a very recent 
transient ring event that sent large amounts of vapor into Saturn's equatorial atmosphere along with some dust, or (2) by small organic-rich 
grains that vaporized upon impact within the INMS instrument, with the bulk of the ring debris characterized by a high dust-to-gas ratio.  
Future mid-infrared observations within the next decade that monitor the presence and/or abundances of HCN, HC$_3$N, CO$_2$ in the aftermath 
of a putative recent ring event could help distinguish between these possibilities.  If HCN and HC$_3$N are not detected with future \textit{JWST} 
or ground-based observations and the observed CO$_2$ abundance remains unchanged, then the dust-grain scenario would seem most likely. 


\section{Acknowledgments}\label{sec:acknow}
We gratefully acknowledge support from the NASA Solar System Workings program grants 80NSSC20K0462 (J.M.) and 80NSSC19K0546 (L.M.), and Cassini Data 
Analysis Program grants 80NSSC19K0902 (T.K., Z.B., J.M.) and 80NSSC19K0903 (J.S.).
L.N.F.~is supported by a European Research Council Consolidator Grant (under the European Union's Horizon 2020 research and innovation programme, 
grant agreement number 723890) at the University of Leicester.  J.H.W. is supported by SwRI internal funds.  

\vspace{24pt}
\appendix{\textbf{Appendix A: Molecular diffusion coefficients}}
\vspace{12pt}

\begin{figure*}[!htb]
\begin{center}
\includegraphics[clip=t,scale=0.50]{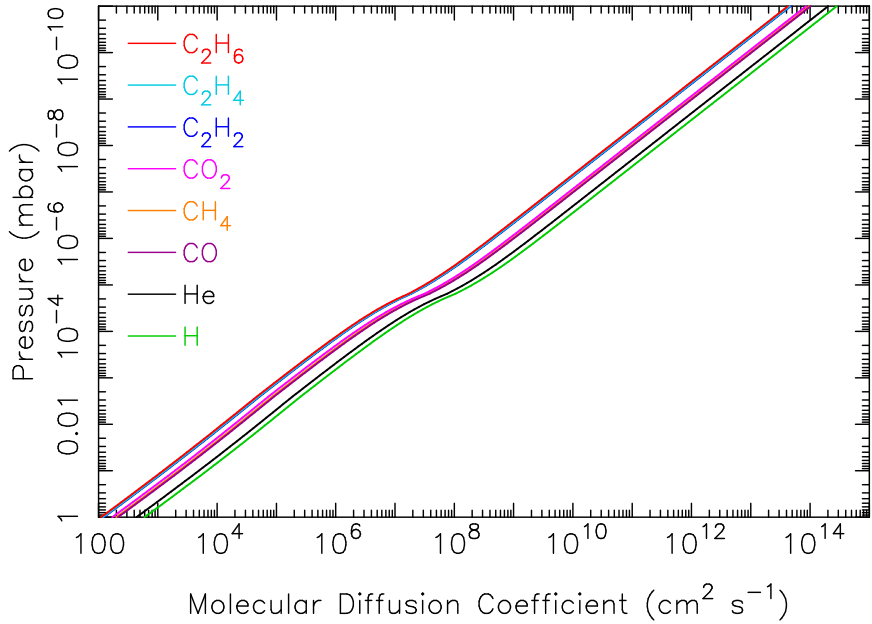}
\end{center}
\caption{Molecular diffusion coefficient profiles for C$_2$H$_6$-H$_2$, C$_2$H$_4$-H$_2$, C$_2$H$_2$-H$_2$, CO$_2$-H$_2$, CH$_4$-H$_2$, CO-H$_2$, 
He-H$_2$, and H-H$_2$ in our 9.2$\deg$ N latitude model.} 
\label{figmolecdif}
\end{figure*}

For H, He, CH$_4$, CO, and CO$_2$ mutual diffusion coefficients with H$_2$, we take values from laboratory data \citep{marrero72}.  
For C$_2$H$_2$, C$_2$H$_4$, and C$_2$H$_6$, we calculate the molecular diffusion coefficients theoretically using Lennard-Jones potentials 
\citep[][pp.~581-583]{reid87}.  The resulting diffusion coefficient profiles for these species are shown in Fig.~\ref{figmolecdif}.  For 
all other species, we use a crude approximation that starts from the CH$_4$-H$_2$ expression and scales based on species mass 
\citep{gladstone96,moses00a}:
\begin{equation}
D_i \ = \ \frac{2.3\scinot17. \, T^{0.765}}{n} \sqrt{ \frac{16.04}{m_i} \left( \frac{m_i + 2.016}{18.059} \right) } \ \ ,  
\end{equation}
where $D_i$ is the molecular diffusion coefficient (cm$^2$ s$^{-1}$), $T$ is the temperature (K), $n$ is the atmospheric 
number density (cm$^{-3}$), and $m_i$ is the mass (amu) of the diffusing species $i$.  
Based on the discussion in \citet{reid87}, the above expression is too crude for many of the heavier hydrocarbons (including C$_6$H$_6$) and 
should be replaced by more robust theorectical expressions in future work.

\vfill \eject

\vspace{24pt}
\appendix{\textbf{Appendix B: Model caveats and sensitivity tests}}
\vspace{12pt}

There are several caveats that need to be emphasized with respect to the models.  First, as is discussed in section~\ref{sec:modinputs}, the 
adopted $K_{zz}$ profiles were developed to fit the CH$_4$, C$_2$H$_2$, and C$_2$H$_6$ retrievals from \textit{Cassini} CIRS and UVIS observations 
at the nearest latitudes and times to the Grand Finale orbits.  Aside from the potential problems associated with the fact that the CIRS and UVIS 
observations used for this purpose were not simultaneous and co-located with the INMS measurements, another potential issue is that these $K_{zz}$ 
profiles were developed in the context of the nominal-chemistry models that do not include the large INMS-derived ring-influx rates.  Because the 
molecules flowing in from the rings affect the profiles 
of CH$_4$, C$_2$H$_2$, and C$_2$H$_6$, it might be reasonable to consider adjusting $K_{zz}$ --- a free parameter in the models --- to better 
fit the observations in the context of these strong-inflow scenarios.  The upper stratospheric $K_{zz}$ can certainly be adjusted to better fit 
the UVIS CH$_4$ data in the models in which the enhanced ultraviolet shielding has caused the CH$_4$ abundance to depart from 
observed values (e.g., Cases E \& F); however, the predicted C$_2$H$_2$/C$_2$H$_6$ ratio in all of the models except Cases B and D is
far removed from what is observed, such that it would be impossible to simultaneously reproduce both C$_2$H$_2$ and C$_2$H$_6$ by simply by changing
$K_{zz}$.  We could, however, adjust $K_{zz}$ in the middle and lower stratosphere to better fit one or the other of these molecules.  
If we were to choose C$_2$H$_2$ for this fitting exercise, we would need to substantially reduce $K_{zz}$ in the middle and lower stratosphere to provide 
more acetylene to better reproduce the CIRS retrievals for Cases A, C, E, and F.  The resulting abundances of other long-lived stratospheric species, 
including C$_2$H$_6$, CO, CO$_2$, H$_2$O, N$_2$, HCN, CH$_3$CN, and HC$_3$N would be correspondingly increased due to the slower diffusion through 
the stratosphere and slower eventual loss out through the bottom boundary of the model.  The model-data mismatch would therefore be worse for these 
species.  By choosing to fit C$_2$H$_6$ for this exercise, we instead need to to increase $K_{zz}$ in the middle and lower stratosphere, which would 
reduce stratospheric C$_2$H$_6$ and better fit the ethane observations for Cases A, C, and E; then, the column abundances of the above-mentioned 
long-lived species would also be reduced, potentially improving the comparison with observations.  We will therefore follow this path in adjusting 
$K_{zz}$ for Cases A, C, and E.  
Note that the chemical production rate of 
C$_2$H$_6$ in the Case F model is so low that adjustments to $K_{zz}$ still do not produce enough to explain observations, and any decrease 
in $K_{zz}$ to improve the C$_2$H$_6$ fit would increase the abundance of many ring-derived species beyond their already large over-predictions.
The problems with Case F therefore cannot be resolved by adjusting $K_{zz}$.

\begin{figure*}[!htb]
\vspace{-14pt}
\begin{tabular}{ll}
{\includegraphics[clip=t,scale=0.30]{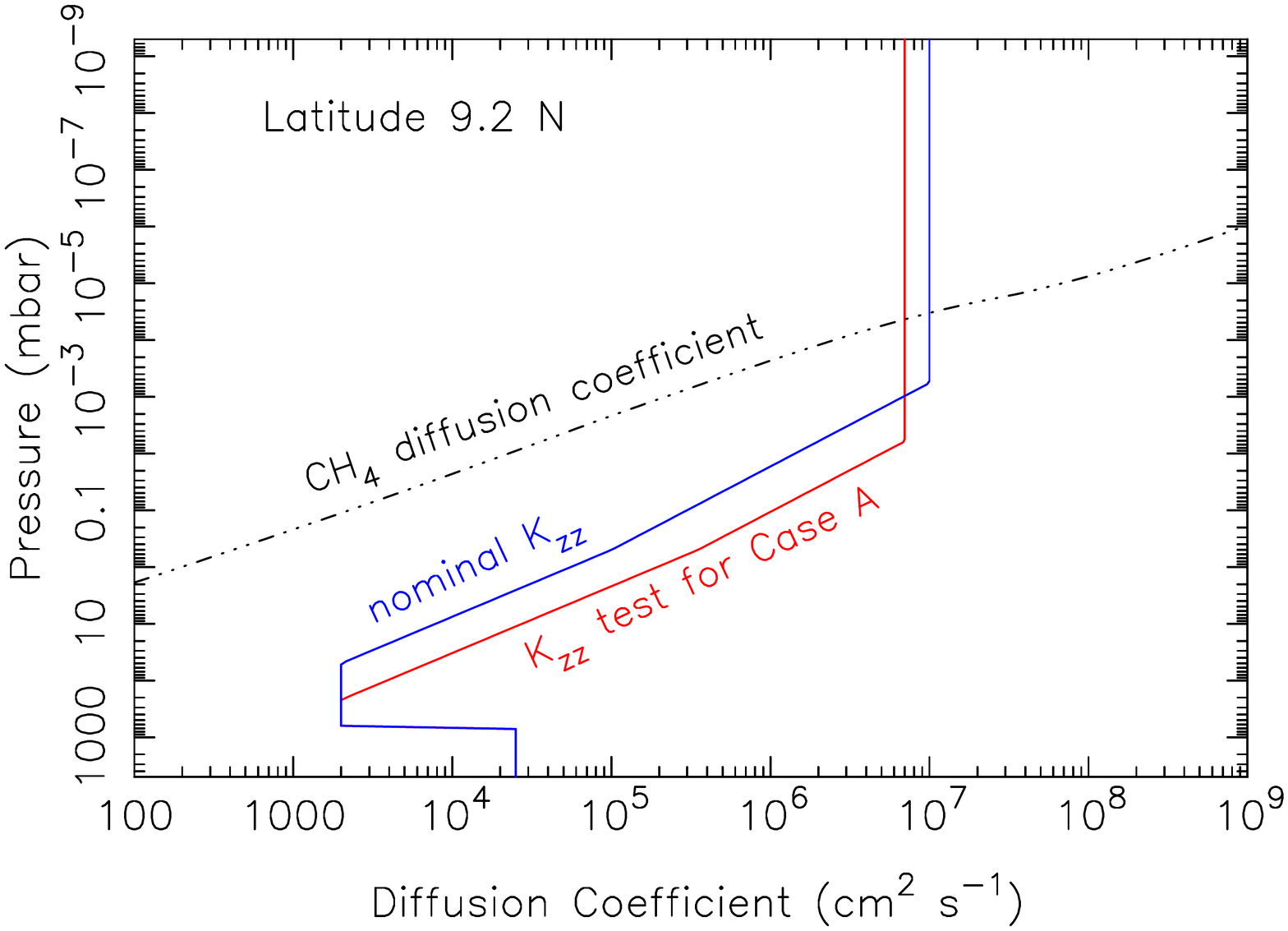}}
&
{\includegraphics[clip=t,scale=0.30]{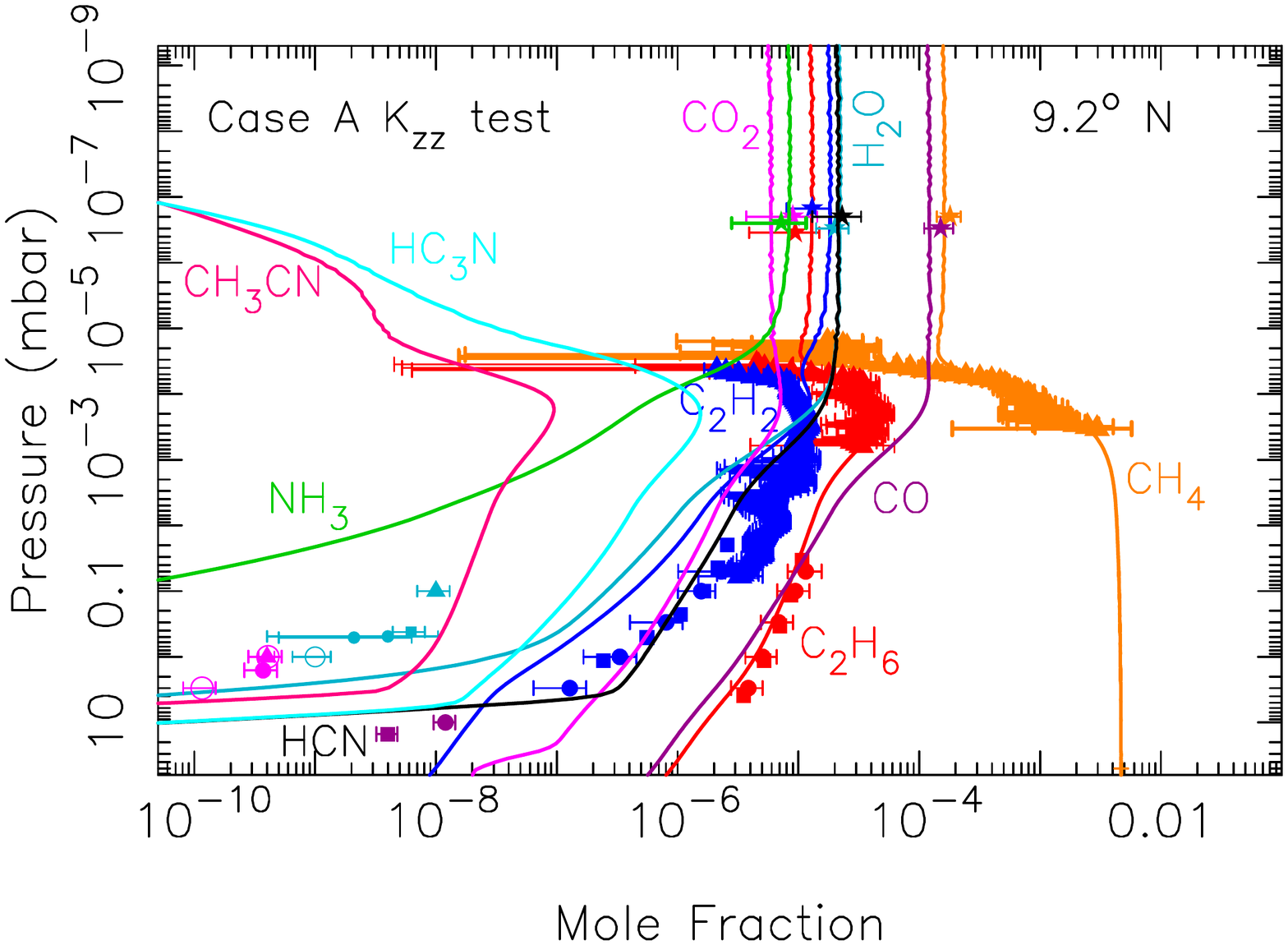}}
\end{tabular}
\caption{Eddy diffusion coefficient profile (Left) and corresponding mixing ratio profiles of key observable species (Right) from an eddy 
diffusion coefficient sensitivity test in which the $K_{zz}$ profile was adjusted (red curve in left figure) to better fit the CH$_4$ and 
C$_2$H$_6$ UVIS and CIRS observations under the assumption that ring vapor is flowing in at the top of the atmosphere with the Case A 
influx rates from Table~\ref{tabinflux}.  Note that although the CH$_4$ and C$_2$H$_6$ observations are well fit by this adjusted $K_{zz}$ 
profile, the stratospheric CO, CO$_2$, H$_2$O, HCN, CH$_3$CN, and HC$_3$N abundances are still much greater than is observed, and the C$_2$H$_2$ 
abundance is smaller than observed.}
\label{figkzztesta}
\end{figure*}

\begin{figure*}[!htb]
\vspace{-14pt}
\begin{tabular}{ll}
{\includegraphics[clip=t,scale=0.30]{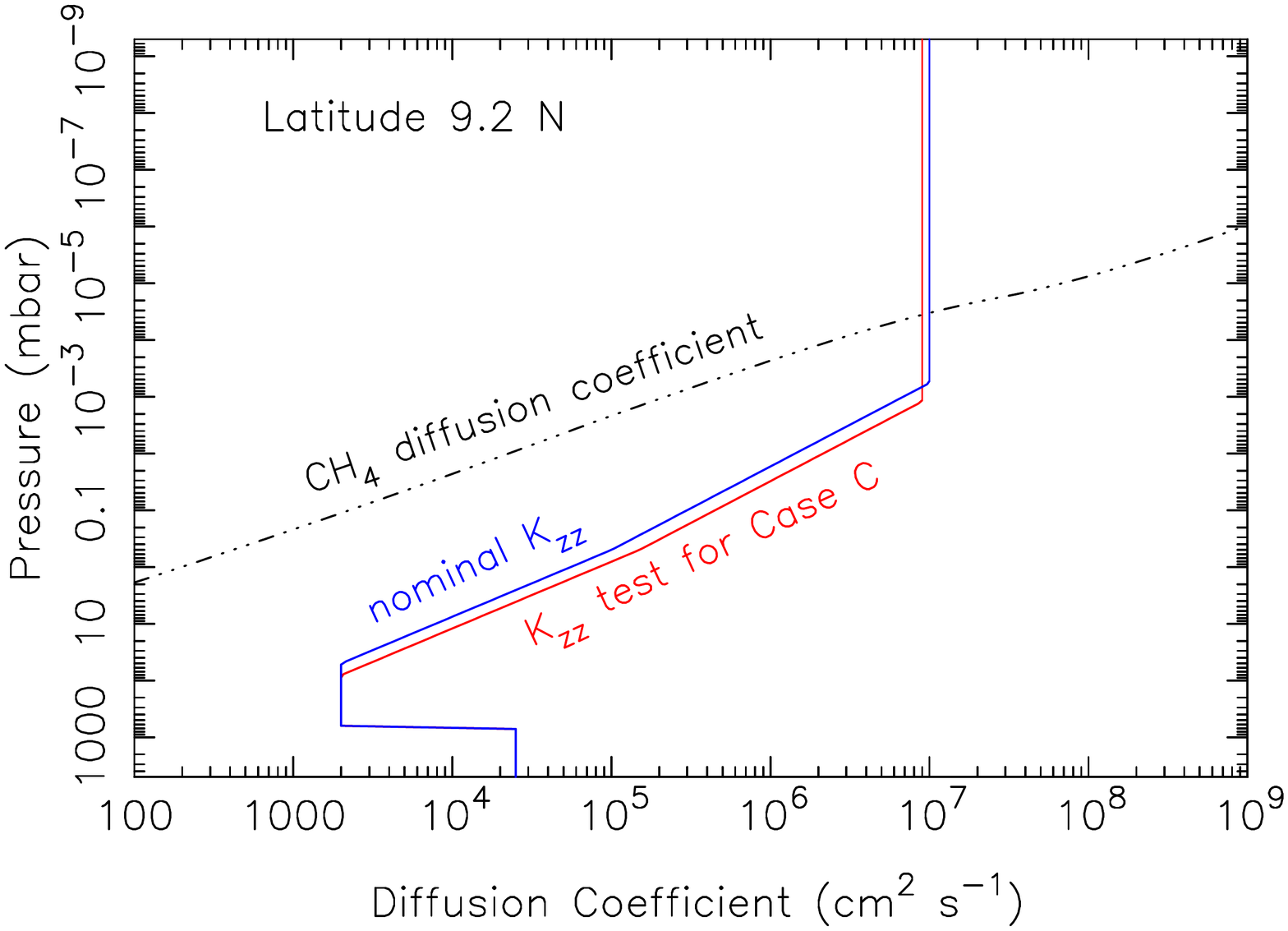}}
&
{\includegraphics[clip=t,scale=0.30]{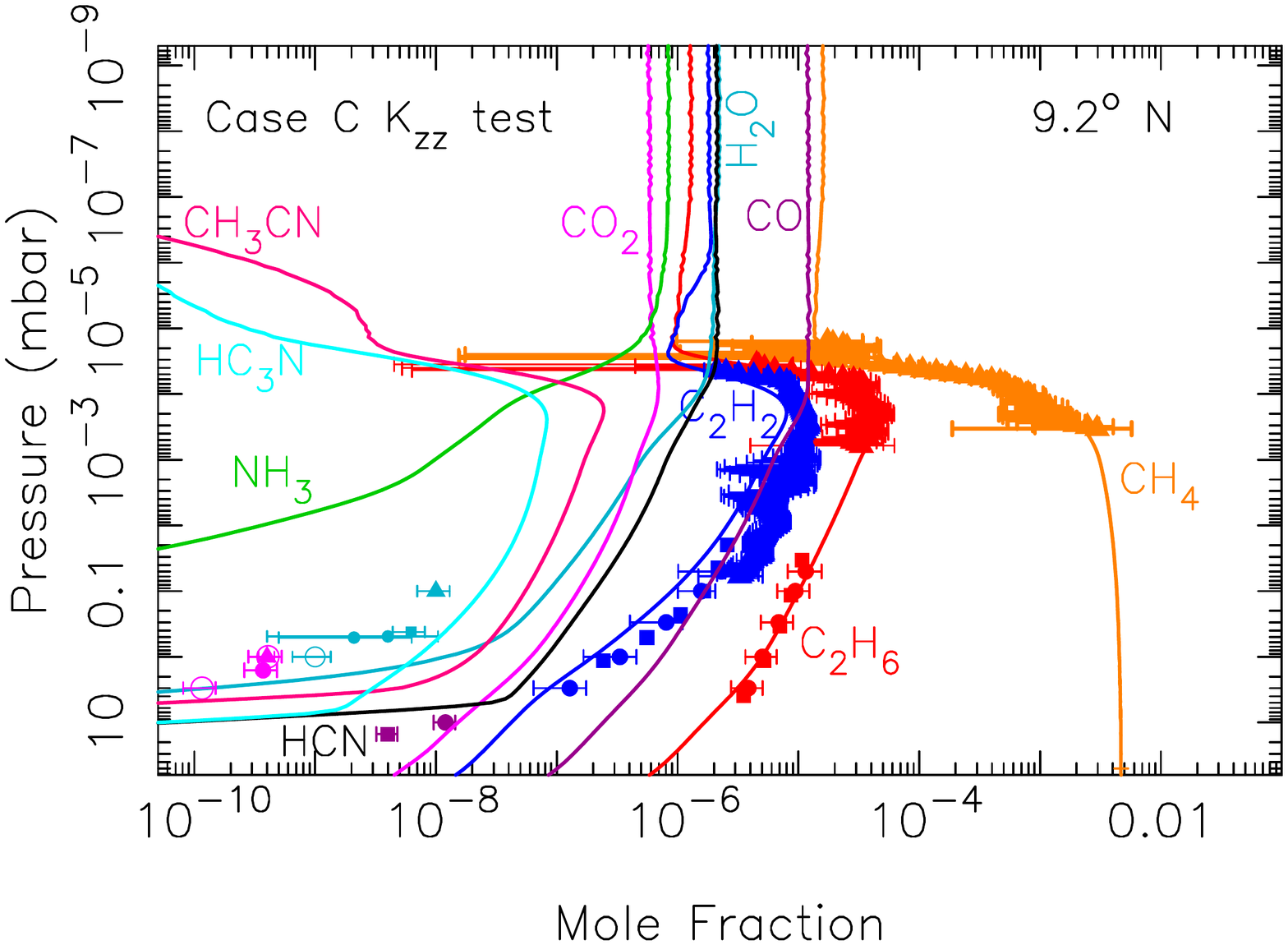}}
\end{tabular}
\caption{Eddy diffusion coefficient profile (Left) and corresponding mixing ratio profiles of key observable species (Right) from an eddy 
diffusion coefficient sensitivity test in which the $K_{zz}$ profile was adjusted (red curve in left figure) to better fit the CH$_4$ and 
C$_2$H$_6$ UVIS and CIRS observations under the assumption that ring vapor is flowing in at the top of the atmosphere with the Case C 
influx rates from Table~\ref{tabinflux}.} 
\label{figkzztestc}
\end{figure*}

To explore whether we can ``hide'' these ring-derived stratospheric molecules through faster diffusion through the stratosphere in Cases A, C, and E, 
we explore this second possibility of adjusting $K_{zz}$ to improve the fit to C$_2$H$_6$ and CH$_4$ observed by CIRS and UVIS, respectively.  
Figures~\ref{figkzztesta} and \ref{figkzztestc} show results for alternative $K_{zz}$ profiles for the Case A and Case C inflow scenarios for 
9.2$\deg$~N latitude.  For the Case A test, the new $K_{zz}$ profile has larger eddy diffusion coefficient values in Saturn's middle and 
lower stratosphere, which helps remove excess stratospheric C$_2$H$_6$ such that the CIRS nadir and limb retrievals for ethane are better reproduced.  
However, a comparison of the new species profiles (colored curves) with the observations (colored data points) in Fig.~\ref{figkzztesta} reveals 
that the predicted CO, CO$_2$, H$_2$O, and HCN abundances are still much greater than has been observed in the stratosphere of Saturn 
\citep{noll90,feuchtgruber99,moses00b,bergin00,cavalie09,cavalie10,cavalie19,fletcher12spire,abbas13}, 
and the model still predicts observable abundances of HC$_3$N (and probably CH$_3$CN) that are inconsistent with CIRS (and millimeter) spectra 
(see section~\ref{sec:obs}).  Moreover, the resulting predicted abundances of C$_2$H$_2$ (and C$_3$H$_4$, C$_4$H$_2$, not shown) are significantly 
smaller than are indicated by infrared observations 
\citep{degraauw97,moses00a,greathouse05,fletcher09,fletcher12spire,fletcher19chap,fletcher21rev,sinclair13,guerlet09,guerlet10,guerlet15,koskinen18}.  
The situation for Case E at 5.5$\deg$~S latitude (not shown) is similar, in that a larger eddy diffusion coefficient in the middle and lower 
stratosphere provides a better fit to the C$_2$H$_6$ retrievals but cannot overcome the predicted significant excess of CO, CO$_2$, H$_2$O,
HCN, HC$_3$N, and CH$_3$CN.

The original Case C model --- the one with horizontal mixing of Case A --- does not as grossly over-predict the stratospheric C$_2$H$_6$ abundance, 
so less of an adjustment to the $K_{zz}$ profile 
is needed to improve the ethane fit (see Fig.~\ref{figkzztestc}).  However, after tweaking $K_{zz}$, the resulting CO, CO$_2$, H$_2$O, HCN, 
HC$_3$N, and probably CH$_3$CN predictions are still greatly over-estimated in comparison with observations and upper limits for these 
species.  It therefore appears that the stratospheric oxygen and nitrogen-bearing species that result from the ring-vapor influx cannot 
flow vertically through the stratosphere and into the deep troposphere fast enough in the equatorial region of Saturn to explain the 
current model-data mismatches.  

A second caveat to the models is that some of the predicted behavior depends on reactions that have uncertain or outdated rate coefficients.  For the major 
observable ring-derived stratospheric species, we identify a few reactions in particular that can affect production or loss of those molecules but for which the 
rate coefficients have been estimated and deserve further study or updates.  One such reaction is C$_3$N + CH$_4$ $\rightarrow$ HC$_3$N + CH$_3$, which affects 
not only HC$_3$N but many species through the production of the important radical CH$_3$.  In the models described in section~\ref{sec:results}, we have 
simply estimated the rate coefficient for this reaction to be $3\scinot-11.$ cm$^{-3}$ s$^{-1}$.  However, we recently have become aware of a measured 
rate coefficient of this reaction from the thesis work of \citet{fournier14}, who have determined a value of 8.165$\scinot-10. \, T^{-0.574} \, \exp{(-2.974/T)}$ 
cm$^{-3}$ s$^{-1}$ in the temperature range 24--300 K.  Our estimated value is slightly smaller than this measured rate coefficient at temperatures relevant 
to Saturn's stratosphere.  When we replace our previous estimate with the \citet{fournier14} expression, we find that the stratospheric column abundances 
of HCN, HC$_3$N, and CH$_3$CN above their condensation regions are relatively unaffected, whereas the main notable effect is a minor $\sim$10\% increase in 
the C$_2$H$_6$ column abundance above 1 mbar under Case A conditions, due to the increased CH$_3$ production.

Other reaction-rate uncertainties involve the production of CH$_2$CN, which is a precursor to the production of CH$_3$CN.  We could find no measured 
rate coefficients for the reaction of triplet ground-state methylene radicals ($^3$CH$_2$) with HCN, but consider that the reaction $^3$CH$_2$ + HCN 
$\rightarrow$ CH$_2$CN + H could potentially be one way to produce CH$_2$CN in Saturn's stratosphere.  This reaction most likely has a significant 
energy barrier that would make it relatively slow at Saturn stratospheric temperatures.  In our models, we estimated the rate coefficient to be 
$5\scinot-11. \, \exp{(-1400/T)}$ cm$^{-3}$ s$^{-1}$, which might still lead to an over-estimate in the rate of this reaction at low temperatures.  
\citet{loison15} include this reaction in their Titan model with an estimated rate coefficient of $1.5\scinot-12. \, \exp{(-3300/T)}$ cm$^{-3}$ s$^{-1}$ 
based on the rate coefficient of reactions of C$_2$H with alkynes.  When we replace our current expression with the one from \citet{loison15} we 
find very little difference in the results for any observable species.  However, we have included the endothermic reaction $^3$CH$_2$ + C$_2$H$_3$CN 
$\rightarrow$ CH$_2$CN + C$_2$H$_3$ in our model, which would have a very high energy barrier and has not been included in Titan models such as 
those of \citet{vuitton19} and \citet{loison15}.  When we omit this reaction, we find that the predicted column abundance of CH$_3$CN above its 
condensation region has been reduced by a factor of $\sim$7 under Case A conditions, along with a $\sim$10\% reduction in the column abundance 
of C$_2$H$_6$ above 1 mbar and a $\sim$3\% increase in the column abundance of HCN.

Photochemical models that consider rate-coefficient uncertainty propagation via Monte Carlo techniques suggest that modeled C$_2$H$_2$ and C$_2$H$_6$ 
mixing ratios are uncertain by a maximum factor of 1.4 for C$_2$H$_6$ and 1.6 for C$_2$H$_2$ in Saturn's stratosphere, with uncertainty factors 
growing with increasing molecular weight to a maximum of $\sim$5 for CH$_3$C$_2$H and $\sim$8 for C$_4$H$_2$ \citep[e.g.,][]{dobrijevic03}.  Similar 
studies for Titan \citep[e.g.,][]{dobrijevic16} find uncertainties for HCN that are similar to that of C$_2$H$_2$ and C$_2$H$_6$, but that grow larger 
for CH$_3$CN, and can be as much as an order of magnitude for HC$_3$N.  Uncertainty factors for H$_2$O, CO, and CO$_2$ on Titan due to rate-coefficient 
uncertainties, however, are relatively small \citep{dobrijevic14,loison15} as a result of a smaller number of better-studied reactions controlling 
the behavior of these species.  Rate coefficient uncertainties are not likely to resolve the enormous over-prediction of CO and CO$_2$ in our models 
compared to observations, although they could potentially affect the predicted abundance of HC$_3$N enough to affect its observability for the Case D 
model, in which horizontal spreading is invoked along with the assumption that only the most volatile species CO, N$_2$, and CH$_4$ are flowing into 
the atmosphere as vapor (see section~\ref{sec:obs}).

\bibliographystyle{cas-model2-names}

\bibliography{ref_inms}



\end{document}